%% file: thesis.tex
\documentstyle[epsfig]{uuthesis}
\title{Hybrid meson decay from the lattice}
\author{Ziwen Fu}
\thesistype{dissertation}
\department{Department of Physics}
\degree{Doctor of Philosophy}
\submitdate{December 2006}
\copyrightyear{2006}
\fourlevels

\def\Dsl{\,\raise.15ex\hbox{$/$}\mkern-13.5mu D} 
\def\vev#1{\langle      #1 \rangle}
\def\Tr{ \mathop{\rm Tr}\nolimits }
\newcommand{\be }{\begin{equation}}
\newcommand{\ee }{\end{equation}}
\newcommand{\bea}{\begin{eqnarray}}
\newcommand{\eea}{\end{eqnarray}}

\def\ssqr#1#2{{\vbox{\hrule height.#2pt
      \hbox{\vrule width.#2pt height#1pt \kern#1pt\vrule width.#2pt}
      \hrule height.#2pt}\kern-.#2pt}}
\def\sqrB{\mathchoice\ssqr64\ssqr64\ssqr{5.1}3\ssqr{4.5}3}

    \def\CS{{\cal S}}

\def\Tr{\mathop{\rm Tr}}
\def\Re{\mathop{\rm Re}}
\def\Im{\mathop{\rm Im}}

\def\LL{\left\langle}	
\def\RR{\right\rangle}	
\def\LP{\left(}		
\def\RP{\right)}	

\def\BE{\begin{equation}}
\def\EE{\end{equation}}
\def\BEA{\begin{eqnarray}}
\def\EEA{\end{eqnarray}}

\def\DPT{\displaystyle}
\def\HSP{\hspace{-0.1cm}+\hspace{-0.1cm}}

\def\ENG#1{  \sqrt{ {\bf k}^2  + M_{#1}^2 } }
\def\ENGG#1{ \sqrt{ {\bf k}^2   \HSP M_{#1}^2 } }
\def\EP#1{ \sqrt{ {\bf (p+k)}^2 + M_{#1}^2 } }
\def\EPP#1{ \sqrt{ {\bf (p\hspace{-0.1cm}+\hspace{-0.1cm}k)}^2 \HSP M_{#1}^2 }}
\def\FZW#1{\frac{e^{-\LP \EP {#1}+\ENG {#1} \RP t }}{\EPP {#1} \ENGG {#1}} }
\def\OTS#1#2{\frac{e^{-\LP \EP {#1}+\ENG {#2} \RP t }}{\EPP {#1} \ENGG {#2}} }

\def\Tt{ t \hspace{-0.1cm} + \hspace{-0.1cm} T}
\def\Rr{ {\bf r}, {\bf R} }
\def\PLUSONE{ \hspace{-0.1cm} + \hspace{-0.1cm}}
\def\MINUSONE{ \hspace{-0.1cm} - \hspace{-0.1cm}}
\def\EQATHR{ \hspace{-0.3cm} &=& \hspace{-0.3cm}}

\newcommand{\etc}{etc.\ }

\renewcommand{\Re}{{\rm Re}}
\renewcommand{\Im}{{\rm Im}}

\begin{document}
\frontmatterformat
\titlepage

  \prefacesection{Abstract}
  \par \input{abstract}
  \ifnoisy \typeout{Abstract.} \fi

\tableofcontents


  \prefacesection{Acknowledgements}
  \par \input{acknowledge}

  \ifnoisy \typeout{Acknowledgements.} \fi

\maintext
\include{Chap1}
\include{Chap2}
\include{Chap3}

\include{Chap4}
\include{Chap5}
\include{Chap6}
\include{Chap7}
\include{Chap8}
\include{Chap9}
\numberofappendices 5
\appendix
\include{appA}
\include{appB}
\include{appC}
\include{appD}
\include{appE}
\bibliographystyle{unsrt}
\bibliography{thesis}

\end{document}

%% file: abstract.tex
Besides the conventional hadrons containing valence quarks and valence antiquarks,
quantum chromodynamics (QCD) suggests the existence of the hybrid hadrons
containing valence gluons in addition to the quarks and antiquarks,
and some experiments may have found some.
A decisive experimental confirmation of its existence, however, is still needed.
At present, lattice simulations have offered the practicable ways of
theoretically guiding us to search for the hybrid states.

In this dissertation, we study the spectroscopy
and the decay rate of the heavy hybrid mesons with a heavy
$b$ quark, a heavy $\bar b$ antiquark and a gluon ($b\overline{b}g$)
to the chosen channels,
and use lattice methods to calculate the transition matrix elements in full QCD.
We are particular interested in the spin-exotic hybrid mesons.

For sufficiently heavy quarks (e.g., $b$ quark),
we use the leading Born-Oppenheimer (LBO) approximation
to calculate the static potential energy at all $b\overline{b}$ separations.
Then, by solving the Schr\"odinger equation with this potential,
we reconstruct the motion of the heavy quarks.
In a similar way we can determine decay rates.
In this dissertation,  we use the lattice method
to calculate the mass of the $f_0$ meson at a single lattice spacing
and light quark mass, namely,  $m_{f_0} = (768 \pm 136)$~MeV.
Most of all we consider the decay channels involving the production of a scalar meson.
We obtain the partial decay rate ($\Gamma$) for the  channel
$H \rightarrow \chi_b + \pi + \pi$, namely,  $ \Gamma = 3.62(98)$~MeV.
All of our results are consistent with those of other researchers.
Knowledge of the masses and the decay rates
should help us  considerably in experimental searches for the hybrid mesons.

%% file: acknowledge.tex
First of all, I would like to thank the MILC collaboration,
which  has developed a parallel 
computer code for the simulation of SU(3) lattice gauge theory. 
Most of the computer code that I used for my Ph.D study 
was adapted or modified from the MILC code.
Moreover, I was a user of MILC collaboration's lattice configurations
during my Ph.D study.

I thank Urs Heller for his data of lattice Coulombic correction term.
I thank  C. Bernard and  Sasa Prelovsek 
for their help in understanding the chiral perturbation theory and the bubble
contribution to the scalar correlator.

Finally, I thank especially my advisor, Carleton DeTar. 
This project is quite difficult. We encountered  many problems.
Throughout, Prof. DeTar always found ways to solve the problem,
and allow my studies to continue. 
Beyond that, Prof. DeTar is not only an expert in physics and math,
but also in computer science.
The computer codes I use in this dissertation were developed
under his guidance.
I could not have finished this dissertation without 
his help, encouragement, and extraordinary patience.
Also Prof. DeTar's personality and enthusiasm for physics 
will inspire me to continue to pursue a career in physics.

An allocation of computer time from the Center for High Performance
Computing at the University of Utah is gratefully acknowledged. 
During the development of this dissertation, I devoted my life to
CHPC's ICEBOX system. I experienced happiness and bitterness 
on this machine. ICEBOX now is retired; I will miss her.

%% file: Chap1.tex
\chapter{ INTRODUCTION }

Besides the conventional hadrons, quantum chromodynamics (QCD)
predicts the existence of the hybrid hadrons, which are hadrons
but also contain the excited gluon fields.
Early results in $e^{+}e^{-}$ annihilation cross section from the CUSB
and CLEO collaborations~\cite{Besson:1984bd,Lovelock:1985nb}
showed that there exists a complex resonance structure
in the energy region above the $\Upsilon(4S)$ (near 10.7 GeV),
where the possible lowest hybrid states ($b\bar{b}g$) are
predicted in Ref.~\cite{Kuti:1998rh}.
However, because it is difficult to differentiate a resonance
from a threshold effect, we just have a ``hint'' of the existence of
the hybrid states,
a decisive experimental confirmation is still needed.

We need some theoretical guidance to search efficiently for the hybrid states.
Due to the nonperturbative nature of the gluons, however,
a perturbative theory in this field is still not available.
Hence, a clear understanding the hybrid states
by analytical methods is very difficult.
At present, the lattice simulations have offered practical ways of
theoretically guiding us to search for the hybrid states.

The heavy hybrid mesons  can be investigated directly
by the leading Born-Oppenheimer (LBO) approximation.
The results from our LBO  approximation can compare with
the corresponding results from other people's experiments.
We focus on the heavy hybrid mesons ($Q\overline{Q}g$), which
consist of a heavy quark ($Q$)
and  a heavy antiquark ($\overline{Q}$) plus a gluon (g).
We study its spectroscopy and the decay amplitude
from the heavy hybrid meson to $(Q\bar{Q})(q\bar{q})$,
and use lattice methods to calculate the transition matrix elements
in full QCD~\cite{McNeile:2002az}.
Finally, we study its phenomenological implications.
We are particular interested in the spin-exotic hybrid meson,
which is the meson with $J^{PC}$ quantum numbers
which  are not permitted in the current quark model.
The lightest one is believed
to be with quantum numbers $J^{PC}=1^{-+}$~\cite{Juge:1999ie},
which does not mingle with a nonhybrid $Q\bar{Q}$
bound state~\cite{McNeile:2002az}.

The spectroscopy of the hybrid mesons (including spin-exotic hybrid
mesons) has been extensively studied over the
past years~\cite{Juge:1999ie,Bali:2000vr}.
The BO approximation was proposed for
the study of the hybrids states
in Refs.~\cite{Hasenfratz:1980jv,Horn:1977rq}.
For sufficiently heavy quarks (e.g., $b$ quark),
the LBO approximation justifies  a treatment in
which the heavy quark ($Q$) and antiquark ($\overline{Q}$)
are regarded as approximately static,
and we assume that the gluon field adjusts rapidly to their movement.
Hence, for various gluonic excitations of the flux tube of $Q\overline{Q}$,
we compute its corresponding static potential energy at $Q\overline{Q}$ separation $R$.
Then by numerically solving the Schr\"odinger equation with this potential,
we reproduces the motion of the heavy quarks.
In a similar way we can calculate the decay rate.

Decay modes are also intensively investigated
in Ref.~\cite{McNeile:2002az}.
However, at present,  our knowledge about the decay channels is still limited.
We make a contribution to the understanding
of the spectroscopy of the hybrid mesons and its  decay model.
There are many new features in our simulation:
(1) We adopt gauge configurations with improved actions,
which include the Symanzik improvements at 1-loop level,
and ``Asqtad'' fermion action~\cite{MILC:ASQ}.
(2) We choose the
improved wave functions for the spin-exotic gluon
channels, which are used to integrate all the decay rate
at all $Q\overline{Q}$ separation $R$.
(3) We consider decay channels
$
H \rightarrow \chi_b + \pi + \pi
$ involving the emission of two pions (here $H$ stands for
hybrid exotic state).
(4) The mass of the dynamical sea quark we choose is
closer to the physical value.

Our project needs three classes of adjustable input parameters:
1) lattice spacing: $a$,
2) light quark masses ($u,d,s$ quarks),
3) the mass of the bottom quark ($b$ quark).
These input parameters are obtained from experiment.
We obtain the masses of $u,d,s$ quarks from
the masses of $\pi, K$ mesons,
and we obtain  the lattice spacing $a$ from the splittings of
$\Upsilon(2S)-\Upsilon(1S)$ and $\Upsilon (1P)-\Upsilon(1S)$
in $\Upsilon$ spectrum.
And from the $B$ meson,
we obtain the mass of the bottom quark~\cite{Aubin:2004wf}.
This project help us gain the knowledge about ``soft gluons.''
Furthermore, we can predict the mass of some  mesons, e.g., the $f_0 $ meson.
Most importantly, we can calculate the partial transition rate
$\Gamma$, which has been poorly understood.
Knowledge of the masses and decay rates
helps us in experimental searches for the hybrid mesons.

In this dissertation, we use the MILC version of the Kogut-Susskind
staggered quark action,
and the gauge action is the L\"uscher Symanzik improved action.
Also we use gauge configurations, which were generated
by the hybrid-molecular dynamics ``R algorithm''~\cite{MILCHMC:87}.
These  configurations contain three flavors: two light quark ($u$, $d$),
and a strange ($s$) quark.
In our simulation, the light quarks are degenerate in mass,
and we choose the bare strange quark mass and up, down  quarks mass to
be close to the corresponding physical values
(i.e., $am_u = 0.005$, and $am_s = 0.05$). From Ref.~\cite{PDBook}
we find the physical mass of the pion meson is about $140$~MeV,
from our lattice calculation,
we get the mass of the pion meson is about $230$~MeV.
The lattice space $a$
is in our simulation of around 0.12 fm~\cite{Aubin:2004wf}.

This dissertation is organized as follows:
In Chapter 2, we provide background material on particle physics.
We first give a brief review of the quark model.
Then we briefly discuss a few conservation laws. Finally, we
discuss about the quarkonium (heavy quark, antiquark states).
In Chapter 3, we provide background material about QCD.
First we give an overview of continuum QCD.
Then we give an overview of lattice QCD.
In Chapter 4, we use the LBO approximation to study
the spectroscopy of the hybrid mesons.
We first study the method of the LBO approximation.
Then we introduce the static potential.
Finally, we calculate hybrid quarkonium states through the LBO approximation.
In Chapter 5, we first briefly review the hybrid states
on the lattice. Then we give an overview of hybrid meson decays.
Finally, we give a overview of the decays from the lattice.
In Chapter 6, we explain our calculational method in detail.
In Chapter 7, we first use time-dependent perturbation theory to
study the $C_{HAB}$ correlator. Then we introduce a
method to extract transition amplitude $x$ directly from  lattice.
In Chapter 8, we give all the lattice numerical simulation results.
First we present the calculation of the mass of the $f_0$ meson.
Then we show the detailed  procedure to obtain
lattice transition amplitude $x$,
and calculate the decay rate.
In Chapter 9,  we first discuss the evidence for the $f_0$ meson
and the hybrid exotic mesons.
Then we discuss hybrid exotic quarkonium decay.
Finally, we summarize our results.

%% file: Chap2.tex
\chapter{Introduction to particle physics}

Particle physics is the study of  the fundamental particles that
make up matter, and how they interact with
each other~\cite{Griffiths:1987tj}.
At present, the ``quarks,'' along with ``leptons,''
are believed to be fundamental, and the quark model is
a popular theory in this field.

In this chapter, we provide background material about particle physics.
We first give a brief review of the quark model.
Then briefly discuss a few conservation laws. Finally we
discuss about quarkonium (i.e., the bound states of heavy quarks
and antiquarks).
We focus on just a few details needed for this dissertation.
Several good textbooks~\cite{Griffiths:1987tj}\cite{Perkins:1982xb}
provide comprehensive background material.
%
\section{Quark model}
\label{s1:QM}

In particle physics, we define a hadron as a subatomic particle
which takes part in the strong nuclear force.
In the current quark model, hadrons are made up of
quarks, antiquarks, and gluons.
Usually hadrons are further classified into
many types (e.g., baryons and mesons, etc).
In this dissertation we are interested only in baryons and mesons.

In 1961 Murray Gell-Mann~\cite{Griffiths:1987tj} introduced the Eightfold way
to arrange the hadrons which were discovered at that time.
The eight lightest baryons ($n$, $p$, $\Sigma^{+}$,
$\Sigma^{-}$, $\Sigma^{0}$, $\Lambda$, $\Xi^{-}$,
$\Xi^{0}$) and the eight lightest mesons ($ K^{0}$,
$K^{+}$, $K^{-}$, $\bar{K}^0$, $\pi^{0}$,
$\pi^{0}$, $\pi^{+}$, $\eta $)
fill a hexagonal array, with two particles at the center,
according to their charge and strangeness.  They are called
a baryon octet, and meson octet, respectively.~\footnote{
See the figures on page 33, 34 in Ref.~\cite{Griffiths:1987tj}.
}

In order to explain and understand the Eightfold way,
in 1964, Murray Gell-Mann and George Zweig independently
proposed~\cite{Griffiths:1987tj} that just three fundamental constituents
(and their antiparticles) combined in different ways could
explain the Eightfold way. Murray Gell-Mann called these
elementary constituents ``quarks,''
and the three types (or ``flavors'') of quarks
were named up ($u$), down ($d$), and strange ($s$),
respectively~\cite{Griffiths:1987tj}.

In the late 1960s the evidence for the existence of ``quarks'' became clear.
In 1974 a new particle $J/\Psi$~\cite{Griffiths:1987tj} was discovered,
which was later understood to be a bound state of a new
quark-antiquark pair ($c\bar{c}$). The new fourth quark
is known as charm ($c$). In 1977 another new meson called the Upsilon
($\Upsilon$)~\cite{Griffiths:1987tj} was discovered at Fermilab,
which was later shown to be the bound state of
the new quark-antiquark pair ($b\bar{b}$). The new fifth quark
was called bottom ($b$). Finally, in 1995,
the evidence for the existence of the ``top'' ($t$) quark was
obtained at Fermilab.
Hence, the current quark model with six ``flavors'' was fulfilled.

In the current quark model, we have the six quarks:
up, down, strange, charm, bottom and top.
By convention we usually label them
by their first letters: $u$, $d$, $s$, $c$, $b$ and $t$.
Moreover, each quarks has its partner
antiquarks. We label them $\bar{u}$, $\bar{d}$, $\bar{s}$,
$\bar{c}$, $\bar{b}$, $\bar{t}$, respectively.
The charges of up, down, strange, charm, bottom (or beauty),
and top (or truth) quarks
are $\frac{2}{3} e $, $-\frac{1}{3} e $, $-\frac{1}{3} e$,
$\frac{2}{3} e $, $-\frac{1}{3} e $, $\frac{2}{3} e$
respectively, where $e$ is the charge of the electron.
For each quark ($q$), the corresponding
antiquark ($\bar{q}$) has the opposite charge.
Hence, the charges of antiup ($\bar{u}$), antidown ($\bar{d}$),
antistrange ($\bar{s}$), anticharm ($\bar{c}$), antibottom ($\bar{b}$),
antitop ($\bar{t}$) quarks are
$-\frac{2}{3} e$, $\frac{1}{3} e$, $\frac{1}{3} e$,
$-\frac{2}{3} e $, $\frac{1}{3} e$, $-\frac{2}{3} e$,
respectively~\cite{Griffiths:1987tj}.

\begin{table}[h]
\vspace{-0.35cm}
\caption{ \label{t:Qgeneration}
Three quark generation. Where  $Q$ is the electric charges,
$D$ is the ``downness,''
$U$ is the ``upness,''
$S$ is the ``strangeness,''
$C$ is the ``charm,''
$B$ is the ``beauty,'' and
$T$ is the ``truth.''
}
\begin{center}\begin{tabular}{|c|c|c|c|c|c|c|c|c|}
\hline
 Flavor& Mass($GeV/c^2$) &   $Q$         &$D$&$U$&$S$&$C$&$B$&$T$ \\
\hline
$d$    &0.08     &$-{1\over 3}$&-1&0& 0&0& 0&0  \\
$u$    &0.004    &$+{2\over 3}$&0 &1& 0&0& 0&0  \\
\hline
$s$    &0.15     &$-{1\over 3}$&0 &0&-1&0& 0&0  \\
$c$    &1.5      &$+{2\over 3}$&0 &0& 0&1& 0&0  \\
\hline
$b$    &4.7      &$-{1\over 3}$&0 &0& 0&0&-1&0  \\
$u$    &176      &$+{2\over 3}$&0 &0& 0&0& 0&1  \\
\hline
\end{tabular} \end{center}  \vspace{-0.4cm}
\end{table}
The six types of quarks (plus their six antiquarks) are classified into
three pairs, which are listed in Table~\ref{t:Qgeneration}.
They are the up-down, the charm-strange,
and the top-bottom multiplets.
The top    block is the first  generation,
the middle block is the second generation, and
the bottom block is the third  generation~\cite{Griffiths:1987tj}.

Like quarks, there are six types of the leptons~\cite{Griffiths:1987tj},
which also are classified into three pairs:
electron-neutrino ($e-\nu_e$),
muon-neutrino ($\mu-\nu_\mu$),
and tau-neutrino ($\tau-\nu_\tau$).
The electron, muon, and tau each
carry a negative charge,
but the three neutrinos ($\nu_e, \nu_\mu, \nu_\tau$) carry no charge.

The quark model assumes~\cite{Griffiths:1987tj}
that the quarks bind with each other to
form the particles.
There are two basic types of the combination.
The first type of the combination is ``baryons,''
which are made up of three quarks. (Similarly,
an antibaryon is made up of three antiquarks).
Listing all the combination of three quarks gives
the baryon decuplet~\cite{Griffiths:1987tj}.
One example of a baryon is the proton $(p)$,
the proton is
made up of  ``$uud$.''~\footnote{
Here we just include the valence quark.
Later in this chapter we will consider the sea quark,}
whose charge is +1.

Another type is the meson,  which is made up of a quark and
an antiquark. Counting all the quark-antiquark combinations
yields the mesons in which we are especially interested.
We list these mesons in Table~\ref{t:Mesonnonet},
\begin{table}[h]
\vspace{-0.35cm}
\caption{ \label{t:Mesonnonet}
The top block is the meson nonet,
the middle block is ``charmed mesons,''
and the bottom block is ``bottom mesons.'' Here we just
include the mesons which we are interest in this dissertation.
Where the $Q$ stand for charge, and $S$ is strangeness,
the $C$ stand for charm, and $B$ is bottom (or beauty).
}
\begin{center} \begin{tabular}{|l|c|c|c|c|c|}
\hline
$q\bar q$  & $Q$  &  $S$  &  $C$  &$B$ &  Meson               \\
\hline
$u\bar u$  & $0 $ &  $0 $ &  $0 $  &$0 $   &  $\pi^0      $     \\
$u\bar d$  & $1 $ &  $0 $ &  $0 $  &$0 $   &$\pi^+        $     \\
$d\bar u$  & $-1$ &  $0 $ &  $0 $  &$0 $   & $\pi^-       $     \\
$d\bar d$  & $0 $ &  $0 $ &  $0 $  &$0 $   &$\eta         $     \\
$u\bar s$  & $1 $ &  $1 $ &  $0 $  &$0 $   &$K^+          $     \\
$d\bar s$  & $0 $ &  $1 $ &  $0 $  &$0 $   &$K^0          $     \\
$s\bar u$  & $-1$ &  $-1$ &  $0 $  &$0 $   &$K^-          $     \\
$s\bar d$  & $0 $ &  $-1$ &  $0 $  &$0 $   &${\bar K}^0   $     \\
$s\bar s$  & $0 $ &  $0 $ &  $0 $  &$0 $   &$\eta'        $     \\
\hline
$c\bar   d$& $1 $ &  $0 $ &  $1 $  &$0 $   &  $D^+        $     \\
$c\bar   u$& $0 $ &  $0 $ &  $1 $  &$0 $   &  $D^0        $     \\
$\bar{c} u$& $0 $ &  $0 $ &  $-1$  &$0 $   &  $\bar{D}^-  $     \\
$\bar{c} d$& $-1$ &  $0 $ &  $-1$  &$0 $   &  $D^-        $     \\
\hline
$u\bar b$  & $1 $ &  $0 $ &  $0 $  &$-1$   &  $B^+        $     \\
$d\bar b$  & $0 $ &  $0 $ &  $0 $  &$-1$   &  $B^0        $     \\
$\bar{d} b$& $0 $ &  $0 $ &  $0 $  &$1 $   &  $\bar{B}^0  $     \\
$\bar{u} b$& $-1$ &  $0 $ &  $0 $  &$1 $   &  $B^- $     \\
\hline
\end{tabular} \end{center}
\vspace{-0.3cm}
\end{table}
the top block is the meson nonet,
the middle block contains ``charmed mesons''~\cite{PDBook},
and the bottom block contains ``bottom mesons''~\cite{PDBook}.
The ``charmed mesons'' and ``bottom mesons'' are heavy.

The quarks and antiquarks carry spin $\frac{1}{2}$.
When a quark and an antiquark bind to form the mesons
in a state of zero orbital angular momentum ($L=0$),
there are spin-$0$ combinations that are called
``pseudoscalar meson'' (i.e., $\pi^+$, $\pi^-$, $\pi^0$, $ K^{0}$,
$K^{+}$, $K^{-}$, $\bar{K}^0$, $\eta$, $\eta'$),
and spin-$1$ combinations that are called ``vector meson''
(i.e., $\rho^+$, $\rho^-$, $\rho^0$,
$ K^{*0}$, $K^{*+}$, $K^{*-}$, $\bar{K}^{*0}$,
$\phi$, $\omega$).
In the language of group theory,
each quark can occur as spin-up ($m_s=\frac{1}{2}$)
or spin-down ($m_s=-\frac{1}{2}$) states,
and two spin states belong to
the fundamental representation of spin $SU(2)$ ({\bf 2}).
Similarly, each antiquark can also occur as
spin-up ($m_s=\frac{1}{2}$)
or spin-down ($m_s=-\frac{1}{2}$) states,
and two spin states belong to
the complex conjugate  fundamental representation of spin $SU(2)$
({$ \bf \bar{2}$}).
We combine two two-dimensional fundamental representations
({\bf 2}, and {$ \bf \bar{2}$}) of spin $SU(2)$
to obtain a spin triplet ({\bf 3}, spin $s=1$)
and a spin singlet ({\bf 1}, spin $s=0$),
\be
{\bf 2} \otimes {\bf \bar{2}} = {\bf 3} \oplus {\bf 1} .
\ee
Hence, the spin singlet state is pseudoscalar mesons
(namely, the total spin $ s=0 $),
and the  spin triplet state is vector mesons
(namely, the total spin $ s=1 $)~\cite{Griffiths:1987tj}.

In addition to the electric charge,
quarks also carry ``color charge''~\cite{Griffiths:1987tj}.
In the language of group theory,
each quark takes ``red''($r$), ``green''($g$), or ``blue''($b$),
and the three color states belong to
the fundamental representation of color $SU(3)$ ({\bf 3}).
Similarly, each antiquark occurs as ``antired''($\bar r$),
``antigreen''($\bar g$), or ``antiblue''($\bar b$),
and three anticolor states belong to
the complex conjugate  fundamental representation of color $SU(3)$
({$ \bf \bar{3}$}).
We combine two three-dimensional fundamental representations
({\bf 3}, and {$ \bf \bar{3}$}) of color $SU(3)$
to form a color octet ({\bf 8}) and a color singlet ({\bf 1}),
\be
{\bf 3} \otimes {\bf \bar{3}} = {\bf 8} \oplus {\bf 1}  .
\label{color:SU3}
\ee
The quark model asserts that all macroscopically isolated particles
must be ``color singlet.''
For example, mesons are the combinations of a quark of color and
a quark of its anticolor to form a neutral color charge.

The force between color charged particles is the ``strong force.''
The carriers of the strong force are gluons~\cite{Griffiths:1987tj},
which carry no electrical charge, but
have both a color and an anticolor charge.
According to Eq.~(\ref{color:SU3}), the gluon can exist as
color octet states or color singlet  state.
However if the gluon of color $SU(3)$ singlet state exists
as mediator, it can also occur as a free particle.
An interesting fact about quarks and gluons is that
we have never been observed a macroscopically isolated  quark
or gluon. For this reason,
the color-charged quarks (or gluons) are
believed to be confined within a hadron~\cite{Griffiths:1987tj}.
Hence, we have eight physical gluon states.

By convention valence quarks for a given hadron are the quarks
(or antiquarks) that give rise to the quantum numbers
of the given hadron. For example:
the proton's quantum numbers are characterized by two ``$u$'' and
one ``$d$'' quark valence quark. At any given momentum
the proton may contains extra $u\bar{u}$ pairs, or $d\bar{d}$ pairs,
or even $s\bar{s}$ pairs, $c\bar{c}$ pairs, etc. We
call these extra quarks sea quarks~\cite{Griffiths:1987tj}.
%
\section{Some conservation laws}
\label{s1:SCL}
The study of the interactions and the decays has resulted in a few
conservation laws. For the strong interactions and decay,
the conserved quantities include
baryon number ($A$),
lepton number ($L$),
upness ($U$),
downness ($D$),
strangeness ($S$),
charm ($\overline C$),
bottom ($B$),
top ($T$),
isospin ($I$),
parity ($P$), and
charge-conjugation ($C$).
These conservation laws, together with the classical conservation
laws (such as the conservation of energy, momentum, charge)
should apply in the strong interactions and decays.

In this section, we just focus on the conservation laws
of parity ($P$), isospin ($I$), and charge-conjugation ($C$),
which are needed for this dissertation.
%
%
\subsection{Parity}
Many physical processes like strong interactions have a property of
parity invariance (or ``mirror symmetry'').
This means that the mirror image of any physical process
can also ocur with the same probability.
In this subsection, we repeat some discussion in Sec.~4.6
in Ref.~\cite{Griffiths:1987tj}.

In physics, the parity basic operator is an inversion operation,
in which the object is flipped to the opposite location
through the origin~\cite{Griffiths:1987tj}.

Quarks have an intrinsic parity, which is defined to be $+1$, and
for an antiquark the parity is $-1$~\cite{Griffiths:1987tj}.
Since the quantum number parity for the composite system is
multiplicative, the intrinsic parity of the baryon octet and decuplet is +1,
and the intrinsic parity of the meson is $-1$~\cite{Griffiths:1987tj}.
For the excited states of the mesons,
the parity is given by~\cite{Griffiths:1987tj}
\be
P = (-1)^{l+1} ,
\ee
where $l$ is the orbital angular momentum of the meson.

The parity is not  conserved by the weak force,  with the result that
all the neutrinos are found to be
``left-handed''~\cite{Griffiths:1987tj}.
%
\subsection{Isospin}
Isospin is a physical term that was introduced to describe
a group of particles that has almost the same mass~\cite{Griffiths:1987tj}.
For example, the doublet of the proton and the neutron is said
to have isospin ${1\over{2}}$ ($I={1\over{2}}$),
with the third component $I_3 = +{1\over{2}}$ for the proton and $-{1\over{2}}$
for the neutron.
Another example is the three pions ($\pi^{+}$,$\pi^{0}$,
$\pi^{-}$), which compose a triplet.
Hence, its isospin is 1. The third components are
$+1$ for the $\pi^{+}$, 0 for $\pi^{0}$,
and $-1$ for $\pi^{-}$, respectively.

In 1932 Heisenberg suggested that
isospin is conserved in strong interactions.
In the language of group theory, we can declare
that strong interactions are invariant
under an internal $SU(2)$ symmetry
(or internal isospin symmetry)~\cite{Griffiths:1987tj}.
Hence, nucleon ($p$, and $n$) is the two-dimensional
representation of the $SU(2)$ ($I=\frac{1}{2}$),
and pion meson ($\pi^{+}$, $\pi^{0}$, and $\pi^{-}$)
is the three-dimensional representation of the $SU(2)$ ($I=1$).

Isospin for the particles is related to other quantum numbers for the particles
by general Gell-Mann-Nishijima formula~\cite{Griffiths:1987tj}
\be
Q = \frac{A+U+D+S+C+B+T}{2} ,
\ee
where
$Q$ is the charge of the particle,
$A$ is baryon number,
$U$ is upness,
$D$ is downness,
$S$ is stangeness,
$C$ is charm,
$B$ is bottom,  and
$T$ is the top~\cite{Perkins:1982xb}.

At the current quark model, the bare masses of the up and  down quarks
are close. Hence, the ``up'' and ``down''
quarks can form an isospin doublet $(I={1\over{2}})$,
and all the other flavors ($s,c,b,t$)
carry isospin zero~\cite{Griffiths:1987tj}.
By convention the third component of isospin of
the up($u$) quark is assigned to be $ 1\over{2} $,
and the third component of isospin of
the down ($d$) quark is assigned to be $ -{1\over{2}} $
(like proton and neutron)~\cite{Griffiths:1987tj}:
\be
  u = \left| { 1\over2} \ \ \ \ { 1\over2 } \RR,  \ \ \ \ \ \
  d = \left| { 1\over2} \ -{1\over2 } \RR.
\ee
The antiup ($\bar{u}$) and antidown ($\bar{d}$) quarks
also form an isospin doublet~\cite{Griffiths:1987tj}:
\be
  \bar u =  \left| { 1\over2} \ -{ 1\over2 }  \RR,   \ \ \ \ \ \
  \bar d = -\left| { 1\over2} \ \ \ \ {1\over2 } \RR.
\ee
By convention the third component of the isospin of
$\bar{u}$ quark is assigned to be $ -{1\over{2}} $
and the third component of the isospin of the $ \bar d $ quark
is assigned to be $ {1\over{2}}$~\cite{Griffiths:1987tj}.

If we add two particles whose isospins are $I={1\over2}$,
we obtain an ``isotriplet''~\cite{Griffiths:1987tj}
\be
  \left| 1  \ \ \ \ 1\RR \, = \, -u\bar d, \ \ \ \ \ \
  \left| 1  \ \ \ \ 0\RR \, = \, \frac{u\bar u-d\bar d}{\sqrt{2}}, \ \ \ \ \ \
  \left| 1   -1\RR       \, = \, d\bar u,
\ee
and an ``isosinglet''
\be
\left| 0 \ \ \ \ 0\RR \, = \, \frac{u\bar u + d\bar d}{\sqrt{2}}.
\label{ISOSIGNET:CH1}
\ee

In the language of group theory, the $u,d$ quarks belong to
the fundamental representation of isospin $SU(2)$ ({\bf 2}),
and the $\bar{u}, \bar{d}$ quarks belong to
the complex conjugate fundamental representation of spin $SU(2)$
({$ \bf \bar{2}$}).
We combine two two-dimensional representations
({\bf 2}, and {$ \bf \bar{2}$}) of spin $SU(2)$
to obtain an ``isotriplet'' ({\bf 3}, isospin $I=1$)
and an ``isosinglet''  ({\bf 1}, isospin $I=0$),
\be
{\bf 2} \otimes {\bf \bar{2}} = {\bf 3} \oplus {\bf 1}  .
\ee

For the peudoscalar mesons the isotriplet is the pion.
For the vector mesons      the isotriplet is the $\rho $ meson.
Hence, the quark content (or the wave function of ``flavor'')
for $\pi^0$(or $\rho^0$) meson is  given by~\cite{Griffiths:1987tj}
\bea
\pi^0, \ \  \rho^0  &=& \frac{u\bar u - d\bar d}{\sqrt{2}} .
\eea

Isospin is connected with a conservation law,
as shown by the strong process
\bea
\ \ \ \ \ \ p + p   & \longrightarrow & d  +  \pi^+   \nonumber \\
I \ \ \  \frac{1}{2}\ \ \frac{1}{2}  &  & 0 \ \ \ \ 1
\eea
Here we assume the isospin of the deuteron ($d$) is consistent with
the observation that there are no other bound
state of two nucleons~\cite{Perkins:1982xb}.
It is obvious that the above process conserves charge,
angular momentum, baryon number, and isospin.
%
\subsection{Charge conjugation}

Charge conjugation is a physical operation that replaces
each particle in a process
with its the corresponding antiparticle~\cite{Griffiths:1987tj}.
It changes the sign of the charge, baryon number, lepton number,
strangeness number, charm number, beauty number,
and truth  number of the particle.
It does not, however, change the mass, energy, momentum,
and spin of the particle~\cite{Griffiths:1987tj}.

We can show that for a system of a quark
(spin-$1\over{2}$ particle) and its antiquark,
the charge conjugation($C$) is given by~\cite{Perkins:1982xb}
\be
C = (-1)^{l+s} ,
\ee
where $l$ is total orbital angular momentum of the meson, and
$s$ is total spin of the meson.

If we consider the ground states ($l=0$) of mesons,
then for the pesudo-scalar mesons ($s=0$), hence $C=+1$;
for the vector mesons ($s=1$), thus  $C=-1$.

One kind of the notation for these states is to indicate
their total angular momentum ($j=l+s $), parity,
and charge conjugation explicitly.
For the ground states ($l=0$) of the pseudoscalar meson,
the total angular momentum is zero (i.e., $j=0+0=0 $),
and they have negative parity and positive charge conjugation ($C$).
Hence
\be
J^{PC} = 0^{-+} .
\ee
For the ground states of the vector mesons, $s=1$,
we obtain $j=1$,
and they have negative parity and negative charge conjugation ($C$).
Hence
\be
J^{PC} = 1^{--} .
\ee

Charge conjugation is conserved for strong and electromagnetic
interactions, but not for weak interactions.
We know that charge conjugation turns a left-handed neutrino into
a left-handed antineutrino, and changes right-handed neutrino
into a right-handed antineutrino.
In weak processes, there exist only left-handed neutrinos
and only right-handed antineutrinos.
Hence, it is obvious that charge conjugation is not conserved in
weak processes~\cite{Griffiths:1987tj,Perkins:1982xb}.
%

\section{Quarkonium}
\label{s1:Quarkonium}
%
In the quark model, we regard all the mesons ($ q_1 \bar{q_2} $)
as two-particle bound states~\cite{Griffiths:1987tj}.
The potential model works well for hydrogen and positronium.
It is reasonable to ask whether the potential model
works for the mesons or not~\cite{Griffiths:1987tj}?

After the discovery of the bound state $c\bar{c}$
(or charmonium states),  potential models of QCD
have become an important tool in understanding
quarkonium (i.e., the bound state of a heavy quark and heavy
antiquark, e.g., $\Upsilon(b\bar{b})$, $c\bar{b}$, etc.)~\cite{Griffiths:1987tj}.

In hydrogen and positronium model,
there exists only  Coulombic potential~\cite{Griffiths:1987tj}.
(In QED, the force between electrical charges
is mediated by the exchange of photons.)
However, for the meson, we know that the force between color
charged particles through gluon exchange is very powerful.
Thus, the short distance behavior of QCD is dominated by this ``strong force.''
Hence, the potential of a meson
should be Coulombic potential at short range~\cite{Griffiths:1987tj}.
However, due to quark confinement,
the potential can increase indefinitely when the distance $r$
between quark and antiquark becomes large.
Here we assume no $q\bar q$ pair production.
At this range, it is standard to choose a linear fit (i.e., $ V(r) \sim r $)
(and it is confirmed in numerical simulation).
Thus, in this dissertation, we choose the overall potential
to be~\cite{Griffiths:1987tj,Perkins:1982xb}.
In Chapter 4, we will note that for the $\Pi_u$ potential, we choose
a different model.
\be
V(r) \ = \ - \frac{k_1}{r} + k_2 r ,
\ee
where $ k_1, k_2 $ are just two constants.

The mesons, which are made up of light quarks (i.e., $u, d,s $),
are intrinsically relativistic~\cite{Griffiths:1987tj}.
The mesons, which are made up of heavy quarks (i.e., $b, t $),
are intrinsically nonrelativistic~\cite{Griffiths:1987tj}.
The classical quantum mechanics
can describe these systems pretty well.
From Fig.~5.7 in~\cite{Griffiths:1987tj}, we can see
that potential model describe the charmonium very well.
For the bound state $b \bar{b} $ (also called bottomonium),
the potential model describes it very well also. Please see
Fig.~5.9 in~\cite{Griffiths:1987tj}.

In this dissertation we study the hybrid quarkonium
(or hybrid meson).
Hybrid mesons are the bound state of $Q\overline{Q}$
with nontrivial excited gluonic components
($Q\overline{Q}g$),
where $Q$ is a heavy quark (e.g., $b$, $t$ quark),
and $\overline{Q}$ is a heavy antiquark.
A hybrid spin-exotic state is a hybrid meson
whose $J^{PC}$ value is  permitted
in the quark model~\cite{McNeile:2002az}.
From the current quark model,
the quantum numbers for $Q\bar{Q}$ composites are
given in Table~\ref{t:exotic}.


Because the quantum numbers $J^{PC}=  1^{-+},\ 0^{+-}$ and   $2^{+-}$
cannot exist for the meson in the quark model,
they are hybrid spin-exotic mesons.
In Chapter 4, we give more details
about hybrid spin-exotic mesons and the potential model.

\vspace{-0.4cm}
\begin{table}[h]
\caption{ \label{t:exotic}
The ${}^{2S+1}L_{J}$ or  $J^{PC}$ for $Q\bar{Q}$. }
\begin{center} \begin{tabular}{|l|l|l|l}
\hline
$L$ &  Singlet($S=0$)     &  Triplet($S=1$)  \\
\hline
0  & $^1$S$_0 (0^{-+})$   &  $^3$S$_1 (0^{--})$  \\
\hline
1  & $^1$P$_1 (1^{+-})$   &  $^3$P$_{0,1,2} (0^{++}, 1^{++}, 2^{++})$  \\
\hline
2  & $^1$D$_2 (2^{-+})$   &  $^3$D$_{1,2,3} (1^{--}, 2^{--}, 3^{--})$  \\
\hline
3  & $^1$F$_3 (3^{+1})$   &  $^3$F$_{2,3,4} (2^{++}, 3^{++}, 4^{++})$  \\
\hline
\end{tabular} \end{center}
\vspace{-0.3cm}
\end{table} 

%% file: Chap3.tex
\setlength{\unitlength}{0.5in}
\newsavebox{\Staple}
\savebox{\Staple}{\begin{picture}(0,0)	
\thicklines
\put(0.0,0.1){\vector(0,1){0.9}}
\put(0.0,1.0){\vector(1,0){0.9}}
\put(0.9,1.0){\vector(0,-1){0.9}}
\end{picture}}
\newsavebox{\FiveStaple}
\savebox{\FiveStaple}{\begin{picture}(0,0)	
\thicklines
\put(0.0,0.1){\vector(0,1){0.9}}
\put(0.0,1.0){\vector(1,1){0.5}}
\put(0.5,1.5){\vector(1,0){0.9}}
\put(1.4,1.5){\vector(-1,-1){0.5}}
\put(0.9,1.0){\vector(0,-1){0.9}}
\end{picture}}
\newsavebox{\SevenStaple}
\savebox{\SevenStaple}{\begin{picture}(0,0)	
\thicklines
\put(0.0,0.1){\vector(0,1){0.9}}
\put(0.0,1.0){\vector(1,1){0.5}}
\put(0.5,1.5){\vector(1,2){0.3}}
\put(0.8,2.1){\vector(1,0){0.9}}
\put(1.7,2.1){\vector(-1,-2){0.3}}
\put(1.4,1.5){\vector(-1,-1){0.5}}
\put(0.9,1.0){\vector(0,-1){0.9}}
\end{picture}}
\newsavebox{\LepageStaple}
\savebox{\LepageStaple}{\begin{picture}(0,0)	
\thicklines
\put(0.0,0.1){\vector(0,1){0.9}}
\put(0.0,1.0){\vector(0,1){0.9}}
\put(0.0,1.9){\vector(1,0){0.9}}
\put(0.9,1.9){\vector(0,-1){0.9}}
\put(0.9,1.0){\vector(0,-1){0.9}}
\end{picture}}
\newsavebox{\Link}
\savebox{\Link}{\begin{picture}(0,0)	
\thicklines
\put(0.0,0.0){\vector(1,0){0.9}}
\end{picture}}
\newsavebox{\Naik}
\savebox{\Naik}{\begin{picture}(0,0)	
\thicklines
\put(0.0,0.0){\vector(1,0){0.97}}
\put(1.0,0.0){\vector(1,0){0.97}}
\put(2.0,0.0){\vector(1,0){0.97}}
\end{picture}}
\newsavebox{\OneFatNaik}
\savebox{\OneFatNaik}{\begin{picture}(0,0)	
\thicklines
\put(0,0){\usebox{\Link}\makebox(0,0)}
\put(1.1,0)+
\put(1.5,0){\usebox{\Naik}\makebox(0,0)}
\put(4.6,0)+
\put(5.1,0){\usebox{\Staple}\makebox(0,0)}
\end{picture}}
\newcommand{\re}{{\rm Re\,}}
\newcommand{\tr}{{\rm Tr\,}}

\chapter{ Lattice QCD}

It is well known that quantum electrodynamics (QED) describes the
interaction of the charged particles and photons. In the same way,
quantum chromodynamics (QCD) is the popular theory of the strong interactions
of the colored particles, which is formulated in terms of the quarks and gluons.
QCD has been very successful in predicting the phenomena
of high energy hadron physics.

Quantum chromodynamics is a well-developed theory, but calculations
are very difficult, particularly at energies below about $1$~GeV,
where the strong coupling constant is of the order of unit,
so perturbation methods fail.
Fortunately, lattice QCD provides a nonperturbative method in this field.
Lattice QCD is a quantum field theory that formulates QCD
on a discrete Euclidean spacetime grid. We explain later.
The main benefits are that many calculations in QCD
take on forms familiar in classical statistical mechanics.
Since lattice QCD does not introduce new
field variables, it inherits the basic characters of QCD.

In this chapter we provide background material about QCD.
We first give an overview of continuum QCD.
Then we give an overview of lattice QCD and some special ways to
incorporate fermion fields.
Finally, we explain how to generate the configurations.
We focus on some details needed for this dissertation.
Several textbooks~\cite{Montvay:1994cy,Rothe:1992nt}
provide comprehensive background material.
%
\section{Overview of continuum QCD}
In this section, we first introduce the full gauge invariant classical
Minkowski Lagrangian for QCD by directly starting from
the Minkowski Lagrangian of a free Dirac field
for a given flavor, and then introducing the gluon field
through the help of the parallel transport.
Then we write the path integral representation
for the physical vacuum expectation values of any observables.
Later we change this path integral
representation to the corresponding path integral
in the Euclidean space-time. Also we associate the path integral
formulation of the partition function
in QCD with the corresponding partition function
of quantum statistical mechanics.

\subsection{Continuum gauge field theory in QCD}
On the lowest level, QCD describes the interactions of the quarks and
gluons. In Chapter 1, we explained that the quarks
($u$, $d$, $s$, $c$, $b$, $t$) and the antiquarks
($\bar{u}$, $\bar{d}$, $\bar{s}$, $\bar{c}$, $\bar{b}$, $\bar{t}$)
are fermions of fractional electric charge and color charge.
Each of the quarks carries a $SU(3)$ color index.
According to the current colored quark model,
the three colors of a given flavor have the same mass,
although different flavors have different masses~\cite{Griffiths:1987tj}.
Hence, the wave function $\psi_c$ of each quark contains a triplet of fields
($c=1,2,3$). The wave function $\bar{\psi_c}$ of each antiquark
also contains a triplet of fields~\cite{Rothe:1992nt}\cite{Griffiths:1987tj}.

Now we introduce the gauge-invariant Lagrangian
in continuum QCD in the Minkowski space.
The starting point is the Minkowski Lorentz invariant Lagrangian
of the free Dirac field
for a given flavor~\cite{Griffiths:1987tj}\cite{Montvay:1994cy}
\be
{\cal L_{F} } = \sum_{c=1}^{3}
\bar{\psi}_c(x)(i \gamma^\mu \partial_\mu - m) \psi_c(x) ,
\label{STARTF:EQ}
\ee
where $c$ is the quark color index, $m$ is the mass of the given
flavor quark, and $\gamma^\mu$ are the Minkowski Dirac matrices.
The Lagrangian in Eq.~(\ref{STARTF:EQ})
is invariant under the $SU(3)$ rotation~\cite{Montvay:1994cy}
\be
\psi(x)     \rightarrow \psi'(x)       = \Lambda \psi(x) , \ \ \ \ \ \
\bar\psi(x) \rightarrow \bar{\psi}'(x) = \bar\psi(x)\Lambda^\dag  ,
\ee
where $\Lambda \in SU(3)$ is independent of $x$.
The above transformation is called a
global gauge transformation~\cite{Montvay:1994cy}.
If $\Lambda $ depends on $x$, that is,
$\Lambda(x) $ is a function of $x$,
the Lagrangian in Eq.~(\ref{STARTF:EQ})
is not invariant under the gauge transformation
\be
\psi(x)     \rightarrow \psi'(x)       = \Lambda(x) \psi(x) , \ \ \ \ \ \
\bar\psi(x) \rightarrow \bar{\psi}'(x) = \bar\psi(x)\Lambda(x)^\dag ,
\ee
where $\Lambda(x)$ is an element of $SU(3)$.
The above transformation is known as
local gauge transformation~\cite{Montvay:1994cy}.
We will modify the Lagrangian in Eq.~(\ref{STARTF:EQ}),
so it it invariant under a local gauge transformation.

For each space-time point $x$,
there is a space $V_x$ allowed values of ${\psi}(x)$ at that point.
How do we compare $\psi(x)$ on one vector space $V_x$
with $\psi(y)$  at an other  vector space $V_y$.
In the gauge theory of QCD it is the parallel transporter
that connects a point on one vector space
with the corresponding point on another vector space.
The parallel transport is generated infinitesimally~\cite{Mack:1981}
along a smooth curve.
First let ${\cal C}_{yx}$ be a smooth curve in space-time from
point $x$ to point $y$.
Then we define a mapping from vector space $V_x$ to vector space
$V_y$, that is,
\be
U({\cal C}_{yx}): V_x \rightarrow V_y ,
\ee
for $U({\cal C}_{yx}) \in SU(3)$ such that  $U({\cal C}_{yx})\psi(x)$
is an element of the vector space $ V_y$.
That is, vector $U({\cal C}_{yx})\psi(x)$ is regarded
as the vector $\psi(x)$ at the point $x$,
parallel transported along the smooth curve ${\cal C}_{yx}$
to the point $y$~\cite{Montvay:1994cy}.
We say $U({\cal C}_{yx})$ is a
``parallel transporter''~\cite{Montvay:1994cy}
along the smooth curve ${\cal C}_{yx}$.

We can prove that~\cite{Montvay:1994cy,Mack:1981},
under the local gauge transformation
\vspace{-0.3cm}
\bea
\psi(x) \rightarrow \psi'(x) = \Lambda(x)\psi(x) , &&
\bar\psi(x)\rightarrow\bar{\psi}'(x) = \bar\psi(x)\Lambda^\dag(x) , \\
\psi(y) \rightarrow \psi'(y) = \Lambda(y)\psi(y) , &&
\bar\psi(y)\rightarrow\bar{\psi}'(y) = \bar\psi(y)\Lambda^\dag(y) ,
\eea
a parallel transporter transforms as~\cite{Montvay:1994cy}
\be
U({\cal C}_{yx}) \rightarrow U'({\cal C}_{yx}) =
\Lambda^{\dag}(y) U({\cal C}_{yx}) \Lambda(x) .
\ee

Now we consider the straight smooth curve ${\cal C}_{x+dx, x}$
from point $x$ to point $x+dx$.
The Lie algebra of  $SU(3)$ is generated by $\DPT \frac{\lambda^a}{2}$,
the eight Gell-Mann matrices satisfy the commutation relations
\be
[\lambda^a,   \lambda^b ] = 2i  f_{abc} \lambda^c  ,
\ee
where $f^{abc}$ are the structure constants of $SU(3)$.
Hence, the corresponding infinitesimal parallel transporter
can be written as~\cite{Montvay:1994cy}
\be
U({\cal C}_{x+dx, x}) =
{\bf 1} + igA^a_{\mu}(x) \frac{\lambda^a }{2} dx^\mu ,
\label{inftansporter:eq}
\ee
where {\bf 1} is the $3 \times 3$ unit matrix,
$g$ is the color coupling constant, which is just introduced conventionally,
$A^a_{\mu}(x)$ is the gluon gauge field, and
$a$ is the gluon color index ($a=1, ... , 8 $).
The parallel transporter in Eq.~(\ref{inftansporter:eq})
has a vector index $\mu$ that gets contracted between
$A^a_{\mu}(x)$ and  $dx^\mu$.
For notational simplicity, we define the gauge field $A_{\mu}(x)$
in the Lie algebra of $SU(3)$~\cite{Montvay:1994cy}
\be
A_{\mu}(x) \equiv igA^a_{\mu}(x)\frac{\lambda^a }{2} .
\ee
Then we can rewrite Eq.~(\ref{inftansporter:eq}) as
\be
U({\cal C}_{x+dx, x}) =
{\bf 1} + A_{\mu}(x) dx^\mu .
\label{inftansporter:ch3}
\ee
Now we perform a longer distance parallel transport
along smooth path ${\cal C}_{yx}$,
where the points $x, x_1, x_2, x_3 \cdots, x_{n-1}, x_n, y$
form a series of infinitesimally separated points
along the path ${\cal C}_{y,x}$.
Then the parallel transport of $\psi(x)$
to the end of the path (i.e., point $y$) in space-time
is constructed  by taking a series of infinitesimal transports along path
${\cal C}_{yx}$~\cite{Montvay:1994cy,Wu:1975es,Yang:1974kj}.
That is,
\bea
U({\cal C}_{y,x}) \psi(x) \hspace{-0.3cm} &=&  \hspace{-0.3cm}
[{\bf 1} + A_{\mu}(x)(x_1^\mu \hspace{-0.1cm}-\hspace{-0.1cm} x^\mu)] \times
[{\bf 1} + A_{\mu}(x_1)(x_2^\mu \hspace{-0.1cm}-\hspace{-0.1cm} x_1^\mu)] \times
\cdots  \times \nonumber \\
\hspace{-0.3cm} & &\hspace{-0.3cm}
[{\bf 1} + A_{\mu}(x_{n-1})(x_n^\mu
\hspace{-0.1cm}-\hspace{-0.1cm} x_{n-1}^\mu)] \times
[{\bf 1} + A_{\mu}(x_{n})(y^\mu
\hspace{-0.1cm}-\hspace{-0.1cm}x_{n}^\mu)] \psi(y) ,
\eea
here $U({\cal C}_{y, x})$ is the parallel transporter
along smooth curve ${\cal C}_{y,x}$.
This product of a series of infinitesimal parallel
transports, with $x$ (the path's beginning) on the left and $y$
(the path's end) on the right,
results in Dyson's formula~\cite{Dyson:1949ha,Montvay:1994cy}
\be
U({\cal C}_{yx}) = P \exp
\left\{  \int_{ {\cal C}_{yx} }  A_{\mu} dx^{\mu}  \right\} ,
\ee
where the symbol $P$ denotes the path ordering of the integrand.
The parallel transporter $U({\cal C}_{yx})$ is
called the Wilson line from point $y$ to point $x$ along path ${\cal C}_{yx}$,
and when the path ${\cal C}_{yx}$ is a closed loop,
the trace of the parallel transporter $U({\cal C}_{yx})$ is
referred to as a Wilson loop by convention.

Since now the parallel transporter can relate different
vector spaces ($V_x$),
we can define the covariant derivative of $\psi(x)$ by the formula
\be
D_\mu\psi(x)dx^\mu = U({\cal C}_{x+dx, x})\psi(x+dx) - \psi(x) .
\ee
After some algebra, we arrive at~\cite{Montvay:1994cy}
\be
D_\mu = \partial_\mu + A_{\mu}  .
\ee
We can prove that, under the local gauge transformation,
the covariant derivative transforms as~\cite{Montvay:1994cy}
\be
D_{\mu} \psi(x)  \rightarrow D_{\mu}^{\prime} \psi'(x) =
\Lambda^{\dag}(x) D_{\mu} \psi(x)  .
\ee
%
Next we introduce the field tensor $F_{\mu \nu}$ through
the commutator of two covariant derivatives~\cite{Montvay:1994cy}:
\be
F_{\mu \nu}(x) \equiv [D_\mu, D_\nu] =
\partial^\mu  A_\nu(x) - \partial^\nu A_\mu(x) + [A_\mu(x), A_\nu(x)] .
\ee
Since the  field tensor $F_{\mu \nu}$ is also an element
in the Lie algebra of color $SU(3)$,
it can be written in the form~\cite{Montvay:1994cy}
\be
F_{\mu \nu}(x)  = igF^a_{\mu \nu}(x)\frac{\lambda^a }{2} ,
\ee
where  $ F_{\mu\nu}^{a} = \partial^\mu  A_\nu^a - \partial^\nu A_\mu^a -
 gf^{abc}A_\mu^b A_\nu^c $ is the gluon field strength tensor.
We can prove that, under the  local gauge transformation,
the field strength tensor transforms as~\cite{Montvay:1994cy}
\be
F_{\mu\nu}(x)  \rightarrow F_{\mu\nu}^{\prime}(x) =
\Lambda^{\dag}(x) F_{\mu\nu}(x) \Lambda(x)  .
\ee

If we replace the ordinary four-derivative $\partial^\mu$
by the covariant derivative $D^\mu$,
the resulting new Lagrangian~\cite{Griffiths:1987tj}
\be
{\cal L_{F} } = \sum_{c=1}^{3}
 \bar{ \psi }_c(x) (i \gamma^\mu D_\mu - m_f ) \psi_c(x)
\label{EQ:InvG}
\ee
is invariant under the local gauge transformations.

Since we have already introduced the gauge field $A_\mu$,
we should add its own free Lagrangian to Eq.~(\ref{EQ:InvG}).
This term is given by the Yang-Mills Lagrangian
(or pure gauge Lagrangian)~\cite{Griffiths:1987tj}
\be
{\cal L_{YM} } = - \frac{1}{4}  F^{\mu \nu}_{a} F_{\mu \nu}^{a} ,
\ee
which is obviously invariant under the local gauge transformations.
Hence, we finally arrive at the full gauge invariant classical
Minkowski Lagrangian for QCD
(here we consider the case of various flavors)
\be
\label{LQCD:eq}
{\cal L} = -\frac{1}{4}F^{\mu \nu}_{a} F_{\mu \nu}^{a} +
           \sum_{f=1}^{n_f} \sum_{c=1}^{3}
           \bar{ \psi }^f_c (i \Dsl - m_f ) \psi^f_c  ,
\ee
where
$\Dsl = \gamma^\mu_M D_\mu$,
$f$ is the quark flavor index,
$ m_f $  is the mass of the quark with flavor $f$,
$\psi^f_c$ and $\bar{\psi}^f_c$ are the spin quark fields,
$ n_f $  is the number of the fermion flavors,
and $ \gamma^\mu_M  $ are the Minkowski Dirac matrices.
For brevity, we have omitted the quark spin index.

\subsection{Continuum QCD formalism}
We calculate the physical vacuum expectation values
of any observables $F(U,\psi,\bar{\psi})$
which are function of the field variables.
For QCD in the Minkowski space~\cite{Rothe:1992nt},
the expectation values can be written as a
path integral~\footnote{
The path integral (PI) was introduced by Feynman
in quantum mechanics in Ref.~\cite{Feynman:1948ur},
For example, we replace the matrix element
$\LL x\left|e^{-iHt}\right|y\RR$ by an infinite-dimensional integration over
all classical paths, which were weighted by the exponential of
$i$ times the classical action $\int Dx e^{iS}$,
where the $\left|y\RR$ is the eigenstate of Hamiltonian operator $H$,
and $S$ is the classical action. Hence, the Hamiltonian operator
has been removed~\cite{Montvay:1994cy,Rothe:1992nt}.
}
\be
\LL 0 \left| F(U,\psi,\bar{\psi}) \right| 0 \RR \, = \, \frac{1}{ Z }
\int {\cal D} U  {\cal D}\psi {\cal D} \bar{\psi} \, F(U,\psi,\bar{\psi}) \,
e^{ \DPT i{\cal S_M} } ,
\label{AMLQCD:EQ}
\ee
where $\psi \, {\rm and} \, \bar{\psi}$ are Grassmann variables~\footnote{
In Ref.~\cite{Montvay:1994cy} the authors give the detailed discussion of
the properties of Grassmann variables. Please see
Ref.~\cite{Montvay:1994cy} for details.
},
$\left|0\RR$  stands for the ground state of the system (or physical vacuum
states), and
\be
\label{MLQCD:DEF}
Z = \int {\cal D}U \ {\cal D}\psi \ {\cal D}\bar \psi \ e^{i {\cal S_M}} .
\ee
Here $\DPT {\cal S_M} = \int d^4x \ {\cal L} $ is the action of QCD,
${\cal L}$ is described by Eq.~(\ref{LQCD:eq}),  and
\bea
{\cal D}U &=& \prod_{x,\mu} d A_\mu(x) \\
{\cal D}\psi \ {\cal D}\bar{\psi} &=&
\prod_{x,\alpha} d       \psi_\alpha(x)
\prod_{y,\beta } d \bar{\psi}_\beta(y) ,
\eea
where ${\psi} \, {\rm and} \, \bar{\psi}$ are Grassmann variables,
and $\alpha, \beta$ are the indices including color, flavor, and spin.

Hence, in Minkowski space
the paths are weighted by a function $e^{i S}$
($S$ is action)~\cite{Rothe:1992nt}, which is oscillating.
Therefore, it is not suitable for numerical calculations.
For a numerical simulation, it is very convenient to
calculate in  imaginary time (i.e., $ t\rightarrow -i\tau $),
where the variable $\tau$ is called ``Euclidean time.''
This means that we change the Minkowski spacetime metric
to  Euclidean spacetime metric,
then use the Minkowski Lagrangian in
Eq.~(\ref{LQCD:eq}) for QCD to create
the corresponding classical Euclidean action for QCD.
First of all, we should understand the difference between these two metrics.
We know that the Minkowski spacetime is a four-dimensional space
(i.e., $x_0, x_1, x_2, x_3$), whose metric is $g_{\mu\nu}$.~\footnote{
The definition of $g_{\mu\nu}$ is in Ref.~\cite{Donoghue:1992dd}. }
Hence, the Minkowski scalar product is
\be
x \ast y \equiv  g_{\mu\nu} x^{\mu} y^{\nu} =
x^{0}y^{0} - x^{1}y^{1} - x^{2}y^{2} - x^{3}y^{3} .
\ee
Euclidean spacetime (or Cartesian space)
is the space of 4-tuples of real numbers (i.e., $x_1, x_2, x_3, x_4$),
where $x_4$ (or $t$) is time and $x_1, x_2, x_3$ is space.
In Euclidean spacetime, we use the metric $\delta_{\mu\nu}$.
Hence, the Euclidean scalar product is
\be
x \cdot y \equiv  \delta_{\mu\nu} x^{\mu} y^{\nu} =
x^{1}y^{1} + x^{2}y^{2} + x^{3}y^{3} + x^{4}y^{4} .
\ee
In fact, Minkowski spacetime can be considered to have a
Euclidean metric but with imaginary time $x^0 = -icx_4$,
where c is the speed of light.
By convention in high energy physics,
we choose c=1 (so, $x^0 = -ix_4$).

Now we  first make following substitutions:
$x_0 \rightarrow -i x_4 $ (or $t \rightarrow -i \tau$)
and $A_0 \rightarrow -i A_4 $,
where $t\equiv x_0$, and $x_4\equiv\tau$.
Second we adopt the Euclidean Dirac matrices $ \gamma^\mu $.
They are related to the  Minkowski Dirac matrices
$\gamma^\mu_M$ through~\cite{Montvay:1994cy}
\vspace{-0.4cm}
\bea
\gamma^\mu_M  &=& i \gamma^\mu  \ \ \ \ \ \ \ \ \    \mu = 1,2,3 \\
\gamma^\mu_M  &=&   \gamma^\mu  \ \ \ \ \ \ \ \ \ \  \mu = 0  .
\eea
%
Therefore, we obtain  the  classical Euclidean action~\cite{Rothe:1992nt}
for QCD~\footnote{
In principle we can relate Minkowski space observable to
Euclidean space through analytic continuation.
This is a trivial for static quantities
such as masses~\cite{Montvay:1994cy,Donoghue:1992dd,MT:90}.
In Chapter 5, we will discuss some problems
(e.g., the energy conservation) due to this continuation.}
\be
  {\cal S} = \int d^4x \left\{
             \frac{1}{4}F_{a}^{\mu \nu} F^{a}_{\mu \nu} +
             \sum_{f=1}^{n_f} \sum_{c=1}^{3}
             \bar{ \psi }^{f}_{c} M_f\psi^{f}_{c}
  \right\} ,
\ee
where
\vspace{-0.5cm}
\bea
M_f   &=& \Dsl + m_f  \\
\Dsl  &=& \gamma^\mu D_\mu  ,
\eea
and $M_f$ is the Dirac operator.
Then Eq.~(\ref{MLQCD:DEF}) can be rewritten as~\cite{Rothe:1992nt}
\bea
\label{ELQCD:EQ}
Z  &=&  \int {\cal D} U  {\cal D}\psi {\cal D} \bar{\psi}
\ e^{ -{\cal S} }  .
\eea
Therefore, Eq.~(\ref{AMLQCD:EQ}) can be rewritten as~\cite{Rothe:1992nt}
\bea
\LL 0 \left| F(U,\psi,\bar{\psi})  \right| 0 \RR &=& \frac{1}{ Z }
  \int {\cal D} U  {\cal D}\psi {\cal D} \bar{\psi}
  \ F(U,\psi,\bar{\psi})  \  e^{ -{\cal S} }  ,
\label{AELQCD:EQ}
\eea
where ${\cal S}$ is the action in Euclidean space.
We note that the weighting factor in path integral now is
$e^{-S}$.

Now we associate the path integral formula of the partition function
described by Eq.~(\ref{ELQCD:EQ}) and Eq.~(\ref{AELQCD:EQ})
with the ones of quantum statistical mechanics.
In quantum statistical mechanics, we define the partition function as
\bea
  Z\left(\tau \right) &=& \Tr  e^{-\tau H } .
\eea
In terms of the eigenvectors and eigenvalues of the operator
$H$ which satisfies $ H \left| n \RR  = E_n \left| n \RR $,
where $ H    $ is the Hamiltonian,
      $ \tau $ is the ``Euclidean time.''
If we consider the following equation~\cite{Montvay:1994cy}
\bea
\Tr(e^{-\tau H} F ) &=& \sum_{n=0}^{\infty}e^{-E_n\tau}
\LL n \left|F \right| n \RR  \\
Z(\tau) &=& \Tr(e^{-\tau H }) \ = \ \sum_{n=0}^{\infty}e^{-E_n\tau}  ,
\eea
We obtain that the thermal expectation values of
any observables ($F$) through~\cite{Montvay:1994cy}
\be
\LL F \RR = \lim_{\tau \rightarrow \infty}
\frac{\Tr \left(F e^{-\tau H}\right)}{Z\left(\tau \right)} .
\ee
Hence, many calculations in QCD take on forms familiar
in classical statistical mechanics.  In Euclidean space
the definition of the partition function described by Eq.~(\ref{ELQCD:EQ})
and the definition of the expectation value described by Eq.~(\ref{AELQCD:EQ})
are consistent with what we define in quantum statistical mechanics.
Therefore, from now on we can reasonably
call the $Z$ in Eq.~(\ref{ELQCD:EQ}) ``partition function.''
%

We usually separate the action $S$ into two parts
\be
{\cal S} = S_g + S_f .
\ee
The first term,  $S_g$, depends on only the gauge fields.
We call this term the pure gauge action (or Yang-Mills action)
\be
  {\cal S}_g = \int d^4x \frac{1}{4} F^{\mu \nu}F_{\mu \nu} .
\ee
The second term, $S_f$, depends on both the gluon and fermion fields.
We refer to this term as fermion action,
\be
  {\cal S}_f = \int d^4x  \bar{ \psi } M  \psi  ,
\label{eq:CFermion}
\ee
where, for brevity, the flavor index and color index have been suppressed.

For $n_f$ degenerate fermion flavors,
the QCD partition function in Euclidean space time is described
by~\cite{Rothe:1992nt}
\be
Z = \int {\cal D}U \, {\cal D}\psi \, {\cal D}\bar \psi \, e^{ -(S_g+S_f) } .
\ee

Integrating the fermion field for degenerate flavors
gives~\cite{Rothe:1992nt}~\footnote{
For the $n_f$ nondegenerate fermion flavors,
\be
\int {\cal D}\psi \, {\cal D}\bar \psi \, e^{ -S_f }  \, = \,
\prod_{f=1}^{n_f} \, \det{M_f}  ,
\label{nodegmf:ch3}
\ee
where $f$ is the index of flavor, and $M_f$ is Dirac operator for flavor $f$.
}
\be
\int {\cal D}\psi \, {\cal D}\bar \psi \, e^{ -S_f }  \, = \,
{(\det{M})}^{n_f} .
\ee
Thus
\be
Z \, = \, \int {\cal D}U  {(\det{M})}^{n_f} \, e^{ -S_g }
  \, = \, \int {\cal D}U  \, e^{ -S_{eff} } ,
\label{StartG:eq}
\ee
where the effective gauge action $ S_{eff}$ is denoted
as~\cite{Rothe:1992nt}\cite{DeGrand:03}
\be
S_{eff}  \, \equiv  \, S_g - n_f \ln{ (\det{M}) } .
\ee
The quark determinant $\det{M}$  generates quark loops.
In the quenched approximation, we set this determinant to a constant.
In effect, in this approximation we ignore the interaction
between fermions and gauge fields.
This approximation introduces a $5-15\%$ error in the lattice calculations of
the light hadron masses~\cite{MILC:98}.
It greatly simplifies the computation.

The expectation value of any physical observable $\cal O$
that depends on the quark and gluon fields is~\cite{Rothe:1992nt}
\be
\vev{{\cal O}} \, =  \,
\frac{\DPT \int{\cal D} U {\cal D}\psi {\cal D}\bar\psi \,
                    {\cal O} \, e^{-S} }
     {\DPT \int{\cal D} U {\cal D}\psi {\cal D}\bar\psi
         \, e^{-S} } .
\label{eq:expvalue}
\ee
The above integral is a Feynman functional integral over all
configurations of the gauge field.
Usually the fermion fields are integrated out analytically,
leading to a numerical path integral over only the gauge fields.
%
%
\section{ QCD on a lattice }
In QED, a perturbative expansion gives quite accurate results for
many physical problems. At very high energies,
a perturbative expansion in QCD also gives quite accurate
results for some physical systems. However in a low energy physical
system, perturbation theory in QCD is not suitable,
due to a large coupling constant.
To overcome this, numerical techniques
(namely,  Monte Carlo techniques and a lattice discretization)
were proposed to study QCD nonperturbatively.

Numerical techniques require discretizing
the continuum QCD theory on a Euclidean space time grid
($a$ is lattice spacing).
At the same time we should retain as many properties of the
continuum QCD action as possible, and we require that
the continuum theory be recovered
in the physical continuum limit (i.e., $a\rightarrow0$).
Such an lattice QCD action is then used to generate a biased sample of
gauge field configurations.
Our observables are calculated on each of these configurations,
and our observables are averaged on these configurations
to get their expectation values.

In this discretization,  we optionally choose periodic boundary
conditions for the gauge fields in space. That means that the first and
last points along any straight line in the lattice are considered
nearest neighbors (e.g., $\phi(x_0) = \phi(x_N)$, where $\phi$ is any
function which depend on lattice site $x$.
However in the time direction,
periodic boundary conditions for the gauge fields are not optional.

In QCD perturbation theory, we encounter various divergent integrals.
We deal with them by modifying the original integral.
A ``regulator'' parameter characterizes the modification.
The physical limit involves removing the regulator.
The lattice is a natural regulator of ultraviolet divergence.
In lattice regularization, we assume that our physical system
is not a continuum space, but a discrete lattice.
The lattice spacing $a$ serves to limit the highest
possible momentum (namely, $ \DPT \frac{\pi}{a} $)
of particles defined on the lattice. Due to this cutoff,
all of the volume integrals of interest are sums:
\be
  \int d^4x =  \sum_{x} a^4 .
\ee
Hence, we never integrate the momentum to infinity.
This shows that, in order to get the theory in the continuum,
after performing the finite sums, we should take the limit
$a\rightarrow 0$.

In this section, we first give an overview of
the pure gauge action in lattice QCD.
Then we give three methods to incorporate the fermion,
namely, the na\"ive fermion, the Wilson fermion, and the staggered fermion.
Finally, we discuss the taste interpretation of
the staggered fermion.
%
\subsection{Pure gauge action on lattice}
In order to explain confinement, in 1974, Wilson~\cite{Wilson:74}
proposed a lattice method. We now called it lattice QCD (LQCD).
From then on, LQCD become a fundamental tool for the studies of
QCD, and it is still in intensive use. The starting point for LQCD
is to define fermion fields only on the sites of a four-dimensional
grid with four spacetime dimensions of finite extent,
$N_x, N_y, N_z, N_t$, and lattice spacing $a$.
Here $N_x, N_y, N_z$ are space dimensions,
and $N_t$ is the time dimension.

A crucial point in his paper~\cite{Wilson:74}  is the introduction of
the gauge link matrices that connect nearest neighbor lattice sites.
Here we follow the procedures in Sec.~3.1.1 and Sec.~3.2.2 in
Ref.~\cite{Montvay:1994cy}.
But in some places we choose different convention.

For the lattice, let  $x $ be  any point on the lattice, and lattice spacing be
$a$. Then $x + a\hat{\mu}$ is the neighboring point in the direction
of the lattice axis $\mu=1,2,3,4$. where $\hat\mu$ is the unit
vector in the $\mu$ direction.
The parallel transporter that moves $A_{\mu} $ along the straight
path line from point $x$ to point $ x + a \hat{ \mu } $
is~\cite{Montvay:1994cy}
\be
U_{\mu}(x) = \exp  \left\{
    \int_{ {\cal C}_{x+a\hat{\mu},x} } A_{\mu} dx^{\mu}
   \right\}
    \approx e^{ a A_{\mu} } .
\ee
Here we make the approximation
\be
\int_{ {\cal C}_{ x+a\hat{\mu}, x } }  A_{\mu}(x)  dx^{\mu}
\longrightarrow A_{\mu}(x)a  .
\ee
We call  $ U_{\mu}(x)$ the link matrix linking
site $x$ to its nearest neighbor
in the $\mu$ direction, $x+a\hat\mu$.
 A link matrix is an element of the gauge group
$SU(3)$ between pairs of nearest neighbor lattice points.
The gauge field configuration is just the set of
all such link matrices between pairs of the nearest neighbor sites.

\begin{figure}[b]
\setlength{\unitlength}{1pt}
\begin{center} \begin{picture}(140,150)
\setlength{\unitlength}{1pt}
\put(0, 50) { \vector(1,0){50} }
\put(50,50) { \line(1,0){50} }
\put(100,50){ \vector(0,1){50} }
\put(100,100){ \line(0,1){50} }
\put(100,150){ \vector(-1,0){50} }
\put(0,150){ \line(1,0){50} }
\put(0,150){ \vector(0,-1){50} }
\put(0,50){ \line(0,1){50} }
\put(0,  40)  { \makebox(0,0){ $x$} }
\put(80, 40) { \makebox(0,0){ $x+a\hat{\mu}$} }
\put(0,  160) { \makebox(0,0){ $x+a\hat{\nu}$} }
\put(90, 160) { \makebox(0,0){ $x+a\hat{\mu}+a\hat{\nu}$} }
\end{picture} \end{center}
\vspace{-1.8cm}
\caption{   \label{PLAQETTE:FIG}
   A plaquette on a hypercubic lattice.
}
\vspace{-0.5cm}
\end{figure}
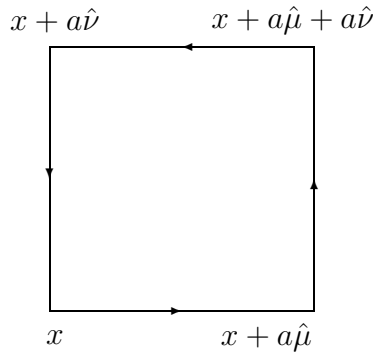

An important property in lattice QCD is that the trace of any
product of link matrices around any closed loop
is gauge invariant.
This property provides a way to translate some gauge invariant quantities
onto the lattice. For example, the smallest closed loop on the lattice is
the unit square (see Fig.~\ref{PLAQETTE:FIG}),
%
which contains four points, $
x,\
x\hspace{-0.1cm}+\hspace{-0.1cm}a\hat\mu, \
x\hspace{-0.1cm}+\hspace{-0.1cm}a\hat\nu, \
x\hspace{-0.1cm}+\hspace{-0.1cm}a\hat\mu
 \hspace{-0.1cm}+\hspace{-0.1cm}a\hat\nu$.
The corresponding product of four link matrices~\cite{Montvay:1994cy},
\be
U_P =
U_\nu(x) U_\mu(x+a\hat{\nu})
U^\dag_\nu(x+a\hat{\mu}) U^\dag_\mu(x)  =
U_{x \nu} U_{x + a\hat\nu,\mu}
U^\dag_{x + a\hat\mu,\nu} U^\dag_{x \mu} ,
\label{plaquette_term}
\ee
is called the plaquette variable in the $\mu - \nu $ plane
at lattice point $x$.

The Wilson gauge action, expressed in terms of the plaquettes,
is~\cite{Montvay:1994cy}
\be
S_G(U) = \sum_P \beta \left\{1 - \frac{1}{6} \left( \Tr
          U_P + \Tr U_P^{-1} \right) \right\}
=
\sum_P \beta \left\{1 - \frac{1}{3} \Re \Tr U_P \right\} ,
\label{SGdef:ch3}
\ee
where the sum over $P$ stands for the sum over all plaquettes,
including each plaquette with only one orientation~\cite{Montvay:1994cy},
that is,
\be
\sum_P = \sum_x \sum_{ 1 \le \mu < \nu \le 4 } .
\ee
In Wilson action $\DPT \beta = \frac{6}{g^{2}} $,
where $ g $ is the bare coupling constant.
The Taylor expansion of the link matrix $U_\mu(x) $ is~\cite{Montvay:1994cy}
\be
  U_\mu(x) = 1 + a A_\mu(x) + \frac{a^2}{2} {A_\mu(x)}^2 .
\ee
The Taylor expansion of the gauge fields is~\cite{Montvay:1994cy}
\be
  A_\nu\left(x+a\hat{\mu}\right) = A_\nu\left(x\right) + a
  \Delta^f_\mu A_\nu\left(x\right) ,
\ee
where $\Delta^f_\mu$ is the forward lattice derivative. That means
\be
\Delta^f_\mu A_\nu\left(x\right) =
\frac{A_\nu\left(x+a\hat{\mu}\right)-A_\nu\left(x\right)}{a} .
\ee
If we use the well-known Campbell-Baker-Hausdorf identity~\cite{Montvay:1994cy}
\be
  e^x e^y = e^{x+y + \frac{1}{2}\left[x , y \right] + \cdots}
\ee
and the formula~\cite{Montvay:1994cy}
\be
  F_{\mu\nu}  \equiv ig F_{\mu\nu}^a \frac{\lambda^a }{2}
\ee
\be
F_{\mu\nu} = \Delta_{\mu}^{f} A_{\nu}(x) -
             \Delta_{\nu}^{f} A_{\mu}(x) +
             [A_{\mu}(x), A_{\nu}(x) ]
\ee
\be
  \Tr(F^{\mu\nu} F_{\mu\nu})  =
  -\frac{g^2}{2} F^{\mu\nu}_a F_{\mu\nu}^a  ,
\ee
we find, after some algebra, that as $a \rightarrow 0$,
the action in Eq.~(\ref{SGdef:ch3}) can be changed into~\cite{Montvay:1994cy}
\be
  S_G =  \frac{1}{4}\sum_{x} a^4 F^{\mu\nu}_a F_{\mu\nu}^a
          + {\cal O}\left(a^6\right) .
\ee
Therefore, we indeed reproduce continuum QCD action as $a \rightarrow 0$,
and it is the correct QCD action for the gauge fields in the lattice.

If we add some terms around larger loops than plaquettes,
as in the Symanzik improved gauge action~\cite{Symanzik83},
the $ {\cal O} \left(a^{6}\right) $
error in the gauge action can be eliminated.
The gauge action used in this dissertation
was a Symanzik improved action~\cite{MILC:97},
\be
S_G = \frac{\beta_{imp}}{3}
\hspace{-0.08cm} \left\{
     \sum_{x;\mu<\nu} P_{\mu\nu}
   - \frac{1}{20 u_0^2}(1\hspace{-0.1cm}+\hspace{-0.1cm}0.4805\,\alpha_s)
     \sum_{x;\mu\neq\nu} R_{\mu\nu}
   - \frac{1}{u_0^2} 0.03325 \,\alpha_s
     \sum_{x;\mu<\nu<\sigma}  C_{\mu\nu\sigma}
\hspace{-0.08cm} \right\},
\label{MILCgaugeaction}
\ee
where $P$ is the standard plaquette in the $\mu , \nu$ plane,
$R$ is the real part of the trace of the ordered product of
$SU(3)$ link matrices along
$1\times 2$ rectangles (see Eq.~(\ref{MILCsg1})),
$C$ is the real part of the trace of the ordered product of
$SU(3)$ link matrices around $1\times 1 \times 1 $ paths
(see Eq.~(\ref{MILCsg2})),
\newlength{\latlength}
\setlength{\latlength}{1mm}
\setlength{\unitlength}{\latlength}
\bea
 \label{MILCsg1}
     R_{\mu\nu} &\equiv& \frac{1}{3} \re\tr
	\raisebox{-4\latlength}{\begin{picture}(20,12)
			\put( 0, 0){\vector( 1, 0){7}}
			\put( 0, 0){\line( 1, 0){10}}
			\put(10, 0){\vector( 1, 0){7}}
			\put(10, 0){\line( 1, 0){10}}
			\put(20, 0){\vector( 0, 1){7}}
			\put(20, 0){\line( 0, 1){10}}
			\put(20,10){\vector(-1, 0){7}}
			\put(20,10){\line(-1, 0){10}}
			\put(10,10){\vector(-1, 0){7}}
			\put(10,10){\line(-1, 0){10}}
			\put( 0,10){\vector( 0,-1){7}}
			\put( 0,10){\line( 0,-1){10}}
			\multiput(10, 1)(0,1){9}{\circle*{0.1}}
		\end{picture}}  \\ 
   C_{\mu\nu\sigma} &\equiv& \frac{1}{3} \re\tr
	\raisebox{-4\latlength}{\begin{picture}(16,16)
			\put( 0, 0){\vector( 1, 0){7}}
			\put( 0, 0){\line( 1, 0){10}}
			\put(10, 0){\vector( 2, 1){4}}
			\put(10, 0){\line( 2, 1){6}}
			\put(16, 3){\vector( 0, 1){7}}
			\put(16, 3){\line( 0, 1){10}}
			\put(16,13){\vector(-1, 0){7}}
			\put(16,13){\line(-1, 0){10}}
			\put( 6,13){\vector(-2,-1){4}}
			\put( 6,13){\line(-2,-1){6}}
			\put( 0,10){\vector( 0,-1){7}}
			\put( 0,10){\line( 0,-1){10}}
			\multiput(1,0.5)(1,0.5){6}{\circle*{0.1}}
			\multiput(6,4)(0,1){9}{\circle*{0.1}}
			\multiput(7,3)(1,0){9}{\circle*{0.1}}
		\end{picture}} ,
 \label{MILCsg2}
\eea
and here $\DPT \beta_{imp} =  \frac{10}{g^2} $,
         $\DPT \alpha_s    = -\frac{4\log(u_0)}{3.0684}$.
The  symbol $u_0$ is the tadpole improvement factor
\be
\label{MILCu0}
 u_0 = { \LP {1 \over 3 } \re \Tr \vev P  \RP }^{1 \over 4} .
\ee
In lattice perturbation theory, there exists
ultraviolet (UV) divergence from tadpole-type graphs~\cite{MILC:97}.
This contribution can be partially removed by absorbing them in
lattice coupling constants. The common method is to replace
lattice gauge matrix $U(x)$ by $U(x)/u_0$~\cite{MILC:97}.
In this dissertation, we also use this method.

\subsection{Fermion action on lattice}
\label{Faol:ch3}
The na\"ive approach to formulate the fermion action is simply
to replace the symmetric derivative in the Dirac operator
with a symmetric difference.
Here we follow the procedure and convention as in
Ref.~\cite{Rothe:1992nt}.
We can change the continuum action in Eq.~(\ref{eq:CFermion}) into
the na\"ive lattice fermion action~\cite{Rothe:1992nt}, that is,
\be
S_N = a^4\sum_x \left\{
      \frac{1}{2a} \sum_{\mu=1}^{4} \bar\psi(x) \gamma^\mu
\left[  U_\mu(x) \psi(x\hspace{-0.1cm}+\hspace{-0.1cm}a\hat\mu)
        -
        U^\dag_{\mu}(x\hspace{-0.1cm}-\hspace{-0.1cm}a\hat\mu)
      \psi(x\hspace{-0.1cm}-\hspace{-0.1cm}a\hat{\mu} )
\right]
+  m \bar\psi_x \psi_x
\right\} ,
\label{eq:naive_fermions}
\ee
where the symmetric difference is denoted as~\cite{Rothe:1992nt}
\be
D_{\mu}=\frac{1}{2a}
\left[
U_{\mu}(x) \psi(x\hspace{-0.1cm}+\hspace{-0.1cm}a\hat\mu)-
U^\dag_\mu(x\hspace{-0.1cm}-\hspace{-0.1cm}a\hat\mu)
\psi(x\hspace{-0.1cm}-\hspace{-0.1cm}a\hat\mu)
\right] .
\ee
This is called the na\"ive fermion action.

Unfortunately, in the continuum limit, the na\"ive action brings out an
unphysical doubling of the fermion modes in each lattice direction.
This mean that for every physical mode at momentum  $ k_\mu = 0 $,
there is a degenerate mode at momentum $ \DPT k_\mu = \frac{\pi}{a}$,
where $ \DPT \frac{\pi}{a} $ is the cutoff momentum of the lattice.
Hence, in four spacetime dimensions, there are 16 degenerate
modes~\cite{Montvay:1994cy}.
Several methods have been proposed to solve this ``doubling problem.''

Wilson added a second order derivative term  to the na\"ive action  to
break this degeneracy. This term includes a coefficient factor $r$,
usually chosen to be 1. For the physical $ k_\mu = 0 $  mode,
this term does not change the continuum limit,
but this term suppresses the remaining 15 $ \DPT k_\mu = \frac{\pi}{a}$
doubler modes~\cite{Montvay:1994cy}.
The Wilson fermion action is~\cite{Rothe:1992nt}
\be
S_{W} = S_{N} - \frac{r}{2a}\sum_{x}a^4 \sum_{\mu=1}^{4}\bar{\psi}(x)
\Biggl\{
       U_{\mu}^{\dagger}(x) \psi(x + a \hat{\mu}) +
       U_{\mu}(x-a\hat{\mu}) \psi(x - a \hat{\mu}) - 2 \psi(x)
\Biggr\} .
\ee
After some algebra, we get~\cite{Rothe:1992nt}
\bea
  S_{W} &=& a^4 \sum_{x} \left\{ (m+\frac{4}{a}) \bar{\psi}(x) \psi(x)
+ \frac{1}{2a}\sum_{\mu=1}^{4} \bar{\psi}(x) [r+\gamma_\mu]
  U_{\mu}(x) \psi(x+a\hat{\mu})  \right. \nonumber \\
 & & \ \ \ \ \ \ \ \ \ \  \left.
  + \frac{1}{2a}\sum_{\mu=1}^{4} \bar{\psi}(x) [r-\gamma_\mu]
  U^{\dagger}_{\mu}(x-a\hat{\mu}) \psi(x-a\hat{\mu})
\right\}  .
\label{eq:wilson_fermions}
\eea
For brevity, we define the hopping parameter $\kappa$
\be
\kappa \equiv \frac{1}{2am+8r} .
\ee
Also we rescale the fermion fields with
\be
\psi \to \psi\sqrt{2\kappa} .
\ee
Then the action can be rewritten as
\be
  S_{W} = \sum_{x y}\bar{\psi}_x M_{x y} \psi_{y} ,
\label{eq:fermion_wilson_action}
\ee
where $M_{x y}$ are the matrix elements of the
fermion matrix~\cite{Rothe:1992nt}, that is,
\be
M_{x y} = \delta_{xy} - \kappa \sum_{\mu=1}^{4}
\Biggl\{ \left[ r+\gamma_\mu\right]  U_{\mu}(x)
\delta_{x,y-a\hat{\mu}}+
\left[r-\gamma_\mu\right]  U^{\dagger}_{\mu}(x-a\hat{\mu})
\delta_{x-a\hat{\mu},y} \Biggr\} .
\label{eq:lattice_fermion_matrix}
\ee
From above equation, we can note that the diagonal term
in $M_{x y}$ is unit.

For the na\"ive fermion action, due to the doubling problem,
a single quark in fact describes 16 identical replicas
of the quark. Each of these replication is
called a ``taste''~\cite{DeGrand:03}.

Kogut and Susskind~\cite{Kogut:83} developed another approach to
solve the doubling problem called the staggered fermion Kogut Susskind action.
In this approach, the 16-fold doubling
problem of the na\"ive fermion action given  in
Eq.~(\ref{eq:naive_fermions}) can be reduced to a factor of four
if we keep only a single fermion field component on
each lattice site.
This procedure reduces the number of tastes per quark fermion
from 16 to 4.
We use another trick (discussed later in Sec.~3.3.2)
to reduce the remaining 4 fermions tastes to 1 in the determinant.

To keep just one single fermion field component on each lattice
site, we do spin-diagonalization~\cite{Montvay:1994cy}.
Staggered fermions $\chi$ are defined by the local transformation.
\be
\psi(x) = \Gamma_x \chi(x)  \qquad
\bar \psi(x) = \bar \chi(x)  \Gamma_x^\dagger \qquad
\Gamma_x = \gamma_1^{x_1}\gamma_2^{x_2}\gamma_4^{x_3}\gamma_4^{x_4} .
\label{eq:STAGGdef}
\ee
where $ \Gamma_x $ is a $ 4 \otimes 4 $ unitary matrix, which diagonalizes
all the $\gamma$ matrices in the action.
In terms of the staggered fermions $\chi$ the quark action can be
written as~\footnote{
This formula is similar to Eq.~(6.26)~\cite{Rothe:1992nt},
but we use different notation and add a taste index.
}
\bea
\label{eq:SFaction}
\CS_S \hspace{-0.3cm} &=& \hspace{-0.3cm} \sum_{x,t}
\left\{ a m_t \bar{\chi_t}(x) {\chi_t}(x) \hspace{-0.1cm}+\hspace{-0.1cm}
      {1 \over 2} \sum_{\mu=1}^{4}  {\alpha_{x \mu} }
      \Bigg[ \bar\chi_{t}(x) U_{\mu}(x)
             \chi_{t}(x\hspace{-0.1cm}+\hspace{-0.1cm}a\hat\mu)
             \hspace{-0.1cm}-\hspace{-0.1cm}
             \bar{\chi_{t}}(x\hspace{-0.1cm}+\hspace{-0.1cm}a\hat\mu)
             U^{\dag}_{\mu}(x) \chi_{t}(x)
      \Bigg]
\right \}   \nonumber \\
\hspace{-0.3cm} &\equiv& \hspace{-0.3cm}
\sum_{x,y,t} \bar\chi_{t}(x) M^{t}[U]_{x,y} \chi_{t}(y)  ,
\eea
where the $t$ is the taste index of a given flavor,
and the matrix $M^t$ is given by~\cite{Rothe:1992nt}
\be
M^{t}[U]_{x,y} \ = \ a m_t\delta_{xy} +
{1\over 2}\ \sum_{\mu} \alpha_{x \mu}
\Bigg[ U_{\mu}(x) \delta_{x,y-a\hat{\mu}}  -
        U^{\dag}_{\mu}(x-a\hat{\mu})\delta_{x,y+a\hat{\mu} }
\Bigg] .
\label{eq:SFactionM}
\ee
The $\gamma_\mu$ matrices have been replaced by the phase factors
\be
\alpha_{x \mu} = (-)^{ x_1 \ + \ \cdots \ + \ x_{\mu-1}}, \ \ \ \
(\mu=1,2,3,4)  .
\label{eq:SFphases}
\ee
From Eq.~(\ref{eq:SFaction}), and Eq.~(\ref{eq:SFactionM}),
we note that the different spin components of $\chi$ are
decoupled, because the phase factor $\alpha_{x,\mu}$ depends only
on the site index and direction index and does not have a spin index.
Here we drop the spin index on $\chi$,
keeping only the color and taste degrees of freedom
at each site. So this trick reduces the original $16$-fold degeneracy
of na\"ive fermions by a factor of four.

The fermion action used in this dissertation
is the ``Asqtad''  action~\cite{MILC:ASQ}.
\begin{figure}[b]
\setlength{\unitlength}{0.75in}
\begin{center}\begin{picture}(7.5, 1.2)
\put(0.0,0){\usebox{\Link}\makebox(0,0)}
\put(1.0,0){\usebox{\Staple}\makebox(0,0)}
\put(2.0,0){\usebox{\FiveStaple}\makebox(0,0)}
\put(3.0,0){\usebox{\SevenStaple}\makebox(0,0)}
\put(4.5,0){\usebox{\LepageStaple}\makebox(0,0)}
\put(5.6,0){\usebox{\Naik}\makebox(0,0)}
\end{picture}\end{center}
\caption{
   \label{PATHSET}
The simple link, three link staple, five link staple,
seven link staple,  Lepage term, and Naik term of Asqtad action.
}
\end{figure}
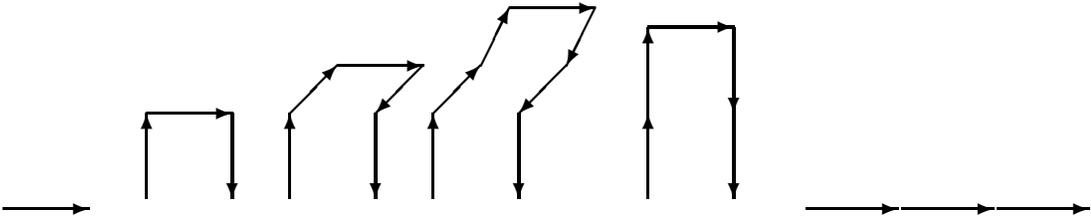
The link paths of ``Asqtad'' action are illustrated in
Fig.~\ref{PATHSET}.
The first term  is the simple link.
The second term is three link staple,
and the third  term is five link staple. The second term and third term
are to reduce the flavor symmetry~\cite{MILC:ASQ}.
The fourth  term (or seven link staple) removes tree level
couplings at the momentum $ \DPT \frac{\pi}{a}$~\cite{MILC:ASQ}.
The fifth  term (or Lepage term) is the correction to the
form factor at small momentum~\cite{MILC:ASQ}.
The sixth term (or Naik term, the third nearest neighbor)
is to improve the quark dispersion relation~\cite{MILC:ASQ}.
If we set the zero momentum coupling to one
and all the other momentum couplings to zero,
and we introduce the tadpole factors
$\DPT (\frac{1}{u_0})^{L-1}$.
where L is the length of path, we obtain the coefficients
$\DPT  \frac{5}{8}          $,
$\DPT  \frac{1}{16} u_0^{-2}$,
$\DPT  \frac{1}{64} u_0^{-4}$,
$\DPT  \frac{1}{384}u_0^{-6}$,
$\DPT -\frac{1}{16} u_0^{-4}$,
$\DPT -\frac{1}{24} u_0^{-2}$,~\cite{MILC:ASQ} respectively.

The fermion action in Eq.~(\ref{eq:SFaction}) is  first term
in the ``Asqtad''  action.
Here is the second term:
\bea
\label{eq:SFaction_2}
\CS_{S2} \hspace{-0.3cm} & = & \hspace{-0.3cm} \sum_{x, t}  \Bigg\{
a m_t \bar{\chi_t}(x) {\chi_t}(x) +
{1 \over 12} \sum_{\mu=1 }^{4} \sum_{\nu \ne \mu }^{4} {\alpha_{x \mu} }
\biggl[ \bar\chi_{t}(x)
U_\nu(x) U_\mu(x\hspace{-0.1cm}+\hspace{-0.1cm}\hat\nu)
    U^\dag_\nu(x\hspace{-0.1cm}+\hspace{-0.1cm}\hat\mu)
\chi_{t}(x\hspace{-0.1cm}+\hspace{-0.1cm}a\hat\mu) \nonumber \\
\hspace{-0.3cm}& &\hspace{-0.3cm} -
         \bar{\chi_{t}}(x\hspace{-0.1cm}+\hspace{-0.1cm}a\hat\mu)
     U_{\nu}(x\hspace{-0.1cm}+\hspace{-0.1cm}\hat\mu)
U^\dag_{\mu}(x\hspace{-0.1cm}+\hspace{-0.1cm}\hat\nu) U^{\dagger}_{\nu}(x)
         \chi_{t}(x) \biggr]
\Bigg\}  .
\eea
The  rest are constructed similarly.
%
%
%
\subsection{Tastes of Kogut-Susskind staggered fermions}
Due to the phase factors $\alpha_{x,\mu}$ the Kogut-Susskind staggered
fermion action has translational invariance under shift by $2a$.
Therefore, in the continuum limit, we can map  a $2^4$ hypercube into
a single lattice point~\cite{Montvay:1994cy}.
For simplicity, we just discuss the free Kogut-Susskind Staggered
fermions.

In position space~\cite{Montvay:1994cy}, we can easily understand
the taste identification. For some purpose it is convenient
to represent the 16 degrees of freedom in a hypercube by
a new field $q$ called a hypercubic field,
which in the continuum limit represents the four spin
components of four degenerate tastes.

Let us first assume that the number of the points in a hypercubic
lattice is even in every direction:
$ L_\mu = 2L^{'}_{\mu} \ (\mu=1,2,3,4) $.
The coordinates of the lattice points lie in the interval
$ 0 \leq x_{\mu} \leq L_{\mu} -1 $.
The coordinates of the hypercubic blocks $ y_\mu $~\cite{Montvay:1994cy}
are denoted by
\be
x_\mu \ =\ 2y_\mu + A_\mu ,
\ee
where $A_\mu=0,1$.
Therefore, it is convenient to divide the lattice into the hypercubes
identified by a
4-vector $y$, that is, $y$ is defined on a lattice of spacing $2a$.
The set of 16 vectors $\{A\}$ identify the points within the hypercube,
and they are called hypercube vectors.
The sum over lattice points can be written as
\be
\sum_x \ =\ {\sum_{y}}^{'} \sum_A  .
\ee

The new fields with definite taste are written as
$q_{y}^{\alpha a} $ and $\bar q_{y}^{\alpha a} $, that is,
\be
     q_y^{\alpha a} = {1 \over 4} \sum_A \Gamma_{A;\alpha a} \ \chi(2y+A)
, \ \ \ \ \ \
\bar q_y^{a \alpha} = {1 \over 4} \sum_A \bar\chi(2y+A) \
\Gamma_{A;a \alpha}^\dag ,
\ee
where $\alpha$ is the Dirac spin index, $a$ is the taste index, and
\be
\Gamma_A \ = \ \gamma_1^{A_1} \gamma_2^{A_2}
               \gamma_3^{A_3} \gamma_4^{A_4} .
\ee
The inverse expressions are
\be
\chi(2y+A)     = \Tr( \, \Gamma_A^\dag q_y \,  ) , \ \ \ \ \ \
\bar\chi(2y+A) = \Tr( \, \bar{q_y} \Gamma_A \, ) .
\ee
Then the free staggered fermion action can be written
in terms of the new fields as
\be
\CS_S = 4m_q {\sum_y} \bar q_y ({\bf 1} \otimes {\bf 1}) q_y  +
 4{\sum_{y,\mu}} \bar q_y  \bigg[
  \gamma_\mu \otimes {\bf 1} \Delta_\mu -
  \gamma_5 \otimes (\gamma_\mu \gamma_5)^T \sqrB_\mu
                             \bigg] q_y ,
\ee
where we define two different lattice derivatives on the block
lattice,
\bea
\Delta_\mu  & \equiv & {1 \over 4} ( f_{y+\hat{\mu}}-f_{y-\hat{\mu}} )
\ \ \ \ \ \ \ \ \longrightarrow  a \partial_\mu f_y \nonumber \\
\sqrB_\mu  & \equiv & {1 \over 4} ( f_{y+\hat{\mu}}+f_{y-\hat{\mu}}
-2f_y)  \longrightarrow  a^2 \partial_\mu^2 f_y  .
\eea
In the continuum limit,  the first term  tends to the usual kinetic term.
The second term is a lattice artifact of order ${\cal O}(a)$,
which disappears in continuum limit~\cite{Montvay:1994cy}.
At finite lattice spacing $a$, the 16 degrees of staggered fermions
in the hypercube are a mixture of spin and taste.

We can write the general bilinear operator as
\be
{\cal O}_{SF} = \bar{q_y} (\gamma_S \otimes \xi_T) q_y
              = \sum_{A,B} \bar \chi(y+A) \ {\rm Tr}
                \big(\Gamma_A^\dag \gamma_S \Gamma_B \xi_T \big) \
              \chi(y+B) ,
\label{eq:STAGop2}
\ee
where $ \xi_T \equiv \Gamma_T^\dagger $.
The matrices $\gamma_S$ determine
the angular momentum and parity of the operator~\cite{Gupta:91}
and the matrices $\xi_T$ label  the taste.
Each of these can be one of the standard 16 elements of the
Dirac Clifford algebra.

The notation $\gamma_S \otimes \xi_T $ stands for the
direct product of the matrix $\gamma_S$ and $\xi_T$, where
the former acts on the Dirac indices, and the later
acts on the taste indices.
The bilinear operators are identified
as follows~\cite{STAGkmnp}\cite{Gupta:91}
\bea
\label{eq:SFoperators}
V_\mu &=& \bar q (\gamma_\mu         \otimes \xi_T) q \nonumber \\
A_\mu &=& \bar q (\gamma_\mu\gamma_5 \otimes \xi_T) q \nonumber \\
P     &=& \bar q (          \gamma_5 \otimes \xi_T) q \nonumber \\
S     &=& \bar q (1                  \otimes \xi_T) q ,
\eea
where $V_\mu$ is a vector, $A_\mu$ is an axial vector,
$P $ is a pseudoscalar,
and $S$ is a scalar meson~\cite{Bowler:88}.

We use these operators as interpolating operators to
create and annihilate mesons.
In the continuum, the mesons are in the taste $SU(4)$ representations {\bf 1}
and {\bf 15}, while on the lattice the action
has a reduced symmetry and these representations break into
many smaller representations~\cite{Gupta:91}.
For example, the continuum {\bf 15}-plet of pion mesons
breaks into seven lattice representations
(four 3-dimensional and three 1-dimensional) on the lattice.
\bea
\label{eq:STAGpions}
{\bf 15} \ \to\ &{}& (\gamma_5\otimes\gamma_{i})                  \oplus
                     (\gamma_5\otimes\gamma_{i}\gamma_4)          \oplus
                     (\gamma_5\otimes\gamma_{i}\gamma_5)          \oplus
                     (\gamma_5\otimes\gamma_{i}\gamma_4\gamma_5)
                     \oplus
\nonumber \\
                &{}& (\gamma_5\otimes\gamma_4)                    \oplus
                     (\gamma_5\otimes\gamma_5)                    \oplus
                     (\gamma_5\otimes\gamma_4\gamma_5)   , \nonumber \\
{\bf 1}  \ \to\ &{}& (\gamma_5\otimes1)  .
\eea

In the first line of Eq.~(\ref{eq:STAGpions}), $i=1,2,3$,
the representations are three-dimensional.
The other representations are one-dimensional~\cite{Gupta:91}.
Hence, when we measure the mass of the pion meson on lattice,
there exists many different types of the mass for pion meson.
In Chapter 8, we can see this in detail.
%
\section{Simulation algorithm}
In this dissertation, we use the configurations
which were generated by the hybrid-molecular dynamics
``R algorithm''~\cite{MILCHMC:87}.
In order to explain this algorithm clearly,
we first introduce the process of
generating gauge configurations.
Then we discuss the hybrid-molecular dynamics ``R algorithm.''
%
\subsection{Numerical simulation and upadating process}
\label{NSU:ssec}
One of the main aims of numerical simulations in lattice QCD (LQCD)
is to calculate the expectation values of some observables
$F[\phi]$ which are the functions of the field variables
$[\phi] \equiv \{ \phi_{x\alpha} \}$,
where  $ { \phi_{x\alpha}}$ is a real field component with index
$\alpha$ at lattice site $x$, where the index $\alpha$
includes flavor, color, spin~\cite{Montvay:1994cy}.
For brevity  we often omit $\alpha $.
For simplicity we assume that the variables for the fermions
are integrated out analytically.
\be
\LL F[\phi] \RR  =
\frac{ \DPT \int[d\phi] \ e^{-S[\phi]} \ F[\phi] }
     { \DPT \int[d\phi] \ e^{-S[\phi]}           } ,
\label{eq:AVE}
\ee
where $e^{-S[\phi]}$ is an action suitable for simulation.
The above equation involves a multidimensional integral over
all the field variables $[\phi]$ on the lattice.
But the number of the integration variables in
$ \DPT  [d\phi] = \prod_{x,\alpha} d \phi_{x\alpha} $
is typically very large (in this dissertation, it is over $10^6$).
Hence, a Monte Carlo integration is used to
calculate the integral efficiently~\cite{Montvay:1994cy}.

For a given lattice with field variables $[\phi]$,
a configuration (i.e., $[\phi]$)
is just one point in the space (or domain) of the field variables.
Fortunately, due to the factor $e^{-S[\phi]}$,
only those paths for which the action $S[\phi]$
has the values close to minimum dominate the
path integrals~\cite{Montvay:1994cy}.
Hence, the efficient Monte Carlo simulation uses an importance
sampling~\cite{Montvay:1994cy}\cite{Rothe:1992nt}, which  generates a
series of configurations with the density $ W[\phi]$ proportional to
the Boltzmann factor $e^{-S[\phi]}$~\cite{Montvay:1994cy}. That is,
\be W[\phi] \propto e^{-S[\phi]}  .
\ee
We can call this series of configurations an ensemble of configurations.
Then we can use these  configurations to calculate  the integral like
Eq.~(\ref{eq:AVE})~\cite{Montvay:1994cy}.

The configurations are generated $\{ [\phi_n], \ 1 \le n \le N \} $
in a sequential manner
by creating from one configuration to the previous configuration,
$[\phi] \rightarrow [\phi']$, with a given transition probability.
The transition probability, $P([\phi]\rightarrow [\phi'])$,
specifies the probability that the configuration $\phi$
becomes to a configuration $\phi^\prime$.
The above procedure of creating configurations is
called ``updating''~\cite{Montvay:1994cy}.
For the numerical simulation, we have a variety of schemes
to get from one configuration to others.

As in statistical mechanics,
we want the density $W\left[\phi\right]$  of the configuration.
We also set the equilibrium density, $W_c\left[\phi\right]$,
to be proportional to the $e^{- S[\phi]}$,
where $S[\phi]$ is the action for a configuration.
To achieve this result, a single step causes the density $W$ to evolve
as follow~\cite{Montvay:1994cy}
\be
  W'\left[\phi'\right] = \sum_{\left[\phi\right]}
  P\left(\left[\phi'\right] \leftarrow \left[\phi\right]\right)
    W\left[\phi\right] .
\ee
The transition probability $P$ satisfies~\cite{Montvay:1994cy}
\bea
 \sum_{\left[\phi'\right]}  P\left(\left[\phi'\right] \leftarrow
  \left[\phi\right]\right)
  &=& 1  \\
  P\left(\left[\phi'\right] \leftarrow \left[\phi\right]\right)
  &>& 0  .
\eea
The normalization condition for the density $W$ is~\cite{Montvay:1994cy}
\be
  \sum_{\left[\phi'\right]}  W\left[\phi'\right] = 1  ,
\ee
and we require the detailed balance condition~\cite{Montvay:1994cy}
\be
  P\left(\left[\phi'\right] \leftarrow \left[\phi\right]\right)
W_c\left[\phi\right] = P\left(\left[\phi\right] \leftarrow
  \left[\phi'\right] \right) W_c\left[\phi'\right] .
\label{eq:detailed_balance}
\ee

Different choices of the transition probability
$P([\phi]\rightarrow [\phi'])$ generate
different ensembles.
Here we discuss one common updating algorithm:
the Metropolis method~\cite{Montvay:1994cy}\cite{Rothe:1992nt}.
Briefly, the Metropolis proceeds
as follows~\cite{Montvay:1994cy}\cite{Rothe:1992nt}:
we first randomly choose a possible new configuration
(i.e., $[\phi^\prime]$) from $N$ possible new configurations,
then calculate the new action ($S[\phi^\prime]$). If the new
action is less than the original action($S[\phi]$), we accept the
configuration $[\phi^\prime]$.
If the new action $S[\phi^\prime]$ is greater than
the original action $S[\phi]$,
we accept the configuration $[\phi^\prime]$
with probability $ \DPT \frac{ e^{-S[\phi']}}{ e^{-S[\phi]} }$.
We can simplify this algorithm for the transition probability
$P([\phi]\rightarrow [\phi'])$ as follows~\cite{Montvay:1994cy}\cite{Rothe:1992nt}
\be
  P\left(\left[\phi'\right] \leftarrow \left[\phi\right] \right)=
  N^{-1} \min\left\{1, \frac{W_c\left[\phi'\right]}
    {W_c\left[\phi\right]}\right\} .  \label{eq:metropolis}
\ee
This method results in
$\DPT W_c\left[ \phi\right] \propto e^{-S\left[ \phi \right] }$,
which means that the probability that any particular gauge configuration
is included in the sample (or an ensemble of gauge configuration)
is proportional to $\DPT e^{-S[\phi]}$.

With an ensemble of the sample gauge configurations,
the average value of any operator (${\cal O}$) is simply calculated
according to its average~\cite{Rothe:1992nt}
\be
\overline{\cal O} \ = \ {1 \over N} \sum_{n=1}^{N} {\cal O}[\phi_{n}] +
O\big( \frac{1}{\sqrt{N}} \big)  .
\label{eq:expvalue2}
\ee
If we have a large number of the sample configurations (i.e., $N$ large),
the average value that we measure is very close
to the operator expectation value $\vev{{\cal O}}$.
%
%
%
\subsection{Hybrid-molecular dynamics ``R algorithm''}
The Metropolis method is simple. However, it usually updates
the gauge link matrix locally
(i.e., each step updates one gauge link matrix).
Hence, we need a more efficient algorithm to
update the gauge link matrix globally (i.e.,
each step updates  the entire lattice).
The hybrid-molecular dynamics (HMD) ``R algorithm''~\cite{MILCHMC:87}
is a example. It was used to generate the
configurations for this dissertation.

The basic idea of the HMD ``R algorithm'' can be stated as follow:
by introducing a set of conjugate momenta $H$ for
the gauge link matrices $U$, the Euclidean path integral
associated with QCD theory can be rewritten in the form of
a partition function for a classical statistical mechanical system
in space-time, and the motion of  gauge link matrix
($U$) is governed by a classical Hamiltonian in the simulation time.
Therefore, the configurations are generated in a deterministic way.

In Sec.~\ref{NSU:ssec}, we assume that the variables for the fermions
are integrated out analytically. In some cases,
we cannot analytically integrate out the variables for fermions.
Hence, we consider the general case.
The starting point for this algorithm~\cite{MILCHMC:87} is
to consider the following path integral for the partition function
for $N_f$ degenerate staggered fermion flavors
in Eq.~(\ref{StartG:eq}).~\footnote{
In Sec.~\ref{Faol:ch3}, when we discuss Kogut-Susskind fermions,
we keep only a single fermion field component on each lattice site.
This procedure reduces the number of tastes from 16 to 4.
The factor $\DPT \frac{1}{4}$
(fourth root trick~\cite{Giedt:2005he})
is used to reduce the remaining four fermions tastes to
one taste in the determinant.
In this dissertation,  we use the configurations~\cite{Aubin:2004wf}
with three flavors: two light quarks ($u$, $d$),
and a strange ($s$) quark with $m_u = m_d \not= m_s$.
Then Eq.~(\ref{dsff:ch3}) becomes
\be
 Z = \int {\cal D}U \, e^{-S_g} \,
\left [ \det(M_u(U)) \right ] ^{ \frac{1}{2} }
\left [ \det(M_s(U)) \right ] ^{ \frac{1}{4} }  ,
\label{ndsff:ch3}
\ee
where $M_u[U]_{x,y}$ is the quark matrix for $u$ or $d$ quark, and
      $M_s[U]_{x,y}$ is the quark matrix for $s$ quark.
Hence, only two masses of KS quarks are included
(i.e., the square root of $u$ or $d$ determinant, and
       the fourth root of $s$ determinant).
}
\bea
  Z &=& \int {\cal D}U \, e^{-S_g} \,
  \left [ \det(M(U)) \right ] ^{ \frac{N_f}{4} }  \nonumber \\
    &=& \int {\cal D}U \, e^{-S_eff }  ,
\label{dsff:ch3}
\eea
where, as we explain below,
\be
S_{eff}  \ = \ S_g - \frac{N_f}{4}\ln \det[M(U)^{\dagger}M(U) ]  .
\label{EDNF2:eq}
\ee
Here $S_g(U)$ is the pure gauge action on lattice,
and the quark matrix $M[U]_{x,y}$ is given by Eq.~(\ref{eq:SFactionM}).
From Eq.~(\ref{eq:SFactionM}), we obtain
$\det[M(U)] =\det[M(U)^{\dagger}] $.
The factor of  $\DPT \frac{1}{4}$
comes from the fact that $\det[M(U)]$ stands for four tastes
per quark fermion.
Here, in Eq.~(\ref{EDNF2:eq}), we considered
the fact that the determinants of
the submatrices on the even sites of $M(U)^{\dagger}M(U)$
and the submatrices on the odd sites of $M(U)^{\dagger}M(U)$
are equal~\cite{Polonyi:1984zt}.

In order for the matrix  $U_{x, \mu}$ still to be an element of
$SU(3)$, the molecular dynamics
equation~\footnote{
The general Hamiltonian equations of motion are
\[
\dot{\phi}_{i} =
\frac{\partial H\lbrack \phi,\pi \rbrack }{\partial \pi_i }  ,
\ \ \ \ \dot{\pi}_{i} =
-\frac{\partial H\lbrack \phi,\pi \rbrack }{\partial \phi_i } ,
\]
where $\phi_i$ is field variable, $\pi_i$ is the momenta
cannonically conjugate to $\phi_i$, and $H\lbrack \phi,\pi \rbrack$
is the hamiltonian.
}
of the motion for $U_{x, \mu}$ at site $x$ should obey
\cite{MILCHMC:87}
\be
\dot{U}_{x, \mu}(t) = i {H}_{x,\mu}(t) {U}_{x,\mu}(t)  ,
\label{Motion:UM}
\ee
where $H_{x,\mu}(t)$ is a traceless Hermitian matrix, and
$\dot{U}_{x,\mu}(t)$ is the derivative of $U_{x,\mu}(t)$
with respect to the simulation time ($t$).
Hence, as we promised above,  we introduce the auxiliary field
$H$ conjugate to the gauge field $A$
into the partition function~\cite{MILCHMC:87}
\bea
  Z &=& \int {\cal D}U {\cal D}H \, e^{- \frac{1}{2} \Tr(H^2) - S_g(U) +
        \frac{N_f}{4} \Tr \ln [ M(U)^{\dagger}M(U) ]  }
\nonumber \\
    &=& \int {\cal D}U {\cal D}H \, e^{ -\mathcal{H} }  ,
\eea
leading to the effective Hamiltonian $\mathcal{H}$~\cite{MILCHMC:87}
\be
{\mathcal{H}} = \frac{1}{2} \Tr(H^2) + S_g(U) -
        \frac{N_f}{4} \Tr \ln [ M(U)^{\dagger}M(U) ] ,
\label{EffH:eq}
\ee
and $\DPT \Tr(H^2) = \sum_{x,\mu} \Tr(H_{x,\mu}^2)$.

Now we briefly illustrate the algorithm to generate
a set of gauge field configurations
with a probability distribution
proportional to $e^{ -\mathcal{H} }$, that is,
\be
P[U,H] \propto e^{ -\mathcal{H} } .
\ee
To this end, we adopt two types of updating steps.
First,  the $H$ field is initialized by the
heat bath method~\cite{Duane:1985hz}\cite{Duane:1986iw},
that is, we construct a random $H$ field~\cite{MILCHMC:87}
\be
H_{x,\mu} =  \lambda_{a} h_{x,\mu}^{a} ,
\ee
by setting each $h_{i,\mu}^{a}$ to
a complex Gaussian random number with standard deviation
$<|h_{x,\mu}^{a}|^2> = 1 $.
Thus, the $H$ field is first updated  globally.

Second,  the $U$ field is updated by the equation of the molecular
dynamics. From Eq.~(\ref{Motion:UM}), if we first update
the $H_{x,\mu}$, then we can easily obtain $U$.
Since the $H_{x,\mu}$ is updated globally, the $U$ field is updated
at the whole lattice.
Because it is very difficult to calculate the inverse of
the matrix $M(U)^{\dagger}M(U)$,
it is standard to use  a noisy estimator
method~\cite{Batrouni:1985jn}\cite{Duane:1986iw}
to compute it.
The integration of the equations of motion proceeds by the leapfrog
method.
It generates the sequence $U(t+i\Delta t)$  and  $H(t+i\Delta t)$,
where $i=1,2,\cdots,N$. we take about $\DPT N = \frac{1}{\Delta t}$
molecular-dynamics steps.
The result is called a molecular dynamics trajectory.
If we repeat two types of updating steps
(namely, the heat bath steps and molecular dynamics steps) many times,
we obtain  an ensemble of configurations.

In hybrid-molecular dynamics ``R algorithm,'' since
$ P[U,H] \propto e^{-\mathcal{H}} $, the expectation value
of an observable $F[U]$ can be written in the form
\be
\LL F[ U ] \RR  =
\frac{ \DPT \int[dU][dH] \ e^{-\mathcal{H}} \ F[U] }
     { \DPT \int[dU][dH] \ e^{-\mathcal{H}}        } .
\label{eq:AVEH}
\ee
Because $F[U]$ does not depend on the momentum $H$, and
$\DPT \int[dH] \ e^{-\mathcal{H}} \propto e^{-S_{eff}} $.
Eq.~(\ref{eq:AVEH}) can be rewritten in the form
\be
\LL F[ U ] \RR  =
\frac{ \DPT \int[dU] \ e^{-S_{eff}} \ F[U] }
     { \DPT \int[dU] \ e^{-S_{eff} } } ,
\label{eq:AVEH2}
\ee
which mean that the probability distribution
$P[U] \propto e^{ - S_{eff} }$.

This algorithm
is applicable to any kind of fermions~\cite{MILCHMC:87}.
For example, in this dissertation,
we use configurations~\cite{Aubin:2004wf}
with three flavors: two light quarks ($u$, $d$),
and a strange ($s$) quark.
In our simulation, the light quarks are degenerate in mass,
and we choose the different mass for strange quark.
Hence, we replace the effective Hamiltonian $\mathcal{H}$
described by Eq.~(\ref{EffH:eq}) by
\be
{\mathcal{H}} = \frac{1}{2} \Tr(H^2) + S_g(U) -
        \frac{1}{2} \Tr \ln [ M_l(U)^{\dagger}M_l(U) ] -
        \frac{1}{4} \Tr \ln [ M_s(U)^{\dagger}M_s(U) ] ,
\ee
where $M_l(U)$ is the quark matrix for the light   quark,
and   $M_s(U)$ is the quark matrix for the strange quark.
The calculation is similar to what we discussed before.
The only difference is that we generate two
gaussian random vectors, and use two conjugate-gradient~\footnote{
In mathematics, the conjugate gradient
is an algorithm for the numerical solution of
linear systems.
For example, we want to solve the following linear system
$$
Ax = b ,
$$
where $A$ is the $n \times n$ matrix, which is symmetric and positive definite,
$b$ is $1 \times n$ matrix, and $x$ is the solution.
The conjugate gradient method is an iterative method.
In this dissertation, we solve many systems like this.
}
calculations (i.e., one for the light quark, one for the strange quark)
for each molecular dynamics step.

From the above algorithm, we can see that there are
two sources of systematic errors.
The first error is due to the nonzero time step (i.e., $\Delta t \ne 0 $)
in the integration, which results in the quadratic dependence of
the error~\cite{MILCHMC:87}
(i.e., the error is of the order $(\Delta t)^2$).
This is because that we integrate the molecular dynamics equation
(i.e., Eq.~(\ref{Motion:UM}), Eq.\,(46) and Eq.\,(47)
in Ref.~\cite{MILCHMC:87})
with an error of order  $(\Delta t)^3$ at each time step.
The second error comes from the conjugate gradient residual~\cite{MILCHMC:87}
(i.e., the required accuracy of conjugate gradient calculation),
which we choose in the conjugate gradient calculation.
If we choose the conjugate gradient residual to be small,
we will take too much $CPU$ time.
Otherwise, if we prefer it to be the large,
the accuracy is not what we want.
So in the simulation, we choose a suitable value
for conjugate gradient residual.

Another error comes from our measurements.
Because the conjugate gradient calculation consumes
too much $CPU$ time,
we measure the potential (or the spectrum)
at the physical time separation $\delta t$.
In our simulation, we make a measurement
at intervals of six simulation time units
(namely, $\delta t = 6 \Delta t $).
So our measurements are not entirely statistically independent
(i.e., our measurement is autocorrelated).
We can reduce the effect of this autocorrelation of measurements
by blocking several consecutive
measurements before fitting~\cite{Bernard:2001av}.
In the latter chapter, when we fit the mass of the $f_0$ meson,
we will consider the effect of autocorrelation.

%% file: Chap4.tex
\newsavebox{\HydrogenFig}
\savebox{\HydrogenFig} {\begin{picture}(0,0)	
\setlength{\unitlength}{1pt}
\put(0,   50) { \line(1,0){160} }
\put(0,   50) { \line(3,2){110} }
\put(160, 50) { \line(-2,3){49} }
\put(160, 50) { \circle*{8} }
\put(0, 50)   { \circle*{8} }
\put(110, 123){ \circle*{4} }
\put(80,  42)  { \makebox(0,0){ $R$} }
\put(0,   42)  { \makebox(0,0){ $+e$} }
\put(160, 42)  { \makebox(0,0){ $+e$} }
\put(110, 130) { \makebox(0,0){ $-e$} }
\put(60,  100) { \makebox(0,0){ $R_1$} }
\put(133, 100) { \makebox(0,0){ $R_2$} }
\put(0,   32)  { \makebox(0,0){ Proton 1} }
\put(160, 32)  { \makebox(0,0){ Proton 2} }
\put(110, 140) { \makebox(0,0){ electron} }
\end{picture}}
%
\newsavebox{\GluonFig}
\savebox{\GluonFig} {\begin{picture}(0,0)	
\setlength{\unitlength}{1pt}
\put(0,   50)   { \line(1,0){160} }
\put(0,   50)   { \line(3,2){110} }
\put(160, 50) { \line(-2,3){49} }
\put(160, 50) { \circle*{8} }
\put(0, 50) { \circle*{8} }
\put(110, 123) { \circle*{4} }
\put(80,  42)  { \makebox(0,0){ $R$} }
\put(0,   38)  { \makebox(0,0){ $Q$} }
\put(160, 38)  { \makebox(0,0){ $\bar{Q}$} }
\put(110, 130) { \makebox(0,0){ $g$} }
\put(60,  100) { \makebox(0,0){ $R_1$} }
\put(133, 100) { \makebox(0,0){ $R_2$} }
\put(0,   28)  { \makebox(0,0){ Quark} }
\put(160, 28)  { \makebox(0,0){ Antiquark} }
\put(110, 142) { \makebox(0,0){ gluon} }
\end{picture}}
%
%
%
\newsavebox{\CorAA}
\savebox{\CorAA}{\begin{picture}(0,0)	
\setlength{\unitlength}{1pt}
\put(50,50) { \line(0,1){100} }
\put(50,50) { \line(1,0){100} }
\put(50,150){ \line(1,0){100} }
\put(150,50){ \line(0,1){100} }
\put(100, 40)  { \makebox(0,0){ $ \Sigma_{g}^+$  }}
\put(107, 158) { \makebox(0,0){ $ \Sigma_{g}^+$  }}
\put(180, 50) { \makebox(0,0){ $ t   $ }}
\put(180, 150){ \makebox(0,0){ $ t+T $ }}
\put(180, 40)    { \makebox(0,0){ $ Source $ } }
\put(180, 160)   { \makebox(0,0){ $ Sink $ } }
\put(50,  40)  { \makebox(0,0){ $ Q $  }}
\put(150, 40)  { \makebox(0,0){ $ \overline{Q} $ }}
\put(0,0)    { \vector(2,1){100} }
\put(0,0)    { \vector(2,3){100} }
\put(45, 14)   { \makebox(0,0){ $ ({\bf r},t)   $ } }
\put(90, 100)  { \makebox(0,0){ $ ({\bf r},t+T) $ } }
\put(-5, -5)   { \makebox(0,0){ $ Origin $ } }
\multiput(155, 50 )(2,2){8} {\circle*{0.15}}
\multiput(155, 150)(2,2){8}{\circle*{0.12}}
\multiput(55, 50 )(2,2){8} {\circle*{0.12}}
\put(120,60)   { \vector(1 ,0){40} }
\put(100,60)   { \vector(-1,0){40} }
\put(160,120)  { \vector(0, 1){40} }
\put(160,100)  { \vector(0,-1){40} }
\put(110, 60)   { \makebox(0,0){ $ {\bf R} $ } }
\put(160, 110)  { \makebox(0,0){ $ {\bf T} $ } }
\put(-20,50)  { \vector(1,1){15} }
\put(-20,50)  { \vector(1,0){30} }
\put(10, 44)   { \makebox(0,0){ $ z $ } }
\put(-5, 68)     { \makebox(0,0){ $x$ or $y$ } }
\end{picture}}
%
%
\newsavebox{\BENDOPER}
\savebox{\BENDOPER}{\begin{picture}(0,0)	
\setlength{\unitlength}{1pt}
\put(50,  50) { \line(1,1){25} }
\put(150, 50) { \line(1,1){25} }
\put(75,  75) { \line(1,0){100} }
\put(180, 50) { \line(-1,-1){25} }
\put(280, 50) { \line(-1,-1){25} }
\put(155, 25) { \line(1,0){100} }
\put(165, 50) { \makebox(0,0){ $ - $ }}
\put(50,  42)  { \makebox(0,0){ $ Q $  }}
\put(150, 42)  { \makebox(0,0){ $ \overline{Q} $ }}
\put(180, 58)  { \makebox(0,0){ $ Q $  }}
\put(280, 58)  { \makebox(0,0){ $ \overline{Q} $ }}
\put(-20,50)  { \vector(1,1){15} }
\put(-20,50)  { \vector(1,0){30} }
\put(10, 44)   { \makebox(0,0){ $ z $ } }
\put(-5, 68)     { \makebox(0,0){ $x \ {\rm or} \ y$ } }
\end{picture}}
%
%
\newsavebox{\Corhh}
\savebox{\Corhh}{\begin{picture}(0,0)	
\setlength{\unitlength}{1pt}
\put(50,50) { \line(0,1){100} }
\put(150,50){ \line(0,1){100} }
\put(0,0)    { \vector(2,1){100} }
\put(0,0)    { \vector(2,3){100} }

\qbezier(55,50)(100,80)(155,50)
\qbezier(55,50)(100,20)(155,50)
\qbezier(55,150)(100,180)(155,150)
\qbezier(55,150)(100,120)(155,150)

\put(100, 54)  { \makebox(0,0){ $ \Pi_{u}$ }}
\put(107, 144) { \makebox(0,0){ $ \Pi_{u}$ }}
\put(45, 14)   { \makebox(0,0){ $ ({\bf r},t)   $ } }
\put(90, 100)  { \makebox(0,0){ $ ({\bf r},t+T) $ } }
\put(180, 50)  { \makebox(0,0){ $ t   $ }}
\put(180, 150) { \makebox(0,0){ $ t+T $ }}
\put(180, 40)  { \makebox(0,0){ $ Source $ } }
\put(180, 160) { \makebox(0,0){ $ Sink $ } }
\put(-5, -5)     { \makebox(0,0){ $ Origin $ } }
\put(50,  40)  { \makebox(0,0){ $ Q $  }}
\put(150, 40)  { \makebox(0,0){ $ \overline{Q} $ }}
\put(-20,50)  { \vector(1,1){15} }
\put(-20,50)  { \vector(1,0){30} }
\put(10, 44)   { \makebox(0,0){ $ z $ } }
\put(-5, 68)     { \makebox(0,0){ $x$ or $y$ } }
\end{picture}}
\newsavebox{\BENDSIM}
\savebox{\BENDSIM}{\begin{picture}(0,0)	
\setlength{\unitlength}{0.6pt}
\qbezier(-80, 0)(-30, 30 )(20,0)
\qbezier(-80, 0)(-30,-30)(20,0)
\put(50,  0) { \line(1,1){25} }
\put(150, 0) { \line(1,1){25} }
\put(75,  25) { \line(1,0){100} }
\put(180, 0) { \line(-1,-1){25} }
\put(280, 0) { \line(-1,-1){25} }
\put(155, -25) { \line(1,0){100} }
\put(35,  0) { \makebox(0,0){ \large $\equiv$ }}
\put(165, 0) { \makebox(0,0){ $ - $ }}
\end{picture}}
%
%
%
%
\newsavebox{\BendPiu}
\savebox{\BendPiu}{\begin{picture}(0,0)	
\setlength{\unitlength}{0.8pt}
\put(0, 0)  { \line(0,1){50} }
\put(50, 0) { \line(0,1){50} }
\put(0,    0    ) { \line(-1,-1){12.5} }
\put(50,   0    ) { \line(-1,-1){12.5} }
\put(-12.5,-12.5) { \line(1,0){50} }
\put(0,  50)    { \line(-1,-1){12.5} }
\put(50, 50)    { \line(-1,-1){12.5} }
\put(-12.5,37.5) { \line(1,  0){50  } }
\put(0,  -7)  { \makebox(0,0){ \scriptsize    $ Q $  }}
\put(50, -7)  { \makebox(0,0){ \scriptsize    $ \overline{Q} $ }}
\put(70,  0)  { \line(0,1){50} }
\put(120, 0)  { \line(0,1){50} }
\put(70,  0)  { \line(1,1){12.5} }
\put(120, 0)  { \line(1,1){12.5} }
\put(82.5, 12.5)  { \line(1,0){50} }
\put(70,  50) { \line(1,1){12.5} }
\put(120, 50) { \line(1,1){12.5} }
\put(82.5, 62.5) { \line(1,0){50} }
\put(70,  -7)  { \makebox(0,0){ \scriptsize    $ Q $  }}
\put(120, -7)  { \makebox(0,0){ \scriptsize    $ \overline{Q} $ }}
\put(150, 0)  { \line(0,1){50} }
\put(200, 0) { \line(0,1){50} }
\put(150,    0    ) { \line(-1,-1){12.5} }
\put(200,   0    ) { \line(-1,-1){12.5} }
\put(137.5,-12.5) { \line(1,0){50} }
\put(150,  50) { \line(1,1){12.5} }
\put(200, 50) { \line(1,1){12.5} }
\put(162.5, 62.5) { \line(1,0){50} }%
\put(150,  -7)  { \makebox(0,0){ \scriptsize    $ Q $  }}
\put(200, -7)  { \makebox(0,0){ \scriptsize    $ \overline{Q} $ }}
\put(230,   0)  { \line(0,1){50} }
\put(280,   0)  { \line(0,1){50} }
\put(230,   0)  { \line(1,1){12.5} }
\put(280,   0)  { \line(1,1){12.5} }
\put(242.5, 12.5)  { \line(1,0){50} }
\put(230,   50)    { \line(-1,-1){12.5} }
\put(280,   50)    { \line(-1,-1){12.5} }
\put(217.5, 37.5) { \line(1,  0){50  } }
\put(230,  -7)  { \makebox(0,0){ \scriptsize    $ Q $  }}
\put(280, -7)  { \makebox(0,0){ \scriptsize    $ \overline{Q} $ }}
\put(60, 25)  { \makebox(0,0){ $ + $ }}
\put(135, 25) { \makebox(0,0){ $ - $ }}
\put(215,  25)   { \makebox(0,0){ $ - $ } }
\end{picture}}
%
%
%
%
%
\chapter{Leading Born-Oppenheimer (LBO) }
The heavy hybrid mesons can be investigated directly by
the numerical simulation, and  by  the Born-Oppenheimer (BO)
approximation, which was proposed for the study of the hybrids states
in Refs.~\cite{Hasenfratz:1980jv,Horn:1977rq}.
Here we focus on the leading order in the BO approximation,
and ignore the higher-order terms.
We use the Leading Born-Oppenheimer (LBO) approximation
to study the spectroscopy of the heavy hybrid mesons
($Q\overline{Q}g$), which consist of a heavy quark ($Q$)
and a heavy antiquark ($\overline{Q}$) plus a gluon.
We are particularly interested in the spin-exotic hybrid mesons with $J^{PC}$
quantum numbers, which is not permitted in the quark model.
The lightest one is believed
to be with $J^{PC}=1^{-+}$~\cite{Juge:1999ie},
which does not mingle with a nonhybrid $Q\bar{Q}$
bound state~\cite{McNeile:2002az}.

In this chapter, we use the LBO
approximation to study the spectroscopy of the hybrid mesons.
We first introduce some notations from molecular physics.
Then we study the LBO approximation.
Next we review the APE smearing algorithm, and introduce the static potentials.
Finally, we calculate the hybrid quarkonia through the LBO approximation.
%
\section{Notation from molecular physics}

We borrow the standard notation from diatomic molecular physics
to classify the static potential (i.e., gluonic energies).
Here we describe the gluon wave function  in analogy with
the electron wave function~\cite{McNeile:2002az,Juge:1999ie}.
The symmetries of the gluons include:
\vspace{-0.3cm}
\begin{itemize}
\item $J_{z}:$
The magnitude (denoted by $\Lambda$) of eigenvalues of the projection
of the projection ${\bf J_g \cdot n} $ of the total angular momentum
${\bf J_g}$ of the gluon field onto the separation axis ${\bf R}$
of the quark $Q$  and antiquark $\overline{Q}$ with
unit vector $ {\bf n} = {\bf R}/R$.
The states with $\Lambda=0,1,2,\dots$ are typically denoted
by the capital Greek letters $\Sigma, \Pi,
\Delta, \dots$, respectively.
\item $CP:$ The combined operations of the charge conjugation ($C$)
and spatial inversion ($P$) about the midpoint between
quark $Q$  and antiquark $\overline{Q}$ is also a symmetry.
Its eigenvalue is usually defined by $\eta$.
States which are even ($\eta = 1 $) or odd ($\eta = -1$)
under the above parity-charge-conjugation operation ($CP$)
are denoted by the subscripts $g$, and $u$, respectively.
\item $R:$ For the $\Sigma$ states ($\Lambda = 0$),
there is an extra symmetry.
The wave function of the $\Sigma$ states which are even
(odd) under reflection in any plane passing through the separation
axis (i.e., the plane contains the separation axis) are
specified by a superscript $+(-)$.
\end{itemize}
For the states with $\Lambda \not= 0$, the energy of the gluons
is not changed by operation $R$.
Hence, those states are degenerate (namely, $\Lambda$ doubling).
We can list the low-lying levels of the gluon field as
$\Sigma_g^+$, $\Sigma_g^-$, $\Sigma_u^+$, $\Sigma_u^-$,
$\Pi_g$, $\Pi_u$, $\Delta_g$, $\Delta_u$, etc~\cite{Juge:1999ie}.
In this dissertation, we only consider the states $\Sigma_g^+$ and $\Pi_u $.

\section{Born-Oppenheimer approximation}
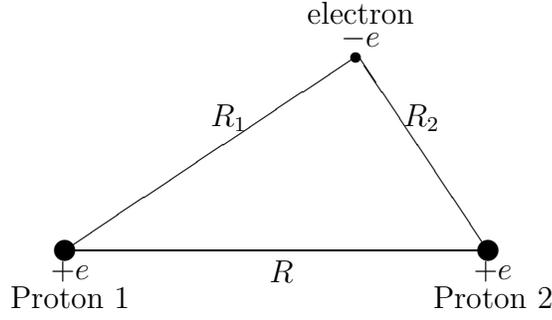
\begin{figure}[t]
\setlength{\unitlength}{1pt}
\begin{center}\begin{picture}(180, 125)
\setlength{\unitlength}{1pt}
\put(0.0,0){\usebox{\HydrogenFig} }
\end{picture} \end{center}
\vspace{-1.5cm}
\caption{ \label{hydrogen:FIG}
    The hydrogen molecule ion, $H_2^{+}$
}
\end{figure}

\vspace{0.2cm}
The hydrogen molecule ion, $H_2^{+}$, consists of a single electron
in the Coulombic field of two protons (see Fig.~\ref{hydrogen:FIG}).
Here we make the approximation that the nuclei (i.e., two protons) are at rest.
Since the nuclear motion is much slower than the motion of the electron,
the electronic wave function depends on the nuclear positions.
The Hamiltonian is
\be
H = -\frac{1}{2m}\bigtriangledown^2
- \ e^2 \LP \frac{1}{R_1}+\frac{1}{R_2} \RP ,
\ee
where $R_1$ and $R_2$ are the distances to the electron from
the respective protons, and $e$ is the charge of
the electron~\cite{DavidJG:94}.

For each fixed $R$, we can use some methods
(e.g., the trial wave function in the variational principle)~\cite{DavidJG:94}
to solve for the motion of the electron,
obtaining the ground state energy $V(R)$ of the system of $H_2^{+}$
as a function of $R$.
We solve the Schr\"odinger equation for the motion of the nuclei
using $V(R)$ as the interacting potential. Then we can recover information
about the positions and motions (e.g., rotation, vibration) of the nulcei~\cite{DavidJG:94}.
In molecular physics this method is called the BO approximation.

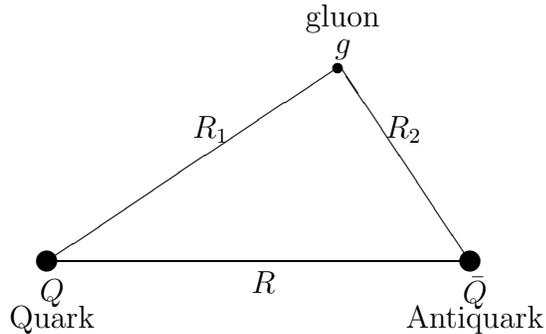
\begin{figure}[b]
\setlength{\unitlength}{1pt}
\begin{center}\begin{picture}(180, 130)
\setlength{\unitlength}{1pt}
\put(0,0){\usebox{\GluonFig} }
\end{picture} \end{center}
\vspace{-1.2cm}
\caption{   \label{ghydrogen:FIG}
The hybrid meson,  $Qg\bar{Q}$
}
\end{figure}

The heavy hybrid mesons 
can be studied similarly with the BO approximation (see Fig.~\ref{ghydrogen:FIG}).
The slow heavy quarks $Q\overline{Q}$ correspond to
the nuclei in the diatomic molecule
while the fast gluons correspond to
the electrons~\cite{Hasenfratz:1980jv}.

To use the BO method, we first consider the quark $Q$ and
antiquark $\overline{Q}$ as the spatially fixed color field sources
and use lattice methods to determine the energy levels of
the gluons as a function of the $Q\overline{Q}$ separation $R$.
Each of these energy levels defines a static potential $V_{Q\overline{Q}}(R)$~\cite{Juge:1999ie}.
At the second step in BO method, the heavy quark motion is then
reproduced by solving the Schr\"odinger equation in each of these static potentials~\cite{Juge:1999ie}.
Conventional quarkonia are based on the static potential ($\Sigma^+_g$),
and the hybrid quarkonium states should come  from the
excited potentials (i.e., $\Pi_u$, etc).

In this chapter, we present the results for the spectrum~\cite{Juge:1999ie}
of the gluonic excitations in the presence of a static quark-antiquark pair.
Using our potentials, we also determine the hybrid quarkonium spectrum.
For comparison with experiments, we also present the conventional
quarkonium spectrum, which comes from $\Sigma_g^+$ potential.
Before we evaluate the static potentials, we fattened~\cite{MILC:ASQ}
the spatial gauge links by
``APE smearing''~\cite{Falcioni:APE}\cite{Parisi:APE}.
Hence, we first discuss the APE smearing.

\section{APE smearing}
To reduce the mixings of our operators with its excited states
the gluonic lattice operators are constructed from
the iteratively-smeared spatial links (i.e., the fat link).
We used the APE smearing algorithm described in
Refs.~\cite{Falcioni:APE,Parisi:APE}
to fatten the spatial gauge link matrix $U_{\mu}(x), \ \mu = 1,2,3$.
The APE smearing averages a link matrix $U_{\mu}(x)$ on the lattice
with its nearest neighbors.
This means that the smearing procedure is to replace the ordinary link matrix
$U_{\mu}(x)$ with the weighted sum of the link and $\alpha$
times its four staples (i.e., three link paths), that is,
\bea
U_{\mu}(x)\ \rightarrow\ U_{\mu}'(x) & = & (1-\alpha) U_{\mu}(x)
+ \frac{\alpha}{4}  \sum_{\nu=1 \atop \nu\neq\mu}^{3}
  \Bigg\{ U_{\nu}(x) U_{\mu}(x+\hat\nu a) U_{\nu}^{\dag}(x+ \hat\mu a)
          \nonumber \\
& &   + U_{\nu}^{\dag}(x-\hat\nu a)	
        U_{\mu}(x-\hat\nu a)
        U_{\nu}^{\dag}(x +\hat\mu a)
  \Bigg \} ,
\label{SMEARING:APE}
\eea
where $\alpha$ is the smearing fraction. In our dissertation, we
choose  smearing fraction $\DPT \alpha = \frac{8}{13}$
in order to get a satisfactory ground state
enhancement~\cite{Legeland:APE,Legeland:1997rs,Engels:1996dz,Albanese:1987ds}.

\begin{figure}[t]
\setlength{\unitlength}{1pt}
\begin{center}\begin{picture}(380,65)
\setlength{\unitlength}{1pt}
\thicklines
\put(00,20){\vector(0,1){40}}
\put(26,40){\makebox(0,0){{\huge $\rightarrow $}}}
\put(100,20){\vector(0,1){40}}
\put(70,40){\makebox(0,0){{\large $(1-\alpha)$}}}
\put(116,40){\makebox(0,0){{\Large $+$}}}
\put(135,40){\makebox(0,0){{\large $\DPT \frac{\alpha}{4}$}}}
\put(150,40){\oval(10,80)[l]}
\put(200,20){\vector(-1,0){40}}
\put(160,20){\vector(0,1){40}}
\put(160,60){\vector(1,0){40}}
\put(210,40){\makebox(0,0){{\Large $+$}}}
\put(220,20){\vector(1,0){40}}
\put(260,20){\vector(0,1){40}}
\put(260,60){\vector(-1,0){40}}
\put(275,40){\makebox(0,0){{\Large $+$}}}
\put(310,20){\vector(-1,-1){20}}
\put(290,0){\vector(0,1){40}}
\put(290,40){\vector(1,1){20}}
\put(320,40){\makebox(0,0){{\Large $+$}}}
\put(330,20){\vector(1,1){20}}
\put(350,40){\vector(0,1){40}}
\put(350,80){\vector(-1,-1){20}}
\put(362,40){\oval(10,80)[r]}
\end{picture}\end{center}
\vspace{-0.4cm}
\caption{Visualization of a smearing iteration.}
\label{FigI}
\end{figure}
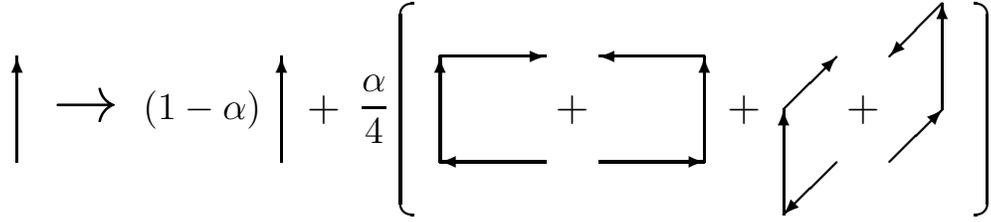

We construct the new  gauge link $U_{\mu}'(x)$ as follows:
Step 1: Calculate the new  gauge link $U_{\mu}'(x)$ by
Eq.~(\ref{SMEARING:APE}) and  project back to $SU(3)$.
Step 2: After we do the step 1 for every link, we replace
the old gauge link $U_{\mu}(x)$ by the new  gauge link $U_{\mu}'(x)$.
Step 1 and Step 2 define a single smearing sweep.
One such a smearing iteration step is shown in Fig.~\ref{FigI}.
By convention, the new  gauge link $U_{\mu}'(x)$ is
called a ``fat link''~\cite{Blum:1996uf}.
In this dissertation, before we evaluate the correlators,
we take 20 smearing sweeps.
%
\section{Computation of the static potentials}
The first step in the LBO approximation
is to determine the energy levels of the gluons as a function of $R$, 
where heavy quark ($Q$) and antiquark ($\overline{Q}$) are
fixed at lattice sites with separation distance $R$ apart.
At this point in the LBO approximation, the quark $Q$ and
antiquark $\overline{Q}$ which provide the color field
are considered to be approximately static~\cite{Juge:1999ie}.

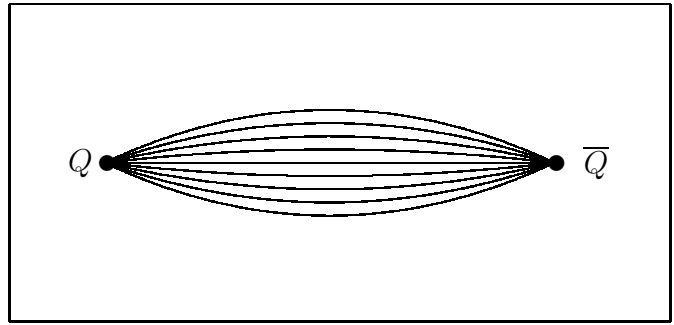
\begin{figure}[b]
\setlength{\unitlength}{1pt}
\begin{center} \begin{picture}(250,108)
\setlength{\unitlength}{1pt}
\put(0,0)   { \line(1,0){250} }
\put(0,0)   { \line(0,1){120} }
\put(0,120) { \line(1,0){250} }
\put(250,0) { \line(0,1){120} }
\put(37,  60) { \circle*{6} }
\put(207, 60) { \circle*{6} }
\put(25,  60) { \makebox(0,0){ $Q$} }
\put(220, 60) { \makebox(0,0){ $\overline{Q}$} }
\qbezier(40,60)(125,100)(210,60)
\qbezier(40,60)(125,20)(210,60)
\qbezier(40,60)(125,90)(210,60)
\qbezier(40,60)(125,30)(210,60)
\qbezier(40,60)(125,80)(210,60)
\qbezier(40,60)(125,40)(210,60)
\qbezier(40,60)(125,70)(210,60)
\qbezier(40,60)(125,50)(210,60)
\qbezier(40,60)(125,60)(210,60)
\end{picture} \end{center}
\vspace{-0.3cm}
\caption{   \label{fluxtube:FIG}
The distribution of color electric field between static color
charges.
}
\end{figure}

It is possible to describe the spatial distribution of the color field
between the static quarks $Q\overline{Q}$.
There are many papers~\cite{Kuti:1998rh} that study the physical properties
of the color flux between static quarks.
We are very interested in the transverse extent of
the color flux tube (see Fig.~\ref{fluxtube:FIG})
and the physical properties of the color fields
(i.e., electric field or magnetic field).
The ground state energy of the color electric field between static quark sources
can be interpreted as the origin of the confining interquark potential
of the heavy $Q\overline{Q}$ pairs.
The hybrid $Q\overline{Q}$ potentials which bind
quark-antiquark ($Q\overline{Q}$) pairs into the
heavy hybrid $Q\overline{Q}g$ molecules are defined by
the excitation content of the confining color flux tube~\cite{Juge:1999ie}.

In this section, we first explain  Monte Carlo estimates of
the correlators, which are used to calculate the static potential of
the $\Sigma^+_g$ and $\Pi_u$ respectively.
Then we discuss the procedure to fit these correlators.
The fitting method we use is the standard
Levenberg-Marquardt~\cite{nr:2000} fitting procedure.
%
\subsection{$C_{A A}$ Correlator}
\begin{figure}[b]
\setlength{\unitlength}{1pt}
\begin{center}\begin{picture}(200,150)
\setlength{\unitlength}{1pt}
\put(0.0,0){\usebox{\CorAA} } 
\end{picture} \end{center}
\vspace{-0.2cm}
\caption{   \label{CAA:FIG}
An illustration of the correlator from $A$ to $A$.
}
\end{figure}
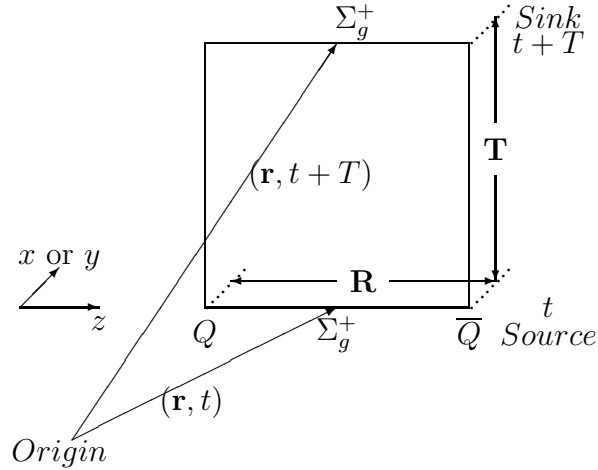

To determine the potential energy of the $Q\overline{Q}$ system we first need
the correlator from the $\Sigma_{g}^+$ (A) state to
the $\Sigma_{g}^+$ (A) state.
On a starting time slice $t$,
the quark $Q$ and antiquark $\overline{Q}$ are fixed
at lattice sites with separation distance of $R$
whose midpoint is at space point {\bf r} (see Fig.~\ref{CAA:FIG}).
The gluonic operator at time slice $t$ is the ground state,
which is the parallel transporter
for a straight link path along the quark and
antiquark ($\Sigma_{g}^{+}$).
Before we evaluate the $C_{A A}$ correlator,
we use the APE smearing method to ``fatten'' spatial links,
which means, in effect,
that the parallel transporter here is not the conventional
parallel transporter using the link matrix, and it is replaced by an
weighted average over the link paths connecting the points~\cite{MILC:ASQ}
(see details in Ref.~\cite{Blum:1996uf}).
The gluonic operator at the source is
\be
{\cal O}_{Source}({\bf r},{\bf R},t)
\equiv \Sigma_{g}^{+}({\bf r}, {\bf R}, t)
\equiv  U({\bf r} \MINUSONE \frac{\bf R}{2} \PLUSONE t{\bf \hat{4} } ,
  {\bf r} \PLUSONE \frac{\bf R}{2} \PLUSONE t{ \bf \hat{4}} ) ,
\ee
where {$\bf \hat{4}$ } is the unit vector in the time direction.

At the final time slice $t_f$ (i.e.,  $t_f \equiv t+T$),
the gluonic operator at time slice $t_f$ is the parallel transporter,
which is a straight link path along the quark $Q$ and antiquark $\overline{Q}$.
The gluonic lattice operator at the sink can be written similarly as
\be
{\cal O}_{Sink}({\bf r},{\bf R}, \Tt)  \equiv
\Sigma_{g}^{+}({\bf r},{\bf R}, \Tt)   \equiv
U({\bf r} \MINUSONE \frac{\bf R}{2} \PLUSONE t_f{\bf \hat{4}} ,
  {\bf r} \PLUSONE  \frac{\bf R}{2} \PLUSONE t_f{\bf \hat{4}} ) .
\ee
Therefore, the timeslice correlator $C_{A A}(T,{\bf R})$
from the $\Sigma_{g}^{+}$ state to the
$\Sigma_{g}^{+}$ state can be written as
\bea
C_{A A} (T, {\bf R} )
\hspace{-0.3cm} &=& \hspace{-0.3cm} \frac{1}{\Omega} \sum_{{\bf r},t}
\LL   \Tr \left [
         {\cal O}_{Sink}^{\dag}({\bf r}, {\bf R}, \Tt) \
         {\cal O}_{Source}({\bf r},{\bf R},t)
\right ] \RR  \nonumber \\
\EQATHR
\frac{1}{\Omega} \sum_{{\bf r},t}
\LL   \Tr \left [
U( {\bf r} \MINUSONE \frac{\bf R}{2} \PLUSONE t_f{\bf \hat{4}}, \
   {\bf r} \PLUSONE  \frac{\bf R}{2} \PLUSONE t_f{\bf \hat{4}} )^{\dag}
U( {\bf r} \MINUSONE \frac{\bf R}{2} \PLUSONE t{\bf \hat{4}}, \
   {\bf r} \PLUSONE \frac{\bf R}{2}  \PLUSONE t{\bf \hat{4}} )
\right ] \RR  \nonumber \\
\EQATHR
\frac{1}{\Omega} \sum_{{\bf r},t}
\LL    \Tr \left [
       \Sigma_{g}^{+}({\bf r}, {\bf R}, \Tt)^\dag
        \ \Sigma_{g}^{+}({\bf r}, {\bf R}, t  )
    \right ] \RR ,
\label{EQ.CAA}
\eea
where $\Omega = N_xN_yN_zN_t$ is the total lattice volume,
$ N_x,N_y,N_z$ are space dimensions, and
$ N_t$ is time dimension.

A few algebraic steps relates the correlator to
the energy $V_{\Sigma_g^+}(R)$ of the lowest state excited by the
operator.
\bea
C_{AA}(T, {\bf R}) &=& \frac{1}{\Omega} \sum_{{\bf r},t}
\Tr \LL 0 \left| {\Sigma_g^+}( {\bf r}, t+T )
                 {\Sigma_g^+}^\dag( {\bf r}, t )
 \right| 0 \RR
\nonumber \\ &=&
\frac{1}{\Omega} \sum_{{\bf r},t}
\LL 0 \left| {\Sigma_g^+}( {\bf r,R}, t+T ) \right| A \RR
\LL A \left| {\Sigma_g^+}^\dag( {\bf r,R}, t ) \right| 0 \RR
\nonumber \\ &=&
\frac{1}{\Omega} \sum_{{\bf r},t}   e^{ -V_{\Sigma_g^+}(R) T }
\LL 0 \left| {\Sigma_g^+}( {\bf r,R}, 0 ) \right| A \RR
\LL A \left| {\Sigma_g^+}^\dag( {\bf r,R}, 0 ) \right| 0 \RR
\nonumber \\ &=&
Z_a^2 \ e^{ -V_{\Sigma_g^+}(R) T } ,
\label{ch4:a}
\eea
where  $ \left| A \RR $ is the eigenstate of the operator
$\Sigma_g^+$,
and
\be
Z_a \equiv  \LL 0 \left| {\Sigma_g^+}(\Rr, 0 ) \right| A \RR .
\ee
We have assumed that the contributions of higher excited states are
negligible. This approximation is valid for large time $T$.

Since the gauge configurations are gauge fixed to the temporal
gauge ($A_4(x) = 0 $) before we estimate
the correlator $C_{AA}$,
the time-link link variables are unit matrices.
Therefore, the product of parallel transporters  in Eq.~(\ref{EQ.CAA})
is equal to the parallel transporter $U({\cal C}_{R,T})$
along the closed rectangular loop
$ {\cal C}_{R,T} $ of the side-lengths $R$ and $T$
(see Fig.~\ref{CAA:FIG}), that is,
$C_{A A} (T, R) = W( {\cal C}_{R,T} )$, the standard  Wilson loop.
%
\subsection{$C_{HH}$ Correlator}
\label{SEC.chh}
\begin{figure}[b]
\setlength{\unitlength}{1pt}
\begin{center}\begin{picture}(200,150)
\put(0.0,0){\usebox{\Corhh} }
\end{picture} \end{center}
\caption{ \label{CHH:FIG}
   An illustration of the correlator from $H$ to $H$.
}
\vspace{-0.5cm}
\end{figure}
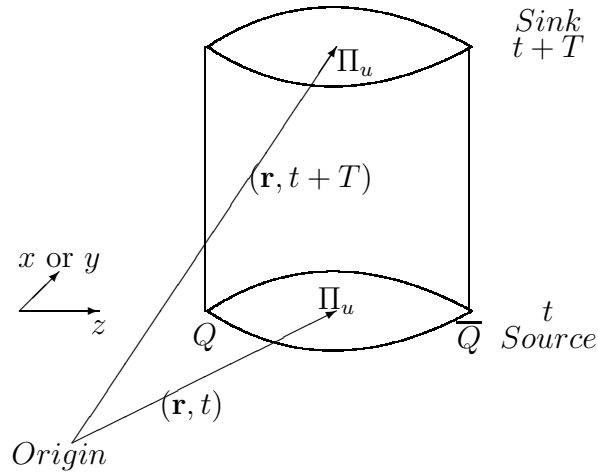

In this section, we calculate the correlator
from the $\Pi_{u}$ state to the $\Pi_{u}$ states
(i.e., $H$ state to $H$ state).
As before, on a starting time slice $t$,
the quark $Q$ and antiquark $\overline{Q}$ are fixed
at lattice sites with separation distance of $R$
whose midpoint is at space point {\bf r} (see Fig.~\ref{CHH:FIG}).
The gluonic operator at time slice $t$ is the
$\Pi_{u}$ lattice operator, which traditionally is
the parallel transporter along a U-shaped path  $\sqcap$ minus
the parallel transporter along the same U-shaped path  $\sqcup$
(i.e., $\, U( {\cal C}_\sqcap) - U( {\cal C}_\sqcup$))
in the opposite transverse direction (see Fig.~\ref{FIG.CreationP}).
The symbol $\, \sqcap \,$ denotes the U-shaped path
from the quark $Q$ to the antiquark $\overline{Q}$ whose transverse extent
can be in the positive $x$ or $y$ direction,
and $\, \sqcup\,$ denotes the U-shaped path
from the quark $Q$ to the antiquark $\overline{Q}$ whose transverse extent
can be in the negative $x$ or $y$ direction.
The gluonic lattice operator at the source can be written as
\be
{\cal O}_{Source}({\bf r},{\bf R},t)  \equiv \Pi_{u}({\bf r},{\bf R},t)
\equiv \setlength{\unitlength}{1pt}
\begin{picture}(120,20)	
\setlength{\unitlength}{0.7pt}
\qbezier(0, 3)(50, 33)(100,3)
\qbezier(0, 3)(50,-27)(100,3)
\end{picture}
\hspace{-1.7cm}.
\ee
For notational simplicity, we define a symbol
\be
\begin{picture}(100,10)
\setlength{\unitlength}{0.6pt}
\put(80, 0){\usebox{\BENDSIM}}
\end{picture}
\hspace{-2.1cm}.
\label{FIG.SIMP}
\ee
\vspace{0.05cm}

At the final time slice $t_f$ (i.e.,  $t_f \equiv t+T$),
the gluonic operator at time slice $t_f$  can be written as
\be
{\cal O}_{Sink}({\bf r},{\bf R}, \Tt)  \equiv \Pi_{u}({\bf r},{\bf
R},t+T) .
\ee

\begin{figure}[t]
\setlength{\unitlength}{1pt}
\begin{center}\begin{picture}(320,60)
\put(0.0,0){ \usebox{\BENDOPER} }
\end{picture} \end{center}
\vspace{-1.2cm}
\caption{ \label{FIG.CreationP}
   The creation operator for the $\Pi_u$ state.
}
\end{figure}
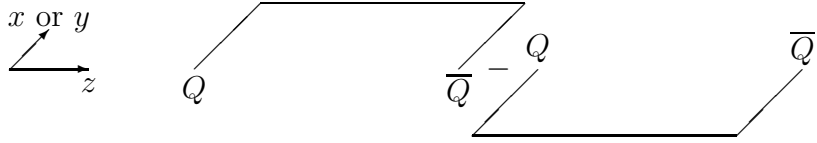
Hence, the timeslice correlator can be written as
\bea
C_{HH}(T, {\bf R} )
&=& \frac{1}{\Omega} \sum_{{\bf r},t}
    \LL \Tr \left [
         {\cal O}_{Sink  }^{\dag}({\bf r}, {\bf R}, \Tt) \
         {\cal O}_{Source}({\bf r}, {\bf R}, t)
                     \right ] \RR  \nonumber\\
&=& \frac{1}{\Omega} \sum_{{\bf r},t}
    \LL \Tr \left [  \Pi_{u}({\bf r}, {\bf R}, \Tt) \
                 \Pi_{u}({\bf r}, {\bf R},t) \RR
\right ]
\eea
%
Following steps similar to Eq.~(\ref{ch4:a}), we calculate the energy
of the lowest $\Pi_u$ state $V_{\Pi_u}(R)$ through
\bea
C_{HH}(T, {\bf R}) &=& \frac{1}{\Omega} \sum_{{\bf r},t}
\Tr \LL 0 \left| {\Pi_u}( {\bf r}, t+T )
                 {\Pi_u}^\dag( {\bf r}, t )
 \right| 0 \RR
\nonumber \\ &=&
\frac{1}{\Omega} \sum_{{\bf r},t}
\LL 0 \left| {\Pi_u}( {\bf r,R}, t+T ) \right| H \RR
\LL H \left| {\Pi_u}^\dag( {\bf r,R}, t ) \right| 0 \RR
\nonumber \\ &=&
\frac{1}{\Omega} \sum_{{\bf r},t}   e^{ -V_{\Pi_u}(R) T }
\LL 0 \left| {\Pi_u}( {\bf r,R}, 0 ) \right| H \RR
\LL H \left| {\Pi_u}^\dag( {\bf r,R}, 0 ) \right| 0 \RR
\nonumber \\ &=&
Z_h^2 \ e^{ -V_{\Pi_u}(R) T } ,
\label{ch4:h}
\eea
where  $ \left| H \RR $ is the eigenstate of the operator $\Pi_u$, and
\be
Z_h \equiv  \LL 0 \left| {\Pi_u}(\Rr, 0 ) \right| H \RR .
\ee
We have assumed that the contributions of higher excited states are
negligible. This approximation is valid for large time $T$.

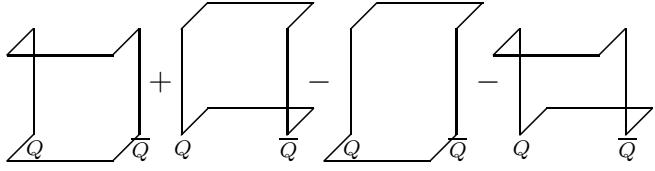
\begin{figure}[t]
\setlength{\unitlength}{0.8pt}
\begin{center}\begin{picture}(300,50)
\put(0.0,0){\usebox{\BendPiu} }
\end{picture} \end{center}
\caption{ \label{BENDLOOP:FIG}
   The bent loop for  $\Pi_u$ states.
}
\end{figure}

Since the gauge configurations are gauge fixed to the temporal
gauge ($A_4(x) = 0 $) before we estimate the correlator $C_{HH}$,
the time-link link variables are unit matrices.
Hence, in our simulation, we actually measure
four bent Wilson loops (see Fig.~\ref{BENDLOOP:FIG}).
%
\subsection{The results for the static potentials}
In this dissertation, we want to reduce the mixings of our
operators with their higher lying excited states.
According we take 20 APE smearing sweeps.
The correlators of $C_{AA}(t,R)$ and $C_{HH}(t,R)$
are then fitted by using a single exponential term.
Our best fit values are obtained by using the correlated $\chi^2$ method.
For the correlator of $C_{AA}(t,R)$, our fitting range of
minimum distance $t_{\rm min}$ to $t_{\rm max}$ of the source-sink
separation is listed in Table~\ref{sigma_TABLE:ch4},
where the $a$ is the lattice spacing.
And the average
least $\chi^2$ per number of degrees of freedom for different $R$ is
approximately close to one. We list some of the fitting
information in Table~\ref{sigma_TABLE:ch4}.

\begin{table}[b]
\vspace{-0.3cm}
\caption{ \label{sigma_TABLE:ch4}
Fitting information for static potential $V_{\Sigma_g^+}$.
The first column is the separation distance,
the second column, static potential,
the third column, the statistical error of the
corresponding static potential, and the remaining columns are $\chi^2$
per number of degrees of freedom, and the time range
for the chosen fit.
}
\begin{center}\begin{tabular}{|c|c|c|c|c|}
\hline
$ R/a   $  & $aV_{\Sigma_g^+}$ & Error for $a V_{\Sigma_g^+}$ &
$\chi^2/D $  & Fitting range    \\
\hline
1 &       0.5396 &  7.999040e-04  &   0.2/3  & 8-12 \\
\hline
2  &      0.8340 &  5.111520e-03  &   5.3/3  & 7-11 \\
\hline
3   &     0.9927 &  7.192000e-03  &   1.0/3  & 6-10\\
\hline
4   &     1.0996 &  1.826420e-02  &   0.7/3  & 6-10\\
\hline
5  &      1.2234 &  1.291270e-02  &   0.9/3  & 5-9 \\
\hline
6  &      1.3007 &  2.307760e-02  &   2.2/3  & 5-9 \\
\hline
7  &      1.3912 &  4.568450e-02  &   0.2/3  & 5-9 \\
\hline
8  &      1.5070 &  7.961440e-02  &   3.8/3  & 5-9 \\
\hline
9  &      1.6254 &  3.601540e-02  &   3.5/3  & 4-8 \\
\hline
10 &      1.7338 &  5.708190e-02  &   6.6/3  & 4-8 \\
\hline
11 &      1.8307 &  1.003580e-01  &   2.3/3  & 4-8 \\
\hline
12 &      1.8838 &  2.552650e-02  &   2.5/3  & 3-7 \\
\hline
\end{tabular}
\vspace{-0.5cm}
\end{center}\end{table}

\begin{table}[ptbh!]
\vspace{-0.3cm}
\caption{ \label{pi_TABLE:ch4}
Fitting information for static potential $V_{\Pi_u}$.
The first column is the separation distance,
the second column, static potential,
the third column, the statistical error of the corresponding static potential,
and the remaining columns are
$\chi^2$ per number of degrees of freedom,
and the time range for the chosen fit.
}
\begin{center}\begin{tabular}{|c|c|c|c|c|}
\hline
$ R/a $ & $aV_{\Pi_u}$ & Error for $aV_{\Pi_u}$ &
$\chi^2 /D $  & Fitting range    \\
\hline
1  &	1.6602 &  6.8816e-03& 	1.0/3& 3-7 \\
\hline
2 &	1.6430&   6.1839e-03& 	0.7/3& 3-7 \\
\hline
3&	1.6442&   6.0712e-03& 	3.3/3& 3-7 \\
\hline
4&	1.6586&   6.5216e-03& 	4.3/3& 3-7 \\
\hline
5&	1.6996&   7.6446e-03& 	2.0/3& 3-7 \\
\hline
6&	1.7279&   9.1389e-03& 	5.0/3& 3-7 \\
\hline
7&	1.7976&   1.1532e-02& 	4.9/3& 3-7 \\
\hline
8&	1.8317&   1.5096e-02& 	0.9/3& 3-7 \\
\hline
9&	1.9373&   2.1053e-02& 	1.2/3& 3-7 \\
\hline
10&	1.9626&   3.0135e-02& 	2.1/3& 3-7 \\
\hline
11&	2.0732&   4.1658e-02& 	3.3/3& 3-7 \\
\hline
12&	2.1580&   6.1574e-02& 	3.4/3& 3-7 \\
\hline
\end{tabular}
\vspace{-0.5cm}
\end{center}\end{table}

For the correlator of $C_{HH}(t,R)$, our fitting range of
minimum distance $t_{\rm min}$ to $t_{\rm max}$ of the source-sink
separation is listed in Table~\ref{pi_TABLE:ch4}.
And the average $\chi^2$ per number of degrees of freedom
for different $R$ is approximately close to one.
We list some of the fitting
information in Table~\ref{pi_TABLE:ch4}.

\epsfverbosetrue
\begin{figure}[b]
\begin{center}
\leavevmode
\epsfxsize=4.0in\epsfbox{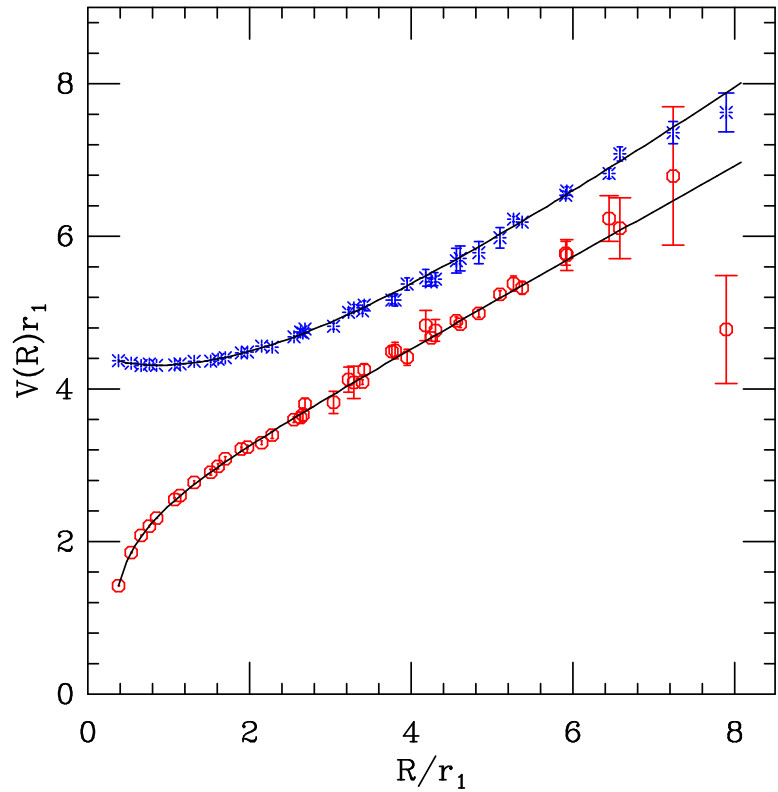}
\end{center}
\vspace{-0.5cm}
\caption{ \label{fig:sigma}
The static ground potential $V_{\Sigma_g^+}(R)$
and first gluonic excitations potential $V_{\Pi_u}(R)$ in terms of
the scale  parameter $r_1$ with respect to
the quark-antiquark separation $R$.
}
\end{figure}
%
%
\subsection{Setting the scale}
This scale parameter $r_1$ depends only weakly on the
valence quark mass and can be calculated
accurately~\cite{Bernard:2000gd}\cite{Bernard:2001av}.
Hence, we prefer to use it to rescale our data.
The definition of this parameter and computation of this
quantity are shown in Ref.~\cite{Bernard:2000gd}\cite{Bernard:2001av}.
The scale parameter $r_1$ is determined through
\be
\left. R^2 \ \frac{d V_{\Sigma_g^+}(R)} {d R} \right|_{R=r_1} = 1.0 .
\ee
From Ref.~\cite{Aubin:2004wf}, we obtain the following value for $r_1$:
\bea
r_1      &=& 0.317(7) \ {\rm fm} \ \ \ \ {\rm or} \\
r_1^{-1} &=& 621(14)  \ {\rm MeV} .
\label{r1:eq1}
\eea
From Table II in Ref.~\cite{Aubin:2004wf}, we get
\be
\frac{r_1}{a} = 2.632(13) .
\ee
Therefore, we obtain~\cite{Aubin:2004wf}
\be
a^{-1} = 1.634(45) {\rm GeV} .
\ee
Our fitting results for the static ground potential
$V_{\Sigma_g^+}(R)$ and first gluonic excitations
potential $V_{\Pi_u}(R)$
are shown in Fig.~\ref{fig:sigma}.
The results are expressed in terms of the scale parameter $r_1$.

\subsection{Fitting procedure}
The static ground potential $V_{\Sigma_g^+}(R)$
is shown as the low part (the octagon one) in Fig.~\ref{fig:sigma},
the solid curve for $V_{\Sigma_g^+}(R)$ potential
in Fig.~\ref{fig:sigma} is a fit to the data
by using a Coulombic plus linear form and lattice Coulombic
potential~\cite{Booth:1992bm}, that is,
\be
V_0 + {e_c}  \left[ \frac{1}{R} \right]   + \sigma R ,
\label{sigf:eq}
\ee
where $V_0$, $e_c$, and $\sigma$ are
three fitting parameters,
and $\left[ \frac{1}{R} \right]$ is the lattice Coulombic potential,
which is calculated in free field theory for the Symanzik improved gauge
action~\cite{Weisz:1982zw}\cite{Bonnet:2001uh}\cite{Booth:1992bm} as
\be
\left[ \frac{1}{R} \right] =
\int_{-\pi}^{\pi} \frac{d^3k}{8\pi^2}
\frac{ e^{i {\bf k} \cdot {\bf R} } }
     { \DPT          \sum_{i=1}^{3} \sin^2( \frac{k_i}{2} )
       + \frac{1}{3} \sum_{i=1}^{3} \sin^4( \frac{k_i}{2} ) } .
\label{colcorr:ch4}
\ee
This lattice correction term is used to correct the
rotational symmetry of the lattice~\cite{Booth:1992bm}.
In practice we employ it at distances less than $3a$.
The $V_0$ can be interpreted as a self-energy.
Since only the tree-level coefficients are needed,
we use Heller's data for the lattice correction term
$ \DPT \left[ \frac{1}{R} \right]$.
In Table~\ref{SIr_TABLE:ch4}, we list some of them.

\begin{table}[b]
\vspace{-0.4cm}
\caption{  \label{SIr_TABLE:ch4}
$\left[\frac{1}{R}\right]$ for separate distance $R$.
The first column is the
separation distance, the second column is $\left[\frac{1}{R}\right]$,
and the remaining column is $\frac{1}{R}$.
}
\begin{center}\begin{tabular}{|c|c|c|}
\hline
$R$     & $\left[\frac{1}{R}\right]$ & $\frac{1}{R}$   \\
\hline
1.000000000  &   1.038940053  &   1.000000000 \\
\hline
2.000000000  &   0.512582839  &   0.500000000 \\
\hline
3.000000000  &   0.332863472  &   0.333333333 \\
\hline
4.000000000  &   0.247627405  &   0.250000000 \\
\hline
5.000000000  &   0.197306796  &   0.200000000 \\
\hline
6.000000000  &   0.163888673  &   0.166666667 \\
\hline
7.000000000  &   0.140048006  &   0.142857143 \\
\hline
8.000000000  &   0.122177437  &   0.125000000 \\
\hline
\end{tabular}
\end{center}\end{table}

Our fitting gives an acceptable result
with a $\chi^2$ of 46.9 for  39 degrees of freedom, with
\vspace{-0.3cm}
\bea
  r_1 V_{0}    &=&   2.2731(80) \nonumber \\
  e_{c}        &=&  -0.3927(27) \nonumber \\
  \sigma r_1^2 &=&   0.5870(39) \nonumber ,
\eea
where we choose all the data within the fitting range of r
from $ \DPT \frac{ R_{min} }{r_1} = 0.0     $
to   $ \DPT \frac{ R_{max} }{r_1} = 7.89688 $.
Since $r_1 V_{0} = 2.2731(80)$,
we obtain the self energy $V_0$,
that is,
\be
V_0 = 1.4116(45) \ {\rm GeV} .
\ee

The first gluonic excitations potential $V_{\Pi_u}(R)$
is illustrated as the upper part in Fig.~\ref{fig:sigma}.
From Fig.~\ref{fig:sigma} we note that at the short distance $R$
the force between $Q\overline{Q}$ is repulsive
for the $\Pi_u$ potential,
but the force between $Q\overline{Q}$
is attractive for the $V_{\Sigma_g^+}$ potential.
This can be explained by the color factor~\cite{Griffiths:1987tj}.
The $\Sigma_g^+$ potential ($V_{\Sigma_g^+}$) is
the singlet static potential in which the quarks $Q\overline{Q}$
state is in the color singlet representation.
The $\Sigma_g^+$ potential behaves like Eq.~(\ref{sigf:eq}).
On the other hand the $\Pi_u$  potential is a hybrid static potential
in which the $Q\overline{Q}$ state is in the color octet
representation.
We know that the color factor
$\DPT f_c = -\frac{1}{6}$~\cite{Griffiths:1987tj} for
color octet configuration of $Q\overline{Q}$,
and $\DPT f_c = \frac{4}{3}$~\cite{Griffiths:1987tj}
for color singlet configuration of $Q\overline{Q}$.
At short distance $R$,
the $\Pi_u$  potential $V_{\Pi_u}$
should behave like $\DPT
\frac{\alpha_s}{6R}$~\cite{Brambilla:1999xf}.
That is,
\vspace{-0.3cm}
\bea
 V_{\Pi_u} &=& \frac{\alpha_s}{6R},  \ \ \ \  R \to 0 ,
\label{Rep:eq2}
\eea
where $\alpha_s$ is strong coupling constant.
Eq.~(\ref{sigf:eq}) and Eq.~(\ref{Rep:eq2}) can
help explain why at short distance $R$,
the force between $Q\overline{Q}$ is weakly
repulsive for the $V_{\Pi_u}$  potential,
but the force between $Q\overline{Q}$
is attractive for the $V_{\Sigma_g^+}$ potential.

The solid curve for the $V_{\Pi_u}(R)$ potential
in Fig.~\ref{fig:sigma} is a fit to the data
by using the equation
\be
e_{\pi} \left[\frac{1}{R}\right] + c_0 + \sqrt{b_0 + b_1 R + b_2 R^2} ,
\ee
where $e_{\pi}$, $c_0$, $b_0$, $b_1$, and $b_2$ are
five fitting parameters. It gives an acceptable fit
with a $\chi^2$ of 44 with 37 degrees of freedom, with
\bea
  c_{0} r_1     &=&   2.018  \pm 0.211  \nonumber \\
  b_{0} r_1^2   &=&   5.313  \pm 0.866  \nonumber \\
  b_{1} r_1^3   &=&  -0.761  \pm 0.101  \nonumber \\
  b_{2} r_1^4   &=&   0.561  \pm 0.028  \nonumber \\
  e_{\pi}       &=&   0.036  \pm 0.017 .
\eea

For all $R$,
in Ref.~\cite{Juge:1999ie}, the authors confirmed that
the first-excited potential of the color flux tube is the $\Pi_u$
potential. Hence, we expect that the lowest lying hybrid mesons should
come from the $V_{\Pi_u}(R)$  potential.

%
\section{Hybrid quarkonium}
The second step in the LBO approximation is to reproduce
the heavy quark motion by solving the Schr\"odinger equation
with the potentials of  $V_{\Sigma_g^+}(R)$ or $V_{\Pi_u}(R)$.
For notational simplicity, we adopt the spherical coordinates.
The reduced mass $\mu$ of the two-body system is
\be
\label{REDUCE:eq}
\mu  = \frac{M_Q M_{\overline{Q}}} {M_Q + M_{\overline{Q}}} = \frac{M_Q}{2} ,
\ee
where $M_Q$ is the mass of the heavy quark $Q$, 
      $M_{\overline{Q}}$ is the mass of the heavy antiquark $\overline{Q}$,
and we consider $M_Q = M_{\overline{Q}}$
The reduced radial Schr\"odinger equation of this system is given
by~\cite{Juge:1999ie,Landau:Lif}
\be
\frac{d^2u(R)}{dR^2}+
2\mu \left [ E-V_{\rm eff}(R) \right ] \, u(R)=0 ,
\label{eqn:schrodinger}
\ee
where
$\DPT
V_{\rm eff}(R) = V_{Q\overline{Q}}(R) +
\frac{ \left\langle {\bf L}_{Q\overline{Q}}^2 \right\rangle }
{\DPT 2\mu R^2}$ is the effective potential~\cite{Juge:1999ie},
${\bf L}_{Q\overline{Q}}$ is the orbital angular momentum
of the quark $Q$ and antiquark $\overline{Q}$
(discussed below in detail)~\cite{Juge:1999ie}, and
\be
u(R) = \rho(R) \, R ,
\ee
where  $\rho(R)$ is the radial wave function,
and $V_{Q\overline{Q}}(R)$  is either $V_{\Sigma_g^+}(R)$ or $V_{\Pi_u}(R)$~\cite{Juge:1999ie}.

The total angular momentum of the hybrid meson
($Q\overline{Q}g$) is denoted by~\cite{Juge:1999ie}
\be
{\bf J}={\bf L}+{\bf S} ,
\ee
where ${\bf S}$ is the sum of the spins
of the quark $Q$ and antiquark $\overline{Q}$,
and ${\bf L}$ is the total orbital angular momentum of the hybrid meson.
The orbital factor ${\bf L}$  is given by~\cite{Juge:1999ie}
\be
{\bf L}={\bf L}_{Q\overline{Q}} + {\bf J}_g ,
\ee
where ${\bf J}_g$ is the total angular momentum of the gluons.
We choose the vector ${\bf n}$ to be unit vector along the axis
of the quark $Q$ and antiquark $\overline{Q}$ (or $z$ axis).
For the gluon state with $\Lambda$, we write the mean value of
${\bf J}_g$ as~\cite{Landau:Lif}
\be
 \left\langle {\bf J}_g \right\rangle  = {\bf n} \Lambda .
\ee
Since classically,
${\bf L}_{Q\overline{Q}}= {\bf R} \times {\bf P}_{Q\overline{Q}}$, it
should be perpendicular to the ${\bf n}={\bf R}/R$ on average
(i.e., $ \left\langle
               {\bf L}_{Q\overline{Q}} \cdot {\bf n}
         \right\rangle  = 0 $ ).
Hence, we deduce that~\cite{Landau:Lif}
$\left\langle ( {\bf L} - {\bf J}_g ) \cdot {\bf n} \right\rangle =0$,
or,
\be
\left\langle {\bf L}   \cdot {\bf n} \right\rangle =
 \left\langle {\bf J}_g \cdot {\bf n} \right\rangle = \Lambda .
\label{eq:Lambdag}
\ee
In the LBO approximation, the eigenvalue $L(L+1)$
of ${\bf L}^2$ and the eigenvalue $S(S+1)$ of ${\bf S}^2$ are also good quantum numbers.
Hence, the total orbital angular momentum of
the quark $Q$ and antiquark $\overline{Q}$
is written as~\cite{Juge:1999ie}
\bea
\left\langle {\bf L}_{Q\overline{Q}}^2 \right\rangle
&=&
\left\langle { \bf L    }^2 \right\rangle +
\left\langle {{\bf J}_g }^2 \right\rangle - 2
\Big\langle  { \bf L    } \cdot {\bf J}_g  \Big\rangle  \\
&=&
L(L+1) + \left\langle {\bf J}_g^2 \right\rangle - 2\Lambda^2 .
\label{Jgn:ch4}
\eea
For the potential of $V_{\Sigma_g^+}$, there are no gluon excitations,
so $\left\langle {\bf J}_g^2 \right\rangle = 0$.
For the $V_{\Pi_u}$ potential, we have $ \Lambda = 1 $, so
$\left\langle {\bf J}_g \cdot {\bf n} \right\rangle = 1$~\cite{Juge:1999ie}.
Hence, $J_g$ can choose any integer number which is larger than 1
(namely, $J_g=1,2,3, \cdots $).  From Eq.~(\ref{Jgn:ch4}),
we note that if the system of the hybrid meson have more than one gluon,
the energy level will be higher than that of the system with just one gluon~\cite{Juge:1999ie}.
However, here we focus on the decay of the lightest exotic hybrid state.
Therefore, we consider the system with just
one gluon state~\cite{Juge:1999ie}
(namely, $\left\langle {\bf J}_g^2 \right\rangle = 2 $).

We know that $\Lambda$ is the $z$-component of the angular momentum of the gluon. 
By convention if $\Lambda$ is positive, we call it
a left-handed gluon state ($\vert {\rm left}\rangle$)
which is not an eigenstate of a $PC$.
If $\Lambda$ is negative, we call it
a right-handed gluon state ($\vert {\rm right}\rangle$)
which is also not a $PC$ eigenstate.
However, if as we discuss in more detail in Appendix B,
the linear combinations of left-handed and
right-handed gluons states are
$\vert {\rm left}\rangle + \epsilon\ \vert {\rm right}\rangle$~\cite{Juge:1999ie},
where $\epsilon = \pm 1$. are $PC$ eigenstates.
For $\Sigma_g^+$, since $\Lambda=0$,
we have $\epsilon = 1$.
Let $\eta = \pm 1$  be the $PC$ quantum number of the gluons,
which correspond to the subscripts $g(u)$.
In Appendix A, we prove that the parity and charge conjugation
of the hybrid meson are given in the form~\cite{Juge:1999ie}
\vspace{-0.3cm}
\bea
\label{appb:eq:1}
P &=& \epsilon \       (-1)^{L+1} \\
C &=& \epsilon \ \eta\ (-1)^{L+S} .
\label{appb:eq:2}
\eea
%

\epsfverbosetrue
\begin{figure}[b]
\begin{center}
\leavevmode
\epsfxsize=4.0in\epsfbox{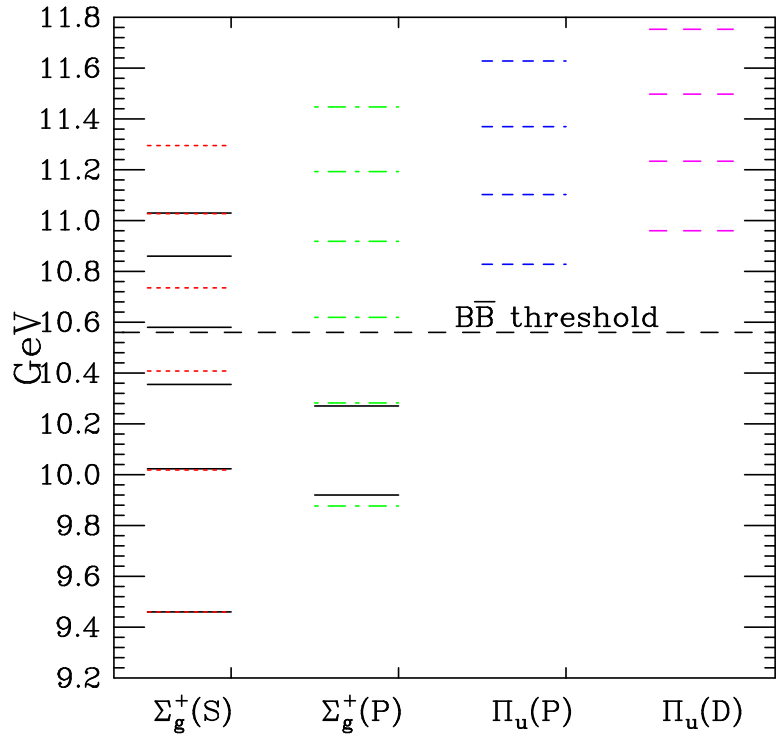}
\end{center}
\vspace{-0.3cm}
\caption{  \label{fig:spectrum}
Spin-averaged $b\bar b$ spectrum in the $LBO$ approximations.
Solid lines specify the experimental measurements.
Short dotted, short dashed lines indicate the $S$ and $P$ state masses
obtained by solving the Schr\"odinger equation with $V_{\Sigma_g^+}$
by using $r_1^{-1}= 0.621$ GeV  and $M_{b1}=3.88$ GeV for the $b$ quark mass,
respectively. Long dashed, dot-dashed lines indicate the hybrid quarkonium
states obtained from the $V_{\Pi_u}$ $(L=1,2)$ potentials respectively.
}
\end{figure}

In general the potentials evaluated from Wilson loops in lattice simulations
are built up by the same diagrams as the self-energy
for the Green's function~\cite{Morningstar:1993de}.
The self energy $V_0$ renormalizes the quark mass so we absorb it
into the definition of $M_Q$.
The difference between $V_0$ and $c_0$ is physically significant,
however we should remove the unwanted self-energy contributions.
In practice we fit our data for the $\Sigma_g^+$ potential
to the form in Eq.~(\ref{sigf:eq}).
The constant $V_0$ is our approximate estimate of the self-energy contributions.

For the $\Pi_u$ state, we have $\Lambda=1$, from
Eq.~(\ref{eq:Lambdag}), we obtain that $L \geq 1$.
The spectra of the heavy hybrid meson ($b\overline{b}g$) built on
the potentials of the $V_{\Sigma_g^+}$ and $V_{\Pi_u}$ are shown
in Fig.~\ref{fig:spectrum}. The energy levels
for each state are listed in Table~\ref{SPECT_TABLE:ch4}.

\begin{table}[t]
\vspace{-0.3cm}
\caption{  \label{SPECT_TABLE:ch4}
Quarkonium and hybrid quarkonium as computed.
The first  column gives the first six  energy levels for $\Sigma^+_g(S)$,
the second column, the first six  energy levels for $\Sigma^+_g(P)$,
the third  column, the first four energy levels for $\Pi_u(P)$,
the fourth column, the first four energy levels for $\Pi_u(D)$.
}
\begin{center}\begin{tabular}{|c|c|c|c|}
\hline
$\Sigma^+_g(S) \ {\rm GeV}$ & $\Sigma^+_g(P) \ {\rm GeV}$ &
$\Pi_u(P) \ {\rm GeV}$  &  $\Pi_u(D) \ {\rm GeV}$   \\
\hline
9.4603   &  9.87656  &  10.8283 &  10.96    \\
\hline
10.0183  & 10.2821   &  11.1025 &  11.2332  \\
\hline
10.4076  & 10.6194   &  11.3694 &  11.4972  \\
\hline
10.7349  & 10.9186   &  11.628  &  11.7525  \\
\hline
11.0269  & 11.1924   &          &           \\
\hline
11.2951  & 11.4475   &          &           \\
\hline
\end{tabular}
\vspace{-0.2cm}
\end{center}\end{table}

From Ref.~\cite{El-Khadra:1997hq}, we know that in the heavy quark limit,
the Hamiltonian operators for $\Sigma_g^+$, $\Pi_u$ in our LBO approximation is
\bea
H_{\Sigma_g^+} &=& 2M_{b0} + \frac{p^2}{2M_{b1}} +
V_{\Sigma_g^+}(R)  \\
H_{\Pi_u}      &=& 2M_{b0} + \frac{p^2}{2M_{b1}} + V_{\Pi_u}(R) ,
\eea
where $M_{b0}$ is called the rest mass,
  and $M_{b1}$ is referred as
the kinetic mass~\cite{El-Khadra:1997hq}.
The heavy quark rest mass $M_{b0}$ was determined
by matching the known mass of the $\Upsilon(1S)$.
\be
M_{\Upsilon(1S)} = 2M_{b0} + E_0 ,
\ee
where the $E_0$ is the energy of the ground state in
the potential of $\Sigma_g^+$.
From our LBO approximation, we obtain $E_0 = 0.1711 \, {\rm GeV}$.
We know $M_{\Upsilon(1S)} = 9.4603 \, {\rm GeV}$. Hence, we estimate
\be
M_{b0} = 4.64 \, {\rm GeV} .
\ee

The heavy quark mass $M_b$ is tuned by using a directly measurable physical quantity. 
A conventional choice is to use the experimentally known $\Upsilon$ spectrum, namely,
the splitting of 2S-1S ($m_E$(2S-1S)) and the splitting of 1P-1S ($m_E$(1P-1S)).
From the LBO approximation, we obtain the calculated
splitting of 2S-1S ($m_C$(2S-1S)) and
the splitting of 1P-1S ($m_C$(1P-1S)).
By adjusting the kinetic mass of the $b$ quark ($M_{b}$),
we make both the difference of splitting of 2S-1S
(namely, $m_C$(2S-1S) - $m_E$(2S-1S))
and the difference of splitting of 1P-1S
(namely, $m_C$(1P-1S) - $m_E$(1P-1S))
as small as possible. Then we obtain the kinetic mass $M_{b}$.
In practice we minimize this formula
\be
f(M_b) =
\sqrt{\big[ m_C({\rm 2S \MINUSONE 1S}) - m_E({\rm 2S \MINUSONE 1S}) \big]^2
     +\big[ m_C({\rm 1P \MINUSONE 1S}) - m_E({\rm 1P \MINUSONE 1S}) \big]^2 }
\ee
to obtain the mass of the $b$ quark ($M_{b}$). In our simulation we estimate
\be
M_{b1} = 3.88 \, {\rm GeV} .
\ee

From Fig.~\ref{fig:spectrum}, we note that below the
$B\overline{B}$ threshold, our LBO results are consistent with
the spin-averaged experimental measurements,
and above the $B\overline{B}$ threshold,
our LBO results are not in good agreement with
the spin-averaged experimental measurements.
This disagreement probably comes from the fact that
above the $B\overline{B}$ threshold, the $\Upsilon(4S)$ and
$\Upsilon(5S)$, etc, can decay to $B\overline{B}$~\cite{PDBook}
and our model does not take those decays into accounts.
However due to their exotic properties and heavy quark kinematics, our
hybrid states decay only weakly to $B\overline{B}$ (discussed in Chapter 5).
Therefore, our hybrid energy levels are presumably insensitive to
this effect.
This disagreement also comes from the fact that
in our LBO approximation we ignore the high order terms.

From Fig.~\ref{fig:spectrum}, we note that the mass of
the lowest-lying hybrid from the $V_{\Pi_u}$ is approximate 10.83 GeV.
In fact we can obtain the precise value of lowest-lying hybrid $m_H$
from the LBO approximation based on the potential of $V_{\Pi_u}$, namely,
\be
 m_H = 10.8283 \, {\rm GeV} .
\label{mH:eq1}
\ee
Above $11$~GeV, the LBO approximation based on $V_{\Pi_u}$
predicts several hybrid states with approximately
uniform energy level separations about ${\rm 200 - 300~MeV}$.
This is a distinct feature of the hybrid quarkonium spectrum.

\begin{figure}[t]
\begin{center}
\leavevmode
\epsfxsize=4in\epsfbox{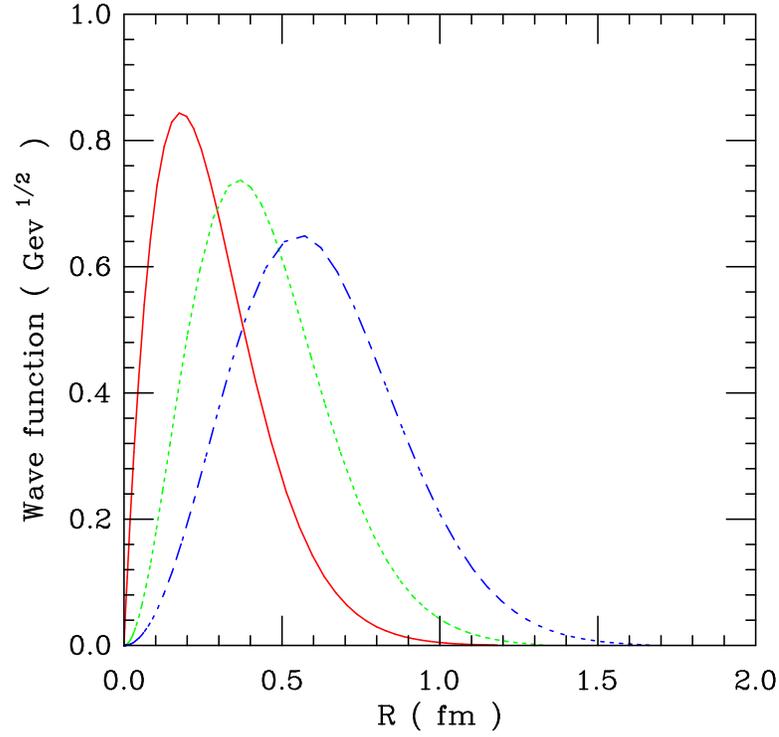}
\end{center}
\vspace{-0.3cm}
\caption{ \label{fig:wf}
Reduced radial wave functions versus quark-antiquark separation $R$.
The solid and dotted curves specify the reduced radial wave functions
for the $1S$ and $1P$ states of $\Sigma_g^+$, respectively,
and the short dashed curve specifies the reduced radial
wave function of $\Pi_u$ hybrid state,
where all the wave function are normalized.
}
\end{figure}

The reduced radial wave functions for the $1S$ and $1P$
conventional quarkonium and that of the lowest-lying $\Pi_u$
hybrid quarkonium state are shown in Fig.~\ref{fig:wf}.
\epsfverbosetrue
For notational convenience,
we call the reduced radial wave function for the $1S$
conventional quarkonium state
$u_{\eta_b}(R)$, the reduced radial wave function for the $1P$
conventional quarkoniun state $u_{\chi_b}(R)$,
and the reduced radial wave function for the lowest-lying $\Pi_u$
hybrid quarkonium state $u_{H}(R)$.

From Fig.~\ref{fig:wf}
we note another distinct feature of the heavy hybrid states.
It is the shape of the reduced radial wave function.
The hybrid reduced radial wave functions all vanish
at the origin ($R=0$). This can be explained
by the repulsive Coulombic term in the potential
$V_{Q\overline{Q}}(R)$ and the special centrifugal
term $V_{\rm eff}(R)$ in the
LBO Schr\"odinger equation~\cite{Hasenfratz:1980jv}.
As depicted in Fig.~\ref{fig:wf},
we also note that the hybrid Schr\"odinger state have
a broader reduced radial wave function than
those of the conventional quark-antiquark states.
In Chapter 5, we use these  LBO results
to develop the decay model for the hybrid state.
In Chapter 8, we use these results to calculate the decay rate.

%% file: Chap5.tex
\chapter{Hybrid meson decay channel}
Most hadronic states are unstable under strong interactions.
A comprehensive understanding of the hadronic decay from QCD involves
developing the physical models to deal with the unstable hadrons and
to extract their hadronic transition matrix elements.
The physical models for studying the strong decays of the hadrons have been
studied for a long time~\cite{Geiger:1994kr}\cite{Page:1998gz}.
Here we concentrate on the hadronic decay in the static limit,
and employ lattice methods to calculate its transition matrix elements.

For the decays of the hybrid mesons,
we study the creation of a light quark-antiquark
state from the gluonic field of the hybrid meson,
then in the heavy quark limit we discuss its open decay channels with the emphasis
on the $\chi_b S$ channel, where $S$ is a scalar meson.
This scalar channel ($1^{-+}$) leads to $\pi^+\pi^-$, $\pi^0\pi^0$,
or $K^+K^-$, $K^+K^-$.

%
%
\section{Hybrid states on the lattice}
In Chapter 4, we studied the properties of the physical hybrid states
by solving the Schr\"odinger equation with these static potentials
in the LBO approximation.
From Fig.~\ref{fig:spectrum}, we can see that the ground
state of the color flux is in the state of $\Sigma^+_g$.
The first excited state is confirmed to be in the state of $\Pi_u$
in Ref.~\cite{Juge:1999ie}, and we show this result in Fig.~\ref{fig:spectrum}.
We found that the mass of a lightest hybrid spin exotic with
$J^{PC}=1^{-+}$ is at $m_H=10.83$~GeV,
about $1.37$~GeV  heavier than the mass of the $\Upsilon(1S)$.
Above this energy level there are many hybrid states,
including the spin exotic states.

\section{Hybrid meson decay model}
In this section, we give a brief review of the possible decay channels of
a hybrid meson through the creation of a light quark-antiquark pair
($q\overline{q}$) in the heavy quark limit.
The jointed symmetry of the $q\overline{q}$
and the final gluonic field is identical with the symmetry of
the initial representation of the gluonic field~\cite{McNeile:2002az}.

For a spin-exotic hybrid meson which contains an excited
gluonic field $\Pi_u$ with $\Lambda=1$ about the separation axis
between $Q\overline{Q}$,
we obtain, from Eq.~(\ref{eq:Lambdag}),  that the orbital angular
momentum of the $Q\overline{Q}$  pair should be
greater than or equal to one ($L \geq 1 $)~\cite{McNeile:2002az}.
To make the problem simple, we assume that the orbital angular
momentum of the $Q\overline{Q}$ is in the $P$-wave (namely, $L=1$).

In practice, the $\Pi_u$ state corresponds
to an excited gluonic field with $L^{PC}=1^{+-}$ or $L^{PC}=1^{+-}$~\cite{Michael:1999ge}.
With the consideration of the spin of the heavy quark $Q\overline{Q}$,
we obtain a set of eight degenerate states~\cite{McNeile:2002az}.
From the discussion in Chapter 2,
we noted that, $J^{PC}=  1^{-+},\ 0^{+-}$ and $2^{+-}$
are the spin-exotic hybrid state~\cite{Michael:1999ge}.
Therefore, the spin-exotic hybrid meson has the heavy quark-antiquark
pair in a spin triplet ($S = 1 $).
In any strong decay this is also conserved.

In Chapter 4, we estimated the energy of the low
hybrid state, namely,  $E_H=10.83$~GeV.
Therefore, if only considering the energy conservation,
The spin-exotic hybrid meson with quantum number $J^{PC}=1^{-+}$
have open decay channels: $ B\overline{B}$, \ $B\overline{B^*}$,
$B^* \overline{B^*}$, $\eta_b\eta$, $\eta_b \eta'       $,
$\chi_{b} \eta     $, $\chi_b S  $, $\Upsilon(1s) \omega$
and $ \Upsilon(1s) \phi$~\cite{McNeile:2002az}.

Page proposed one non-relativistic
symmetrization selection rule~\cite{Page:1996rj}, namely that
the hybrid decay does not lead to two identical states:
\be
H \not \to X + Y \, .
\ee
That is, the decay is suppressed if $X$ and $Y$ have
an identical internal structure in all respects,
as well as an identical radial and gluonic excitation
(namely, each meson has orbital angular momentum $L=0$ (S-wave)).
From this rule we can first exclude the decays to
$B\overline{B}$, $B\overline{B^*}$ and $B^* \overline{B^*}$.

For the hybrid meson with $J^{PC}=1^{-+}$,
we consider its decay to a $(Q\overline{q})(\overline{Q}q)$.
If each heavy-light meson $Q\overline{q}$ has $L=0$,
it is impossible, because the spin triplet state has even $CP$,
and the spin singlet state has $\Lambda=0$~\cite{McNeile:2002az}.
If one or two heavy-light mesons have a nonzero orbital
excitation($L \geq 1$), it is permitted from the consideration of the symmetry.
but not permitted energetically.
From Ref.~\cite{PDBook},
we note that experimental observation mass of the excited (L=1) $B$ meson
is $5.732$~GeV, and the mass of  nonexcited (L=0) $B$ meson is $5.279$~GeV.
Therefore, the hybrid exotic meson with mass $E_H = 10.83$
cannot decay to these two $B$ mesons if one or two $B$ mesons
have a nonzero orbital excitation ($L \geq 1$).

On the other hand, the decays of a hybrid meson with $J^{PC}=1^{-+}$ to
$(Q\overline{Q})(q\overline{q})$ are
allowed energetically~\cite{McNeile:2002az},
because the excitation energy in the gluonic field is enough to
produce a light quark-antiquark $q\overline{q}$ pair
in a flavor singlet state~\cite{McNeile:2002az},
for example, $\eta$, $\omega$ and scalar ($S$) channels.
However, as we have  discussed, its orbital angular momentum is conserved.
Hence, the decays to the quarkonia of $Q\overline{Q}$
are only into a $\chi_b$ state (P-wave state) in the static limit.

For the decays from a hybrid meson with the $\Pi_u$ representation
to a gluonic field with the $\Sigma^+_g$ representation,
the $q\overline{q}$ should have
a $\Pi_u$ representation with  $\Lambda=1$ and $CP=-1$.
Keep in mind that this orbital wave function of the $q\overline{q}$ of separation axis
between $Q\overline{Q}$ has no spin component.
Therefore, if the $q\overline{q}$ is the scalar meson with $J^{PC}=0^{++}$,
its orbital wave function should have odd $CP$~\cite{McNeile:2002az}.

%% file: Chap6.tex
\newsavebox{\CorHH}
\savebox{\CorHH}{\begin{picture}(0,0)	
\setlength{\unitlength}{1pt}
\put(50,50) { \line(0,1){100} }
\put(150,50){ \line(0,1){100} }
\put(0,0)    { \vector(2,1){100} }
\put(0,0)    { \vector(2,3){100} }
\put(50,  50) { \line(1,1){25} }
\put(150, 50) { \line(1,1){25} }
\put(75,  75) { \line(1,0){100} }
\put(50,  50) { \line(-1,-1){25} }
\put(150, 50) { \line(-1,-1){25} }
\put(25,  25) { \line(1,0){100} }
\put(50,  150) { \line(1,1){25} }
\put(150, 150) { \line(1,1){25} }
\put(75,  175) { \line(1,0){100} }
\put(50,  150) { \line(-1,-1){25} }
\put(150, 150) { \line(-1,-1){25} }
\put(25,  125) { \line(1,0){100} }
\put(100, 54)  { \makebox(0,0){ $ \Pi_{u}$ }}
\put(107, 144) { \makebox(0,0){ $ \Pi_{u}$ }}
\put(45, 14)   { \makebox(0,0){ $ ({\bf r},t) $ } }
\put(85, 100)  { \makebox(0,0){ $ ({\bf r},t+T) $ } }
\put(180, 50)  { \makebox(0,0){ $ t   $ }}
\put(180, 150) { \makebox(0,0){ $ t+T $ }}
\put(180, 40)  { \makebox(0,0){ $ Source $ } }
\put(180, 160) { \makebox(0,0){ $ Sink $ } }
\put(-5, -5)     { \makebox(0,0){ $ Origin $ } }
\put(50,  40)  { \makebox(0,0){ $ Q $  }}
\put(150, 40)  { \makebox(0,0){ $ \overline{Q} $ }}
\put(-20,50)  { \vector(1,1){15} }
\put(-20,50)  { \vector(1,0){30} }
\put(10, 44)   { \makebox(0,0){ $ z $ } }
\put(-5, 68)     { \makebox(0,0){ $x$ or $y$ } }
\end{picture}}
\newsavebox{\CorBB}
\savebox{\CorBB}{\begin{picture}(0,0)	
\setlength{\unitlength}{1pt}%
\put(50,50) { \line(0,1){100} }
\put(50,50) { \line(1,0){100} }
\put(50,150){ \line(1,0){100} }
\put(150,50){ \line(0,1){100} }
\put(0,0)    { \vector(2,1){100} }
\put(0,0)    { \vector(2,3){100} }
\put(100,150){ \vector(-1,1){50} }
\put(0, 0)   { \vector(1,4){50} }
\put(0, 0)   { \vector(4,3){120} }
\put(100,50){ \vector( 1,2){20 } }
\put(100, 42)  { \makebox(0,0){ $ \Sigma_{g}$  }}
\put(107, 144) { \makebox(0,0){ $ \Sigma_{g}$  }}
\put(180, 50) { \makebox(0,0){ $ t   $ }}
\put(180, 150){ \makebox(0,0){ $ t+T $ }}
\put(40, 14)     { \makebox(0,0){ $ {\bf r} $ } }
\put(75, 100)    { \makebox(0,0){ $ {\bf r} $ } }
\put(85, 182)    { \makebox(0,0){ $ {\bf s'}-{\bf r} $ } }
\put(20, 100)    { \makebox(0,0){ $ {\bf s'}$ } }
\put(50, 205)    { \makebox(0,0){ $ \bar\Psi\Psi $ } }
\put(126, 97)    { \makebox(0,0){ $ \bar\Psi\Psi $ } }
\put(40,  35)     { \makebox(0,0){ $ {\bf s}$ } }
\put(128, 72)    { \makebox(0,0){ $ {\bf s}-{\bf r} $ } }
\put(180, 40)    { \makebox(0,0){ $ Source $ } }
\put(180, 160)   { \makebox(0,0){ $ Sink $ } }
\put(-5, -5)     { \makebox(0,0){ $ Origin $ } }
\put(50,  43)  { \makebox(0,0){ $ Q $  }}
\put(150, 43)  { \makebox(0,0){ $ \overline{Q} $ }}
\end{picture}}
\newsavebox{\CorHAB}
\savebox{\CorHAB}{\begin{picture}(0,0)	
\setlength{\unitlength}{1pt}%
\put(50,50) { \line(0,1){100} }
\put(50,150){ \line(1,0){100} }
\put(150,50){ \line(0,1){100} }
\put(0,0)    { \vector(2,1){100} }
\put(0,0)    { \vector(2,3){100} }
\put(100,150){ \vector(-1,1){50} }
\put(0, 0)   { \vector(1,4){50} }
\put(50,  50) { \line(1,1){25} }
\put(150, 50) { \line(1,1){25} }
\put(75,  75) { \line(1,0){100} }
\put(50,  50) { \line(-1,-1){25} }
\put(150, 50) { \line(-1,-1){25} }
\put(25,  25) { \line(1,0){100} }
\put(100, 54)  { \makebox(0,0){ $ \Pi_{u}$ }}
\put(107, 144) { \makebox(0,0){ $ \Sigma_{g}$ }}
\put(180, 50)  { \makebox(0,0){ $ t   $ }}
\put(180, 150) { \makebox(0,0){ $ t+T $ }}
\put(45, 14)   { \makebox(0,0){ $ ({\bf  r},t) $ } }
\put(85, 100)  { \makebox(0,0){ $ ({\bf  r},t+T) $ } }
\put(78, 187)  { \makebox(0,0){ $ ({\bf s}-{\bf r} )$ } }
\put(20, 100)  { \makebox(0,0){ $ ({\bf s},t+T)   $ } }
\put(50, 205)  { \makebox(0,0){ $ \bar\Psi\Psi $ } }
\put(180, 40)  { \makebox(0,0){ $ Source $ } }
\put(180, 160) { \makebox(0,0){ $ Sink $ } }
\put(-5, -5)   { \makebox(0,0){ $ Origin $ } }
\put(50,  40)  { \makebox(0,0){ $ Q $  }}
\put(150, 40)  { \makebox(0,0){ $ \overline{Q} $ }}
\put(-20,50)  { \vector(1,1){15} }
\put(-20,50)  { \vector(1,0){30} }
\put(10, 44)   { \makebox(0,0){ $ z $ } }
\put(-5, 68)     { \makebox(0,0){ $x$ or $y$ } }
\end{picture}}
\newsavebox{\AEPiu}
\savebox{\AEPiu}{\begin{picture}(0,0)	
\setlength{\unitlength}{1pt}
\put(0, 0)  { \line(0,1){50} }
\put(50, 0) { \line(0,1){50} }
\put(0,    0    ) { \line(-1,-1){12.5} }
\put(50,   0    ) { \line(-1,-1){12.5} }
\put(-12.5,-12.5) { \line(1,0){50} }
\put(0,  50) { \line(1,  0){50  } }
\put(0,  -7)  { \makebox(0,0){ \scriptsize    $ Q $  }}
\put(50, -7)  { \makebox(0,0){ \scriptsize    $ \overline{Q} $ }}
\put(70,  0)  { \line(0,1){50} }
\put(120, 0)  { \line(0,1){50} }
\put(70,  0)  { \line(1,1){12.5} }
\put(120, 0)  { \line(1,1){12.5} }
\put(82.5, 12.5)  { \line(1,0){50} }
\put(70,  50) { \line(1,0){50} }
\put(70,  -7)  { \makebox(0,0){ \scriptsize    $ Q $  }}
\put(120, -7)  { \makebox(0,0){ \scriptsize    $ \overline{Q} $ }}
\put(60, 25)  { \makebox(0,0){ $ - $ }}
\end{picture}}
%
%
%
%
\newsavebox{\Corhab}
\savebox{\Corhab}{\begin{picture}(0,0)	
\setlength{\unitlength}{1pt}%
\put(50,50) { \line(0,1){100} }
\put(50,150){ \line(1,0){100} }
\put(150,50){ \line(0,1){100} }
\put(0,0)    { \vector(2,1){100} }
\put(0,0)    { \vector(2,3){100} }
\put(100,150){ \vector(-1,1){50} }
\put(0, 0)   { \vector(1,4){50} }
\qbezier(55,50)(100,80)(155,50)
\qbezier(55,50)(100,20)(155,50)
\put(100, 54)  { \makebox(0,0){ $ \Pi_{u}$ }}
\put(107, 144) { \makebox(0,0){ $ \Sigma_{g}$ }}
\put(180, 50)  { \makebox(0,0){ $ t   $ }}
\put(180, 150) { \makebox(0,0){ $ t+T $ }}
\put(45, 14)   { \makebox(0,0){ $ ({\bf r},t)   $ } }
\put(90, 100)  { \makebox(0,0){ $ ({\bf r},t+T) $ } }
\put(78, 187)  { \makebox(0,0){ $ ({\bf s}-{\bf r},0) $ } }
\put(20, 100)  { \makebox(0,0){ $ ({\bf s},t+T)       $ } }
\put(50, 205)  { \makebox(0,0){ $ \bar\Psi\Psi $ } }
\put(180, 40)  { \makebox(0,0){ $ Source $ } }
\put(180, 160) { \makebox(0,0){ $ Sink $ } }
\put(-5, -5)   { \makebox(0,0){ $ Origin $ } }
\put(50,  40)  { \makebox(0,0){ $ Q $  }}
\put(150, 40)  { \makebox(0,0){ $ \overline{Q} $ }}
\put(-25,50)  { \vector(1,1){15} }
\put(-25,50)  { \vector(1,0){30} }
\put(5, 44)   { \makebox(0,0){ $ z $ } }
\put(-10, 68)     { \makebox(0,0){ $x$ or $y$ } }
\end{picture}}
\newsavebox{\CreationH}
\savebox{\CreationH}{\begin{picture}(0,0)	
\setlength{\unitlength}{1pt}
\put(50,  70) { \line(0,1){40} }
\put(150, 110) { \vector(0,-1){40} }
\put(50,  110) { \line(1,0){100} }
\put(50,  50) { \line(0,-1){40} }
\put(150, 10) { \vector(0, 1){40} }
\put(50,  10) { \line(1,0){100} }
\put(210,  70) { \line(1,1){25} }
\put(335,  95) { \vector(-1,-1){25} }
\put(200,  50) { \line(-1,-1){25} }
\put(275,  25) { \vector(1, 1){25} }
\put(235,  95) { \line(1,0){100} }
\put(175,  25) { \line(1,0){100} }

\put(50,  60)  { \makebox(0,0){ $ Q $  }}
\put(150, 60)  { \makebox(0,0){ $ \overline{Q} $ }}
\put(210,  60)  { \makebox(0,0){ $ Q $  }}
\put(310, 60)  { \makebox(0,0){ $ \overline{Q} $ }}
\end{picture}}
%
%
%
\chapter{Our calculational method}
\label{ch:ocm}
%
In Chapter 5  we discussed the possible decay channels
of a hybrid meson in the heavy quark limit.
The goal of this dissertation is to study these hadronic decays.

In this dissertation,  we consider the decay,
\be
H \longrightarrow A \ + \ B \, ,
\ee
where $H$ is unstable spin-exotic hybrid state of the static heavy
quark.
The hybrid meson $H$ is in the  $\Pi_u$ representation.
The $A$  and $B$ stand for two decay products.
The $A$ is a conventional quarkonium state $\chi_b$ in
the $\Sigma_g^+$ representation.
The $B$ is a ${0}^{++} \pi\pi$ state~\cite{McNeile:2002az}.

In Chapter 4, we explained in detail the correlators
$C_{A A}$ and $C_{HH}$, which we used for calculating the static potential
$V_{\Sigma_g^+}(R), \ V_{\Pi_u}(R)$ respectively.
In this chapter,  we discuss in detail the correlators
$C_{B B}$ and $C_{H A B}$,
which are used to calculate the transition amplitude $x$
and the decay rate $\Gamma$ in Chapters 7 and 8.
%
%
\section{$C_{BB}$ correlator}
In this section we calculate the correlator
from B state ($ {0}^{++} \ \pi\pi$ state) to
     B state.
We first give the definition of the $C_{BB}$ correlator.
Then we introduce two methods to calculate it.
Because we use replica trick in our calculation for  the $C_{B B}$ correlator,
we first briefly introduce the replica trick.

\subsection{Replica trick in lattice}
As we discussed in Chapter 3, for the KS staggered fermion,
we reduce the number of the tastes per flavor from 16 to 4.
To further  reduce from 4 to 1, we used the fourth root trick.
At the quark level, the fourth root trick can be achieved
by multiplying each dynamical sea quark loop by a factor of
$ \DPT  \frac{1}{4}$~\cite{Aubin:2003mg}.
In Ref.~\cite{Aubin:2003mg}, the authors apply this method
to staggered chiral perturbation theory (S$\chi$PT).
However in practice, when we use this method to
work at all orders in S$\chi$PT, we cannot make sure that
every time we do it correctly at higher order in S$\chi$PT.
A solution for this is to use the replica trick.~\footnote{
The replica trick is originally proposed
in Refs.~\cite{Edwards-Anderson:1975}
to study the spin glass.
Now it has been widely used in the theory of the disordered systems
such as QCD~\cite{Splittorff:2002eb}.
}
We can state this trick as~\cite{Aubin:2003mg} follows: \\
1) Generalize the number of each quark flavor to
contain $n_i$ quarks ($i\hspace{-0.1cm}=\hspace{-0.1cm}u,d,s$)
(i.e., $n_u$ for up quark, $n_d$ for down quark, $n_s$ for strange quark).
Therefore, we have $ n = \DPT \sum_{i=u,d,s} n_i$ total quark
for all flavors $u,d,s$. \\
2) Calculate physical quantities in which we are interested
(e.g., correlation function) as analytic functions of the $n_i$. \\
3) In the end, set each $n_i = \DPT \frac{1}{4}, \ i=u,d,s$. \\
This trick automatically performs the transition from four tastes to one taste
per flavor for staggered fermion at all orders.
We use this method extensively.
%
%
\subsection{The operator for the $C_{BB}$ correlator}
\begin{figure}[h]
\setlength{\unitlength}{1pt}
\begin{center}\begin{picture}(200,205)
\put(0.0,0){\usebox{\CorBB} } 
\end{picture} \end{center}
\caption{  \label{CBB:FIG}
An illustration of the correlator from $B$ to $B$.
}
\end{figure}
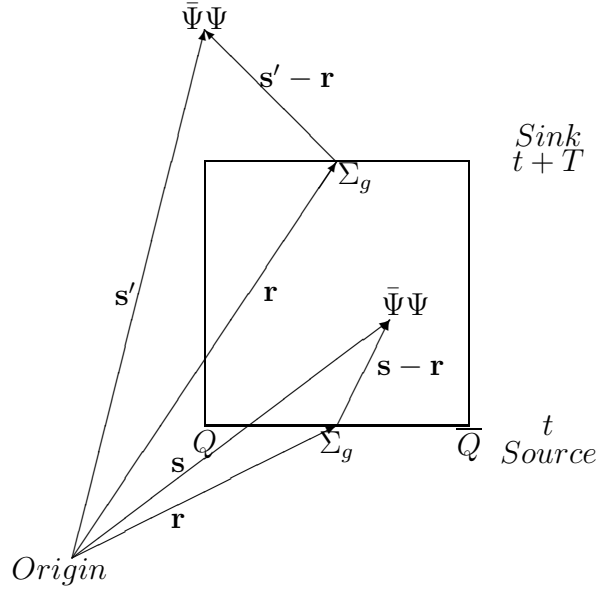

On a starting time slice $t$,
the quark $Q$ and antiquark $\overline{Q}$ are fixed
at lattice sites with separation distance of $R$,
whose midpoint is at space point {$\bf r$}.
Let  ${\bf s}$ be the spatial point of the $\pi\pi$ states
at the source time,
and { $\bf s'$} corresponding position at the sink time
(see Fig.~\ref{CBB:FIG}).
In order to make our $\pi\pi$ state to be $0^{++}$ flavor singlet,
the fermionic lattice operator at the source can be written as
\be
{\cal O}_{Source}({\bf s}, t)  \equiv
\sum_{a, g}
\frac{ \bar u^a_g( {\bf s}, t ) u^a_g( {\bf s},t ) +
       \bar d^a_g( {\bf s}, t ) d^a_g( {\bf s},t ) }{ \sqrt{2n_r} } ,
\ee
where   $g$ is the index of the taste replica,
      $n_r$ is the number of the taste replicas,
and     $a$ is the color index.
At the final time slice $t_f$ ($t_f \equiv t+T$),
the fermionic lattice operator at the sink can be written similarly as
\be
{\cal O}_{Sink}({\bf s'}, t_f)  \equiv
\sum_{ b, g' }
\frac{ \bar u^{b}_{g'}({\bf s'},t_f) u^b_{g'}( {\bf s'},t_f ) +
       \bar d^{b}_{g'}({\bf s'},t_f) d^b_{g'}( {\bf s'},t_f ) }
     { \sqrt{2n_r} } ,
\ee
where $g'$ is the index of the taste replica
and $b$ is the color index.

Therefore, the time slice correlator $C_{BB}$
from $B$ (namely, ${0}^{++} \ \pi\pi$) to $B$
for a chosen momentum $\bf k$
can be evaluated by the formula
\bea
C_{BB}(T, {\bf k} ) \hspace{-0.3cm} &=& \hspace{-0.3cm}
\frac{1}{\Omega} \sum_{{\bf s, s'},t}
\LL \Tr[{\cal O}_{Sink}^{\dag}({\bf s'},t+T) \ {\cal O}_{Source}({\bf
s},t) \RR
e^{ i{\bf k \cdot (s-s')} }
\nonumber\\ \hspace{-0.3cm} &=& \hspace{-0.3cm}
 \frac{1}{ \Omega }\frac{1}{n_r}
\sum_t \sum_{\bf s', s} \sum_{a,b} \sum_{g,g'}
\Biggl\{
\LL \bar u^{b}_{g'}({\bf s'}, \Tt) u^{b}_{g'}({\bf s'}, \Tt) \
    \bar u^{a}_{g }({\bf s}, t)   u^{a}_{g }({\bf s },t) \RR  +
\nonumber\\
\hspace{-0.3cm} &&\hspace{-0.3cm}
\LL \bar u^{b}_{g'}({\bf s'}, \Tt) u^{b}_{g'}({\bf s'}, \Tt) \
    \bar d^{a}_{g }({\bf s}, t)   d^{a}_{g }({\bf s },t) \RR
\Biggl\}  e^{ i{\bf k \cdot (s-s')} } ,
\label{EQ.CBB}
\eea
where $\Omega$ is the total lattice volume.
Here we consider that the $u, d$ quarks are degenerate in mass.
We can simplify the correlator of $C_{BB}$ in Eq.~(\ref{EQ.CBB}) as
\bea
C_{BB}(T, {\bf k} ) \hspace{-0.3cm} &=& \hspace{-0.3cm}
\frac{1}{\Omega}\frac{1}{n_r}
\sum_{t}\sum_{\bf s', s}\sum_{a,b}\sum_{g,g'}
\Biggl\{
\LL \bar u^{b}_{g'}({\bf s'},\Tt) u^{b}_{g'}({\bf s'},\Tt) \
    \bar u^{a}_{g }({\bf s}, t)   u^{a}_{g }({\bf s },t) \RR  +
\nonumber\\ \hspace{-0.3cm} && \hspace{-0.3cm}
\LL \bar u^{b}_{g'}({\bf s'},\Tt) u^{b}_{g'}({\bf s'},\Tt) \
    \bar d^{a}_{g }({\bf s}, t)   d^{a}_{g }({\bf s },t) \RR
\Biggl\}  e^{ i{\bf k \cdot (s-s')} } ,
\label{CBBA:ch6}
\eea
%
After we perform the Wick contractions of the fermion fields,
and sum over the index of the taste replica~\cite{Montvay:1994cy},
we obtain
\bea
C_{BB}(T, {\bf k} )
\label{CBB:EQDCVC}
\hspace{-0.3cm} &=& \hspace{-0.3cm}
\frac{1}{\Omega} \sum_{t} \sum_{ {\bf s'},{\bf s} }
\Bigg\{ \sum_{a,b}
\LL M_{bb}^{-1}({\bf s'},\Tt;{\bf s'},\Tt)
    M_{aa}^{-1}({\bf s},t;   {\bf s},t   ) \RR
e^{ i{\bf k} \cdot ({\bf s }-{\bf s'} )} \Bigg\}
\\
\hspace{-0.3cm}&-&\hspace{-0.3cm}
\frac{2 n_r}{\Omega} \sum_{t} \sum_{ {\bf s'},{\bf s} }
\Bigg\{ \sum_{a,b}
\LL M_{ba}^{-1}({\bf s'},\Tt;{\bf s},t)
       M_{ab}^{-1}({\bf s},t;{\bf s'},\Tt)  \RR
e^{ i{\bf k} \cdot ({\bf s }-{\bf s'} )} \Bigg \}   \nonumber ,
\eea
where $M$ is the fermion matrix.
Therefore, the correlator $C_{BB}(T, {\bf k} )$ contains two parts.
The first term in Eq.~(\ref{CBB:EQDCVC}) is
the disconnected contribution part (i.e., DC part).
We label this term $C_{BB}^{DC}$.
\be
C_{BB}^{DC}(T, {\bf k} ) =
\frac{1}{\Omega} \sum_{t} \sum_{ {\bf s'},{\bf s} }
\Bigg\{ \sum_{a,b}
\LL M_{aa}^{-1}({\bf s},t;   {\bf s},t   )
        M_{bb}^{-1}({\bf s'},\Tt;{\bf s'},\Tt)  \RR
e^{ i{\bf k} \cdot ({\bf s }-{\bf s'} )} \Bigg\} .
\label{DCEQ:BB}
\ee
The second term in Eq.~(\ref{CBB:EQDCVC}) is
the connected contribution for correlator $C_{BB}$.
We label this term $C_{BB}^{CC}$.
\be
C_{BB}^{CC}(T, {\bf k} )  =
\frac{1}{\Omega} \sum_{t} \sum_{ {\bf s'},{\bf s} }
\Bigg\{ \sum_{a,b}
\LL M_{ab}^{-1}({\bf s},t;{\bf s'},\Tt)
        M_{ba}^{-1}({\bf s'},\Tt;{\bf s},t) \RR
e^{ i{\bf k} \cdot ({\bf s }-{\bf s'} )}   \Bigg\}   .
\label{CCEQ:BB}
\ee
%
%
%
\subsection{Direct method}
In our lattice simulation, we use the direct method to calculate
the connected part $ C_{BB}^{CC}(T, {\bf k} ) $.
Eq.~(\ref{DCEQ:BB}) can be changed into
\be
C_{BB}^{CC}(T, {\bf k} )  =
\frac{1}{\Omega} \sum_{t} \sum_{ {\bf s'},{\bf s} }
\Bigg\{  {\Tr}_{c}
\left [ M^{-1}({\bf s},t;{\bf s'},\Tt)
    M^{-1}({\bf s'},\Tt;{\bf s},t) \right ]
e^{ i{\bf k} \cdot ({\bf s }-{\bf s'} )}  \Bigg \}  ,
\ee
where ${\Tr}_{c}$ is trace over the color index.
We know that
\be
 M^{-1}({\bf s},t;    {\bf s'},\Tt)  \equiv
\lbrack M^{-1}({\bf s'},\Tt; {\bf s}, t  )  \rbrack ^{\dagger}
(-1)^{ s' - s }  ,
\ee
where $ s \equiv ( {\bf s}, t), s' \equiv ( {\bf s'}, \Tt)  $,
$ M^{-1}(s; s') $  is $3 \, \times \, 3 $  color matrix, and
$
(-1)^{ s' - s } \equiv (-1)^{ s'_1 + s'_2 + s'_3 - s_1 - s_2 - s_3 +T }
$.
%
Finally, we can rewrite our connected part as
\bea
C_{BB}^{CC}(T, {\bf k} ) \hspace{-0.3cm} &=& \hspace{-0.3cm}
\frac{1}{\Omega} \sum_{t} \sum_{{\bf s'},{\bf s}}  \Bigg\{
(-1)^{ s' - s } e^{ i{\bf k} \cdot ({\bf s }-{\bf s'} )}
\times \nonumber\\
&&\hspace{1.5cm}
{\Tr}_{c} \left[ M^{-1}({\bf s},t; {\bf s'},\Tt)
       M^{-1}({\bf s},t; {\bf s'},\Tt)^{\dag} \right]
\Bigg\} .
\label{CBBCC:ch6}
\eea
%
We use the conjugate gradient method (CG )
to obtain the required matrix element of
the inverse fermion matrix
$ M({\bf s},t;{\bf  s'},\Tt)$ by solving the linear system
\be
\sum_{b} M_{a,b}(     {\bf s''}, t''; {\bf s'}, \Tt)
         M^{-1}_{b,c}({\bf s'},  \Tt; {\bf s }, t  )
= \delta_{a,c} \delta_{t'', t} \delta_{ {\bf s''}, {\bf s} }
\ee
for $M^{-1}_{b,c}({\bf s'},  \Tt; {\bf s }, t  )$,
where the $a,b,c$ are color indices.
Since  we obtain the matrix element of
$ M^{-1}( {\bf s'},\Tt ; {\bf s},t ) $,
we can easily calculate $C_{BB}^{CC}(T, {\bf k} )$
by Eq.~(\ref{CBBCC:ch6}) .
%
%
\subsection{Random source method}
%
With the random source we compute both the $DC$ part and $CC$ part
for the correlator $C_{BB}$.
This method involves relating the desired correlator
to a correlator involving random color fields $\xi$:
\bea
C_{BB}^{RND}(T, {\bf k} ) \hspace{-0.3cm} &=& \hspace{-0.3cm}
\frac{1}{\Omega} \sum_{t} \sum_{ {\bf s'},{\bf s}} \Biggl\{
e^{ i {\bf k \cdot ( s - s' ) } }  \times
\nonumber\\ \hspace{-0.3cm} && \hspace{-0.3cm}
\frac{1}{N_{r}} \sum_{\eta} \sum_{a,a'}
\sum_{{\bf r_{0}' },t_{0}' }
\left[\xi_{a }^{\eta^*}({\bf s},t) M^{-1}_{aa'}({\bf s},t;{\bf r_{0}'},t_{0}')
      \xi_{a'}^{\eta  }({\bf r_{0}'}, t_{0}' ) \right] \times
\nonumber\\ \hspace{-0.3cm} && \hspace{-0.3cm}
\frac{1}{N_{r}} \sum_{\eta'}  \sum_{b,b'} \sum_{{\bf r}',t' }
\left[ \xi_{b}^{\eta'^*}({\bf s'},\Tt) M^{-1}_{bb'}({\bf s'},\Tt;{\bf r'},t')
       \xi_{ b' }^{\eta'}( {\bf r'}, t' ) \right ]
\Biggr\}
\nonumber\\ \hspace{-0.3cm} &=& \hspace{-0.3cm}
\frac{1}{\Omega}  \sum_{t} \sum_{{\bf s'},{\bf s}} \Bigg\{
e^{ i{\bf k} \cdot ({\bf s }-{\bf s'} )}  \times
\nonumber\\ \hspace{-0.3cm} && \hspace{-0.3cm}
\sum_{{\bf r}',t' } \sum_{{\bf r_{0}'},t_{0}' }
\Bigg[
\sum_{a,a'} \sum_{b,b'}
\LP M^{-1}_{aa'}({\bf s },t  ;{\bf r_{0}'},t_{0}' )
    M^{-1}_{bb'}({\bf s'},\Tt;{\bf r'},t' ) \RP  \times
\nonumber\\  \hspace{-0.3cm} && \hspace{-0.3cm}
\frac{1}{N_{r}^{2}} \sum_{\eta} \sum_{\eta'}
\LP \xi_{a}^{\eta ^{*} }( {\bf s},t )
    \xi_{ a' }^{\eta}( {\bf r_{0}'}, t_{0}' )
    \xi_{b}^{\eta'^{*} }( {\bf s'},\Tt)
    \xi_{ b' }^{\eta'}( {\bf r'}, t' ) \RP
\Bigg] \Bigg\}  ,
\label{RandEQ:Cor}
\eea
where $N_{r}$ is the total number of random sources,
$\xi$ is the complex vector of the gaussian random numbers,
$a, a', b, b'$ are color indices, and $\eta, \eta'$ are
the indices of the random sources.
In the MILC code, we simulate the $\xi$ to satisfy
\bea
\label{MILC:xi1}
\langle \Re\xi_{x} \rangle = \frac{1}{N_{r}} \sum_{r} \Re\xi_{x}^{r} = 0
&&
\langle \Im\xi_{x} \rangle = \frac{1}{N_{r}} \sum_{r} \Im\xi_{x}^{r} = 0
\nonumber \\
\langle (\Re\xi_{x})^{2} \rangle =
\frac{1}{N_{r}}\sum_{r}( \Re\xi_{x}^{r} )^{2} = \frac{1}{2} &&
\langle (\Im\xi_{x})^{2} \rangle =
\frac{1}{N_{r}}\sum_{r}( \Im\xi_{x}^{r} )^{2} = \frac{1}{2} ,
\eea
where $\Re\xi$ is the real    part  of $\xi$,
and   $\Im\xi$ is the imagery part  of $\xi$.
We can prove
\bea
\label{appa:eq1}
\frac{1}{N_{r}} \sum_{\eta}
\xi_{a}^{\eta^{*} }(x) \xi_{b}^{\eta}(y)
&=& \delta_{xy}\delta_{ab}        \\
\frac{1}{N_{r}} \sum_{\eta}
\xi_{a }^{\eta^*}(x)  \xi_{b}^{\eta}(y)
\xi_{a'}^{\eta^*}(x') \xi_{b'}^{\eta}(y')
&=&
\delta_{xy }\delta_{ab}  \delta_{a'b'} \delta_{x'y'}  +
\delta_{ab'}\delta_{xy'} \delta_{ba' } \delta_{yx' }  ,
\label{appa:eq2}
\eea
where $a$ is color index, and $\eta$ is the index of random source.

For $ \eta \not= \eta' $,
the correlator in Eq.~(\ref{RandEQ:Cor}) gives the disconnected part,
that is,
\be
C_{BB}^{RND}(T, {\bf k})_{\eta \not= \eta'} =
C_{BB}^{DC }(T, {\bf k} ) \times \LP 1-\frac{1}{N_r} \RP .
\ee

For $ \eta = \eta' $, it gives both the disconnected part
and connected part, that is,
\be
C_{BB}^{RND}(T, {\bf k})_{\eta = \eta'} =
\frac{1}{N_r} \left (
C_{BB}^{DC }(T,{\bf k})+C_{BB}^{CC}(T,{\bf k}) \RP .
\ee
Therefore, we obtain both the disconnected part and the connected part
of the  correlator $C_{BB}$ from this method.
In practice, we compare the result of the connected part from
this method with the result of the connected part
from the direct method to justify our results from this method.
From our calculation we find that
if we choose the $ N_{r} $ large enough, the numbers from two methods agree.

Let
\vspace{-0.3cm}
\bea
\sigma^{\eta}({\bf s},t)      & \equiv &
\sum_{a, a'}  \sum_{{\bf r_{0} }',t_{0}'}
\xi_{a}^{\eta ^{*} }( {\bf s},t )
M^{-1}_{aa'}( {\bf s},t ; {\bf r_{0}'},t_{0}' )
\xi_{a'}^{\eta}( {\bf r_{0}'}, t_{0}' )  \\
\sigma^{\eta'}( {\bf s'},\Tt) &\equiv&
\sum_{b, b'}  \sum_{ {\bf r }',t' }
\xi_{b}^{\eta ^{\prime *} }( {\bf s'},\Tt)
M^{-1}_{bb'}( {\bf s'},\Tt; {\bf r'},t' )
\xi_{ b' }^{\eta' }( {\bf r'}, t' ) .
\eea
Then Eq.~(\ref{RandEQ:Cor}) can be rewritten as
\be
C_{BB}^{RND}(T, {\bf k} ) =
\frac{1}{\Omega} \frac{1}{N_{r}^{2}} \sum_{\eta, \eta'}
\sum_{t} \sum_{ {\bf s'},{\bf s}}   \Bigg\{
\sigma^{\eta}( {\bf s},t) \sigma^{\eta'}( {\bf s'},\Tt)
e^{  i{\bf k} \cdot ({\bf s }-{\bf s'}) }  \Bigg\} .
\ee
If we define the Fourier transform:
$\DPT
\sigma^{\eta}( {\bf k},t )  =
\sum_{\bf s} \sigma^\eta({\bf s},t) e^{ -i{\bf k} \cdot {\bf s} }
$,
then Eq.~(\ref{RandEQ:Cor}) can be rewritten as
\be
C_{BB}^{RND}(T, {\bf k} ) =
\frac{1}{\Omega} \frac{1}{N_{r}^{2}} \sum_{t} \sum_{\eta, \eta'}
\sigma^{\eta}(-{\bf k},t) \sigma^{\eta'}({\bf k},\Tt) .
\ee
From the above discussion, we know that,
for $\eta \not= \eta'$,
the above correlator gives the disconnected part
and for $\eta = \eta'$,
it gives both disconnected and connected parts.
%
%
\section{ $C_{H AB}$ correlator}
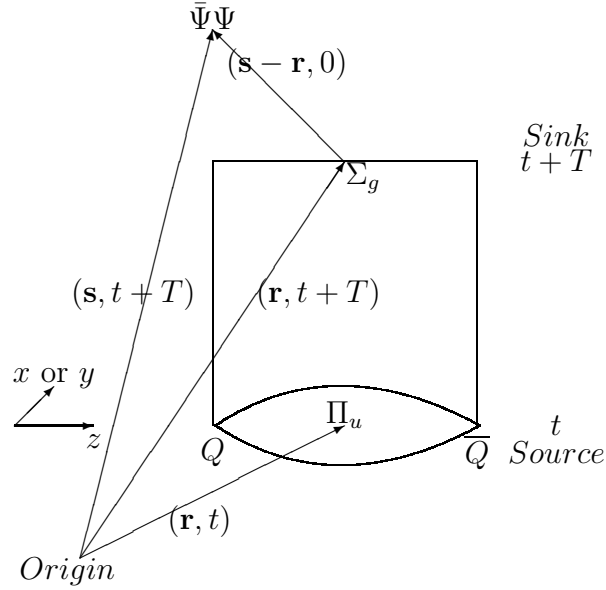
\begin{figure}[t]
\setlength{\unitlength}{1pt}
\begin{center}\begin{picture}(200,205)
\put(0.0,0){\usebox{\Corhab} }
\end{picture} \end{center}
\caption{  \label{CHAB:FIG}
   An illustration of the correlator from $H$ to $AB$.
}
\end{figure}

In this section we show how to calculate the correlator
from the $\Pi_{u}$ state or $H$ to
our two final states $A$ and  $B$  state.
On a starting time slice $t$,
the quark $Q$ and antiquark $\overline{Q}$ are fixed
on lattice sites with a separation distance $R$
whose midpoint is at space point {\bf r} (see Fig.~\ref{CHAB:FIG}).
The gluonic operator at time slice $t$ is  the
$\Pi_{u}$ lattice operator as we described in Sec.~4.4.2.

At the final time slice $\Tt$, the gluonic lattice operator
can be written as
\bea
{\cal O}_{Sink}({\bf r},{\bf R},\Tt)  \hspace{-0.3cm}&\equiv& \hspace{-0.3cm}
\sum_{ {\bf s} } \Sigma_{g}^{+} ({\bf r},{\bf R},\Tt)w({\bf s}-{\bf r})
\times \nonumber\\ \hspace{-0.3cm}&&\hspace{-0.3cm}
\sum_{a } \frac{
\bar u^{a}( {\bf s},\Tt ) u^{a}( {\bf s},\Tt ) \PLUSONE
\bar d^{a}( {\bf s},\Tt ) d^{a}( {\bf s},\Tt ) }{ \sqrt{2} }
\nonumber\\ \hspace{-0.3cm}&=& \hspace{-0.3cm}
\sqrt{2} \sum_{ {\bf s}, a } \Sigma_{g}^{+} ({\bf r},{\bf R},\Tt)
w({\bf s}\MINUSONE{\bf r})
\bar u^{a}( {\bf s},\Tt ) u^{a}( {\bf s},\Tt ) ,
\eea
where we consider that the $u, d$ quarks are degenerate in mass
and their production amplitude are identical.
The $w({\bf s} \MINUSONE {\bf r})$~\footnote{
In Chapter 5, we discussed the $CP$ symmetry of the wave function
$w({\bf s} - {\bf r})$.
}
is the wave function of
$\bar u^a({\bf s},\Tt) u^a({\bf s},\Tt)$.
Here we should note that  wave function
$w({\bf s} - {\bf r})$ is just our choice for the final state.
We assume it is a real function (i.e., $w^{*}=w$).
We should choose a suitable wave function $w({\bf s} \MINUSONE {\bf r})$ to
make sure that the initial and final states are in same symmetries.
With these definitions the time slice correlator can be written as
\bea
C_{H AB}(T, {\bf R} ) &=&
\frac{1}{\Omega} \sum_{{\bf r},t}
\LL \Tr \left [
{\cal O}_{Sink}^\dag({\bf r,R},\Tt){\cal O}_{Source}({\bf r,R},t)
 \right ] \RR  \nonumber\\
&=& \frac{\sqrt{2}}{\Omega} \sum_{{\bf r},t} \sum_{{\bf s}}
\LL \sum_{a} \bar u^a({\bf s},\Tt) u^a({\bf s},\Tt) \RR
w( {\bf s} \MINUSONE {\bf r} )  \times
\nonumber \\ &&
\Tr \left[ \Sigma_g^+({\bf r, R},\Tt) \Pi_u({\bf r, R},t) \right] \nonumber\\
&=& \frac{\sqrt{2}}{\Omega} \sum_{{\bf r},t}
\sum_{{\bf s}} \sum_{a} M^{-1}_{aa}( {\bf s},\Tt ; {\bf s},\Tt )
w( {\bf s} \MINUSONE {\bf r} ) \times \nonumber\\
&& \Tr \left[\Sigma_g^+({\bf r,R},\Tt)\Pi_u({\bf r,R},t) \right] .
\label{EQ.CHAB}
\eea
From the random source method, we can evaluate
$\DPT \sum_{a} \bar u^a({\bf s},\Tt) u^a({\bf s},\Tt)$
from
\bea
\LL \sum_a \bar u^a({\bf s},\Tt) u^a({\bf s},\Tt)
\RR
&=& \sum_a M^{-1}_{aa}( {\bf s},\Tt ; {\bf s},\Tt ) \nonumber\\
&=& \frac{1}{N_{r}} \sum_{ \eta }  \sigma^{\eta}( {\bf s},\Tt ) .
\eea

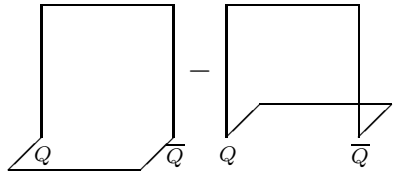
\begin{figure}[b]
\setlength{\unitlength}{1pt}
\begin{center}\begin{picture}(200,40)
\put(45,0){\usebox{\AEPiu} }
\end{picture} \end{center}
\caption{ \label{AEpi:FIG}
   An illustration of the calculation of $\AE$.
}
\end{figure}

Since the configurations are gauge fixed to the temporal
gauge before we estimate the correlator,
the time-link link variables are unit matrices.
Hence, we measure the product of the parallel transporters
in Eq.~(\ref{EQ.CHAB})
(i.e., $\DPT\Tr \left[\Sigma_g^+({\bf r,R},\Tt) \Pi_u({\bf r,R},t)\right]$)
by two bent Wilson loops shown in Fig.~\ref{AEpi:FIG}.
In terms of
\bea
S( {\bf s},\Tt ) &\equiv&
\sum_{a} M^{-1}_{aa}( {\bf s},\Tt ; {\bf s},\Tt )  \\
\AE({\bf r},{\bf R},t,T) &\equiv&
\Tr \left [ \Sigma_{g}^{+}({\bf r},{\bf R},\Tt)
            \Pi_{u}({\bf r},{\bf R},t)
\right]  ,
\eea
we have
\be
C_{HAB}(T, {\bf R} ) =
\frac{ \sqrt{2} }{\Omega} \sum_{{\bf r},t}  \sum_{{\bf s}}
S({\bf s},\Tt) w({\bf s} \MINUSONE {\bf r}) \AE({\bf r},{\bf R},t,T) .
\label{CHAB:ch61}
\ee

We can express the wave function of the $\pi\pi$ state in momentum space,
\be
   w({\bf s}, {\bf r}) \equiv w({\bf s} \MINUSONE {\bf r}) =
   \sum_{\bf p} e^{i{\bf p} \cdot ({\bf r-s})} w({\bf p}) .
\ee
Then we can rewrite Eq.~(\ref{CHAB:ch61}) as
\bea
C_{H AB}( T, {\bf R} )
\hspace{-0.3cm} &=& \hspace{-0.3cm}
\frac{ \sqrt{2} }{\Omega} \sum_{ t, {\bf p} }
\Bigg\{ w({ {\bf p}})
\sum_{{\bf s}} S( {\bf s},\Tt ) e^{ -i {\bf p} \cdot {\bf s} }
\sum_{{\bf r}}
\AE({\bf r},{\bf R},t,T) e^{ i {\bf p} \cdot {\bf r} }
\Bigg\} .
\eea
If we perform the following Fourier transforms,
\bea
\AE(-{\bf p},{\bf R},t,T)  &=&  \sum_{\bf r}
\AE({\bf r},{\bf R},t,T) e^{ i {\bf p} \cdot {\bf r} }  \\
S( {\bf p},\Tt )           &=&  \sum_{{\bf s}}
S( {\bf s},\Tt ) e^{ -i {\bf p} \cdot {\bf s} }  ,
\eea
we arrive at
\be
C_{H AB}( T, {\bf R} ) =  \frac{ \sqrt{2} }{\Omega}
\sum_{t}  \sum_{ {\bf p} }
w({ {\bf p}}) S( {\bf p},\Tt ) \AE(-{\bf p},{\bf R},t,T)  .
\label{chab:eq6}
\ee

Now we discuss the symmetry of the wave function $w({\bf s-r})$.
It depends on the meson, which was created in the decay.
For scalar meson production (with $J^{PC}=0^{++}$),
the $w({\bf s-r})$ must be  in a $\Pi_u$ representation.
This can be achieved by making $w({\bf s-r})$ odd in the $x$ direction
and even in the $y$ and $z$ directions, where $x$ is the direction
of the transverse extent of the $\Pi_u$ state.
Another method to check this symmetry
is to consider the parity,
since $P_x$, $P_y$ and $CP_z$ are conserved in the heavy decay in
heavy quark limit.
Now we consider our $\Pi_u$ potential state, which is
odd under $P_x$ and even under $P_ y$ and  $CP_z$.
The $\Sigma_{g}^{+}$ operator is even under
all three operations.
Therefore, the symmetry requires that the wave function $w$
be odd under the inversion  $P_x$ and
even under $P_y$ and $P_z$ (since $C=+1$ for the scalar meson).
So, we choose the wave function as
\be
w({\bf k, r}) = \sin(k_x \cdot x ) \, \cos (k_y \cdot y ) \, \cos(k_z \cdot
z ) ,
\label{ww:ch6}
\ee
where the ${\bf k}=(k_x, k_y, k_z)$ is the chosen momentum.
It have the desired symmetry.
In terms of plane waves
\bea
w({\bf p})  \hspace{-0.3cm}&=&\hspace{-0.3cm}
\frac{1}{V}\sum_{\bf r} e^{i{\bf p} \cdot {\bf r}} w({\bf r}) \nonumber\\
\hspace{-0.3cm}&=&\hspace{-0.3cm}
\frac{1}{8i}\Bigg\{
 \delta_{ \bf p, k}   \PLUSONE  \delta_{ \bf p, k_1}
\PLUSONE \delta_{ \bf p, k_2} \PLUSONE \delta_{ \bf p, k_3} -
          \delta_{ \bf p, -k}   \MINUSONE \delta_{ \bf p, -k_1}
\MINUSONE \delta_{ \bf p, -k_2} \MINUSONE \delta_{ \bf p, -k_3}
\Bigg\}   ,
\eea
where $ {\bf k_1}  \equiv ( k_x,  k_y, -k_z )$,
${\bf k_2} \equiv ( k_x, -k_y,  k_z )$, and
${\bf k_3} \equiv ( k_x, -k_y, -k_z )$.

We can then rewrite Eq.~(\ref{chab:eq6}) as
\bea
C_{H AB}( T, {\bf R} )_{\bf k} \hspace{-0.3cm}&=&\hspace{-0.3cm}
\frac{ \sqrt{2} }{\Omega}  \sum_{t}  \frac{1}{8i} \Bigg\{
 S( {\bf k }, \Tt) \AE(-{\bf k}, {\bf R}, t,T )
-S( {\bf -k}, \Tt) \AE( {\bf k}, {\bf R}, t,T ) \nonumber\\
\hspace{-0.3cm} &+& \hspace{-0.3cm}
 S( {\bf k_1 }, \Tt) \AE(-{\bf k_1}, {\bf R}, t,T )
-S( {\bf -k_1}, \Tt) \AE( {\bf k_1}, {\bf R}, t,T )
\nonumber\\
\hspace{-0.3cm} &+& \hspace{-0.3cm}
 S( {\bf k_2}, \Tt) \AE(-{\bf k_2}, {\bf R}, t,T )
-S( {\bf -k_2}, \Tt) \AE( {\bf k_2}, {\bf R}, t,T )
\nonumber\\
\hspace{-0.3cm} &+& \hspace{-0.3cm}
 S( {\bf k_3 }, \Tt) \AE(-{\bf k_3}, {\bf R}, t,T )
-S( {\bf -k_3}, \Tt) \AE( {\bf k_3}, {\bf R}, t,T )
\Bigg\}  .
\label{chab:eq6:2}
\eea
We use this formula to calculate the desired correlator $C_{H AB}$.

%% file: Chap7.tex
\chapter{Hybrid meson decay from the lattice}
For notational simplicity and without loss of generality,
in this chapter we consider the generic strong decay
\be
 H \to A + B ,
\ee
where $A$ and $B$ are stable particles.

In Chapter 5, we studied the hybrid meson decay channel.
It is possible to employ some rather restricted conditions
on a lattice to extract the transition matrix elements
($x$)~\cite{McNeile:2002az}.
In this chapter, we first use the time-dependent perturbation theory to
study the $C_{HAB}$ correlator. Then we introduce a
method to extract $x$ directly from lattice~\cite{McNeile:2002az}.
%

\section{Perturbation expansion for $C_{HAB}$ correlator}
In this section we use perturbation theory to study
the correlator $C_{HAB}$ in Sec.~6.3,
that is to calculate the correlator from the $\Pi_{u}$ state
(i.e., H state) to two-body states (namely, AB state).
To use this approach,
we should discuss whether the perturbation theory here is
justified or not.
The condition we need is that
$ \Gamma  \ll \Delta E $, where $ \Gamma $ is the decay rate,
and $ \Delta E = E_H - E_A $.
This is the narrow width approximation.
In Chapter 8,  we will discuss this question.

For notational simplicity we refer to the lattice operator at the source $H$
as (Please see details in Chapter 6),
\be
{\cal O}_{H}( \Rr, t) ,
\ee
and we refer to the lattice operator at the sink $AB$ as
\be
{\cal O}_{AB}( \Rr, \Tt) \equiv
\sum_{\bf s} {\cal O}_{A}( \Rr, \Tt) w({\bf s-r}) {\cal O}_{B}({\bf s}, \Tt) ,
\ee
where
$w({\bf s} \MINUSONE {\bf r})$ is the distribution wave function of the
meson B.

In this notation the  time slice correlator is given by
\bea
C_{HAB}(t, {\bf R} ) &=&  \frac{1}{\Omega} \sum_{{\bf r}, t}
\LL {\cal O}_{AB}^{\dag}(\Rr, \Tt) \ {\cal O}_{H}(\Rr, t)
\RR  ,
\eea
where ${\bf R}$ indicates  that the  correlator $C_{HAB}(t)$ is
calculated at a fixed orientation of the $Q\overline{Q}$,
and ${\bf r}$ is the midpoint of the
$Q\overline{Q}$ (see Fig.~\ref{CHAB:FIG}).
The system can be described by a Hamiltonian $\hat H$, that is,
\be
\hat H = \hat H_0 + V_I ,
\ee
where $\hat H_0$ is the unperturbed Hamiltonian,
and $V_I$ is the transition potential.
In perturbation theory we assume that $V_I$ is numerically small.
Hence,
\be
e^{- \hat{H} t } \simeq
e^{-\hat H_0 t}  +
\sum_{t_1=0}^{t} e^{-\hat H_0(t-t_1)} V_I(t) e^{-\hat H_0 t_1} + \cdots .
\ee
Therefore, we obtain
\bea
C_{HAB}(T, {\bf R} )  \hspace{-0.3cm} &=& \hspace{-0.3cm}
\frac{1}{\Omega} \sum_{ {\bf r}, t }
\LL 0 \left| e^{\hat H(t+T)} {\cal O}_{AB}^\dag(\Rr, 0) e^{-\hat H(t+T)}
\, e^{-\hat H t}{\cal O}_{H}(\Rr,0) e^{-\hat H t} \right| 0 \RR
\nonumber\\  \hspace{-0.3cm} &\approx& \hspace{-0.3cm}
\frac{1}{\Omega} \sum_{{\bf r}, t } \sum_{t_1 =0}^{T-1}
\LL 0 \left| {\cal O}_{AB}^\dag(\Rr,0) \, e^{-\hat H_0 (T-t_1) } \, V_I(t_1) \,
e^{-\hat H_0 t_1} \, {\cal O}_{H}(\Rr, 0) \right| 0 \RR  +
\nonumber\\ \hspace{-0.3cm} && \hspace{-0.3cm}
\frac{1}{\Omega} \sum_{\bf r }
\LL  0 \left| {\cal O}_{AB}^{\dag}(\Rr,0) \, e^{-\hat H_0 T} \, {\cal
  O}_{H}(\Rr,0) \right| 0 \RR
\label{CHAB:CH7:1}
\eea
%
The second term in Eq.~(\ref{CHAB:CH7:1}) is
zero, because the $H_0$ does not, by itself, cause the transition.
Thus we can rewrite Eq.~(\ref{CHAB:CH7:1}) as
\be
C_{HAB}(T, {\bf R} )  =
\frac{1}{\Omega} \sum_{{\bf r}, t } \sum_{t_1 =0}^{T-1}
\LL 0 \left| {\cal O}_{AB}^\dag(\Rr,0) \ e^{-\hat H_0(T-t_1)} \, V_I(t_1) \,
e^{-\hat H_0 t_1} \, {\cal O}_{H}(\Rr, 0) \right| 0 \RR .
\ee
In this dissertation, we model the transition potential as
\be
V_I(t') \equiv \sum_{\bf R'} \sum_{\bf r'} \sum_{\bf s'} {\bar x({\bf R'})}
\, {\cal C}_{\Sigma_g^+}^\dag({\bf R', r'},t')
\, {\cal C}_{\Pi_u}({\bf R', r'},t')
\, a^3\bar\psi\psi({\bf s'},t')
\, w_I( \bf s'- r') ,
\label{V_I:model}
\ee
where $ \bar x({\bf R'}) $ is the overall normalization factor
that controls the transition amplitude,
${\cal C}_{\Sigma_g^+}^\dag$ is a creation operator for the $\Sigma_g^+$
state, ${\cal C}_{\Pi_u}$ is an annihilation operator
for $\Pi_u$ state, and $w_I({\bf s'} \MINUSONE {\bf r'})$
is the distribution wave function of the
$\bar\psi\psi({\bf s'},t')$, which is dictated by the decay.
\bea
C_{HAB}(T, {\bf R} )  \hspace{-0.2cm} &=& \hspace{-0.2cm}
{\bar x({\bf R}) } \, \frac{1}{\Omega}
\sum_{{\bf r},t} \sum_{t' =0}^{T-1} \sum_{\bf R'} \sum_{\bf r'} \sum_{\bf s'}
\bigg\langle 0 \bigg| {\cal O}_{AB}^\dag(\Rr,t') \, e^{-\hat H_0 (T-t') } \,
{\cal C}_{\Sigma_g^+}^\dag({\bf R', r'},t') \nonumber\\
\hspace{-0.2cm}&&\hspace{-0.2cm}
\, {\cal C}_{\Pi_u}({\bf R', r'},0)
\bar\psi\psi({\bf s'},t')  \, w_I( {\bf s'- r'})  \,
e^{-\hat H_0 t'} \, {\cal O}_{H}(\Rr, 0) \bigg| 0 \bigg\rangle .
\eea
If we plug into the operator ${\cal O}_{AB}^{\dag}(\Rr,0)$,
we obtain
\bea
C_{HAB}(T,{\bf R})  \hspace{-0.2cm} &=&  \hspace{-0.2cm}
{ \bar x({\bf R}) } \frac{1}{\Omega}
\sum_{{\bf r},t}\sum_{t'=0}^{T-1}\sum_{\bf R', r'}\sum_{\bf s, s'}
\bigg\langle 0 \bigg| {\cal O}_{\Sigma_g^+}^\dag(\Rr,0)
\bar\psi\psi({\bf s},t\PLUSONE T) w^*({\bf s\MINUSONE r})
e^{-\hat H_0 (T-t') }
\nonumber\\
\hspace{-0.2cm} && \hspace{-0.2cm}
{\cal C}_{\Sigma_g^+}^\dag({\bf R', r'},0)  {\cal C}_{\Pi_u}({\bf R', r'},0)
\bar\psi\psi({\bf s'},t')  \, w_I( {\bf s' \MINUSONE r'})  \,
e^{-\hat H_0 t'} \, {\cal O}_{\Pi_u}(\Rr, 0) \bigg| 0 \bigg\rangle
\nonumber\\  \hspace{-0.2cm}&=&\hspace{-0.2cm}
{\bar x({\bf R}) } \frac{1}{\Omega}
\sum_{{\bf r}, t } \sum_{t' =0}^{T-1} \sum_{\bf s, s'} \sum_{\bf p, q }
Z_a \, e^{ -E_{\Sigma_g^+} (T-t') }
Z_h \, e^{ -E_{\Pi_u} t' }
e^{ -i \bf q \cdot (s - r) }  e^{ i \bf p \cdot (s' - r) }
\nonumber\\  \hspace{-0.2cm}&&\hspace{-0.2cm}
w^*({\bf  q}) \ w_I({\bf p})
\LL 0 \left| \bar\psi\psi({\bf s},\Tt)
             \bar\psi\psi({\bf s'},t'+t) \right| 0 \RR
\nonumber\\  \hspace{-0.2cm}&=&\hspace{-0.2cm}
{\bar x({\bf R})} \frac{1}{N_t}
\sum_{\bf q} \sum_{t} \sum_{t' =0}^{T-1} \sum_{\bf s, s'}
Z_a \, e^{ -E_{\Sigma_g^+} (T-t') }
Z_h \,  e^{ -E_{\Pi_u} t' }
e^{ i \bf q \cdot (s - s') } w^*({\bf q})w_I({\bf q})
\nonumber\\  \hspace{-0.3cm}&&\hspace{-0.3cm}
\LL 0 \left| \bar\psi\psi({\bf s},\Tt)
             \bar\psi\psi({\bf s'},t'+t) \right| 0 \RR ,
\label{aE:chab}
\eea
where $Z_a$ is defined in Eq.~(\ref{ch4:a}),
      $Z_h$ is defined in Eq.~(\ref{ch4:h}), and
we express the wave function of the $\pi\pi$ state in
momentum space,
\be
w({\bf s} \MINUSONE {\bf r}) =
\sum_{\bf p} e^{i{\bf p} \cdot ({\bf r-s})} w({\bf p}) .
\ee
We can rewrite Eq.~(\ref{aE:chab}) as
\bea
C_{HAB}(T, {\bf R})  \hspace{-0.2cm} &=&  \hspace{-0.2cm}
{\bar x}({\bf R}) V
\sum_{\bf q} \ w^*({\bf q}) w_I({\bf q}) \ \sum_{t' =0}^{T-1}
Z_a \, e^{ -E_{\Sigma_g^+} (T-t') }  Z_h \, e^{ -E_{\Pi_u} t' }  \times
\nonumber\\  \hspace{-0.2cm} && \hspace{-0.2cm}
\frac{1}{\Omega} \sum_{ t} \sum_{\bf s, s'}
e^{ i \bf q \cdot ( s-s') }
\LL 0 \left| \bar\psi\psi({\bf s},\Tt)
             \bar\psi\psi({\bf s'},t'+t) \right| 0 \RR ,
\eea
where $V=N_xN_yN_z$.
In Chapter 6, we calculate $C_{BB}(T,{\bf k})$ through
\be
C_{BB}(T,{\bf k}) = \frac{1}{\Omega}
\sum_{t} \sum_{\bf s, s'}
e^{ i \bf k \cdot ( s-s') }
\LL 0 \left| \bar\psi\psi({\bf s},\Tt)
             \bar\psi\psi({\bf s'}, t) \right| 0 \RR  .
\ee
Therefore, we obtain
\be
C_{HAB}(T, {\bf R})  = {\bar x}({\bf R})  V \sum_{\bf q}
w^*({\bf q})w_I({\bf q})
\sum_{t' =0}^{T-1}
Z_a \, e^{ -E_{\Sigma_g^+} (T - t') } \, Z_h \,
       e^{ -E_{\Pi_u} t' }
C_{BB}(T \MINUSONE t', {\bf q} ) ,
\ee
From Eq.~(\ref{ch4:a}) and Eq.~(\ref{ch4:h}),
we have
\bea
C_{AA}(R, T \MINUSONE t') &=&
Z_a^2 \ e^{ -E_{\Sigma_g^+}(R) \times (T-t') }  \\
C_{HH}(R, t')             &=&
Z_h^2 \ e^{ -E_{\Pi_u}(R)      \times t' }  .
\eea
So we can rewrite this result in terms of all the correlators as
\be
C_{HAB}(T, {\bf R} )  \hspace{-1mm} =  \hspace{-1mm} \sum_{\bf q}
w^*({\bf q}) w_I({\bf q}) \, \frac{V}{Z_a Z_h} \sum_{t' = 0}^{T-1}
C_{AA}( {\bf R}, T \MINUSONE t') \, {\bar x}({\bf R}) \,
C_{HH}({\bf R}, t') \, C_{BB}( {\bf q}, T\MINUSONE t' ) .
\label{3pt:aD:CH70}
\ee
In Chapter 6, we discussed the symmetry of the wave function
$w$, and we chose the wave function $w$ as Eq.~(\ref{ww:ch6}).
Then we obtain
\be
C_{HAB}(T, {\bf R, k} )  = \frac{ w_I( {\bf k}) V }{Z_a Z_h}
\sum_{t' = 0}^{T-1} C_{AA}( {\bf R}, T \MINUSONE t') {\bar x}({\bf R})
C_{HH}({\bf R}, t') C_{BB}( T \MINUSONE t', {\bf k} ) ,
\label{3pt:aD:CH7}
\ee
where we consider the cubic symmetry
(namely,
$C_{BB}(T \MINUSONE  t', {\bf R,-k} ) =
 C_{BB}(T \MINUSONE  t', {\bf R, k} ) $, etc).
Here we absorb the
$i$ into $w_I({\bf k})$ (i.e., $ i \, w_I({\bf k}) \to  w_I({\bf k})$).
Note that the wave function $w_I$ must
have the same symmetry as the wave function $w$.
%
%
\section{Decays from the lattice}
%
In practice
it is possible to extract the transition amplitude $x$
directly from the lattice~\cite{McNeile:2002az}.
Here  we describe this approach in principle.
Using a suitable lattice operator to create the hybrid state $H$ at
$t=0$ and annihilate a two-body state with relative
momenta ${\bf k}$ and ${\bf -k}$ at time $t$.
In Sec.~7.1, we used perturbation theory to obtain
the correlator $C_{H-AB}(t)$ from an $H$ state
with mass $E_H$ and a two-body state ($A, B$) with energy $E_A, E_B$,
that is,~\footnote{
This equation is same as Eq.~(\ref{3pt:aD:CH7}), just change $T$ into $t$.
}
\be
C_{HAB}(t, {\bf R, k} )  = \frac{ w_I({\bf k}) V }{ Z_a Z_h } \sum_{t'=0}^{t-1}
C_{AA}({\bf R}, t\MINUSONE t'){\bar x}({\bf k}) \,
C_{HH}({\bf R}, t') \,
C_{BB}(T\MINUSONE t', {\bf k} ) ,
\label{3pt:eq}
\ee
where the summation is over the intermediate $t$-value $t'$,
and $t'$ is an integer.
By obtaining $Z_h$ from the correlator $C_{H-H}(t)$ from $H \to H$,
and $Z_a$ from the correlator $C_{A-A}(t)$ from $A \to A$,
also measuring the correlator $C_{B-B}(t)$ from $B \to B$,
and the correlator $C_{H-AB}(t)$ from $H \to AB$,
we can extract the transition amplitude $\bar x({\bf R})$
in principle, that is,
\be
{\bar x({\bf R}) } w_I({\bf k}) = \frac{Z_a \, Z_h}{V}
\frac{ C_{HAB}(t, {\bf R, k} ) }
{\DPT \sum_{t' = 0}^{t-1}
C_{AA}( {\bf R}, t\MINUSONE t' ) \
C_{HH}( {\bf R}, t' ) \
C_{BB}( t \MINUSONE t', {\bf k}) }  .
\label{xrate:FullQCD}
\ee
If we define
\be
x({\bf k,R}) \equiv {\bar x({\bf R})} \, w_I({\bf k}) \, V ,
\label{x.def:ch7}
\ee
we obtain
\be
{ x({\bf k, R}) } =
\frac{ C_{HAB}(t, {\bf R, k} ) \ Z_a \, Z_h }
{\DPT \sum_{t' = 0}^{t-1}
C_{AA}({\bf R}, t\MINUSONE t') \
C_{HH}({\bf R}, t') \
C_{BB}(t\MINUSONE t', {\bf k}) }  .
\label{xrate:FullQCD2}
\ee
Therefore, from Eq.~(\ref{xrate:FullQCD2}),
we can extract the transition amplitude $x({\bf k, R})$ from the lattice.

In this dissertation we focus on the decay channel to $\pi\pi$
branch, that is,
\be
H \rightarrow \chi_b + \pi + \pi \, ,
\ee
where $\pi\pi$ is in scalar channel.
In the lattice simulation, due to the taste symmetry breaking,
there exist the splittings between pions ($\pi$) of various
tastes~\cite{Aubin:2003mg}\cite{Aubin:2004wf}\cite{Prelovsek:2005rf}.
We classify the pions into five types, namely,
$\pi_5$, $\pi_A$, $\pi_V$,  $\pi_I$, and $\pi_T$ which correspond
the pions with taste pseudoscalar, axial, vector, singlet, and tensor
respectively. In order to calculate the decay channel to $\pi\pi$
branch correctly, we should count these
five channels correctly.
In Ref.~\cite{SCALAR:2006}, the authors predict the
weights for each  taste of the $\pi\pi$ intermediate state
in bubble contribution channel. We list them in
Table~\ref{pipi_weight_TABLE}.
\begin{table}[t]
\vspace{-0.3cm}
\caption{ \label{pipi_weight_TABLE}
The weights for different tastes of the $\pi\pi$ intermediate states.
The first  column is the taste,
the second column is its weight.
}
\begin{center}\begin{tabular}{|c|c|}
\hline
{\rm Taste} &  {\rm Weight}   \\
\hline
I (singlet) &  $-3/4$  \\
\hline
V (vector)  &  $4/4$  \\
\hline
T (tensor)  &  $6/4$  \\
\hline
A (axial)   &  $4/4$  \\
\hline
P (pseudoscalar)  & $1/4$ \\
\hline
\end{tabular}
\vspace{-0.2cm}
\end{center}\end{table}

In our model, the transition amplitude of the Goldstone $\pi_5\pi_5$
can be calculated in lab frame (see details in Appendix D) by the formula
\be
x_{PP}({\bf p}) =
\LL \pi_5({\bf p}_1)\pi_5({\bf p}_2), A \left| V_I^{\rm lab} \right| H \RR ,
\ee
where $V_I^{\rm lab}$ is the interaction potential in lab frame,
which is denoted in Eq.~(\ref{V_I:model:ade}), and the subscript $PP$
specifies the Goldstone $\pi_5\pi_5$ channel,
that is,
\bea
x_{PP}({\bf p}_1, {\bf p}_2 ) &=&
\LL \pi_5({\bf p}_1)\pi_5({\bf p}_2)
\left| \bar\psi\psi( {\bf p} ) \right| 0 \RR
x(p)
\delta^{(3)}({\bf P}_H - {\bf P}_A  - {\bf p}) \nonumber\\
&=& b_{PP}({\bf p}_1, {\bf p}_2) \, x( p) \,
\delta^{(3)}({\bf P}_H - {\bf P}_A  - {\bf p} ) ,
\eea
where $x(p)$ is calculated by Eq.~(\ref{GGG:DDD})
(see details in Appendix D),
${\bf p}={\bf p}_1+{\bf p}_2$,
and
$ b_{PP}({\bf p}_1,{\bf p}_2)  \equiv
\LL \pi_5({\bf p}_1)\pi_5({\bf p}_2)
\left| \bar\psi\psi( {\bf p} ) \right| 0 \RR $
is a constant, which does not depend on ${\bf R}$,
calculated to be~\cite{SCALAR:2006}
\be
b_{PP}( {\bf p}_1, {\bf p}_2 ) =
\frac{1}{a^3}\sqrt{ \frac{B_0^2}{4L^3} }
\frac{1}{ \sqrt{2 E_{\pi_P}({\bf p}_1)} }
\frac{1}{ \sqrt{2 E_{\pi_P}({\bf p}_2)} } ,
\label{b55:chap7}
\ee
where $B_0$ is the coupling constant,
$E_{\pi_P}({\bf p}) = \sqrt{ {\bf p}^2 + M_{U_5}^2 }$
and $M_{U_5}$ is the mass of Goldstone pion.
Please see details in Ref.~\cite{SCALAR:2006} or Appendix D.

Similarly, another channels can be also measured. That is,
\bea
x_{\rm AA}({\bf p}_1, {\bf p}_2 ) &=&
b_{\rm AA}({\bf p}_1, {\bf p}_2 ) \, x( p) \,
\delta^{(3)}{ ({\bf P}_H - {\bf P}_A  - {\bf p} ) } \\
x_{\rm VV}({\bf p}_1, {\bf p}_2 ) &=&
b_{\rm VV}({\bf p}_1, {\bf p}_2 ) \, x( p) \,
                        \delta^{(3)}{ ({\bf P}_H - {\bf P}_A  - {\bf p} ) } \\
x_{\rm II}({\bf p}_1, {\bf p}_2 ) &=&
b_{\rm II}({\bf p}_1, {\bf p}_2 ) \, x( p) \,
                        \delta^{(3)}{ ({\bf P}_H - {\bf P}_A  - {\bf p} ) } \\
x_{\rm TT}({\bf p}_1, {\bf p}_2 ) &=&
b_{\rm TT}({\bf p}_1, {\bf p}_2 ) \, x( p) \,
                        \delta^{(3)}{ ({\bf P}_H - {\bf P}_A  - {\bf p} ) }  ,
\eea
where the subscripts ${\rm AA}$, ${\rm VV}$, ${\rm II}$, and ${\rm TT}$
stand for the axial, vector, singlet, tensor channels respectively.
\bea
b_{AA}( { \bf p}_1, {\bf p}_2 ) &=&
\frac{1}{a^3}\sqrt{ \frac{B_0^2}{4 L^3} }
\frac{1}{ \sqrt{ 2 E_{\pi_A}({\bf p}_1)} }
\frac{1}{ \sqrt{ 2 E_{\pi_A}({\bf p}_2)} } \\
b_{VV}( { \bf p}_1, {\bf p}_2 ) &=&
\frac{1}{a^3}\sqrt{ \frac{B_0^2}{4 L^3} }
\frac{1}{ \sqrt{ 2 E_{\pi_V}({\bf p}_1)} }
\frac{1}{ \sqrt{ 2 E_{\pi_V}({\bf p}_2)} } \\
b_{II}( { \bf p}_1, {\bf p}_2 ) &=&
\frac{1}{a^3}\sqrt{ \frac{B_0^2}{4 L^3} }
\frac{1}{ \sqrt{ 2 E_{\pi_I}({\bf p}_1)} }
\frac{1}{ \sqrt{ 2 E_{\pi_I}({\bf p}_2)} } \\
b_{TT}( {\bf p}_1, {\bf p}_2 ) &=&
\frac{1}{a^3}\sqrt{ \frac{B_0^2}{4 L^3} }
\frac{1}{ \sqrt{ 2 E_{\pi_T}({\bf p}_1)} }
\frac{1}{ \sqrt{ 2 E_{\pi_T}({\bf p}_2)} }  ,
\eea
where $E_{\pi_V}({\bf p})$, $E_{\pi_A}({\bf p}),
$ $E_{\pi_I}({\bf p})$, and $E_{\pi_T}({\bf p})$ are denoted
like $E_{\pi_p}({\bf P})$.

The transition amplitude of the $\pi\pi$ channel
can be calculated by the formula
\be
\LL \pi\pi, A \left| V_I^{\rm lab} \right| H \RR
= x_{\pi\pi}(p) \delta^{(3)}{ ({\bf P}_H - {\bf P}_A  - {\bf p} ) }
\ee
where $V_I^{\rm lab}$ is denoted by Eq.~(\ref{V_I:model:ade}),
and the subscript $\pi\pi$ specifies the $\pi\pi$ channel.
After some algebra, we obtain
\bea
x_{\pi\pi}(p_1, p_2) &=&
b_{\pi\pi}( {\bf p}_1, {\bf p}_2 ) \, x( p) ,
\label{x_pipi:chap7}
\eea
where $ b_{\pi\pi}({\bf p}_1, {\bf p}_2) \equiv
\LL \pi\pi({\bf p}) \left| \bar\psi\psi( {\bf p} ) \right| 0 \RR $
is a constant, which does not depend on ${\bf R}$.
If we consider Table.~\ref{pipi_weight_TABLE}, we obtain
\be
b_{\pi\pi}^2({\bf p}_1, {\bf p}_2) =
\frac{1}{4}b_{PP}^2( {\bf p}_1, {\bf p}_2 ) +
           b_{AA}^2({\bf p}_1, {\bf p}_2) +
           b_{VV}^2({\bf p}_1, {\bf p}_2) -
\frac{3}{4}b_{II}^2({\bf p}_1, {\bf p}_2) +
\frac{6}{4}b_{TT}^2({\bf p}_1, {\bf p}_2).
\label{b_pipi:chap7}
\ee
In continuum limit, we can estimate
\be
b_{\pi\pi}({\bf p}_1, {\bf p}_2) =
\frac{1}{a^3}\sqrt{ \frac{3B_0^2}{4L^3} }
\frac{1}{ \sqrt{2 E_{\pi}({\bf p}_1)} }
\frac{1}{ \sqrt{2 E_{\pi}({\bf p}_2)} },
\ee
where $E_{\pi}({\bf p})$ is the energy of pion  with momentum ${\bf p}$.
In practice, we can rewrite Eq.~(\ref{x_pipi:chap7}) as
\bea
x_{\pi\pi}(p_1, p_2) &=&
\hat x_{\pi\pi}(p) \frac{1}{
\sqrt{ 2 E_{\pi}({\bf p}_1) }
\sqrt{ 2 E_{\pi}({\bf p}_2) } },
\label{x_pipi2:chap7}
\eea
where
\be
\hat x_{\pi\pi}(p) =
\frac{1}{a^3}\sqrt{ \frac{3B_0^2}{4L^3} } \, x( p ) .
\label{xhat_pipi:chap7}
\ee
We can also define a constant, that is.
\be
\hat b_{\pi\pi} =
\frac{1}{a^3}\sqrt{ \frac{3B_0^2}{4L^3} } .
\label{bhat_pipi:chap7}
\ee
In Chapter 8, we show how to
calculate the decay rate from $x_{\pi\pi}(p)$.

%% file: Chap8.tex
\chapter{ Our numerical results }
\label{ch:onr}
In this chapter we give our lattice numerical simulation results.
First we present the calculation of the mass of the $f_0$ meson.
Then we show the detailed  procedure to obtain
the lattice transition matrix element $x$,
and calculate the decay rate.

In this dissertation, we use 520 dynamical MILC fermion configurations
with $N_f = 2+1$ flavors of sea quarks
(namely, $a m_{u,d} = 0.005$, and $a m_{s} = 0.05$).
The coupling constant is $\beta=6.76$.
The lattice dimension is 24$^3$64,
with the lattice spacing $a$  of around 0.12 fm.
All the results we show in this chapter come
from this ensemble of gauge configurations.
%
\section{$f_0$ meson}
Recently there exists a growing interest in the $f_0$ meson.
The analyzes of the $\pi\pi$ scattering phase shift
suggested a brand resonance of the $f_0$ meson with $I=0$
and $J^{PC}=0^{++}$~\cite{Igi:1998gn}.
In this section we first discuss an interpolating
operator for the $f_0$ meson.
Then we give the detailed procedures for determining
its mass.
\subsection{Operator for $f_0$ meson}
We choose an operator with isospin $I=0$  and $J^{PC}=0^{++}$
at the sink and source,
\be
\hat{f_0}(x) \equiv \sum_{r=1}^{n_r}
\frac{ \bar{u}_r(x)u_r(x) + \bar{d}_r(x)d_r(x)}{\sqrt{2n_r}} ,
\ee
where $r  $ is the index of the taste replica,
      $n_r$ is the number of the taste replica, and
for notational simplicity we omit the color index and spin index.

The timeslice correlator for the $f_0$ meson is given by,
\be
C(t, {\bf k} )  = \sum_{\bf x}
\bigg\{
\langle \hat{f_0}({\bf x},t) \hat{f_0}({\bf 0},0) \rangle -
{\langle \hat{f_0} \rangle}^2
\bigg\}
e^{ i {\bf k \cdot x} }  ,
\label{scor0:ch7}
\ee
where the $x=({\bf x},t)$ is lattice position, ${\bf 0}$ is zero vector
(namely, ${\bf 0} = (0,0,0)$),
$\langle \hat{f_0} \rangle $ is the vacuum expectation value,
and ${\bf k}$ is the chosen momentum.

In Chapter 6, we discussed this correlator in detail.
From Ref.~\cite{Aubin:2003mg}, we know that
$n_r=1/4$ for the $1+1+1$ theory.
Thus, we obtained
\be
C(t,{\bf k})  =  C^{CC}(t,{\bf k}) -
\frac{1}{2} C^{DC}(t,{\bf k}) ,
\label{scor:ch8}
\ee
where $C(t,{\bf k})$ is described by Eq.~({\ref{CBB:EQDCVC}),
$C^{CC}(t,{\bf k})$ is described by Eq.~(\ref{CCEQ:BB}), and
$C^{DC}(t,{\bf k})$ is described by Eq.~(\ref{DCEQ:BB}).
The superscript $CC$ stands for the connected contribution.
The superscript $DC$ stands for the disconnected contribution.
%
%
%
%
\subsection{Procedures for determining the $f_0$ mass}
As we discussed in Chapter 6
we used the point sources and  the point sink operators
for the $f_0$  propagator.
For the point source, we just set a $1$ on only one point
on a chosen time slice, and  zero everywhere else.
In order to improve the statistics, when we calculate
the connected part,
we computed correlators from eight source time slices
evenly spread through the lattice
(i.e., only one source time slice was chosen at a time),
and averaged the correlators.
Moreover, for the nonzero momentum mesons
we used a quark source with $1$ and an antiquark source
with $e^{i {\bf k} \cdot {\bf s} }$ on only one point
on the chosen time slices,
where the ${\bf k}$ is our chosen momentum, and
      the ${\bf s}$ is the position of the quark source (see Fig. 6.1).

From Ref.~\cite{SCALAR:2006}, we know, for staggered quarks,
that the meson correlators
have the generic single-particle contribution~\cite{Bernard:2001av}
\be
\label{sfits:ch7}
{\cal C}(t) =
\sum_i A_i \left( e^{-m_i t} + e^{-m_i(N_t-t)} \right) +
\sum_i A_i^{'}(-1)^t
\left(  e^{-m_i' t} + e^{-m_i'(N_t-t)}  \right)  ,
\ee
where the oscillating terms correspond to a particle with
opposite parity.
In our case for the $f_0$ meson correlator,
we assume only one mass with each parity in the fits
of Eq.~(\ref{sfits:ch7}).~\footnote{
In Ref.~\cite{SCALAR:2006} we show that, in our concrete calculation,
our  operator is the state with $I \otimes I$
and  state of $\gamma_0\gamma_5 \otimes \gamma_0\gamma_5$
(its oscillating term).
}
From the discussion in Appendix B, we must consider the
two meson contributions. Therefore, we have
\be
{\cal C}(t) = A_1\left( e^{-m_1 t} + e^{-m_1(N_t-t)} \right) +
       A_2(-1)^t \left( e^{-m_2 t} + e^{-m_2(N_t-t)} \right)
       + B^{S\chi PT}_{f_0}(t) ,
\label{scalar:ch7}
\ee
where $B^{S\chi PT}_{f_0}(t)$ is denoted in Eq.~(D.14). If we
rearrange the terms in Eq.~(D.14), we obtain
\be
B^{S\chi PT}_{f_0}(t) = \frac{B_0^2 }{ 4 L^3 }
\Bigg \{ {\rm f_{Bubble}}(t) + \delta_{V} {\rm f_{Vhairpin}}(t)
        + \delta_A {\rm f_{Ahairpin} }(t)  \Bigg \} ,
\ee
here $B_0$ is the coupling constant,
$\delta_V=a^2\delta_V^{\prime}$ is the hairpin coupling of
a pair of taste-vector mesons,
$\delta_A=a^2\delta_A^{\prime}$ is the hairpin coupling of
a pair of taste-axial mesons,
and
\bea
{\rm f_{Vhairpin}}(t) &\equiv& \frac{1}{\delta_V} {\sum_{\bf k}} \left\{
-4\FZW {U_V}  \right.
\nonumber\\ &&  \hspace{-2.6cm}
+C_{V_{\eta }}^2 \FZW {\eta  V} + C_{V_{\eta'}}^2 \FZW {\eta' V}
\nonumber\\ &&  \hspace{-2.6cm}
-C_{V_{\eta}}C_{V_{\eta'}} \left.
\left[ \OTS {\eta  V} {\eta' V} + \OTS {\eta'  V} {\eta V}  \right]
\right\}  \\
{\rm f_{Bubble}(t) }  &\equiv& {\sum_{\bf k}} \left\{
\frac{1}{9} \FZW {\eta I}  -\FZW {U_I} \right.
\nonumber\\ &&  \hspace{-2.6cm}  \left.
+\frac{1}{16}\sum_{b=1}^{16} \left[ 4\FZW {U_b} + \FZW {K_b} \right]
\right\} ,
\label{btt:a0}
\eea
where $\delta_V$, $\delta_A$, $B_0$, $C_{V_{\eta }}$,
$C_{V_{\eta'}}$, $C_{A_{\eta }}$, $C_{A_{\eta'}}$
are constants, which are given in Appendix B,
$M_{\eta V}$, $M_{\eta A}$, $M_{\eta' V}$, $M_{\eta' A}$ are
given in Appendix B, and for ${\rm f_{Ahairpin}}(t)$, we just require
$V \to A$ in ${\rm f_{Vhairpin}}(t)$.

The tree-level masses of the mesons are~\cite{Aubin:2003mg}
\be
M_{ff^\prime_{~\!b}}^2 = B_0(m_f+m_{f^\prime}) + a^2\Delta_b ,
\ee
where $\bar f f^\prime$ are two flavors which make up,
$b=1,..,16$ are the taste, and
the term of $a^2\Delta_b$ comes from taste symmetry breaking.
Here for brevity, we denote
\be
M_{U_b} \equiv M_{\pi_b}\equiv M_{uu_{~\!\!b}}
\equiv M_{dd_{~\!\!b}}  \equiv M_{ud_{~\!\!b}}, \ \
M_{S_b} \equiv M_{ss_{~\!\!b}}, \ \
M_{B_b} \equiv M_{us_{~\!\!b}}  .
\ee
By convention the $M_{U_P}$ is the mass of the Goldstone pion, and
              the $M_{K_P}$ is the mass of the Goldstone kaon.

All the masses needed for Eq.~(8.7) are well known.
From the mass-squared splittings in Table III of Ref.~\cite{Aubin:2004fs}
also the $aM_{U_P} = 0.1594$, $aM_{K_P} = 0.36523$
in Ref.~\cite{Aubin:2004wf}, we obtain
\bea
aM_{U_A} = 0.2342, \ \ aM_{U_T} = 0.2694, &&\hspace{-0.3cm}
aM_{U_V} = 0.2966, \ \ aM_{U_I} = 0.3205.  \nonumber \\
aM_{K_A} = 0.4036, \ \ aM_{K_T} = 0.4250, &&\hspace{-0.3cm}
aM_{K_V} = 0.4428, \ \ aM_{K_I} = 0.4591.  \nonumber
\eea
The $M_\eta$ is estimated from the Gell-Mann-Okubo formula
\be
m_\eta^2 = \frac{ m_\pi^2 + 2m_S^2}{3} ,
\ee
that is,
\be
aM_{\eta I} = 0.4958 .
\ee

The first term ($J^{PC} = 1^{++}$) in Eq.~(\ref{scalar:ch7})
gives the mass of the $f_0$ meson,
and the oscillating term ($J^{PC} = 1^{+-}$) in Eq.~(\ref{scalar:ch7})
gives the mass of the $\eta_{A}$ meson.
In Ref.~\cite{SCALAR:2006}, we do a fit to all of the correlator data at once.
In this grand fit, the masses of the $f_0$ meson, $a_0$ meson, $\eta_{05}$ meson,
and $\pi_{05}$ meson, the couplings of $\delta_V$, $\delta_A$, and $B_0$ were determined.
We choose all the data within the fitting range of t
from $5$ to $18$ (namely, the minimum distances $t_{ \rm min} = 5 $).
Our fitting gives an acceptable result with a $\chi^2$ of 127 with 114
degrees of freedom, with
\be
  am_{f_0}  =  0.47(7)
\ee
From Ref.~\cite{Aubin:2004wf}, for our set of configurations,
we know
\be
a^{-1} = 1.634(45) \, {\rm GeV} ,
\ee
then gets
\be
  m_{f_0}  =  ( 768 \pm 136 ) \, {\rm MeV} .
\ee
Also we get
\be
  a B_0  =  3.12 \pm 0.42 .
\ee
In next section we use this number to evaluate the decay rate.
%
%
\section{Lattice transition matrix element}
From the discussion of the symmetry of the wave function in Sec.~6.3,
we know the possible minimum momentum is
$\DPT {\bf k}=(1,0,0)\frac{2\pi}{L}$
for the scalar meson creation~\cite{McNeile:2002az}.
In this section, we will show the detailed procedures to
obtain the overall decay rate at this chosen momentum.
Then we will consider the momentum of $\DPT {\bf k}=(1,1,0) \frac{2\pi}{L}$, \
$\DPT {\bf k}=(1,1,1)\frac{2\pi}{L}$, and
$\DPT {\bf k}=(2,0,0) \frac{2\pi}{L}$.
From the data of these four chosen momenta, we can obtain
a decay rate that has physical meaning.
%
%
%
\subsection{The wave function of the excited gluonic states}
In Chapter 4, we used the LBO approximation to solve for the gluonic field around
a static quark-antiquark $Q\overline{Q}$ at separation $R$ to
determine first the static potential. Then we solved the
Schr\"odinger equation with that potential to determine its spectrum and wave function.
From Fig.~\ref{fig:wf}, we note that the potential ($\Pi_u$)
that binds the hybrid meson is comparatively flat.
Hence, the hybrid meson has a broad radial wave function,
which we denote by $u_H$ in Chapter 4.
This observation allows a quick estimate for the decay rate.
For example, in Chapter 5, we justified the decay
from the hybrid meson state to the P-wave $\chi_b$ state
(i.e., we refer to the wave function of this state as $u_{\chi_b}$ in
Chapter 4).
In order to integrate over $R$ easily, we rescale Fig.~\ref{fig:wf}
in terms of lattice spacing $a$, and show it in Fig.~\ref{wfR:ch8}.
We see that the lack of the nodes in the relevant
wave functions results in a quite large spatial wave function
overlap factor. Moreover if we assume the transition rate
is independent of $R$,  we obtain
\be
\int u_{H} \ u_{\chi_b} dR = 0.8643 ,
\ee
where $u_{H}$, $u_{\chi_b}$ are normalized.
From the other side, this big overlap factor justifies the decay
from the hybrid meson state to the P-wave $\chi_b$ state.
Later in this chapter, we calculate the decay rate of this channel.
\begin{figure}[t]
\begin{center}
\epsfxsize=8.0cm\epsfbox{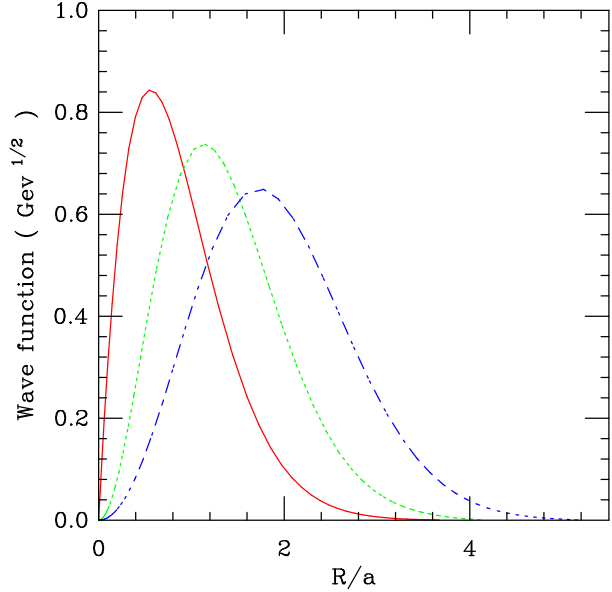}
\vspace{-0.2cm}
\caption{ \label{wfR:ch8}
Reduced radial wave function versus $R$ in units of $a$.
The solid curve, dotted curve, and  short dash curve specify
the reduced radial wave functions for the $1S$, $1P$ states
of $\Sigma_g^+$, and $\Pi_u$ hybrid state, respectively.
}
\vspace{-0.7cm}
\end{center}\end{figure}
%
%
%
%
\subsection{Simulation procedure}
In Chapter 6 we have already discussed how to create operators
for the $\Pi_u$, $\Sigma_{g}^{+}+ S(0^{++})$, and we use
Eq.~(\ref{chab:eq6}) to calculate the correlator $C_{HAB}$.
We also discussed the symmetry of the wave function $w({\bf r})$.

In practice we evaluate the difference of two Wilson loops.
As we discussed in Chapter 6,  we  call this observable $\AE({\bf r})$ and
its spatial Fourier transform $\AE({\bf k})$.
The fermion loop of light quark meson ($f_0$ meson)
is evaluated at each spatial point ${\bf s}$ at time
$t+T$ and its Fourier transform is $S({\bf k})$.
We actually used 20 iterations of $APE$ smearing
for both spatial ends of $\AE$.
The transverse extent of the $\Pi_u$ end is two lattice spacings.
We summed over all possible orientation to improve statistics.
%
\subsection{Our numerical results for decay rate}
For notational simplicity, we focus on the chosen momentum
$\DPT {\bf k}=(1,0,0)\frac{2\pi}{L}$.
For scalar decays, we choose the wave function as in
Eq.~(\ref{ww:ch6}). According to Eq.~(\ref{chab:eq6:2}),
in terms of these components of the momentum, we estimate
the required  correlator $C_{HAB}(t, {\bf R, k})$.
\begin{figure}[b]
\begin{center}
\epsfxsize=8cm\epsfbox{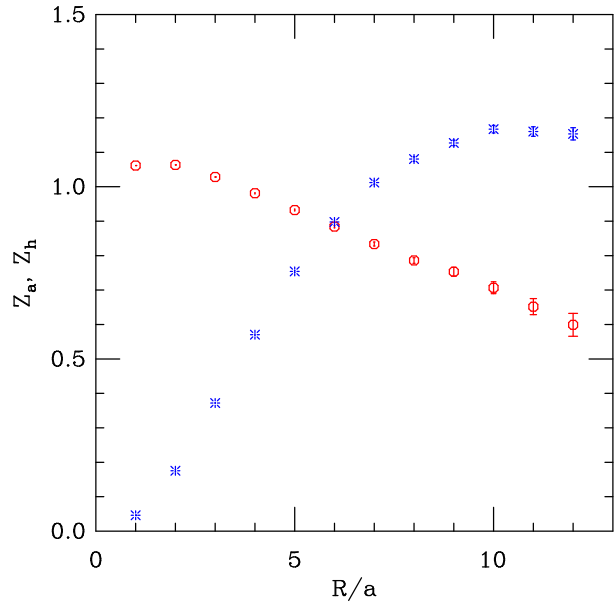}
\vspace{-0.2cm}
\caption{ \label{a_h.fig}
The overlap amplitudes of $Z_a$ and $Z_h$ versus $R$,
the burst one is for $Z_h$, and the octagon one is for $Z_a$.
}
\vspace{-0.7cm}
\end{center}\end{figure}
From Eq.~(\ref{xrate:FullQCD2}), we evaluate the
bare transition amplitude $x({\bf k, R}, t)$
at each $t$ and ${\bf R}$ for fixed momentum ${\bf k}$, that is,
\be
{ x({\bf k, R}, t) } =
\frac{ C_{HAB}(t, {\bf R, k}) \, Z_a \, Z_h }
{\DPT \sum_{t' = 0}^{t-1}
C_{AA}({\bf R}, t \MINUSONE t') \
C_{HH}({\bf R}, t') \
C_{BB}(t \MINUSONE t',{\bf k}) } .
\ee

We measure $Z_h$ from the correlator $C_{H-H}(t)$ from $H \to H$
as explained in Sec.~4.4.2,
and $Z_a$ from the correlator $C_{A-A}(t)$ from $A \to A$
as explained in Sec.~4.4.1.
The $Z_a$ and $Z_h$ values for different $R$ are shown
in Fig.~\ref{a_h.fig}, where the burst  one is for $Z_h$, and
the octagon one is for $Z_a$.

When we evaluate the bare transition amplitude $x({\bf k, R}, t)$,
we use jackknife method.
Here we should point out when we construct the jackknife sample,
we just throw out one measurement.

For illustration, the values of $x({\bf k, R}, t)$
for $ \DPT \frac{R}{a} = 1$ and momentum
$\DPT {\bf k}=(1,0,0)\frac{2\pi}{L}$
at different $t$ are shown in Fig.~\ref{x_1_1.fig}.
For the range from $t=2$ to $t=5$, there exists a plateau for
$x({\bf k, R}, t)$. We fit this plateau to get
$x({\bf k, R}) = 0.3253 \pm 0.0093$
with a $\chi^2$ of 0.03 for 3 degrees of freedom.
\begin{figure}[t]
\begin{center}
\epsfxsize=8cm\epsfbox{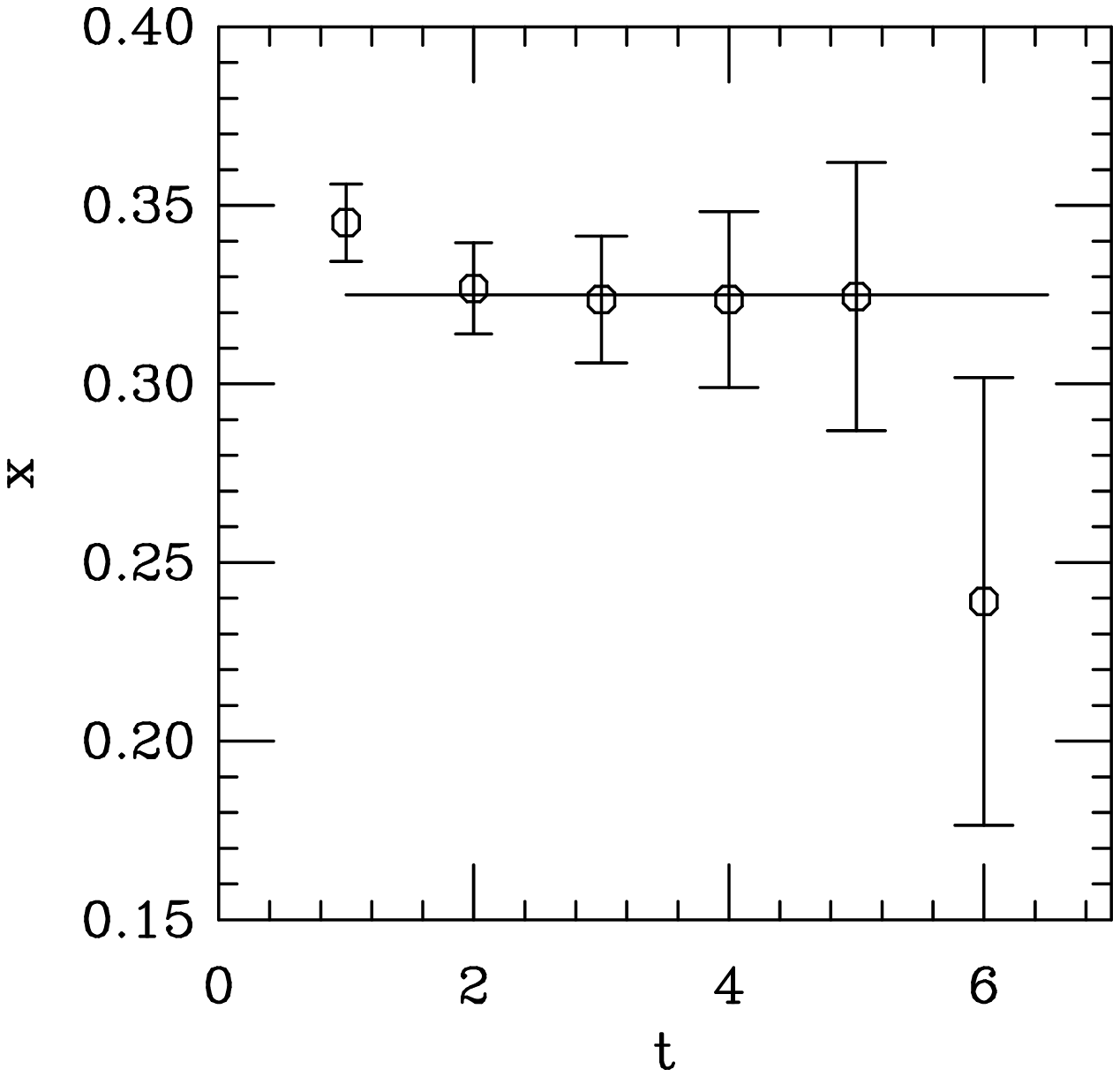}
\vspace{-0.2cm}
\caption{ \label{x_1_1.fig}
The  $x({\bf k, R}, t)$ values for $\DPT \frac{R}{a} = 1$  versus $t$
with $\DPT {\bf k}=(1,0,0)\frac{2\pi}{L}$.
}
\vspace{-0.7cm}
\end{center}\end{figure}

In our lattice simulation we obtain the $x({\bf k, R}, t)$ values
shown in Fig.~\ref{xfig100_1_6} and Fig.~\ref{xfig100_7_12}.
From these data we can extract the values of
$x({\bf k, R})$ for momentum $\DPT {\bf k}=(1,0,0)\frac{2\pi}{L}$
at all separations $R$.
In Table~\ref{x_TABLE}, for $\DPT {\bf k}=(1,0,0)\frac{2\pi}{L}$
we list all the values of $x({\bf k, R})$ for
different $R$, fitting range, etc.
Also we plot the values of $x({\bf k, R})$ in Fig.~\ref{satu:ch7}.
\begin{figure}[ptbh!]
\begin{center}
\epsfxsize=8cm\epsfbox{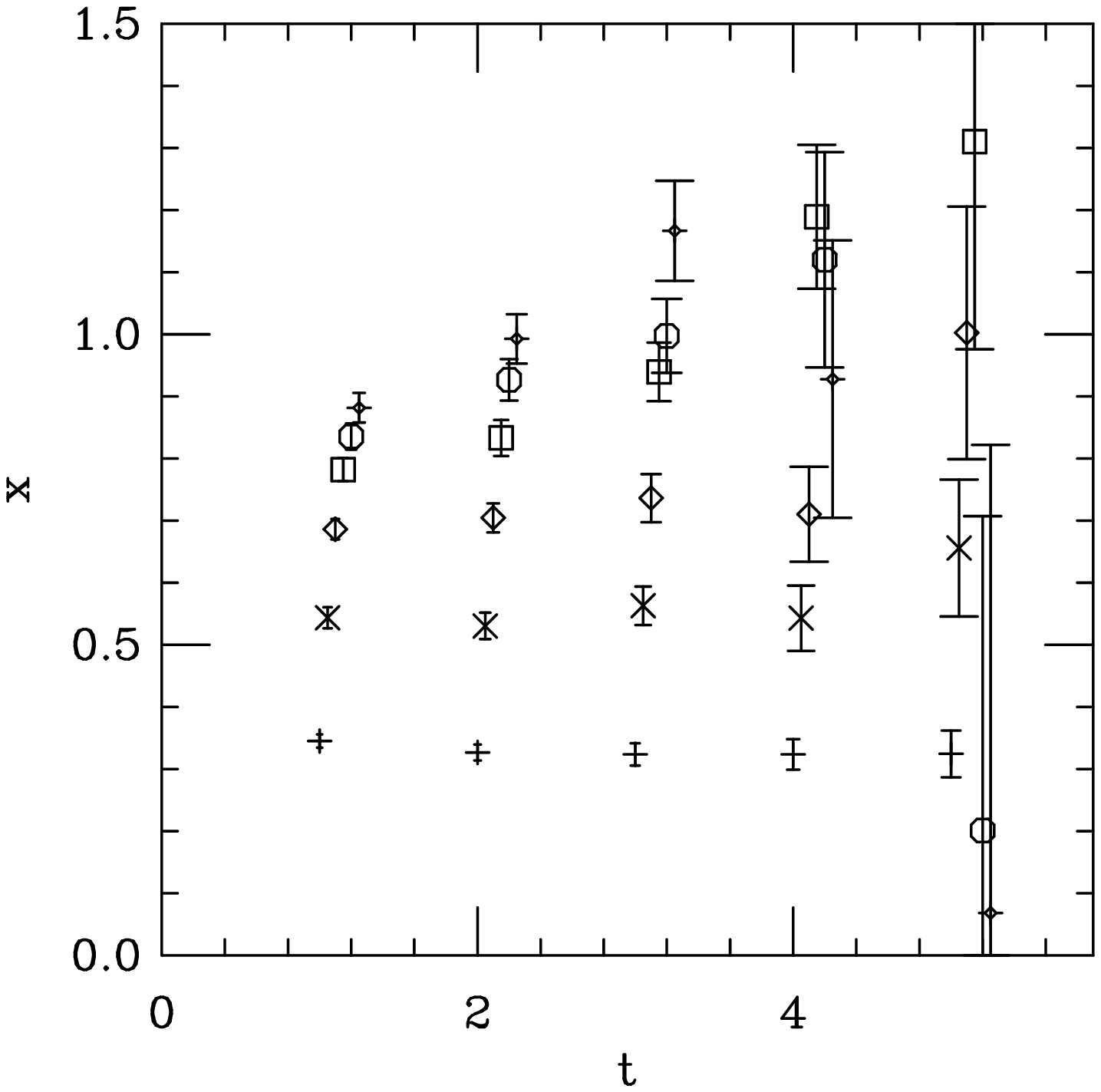}
\caption{ \label{xfig100_1_6}
The $x({\bf k, R}, t)$ values with momentum
$\DPT {\bf k}=(1,0,0)\frac{2\pi}{L}$ versus $t$.
Here values for $\DPT \frac{R}{a} = 1, 2, \cdots, 6$  are
represented by the  plus sign, cross, diamond,
square, octagon, and fancy diamond,  respectively.
}
\vspace{-0.3cm}
\end{center}\end{figure}
\begin{figure}[ptbh!]
\begin{center}
\vspace{-0.3cm}
\epsfxsize=8cm\epsfbox{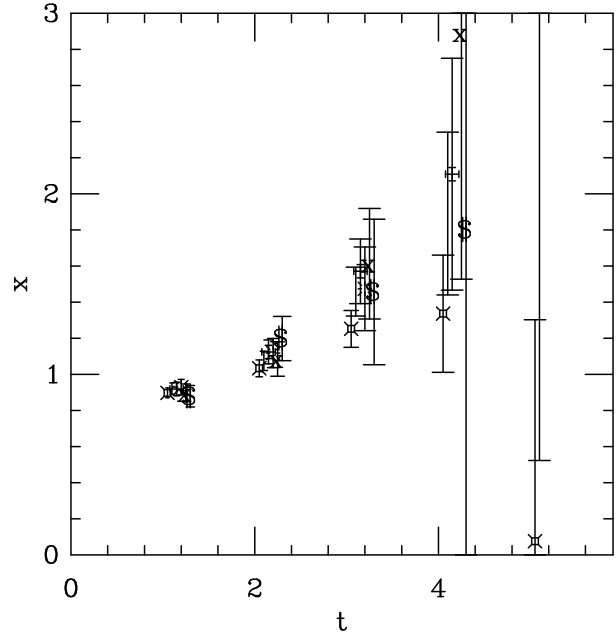}
\caption{ \label{xfig100_7_12}
The $x({\bf k, R}, t)$ values with momentum
$\DPT {\bf k}=(1,0,0)\frac{2\pi}{L}$  versus $t$.
Here values for $\DPT \frac{R}{a} = 7, 8, \cdots, 12$ are
represented by  fancy square, fancy cross, fancy plus,
burst, cross, and dollar sign, respectively.
}
\vspace{-0.3cm}
\end{center}\end{figure}
\begin{table}[ptbh!]
\caption{ \label{x_TABLE}
The $x({\bf k, R})$ values with momentum
$\DPT {\bf k}=(1,0,0)\frac{2\pi}{L}$.
The first column is the separate distance, and
Column two is the transition amplitude.
The remaining columns are the time range for the chosen fit,
$\chi^2$ and number of degrees of freedom for the fit.
}
\begin{center}
\begin{tabular}{|l|c|c|c|l|}
\hline
$R/a$     &  $ax$          & Fitting range &  $\chi^2/D$  \\
\hline
1          & 0.3215(105)    &2--5  &  0.0266/3      \\
\hline
2          & 0.5430(196)  &2--5  &  1.81/3      \\
\hline
3          & 0.7170(272)  &2--5  &  2.51/3       \\
\hline
4          & 0.8636(303)  &2--3  &  3.71/1       \\
\hline
5          & 0.9520(382)  &2--4  &  2.02/2       \\
\hline
6          & 1.1328(826)  &3--5  &  3.01/2       \\
\hline
7          & 1.270(119)  &3--5  &  0.987/2     \\
\hline
8          & 1.511(163)   &3--5  &  1.07/2       \\
\hline
9          & 1.615(189))  &3--5  &  1.36/2      \\
\hline
10         & 1.567(240)   &3--5  &  3.54/2     \\
\hline
11         & 1.672(297)   &3--5  &  0.893/2   \\
\hline
12         & 1.465(395)   &3--5  &  0.16/2      \\
\hline
\end{tabular}
\end{center}\end{table}
%
\begin{figure}[ptbh!]
\begin{center}
\epsfxsize=8cm\epsfbox{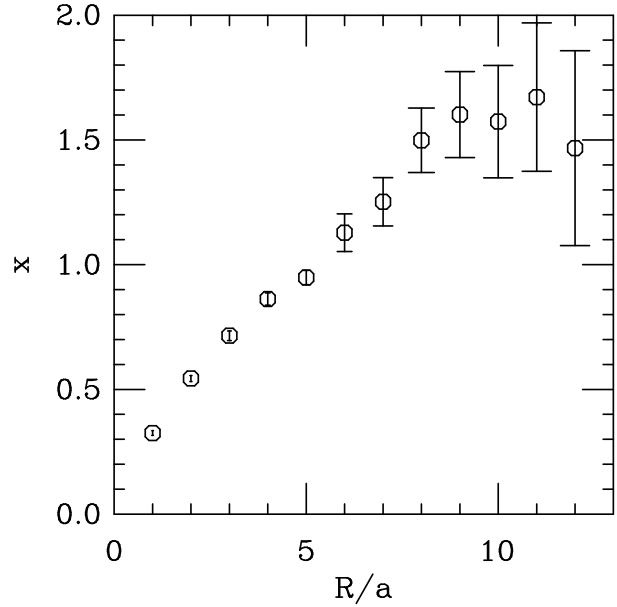}
\caption{
The $x({\bf k, R})$ values with momentum
$\DPT {\bf k}=(1,0,0)\frac{2\pi}{L}$ versus $R$.
}
\label{satu:ch7}
\end{center} \end{figure}
From Fig.~\ref{satu:ch7}, we note
an increase of $x({\bf k}, R)$  with respect to $R$.
and a hint of saturation at large $R$.
One approach to interpret the $R$-dependence of  $x({\bf k}, R)$
is  by noting that the
scalar  meson wave function $w$ has a node at the center of the $\Pi_u$ state
in the transverse  direction, since the $w$ is odd in this direction,
and therefore it is sensitive to the transverse width of
the excited gluonic flux~\cite{Green:1996be} in the $\Pi_u$ state.
Just as we find, this can explain the increase~\cite{Green:1996be}
with longitudinal extent $R$ leading  to a plateau.

Now we have the bare transition amplitude values $x({\bf k},R)$.
These values depend on $R$. In Appendix D, we develop a formula
to integrate out $R$.
From Eq.~(\ref{WLY:FZW1}), we obtain the overall transition amplitude
for fixed momentum ${\bf k}$ (i.e., $\tilde x({\bf k})$),
\be
\tilde  x({\bf k}) = \frac{1}{2} \int u_{H}(R)
\ x(R, {\bf k}) \
u_{\chi_b}(R) \ dR .
\label{Detar:chap8}
\ee
We obtained the values of $u_{H}(R)$ and $u_{\chi_b}(R)$
from the LBO treatment in Chapter 4,  and we also plot them in
Fig.~\ref{wfR:ch8}.
From Fig.~\ref{wfR:ch8}, we note that when $R > 3a$,
the value of $u_{\chi_b}(R)$ is approximate close to zero.
Hence, only values of $x(R, {\bf k})$ for $R=1a, 2a, {\rm and} \, 3a$
mainly contribute to the $\tilde  x({\bf k})$ value.

From the values of $x(R, {\bf k})$, $u_{H}(R), u_{\chi_b}(R)$,
we can get an overall $x$ value
for momentum $\DPT {\bf k}=(1,0,0)\frac{2\pi}{L}$
by Eq.~(\ref{Detar:chap8}), that is,
\be
a\tilde x(100) = 0.1693(60)  ,
\ee
where the $100$ specifies the chosen momentum.

From the same procedure, we can get  overall $x$ value
for momentum $\DPT{\bf k}=(2, 0, 0) \frac{2\pi}{L}$, that is,
\be
a\tilde x(200) = 0.2007(50)  ,
\ee
the overall  $x$ value for momentum $\DPT {\bf k}=(1, 1, 0) \frac{2\pi}{L}$,
that is,
\be
a\tilde x(110) = 0.1482(41)  ,
\ee
and  the overall  $x$ value for momentum $\DPT{\bf k}=(1,1,1) \frac{2\pi}{L}$,
that is,
\be
a\tilde x(111) = 0.1115(28)  .
\ee
For concreteness,
in Tables~\ref{x_TABLE_200},~\ref{x_TABLE_110}, and~\ref{x_TABLE_111},
we list the values of $x({\bf k}, R)$ for different $R$
with momentum $\DPT {\bf k}=(2,0,0)\frac{2\pi}{L}$,
$\DPT {\bf k}=(1,1,0)\frac{2\pi}{L}$,
and $\DPT {\bf k}=(1,1,1)\frac{2\pi}{L}$ respectively.
Since when $R > 3a$, the values of
$u_{\chi_b}(R)$ are negligible, and only $x(R, {\bf k})$
of $R=1a,2a,3a$ mainly contribute to the $\tilde  x({\bf k})$ value.
We omitted the results for $R>6a$ in the table.

\begin{table}[t]
\vspace{-0.3cm}
\caption{ \label{x_TABLE_200}
The $x({\bf k, R})$ values with momentum
$\DPT {\bf k}=(2,0,0)\frac{2\pi}{L}$.
The first column is the separate distance, and
Column two is transition amplitude.
The remaining columns are the fitting range,
$\chi^2$ and number of degrees of freedom for the fit.
}
\begin{center}\begin{tabular}{|l|c|c|c|l|}
\hline
$R/a$     &  $ax$          & Fitting range &  $\chi^2/D$  \\
\hline
1  &  0.4007(114)    & 2-5 & 0.952/3 \\
\hline
2  &  0.6433(181)   & 2-5 & 1.8/3 \\
\hline
3  &  0.8515(221)   & 2-5 & 0.658/3 \\
\hline
4  &  1.0001(242)   & 2-5 & 3.85/3 \\
\hline
5  &  1.0006(276)   & 2-4 & 1.51/2 \\
\hline
6  &  1.2160(310)   & 2-4 & 2.52/2 \\
\hline
\end{tabular}
\vspace{-0.6cm}
\end{center}\end{table}
\vspace{-0.2cm}
\begin{table}[h]
\caption{ \label{x_TABLE_110}
The $x({\bf k, R})$ values with momentum
$\DPT {\bf k}=(1,1,0)\frac{2\pi}{L}$.
The first column is the separate distance, and
Column two is transition amplitude.
The remaining columns are the fitting range,
$\chi^2$ and number of degrees of freedom for the fit.
}
\begin{center}\begin{tabular}{|l|c|c|c|l|}
\hline
$R/a$     &  $ax$    & Fitting range &  $\chi^2/D$  \\
\hline
1  & 0.2909(85)   &  2-5  & 0.344/3  \\
\hline
2  & 0.4695(150)  &  2-5  & 0.841/3   \\
\hline
3  & 0.6352(174)  &  2-5  & 3.1 /3   \\
\hline
4  & 0.7512(217)  &  2-4  & 3.35/2  \\
\hline
5  & 0.9205(483)  &  3-5  & 0.0467/2 \\
\hline
6  & 1.0056(620)  &  3-5  & 0.0896/2 \\
\hline
\end{tabular}
\vspace{-0.2cm}
\end{center}\end{table}

From our grand  fit in Ref.~\cite{SCALAR:2006}, we found
\be
a B_0  =  3.12 \pm 0.42 .
\ee
Hence, from Eq.~(\ref{b_pipi:chap7}), we have
\be
a^4 \hat b_{\pi\pi} =  0.0230(31) .
\ee
Therefore, from Eq.~(\ref{WLY:FZW4}),
we obtain the transition amplitute of the $\pi\pi$ channel
for momentum $\DPT{\bf k}=(1, 0, 0) \frac{2\pi}{L}$, that is,
\be
a^5 \hat x(100) =  0.00389(66) ,
\label{x_pipi_100:ch8}
\ee
where the $100$ specifies the chosen momentum  ${\bf n}=(1, 0, 0)$.
From the same procedure, we can get  overall transition amplitude
$x$ for other three momentum. We list all the results of transition
amplitude in Table~\ref{x_TABLE_ALL},
where the first column is the chosen momentum,
and the second column is overall transition amplitude.

\begin{table}[t]
\vspace{-0.3cm}\caption{
The $x({\bf k, R})$ values with momentum
$\DPT {\bf k}=(1,1,1)\frac{2\pi}{L}$
The first column is the separate distance, and
the second column transition amplitude.
The remaining columns are the fitting range,
$\chi^2$ and number of degrees of freedom for the fit.
\label{x_TABLE_111}
}
\begin{center}
\begin{tabular}{|l|c|c|c|l|}
\hline
$R/a$ &  $ax$          & Fitting range &  $\chi^2/D$  \\
\hline
1  &  0.2742(65)   &  2-5  &0.0458/3  \\
\hline
2  &  0.3843(105)   &  2-5  &0.672/3 \\
\hline
3  &  0.4493(117)  &  2-5  &1.85/3  \\
\hline
4  &  0.3425(132)  &  2-4  &5.11/2  \\
\hline
5  &  0.3540(308)  &  3-5  &2.41/2  \\
\hline
6  &  -0.00207(436) &  3-5  &2.79/2  \\
\hline
\end{tabular}
\vspace{-0.5cm}
\end{center}\end{table}
\begin{table}[h]
\caption{
The overall transition amplitude $x$,
The first column is the chosen momentum,
and Column two is overall transition amplitude.
\label{x_TABLE_ALL}
}
\begin{center}\begin{tabular}{|c|c|}
\hline
${\bf k}$ &  $a^5 \hat x$      \\
\hline
$\DPT (1,0,0) 2\pi/L$ &  0.00389(66) \\
\hline
$\DPT (2,0,0) 2\pi/L$ &  0.00461(74) \\
\hline
$\DPT (1,1,0) 2\pi/L$ &  0.00341(55) \\
\hline
$\DPT (1,1,1) 2\pi/L$ &  0.00256(41) \\
\hline
\end{tabular}
\vspace{-0.3cm}
\end{center}\end{table}

From these $\hat x$-values,
in Appendix D we relate the overall LBO decay amplitude $x(p)$
to the lattice transition amplitudes $x_{\pi\pi}({\bf k})$ as
Eq.~(\ref{xpp:fzw:wly}). That is,
we can use the follow formula
\be
\hat x_{\pi\pi}(p) =
\frac{1}{L^3}\sum_{k_x \ne 0, k_y, k_z}\sum_{x,y,z} a^3
N_s j_1(p r) Y_{11}(\theta, \phi) \,
\hat x({\bf k})
\sin(k_x x) \cos(k_y y) \cos(k_z z) .
\ee
This result gives the transition amplitude $x$ at any given momentum $p$.
Here $N_s$ is a normalization factor,
$j_1(pr)$ is the spherical Bessel function, and
$Y_{1,m_s}(\theta,\phi)$ is the spherical harmonic function.
In terms of our measured values, we have
\bea
\hat x_{\pi\pi}(p) &=&  \frac{1}{L^3} \sum_{x,y,z} a^3
N_s j_1(pr) Y_{11}(\theta, \phi) \times
\Bigg\{
\hat x(100) \sin(x) + \hat x(200) \sin(2x) \nonumber \\
&+&
\hat x(110) \sin(x) \cos(y) +  \hat x(111) \sin(x) \cos(y) \cos(z)
\Bigg\} ,
\label{x4comp:ch8}
\eea
where we sum over just a few small values of $k$.
With these values in Table~\ref{x_TABLE_ALL},
we evaluate $x_{\pi\pi}(p)$ at any given momentum $p$.
We show the transition amplitude $x_{\pi\pi}(p)$ versus momentum $p$
in Fig.~\ref{xpg:ch7}.
Since we just measure a few $x's$ with small
momentum. As discussed in Appendix D, the values shown in
Fig.~\ref{xpg:ch7} for high momentum are not reliable.
We test that it is reliable for the small momentum $p$.

\begin{figure}[t]
\begin{center}
\epsfxsize=9cm\epsfbox{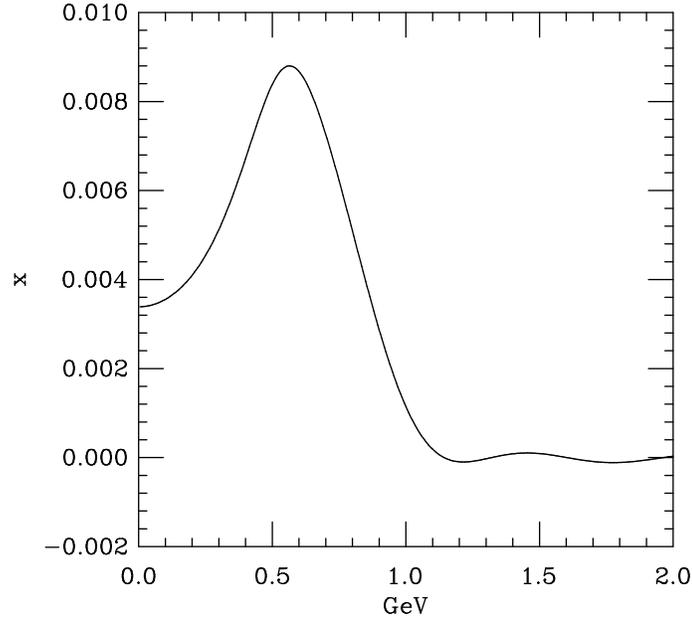}
\vspace{-0.2cm}
\caption{
The transition amplitude $x(p)$  versus momentum $p$.
}
\vspace{-0.6cm}
\label{xpg:ch7}
\end{center} \end{figure}

For the decay
\be
   H \rightarrow \chi_b + \pi + \pi \, ,
\ee
From Fermi's Golden Rule for the transition rate $\Gamma$,
we need to calculate the phase factor. Because the calculation of
phase factor is tedious. We show the detailed procedure in
Appendix E to obtain a formula to evaluate the transition rate $\Gamma$,
\be
\Gamma = 8\pi^3  \left(\frac{La}{2\pi}\right)^6
\int_0^{p_{\rm max}}  dp
\ p^2 \, \hat x_{\pi\pi}(p)^2 \,
\sqrt{ \frac{1}{4} - \frac{m_\pi^2}{ (\Delta M)^2 - p^2 } },
\label{gamma_pipi:ch8}
\ee
where $M_A$ is the mass of the $\chi_b$ state,  $M_H$ is the mass of
the $H$ state, $\Delta M = M_H - M_A$,  and
\be
p_{\rm max} = \sqrt{(\Delta M)^2 - 4 m_\pi^2} .
\ee
Here $m_\pi$ is the mass of pion meson.

For our simulation, we have $M_H = 10.828$~GeV and $M_A = 9.877$~GeV, hence
$\Delta M = 0.951$~GeV, thus $p_{\rm max} = 0.83 \, {\rm GeV}$.
We can estimate $x_{\pi\pi}(p)$ by Eq.~(\ref{x4comp:ch8}).
Hence, by integrating Eq.~(\ref{gamma_pipi:ch8}),
we can obtain overall transition rate
\be
\Gamma = 3.62(98) \, {\rm MeV} ,
\ee
for the decay $H \rightarrow \chi_b + \pi + \pi $.
In Chapter 7, we use the perturbation expansion for $C_{HAB}$
correlator to find a way to
extract the transition amplitude $x$.
To use this method, we should discuss whether it is justified or not.
The condition we need is $ \Gamma  \ll \Delta E $.
Since $\Gamma  \ll \Delta E$,
our lattice method for decay rate is justified. 

%% file: Chap9.tex
\chapter{ Conclusion }
In this chapter, we first discuss the evidence for the $f_0$ meson.
Then we discuss the evidence for light quark hybrid exotic mesons.
Next we discuss hybrid exotic quarkonium decay.
Finally, we summarize our results.

\section{Evidence for $f_0$ meson}
For the past 45 years, the question of the existence and nature
of the $f_0$-meson has attracted the attention of
many authors~\cite{Hatsuda:2001da}.
The subjects of investigation are the internal properties of the
$f_0$-meson and its role as an intermediate particle
in various processes,  both in vacuum and in hot and dense matter.
It appears as a broad enhancement in the
isospin $I=0$, $J=0$ \,
$\pi\pi$ scattering amplitude.
The experimental value of the $f_0$-meson mass
is not accurately determined and lies in a wide interval~\cite{PDBook}
\be
M_{f_0}= 400 - 1200 \, {\rm MeV} ,
\ee
This uncertainty is explained by large values of the decay width of this meson
into two pions~\cite{PDBook}
\be
\Gamma_{f_0} = 600 - 1000 \, {\rm MeV} .
\label{expgsig}
\ee

Recently, the Fermilab experiment E791~\cite{Aitala:2002kr}
showed strong evidence for the existence of an $I=0$, $J=0$ \,
$\pi^+\pi^-$ resonance in charmed $D^+$ meson decay.
The $\pi^+ \pi^-$ resonance in E791 is compatible with the scalar
meson $f_0(500)$. It was also observed in the Cabbibo-suppressed
decay $D^+ \to  \pi^- \pi^+ \pi^+$~\cite{Aitala:2002kr}.
The $D^+ \to f_0(500) \pi^+$ decay contribution is dominant,
being responsible for approximately half of the decays
thought the particular resonant sequence:
$D^+ \to \pi + f_0 \to \pi^-\pi^+\pi^+$ decay~\cite{Aitala:2002kr}.
The measured mass and width of the $f_0$  meson are
$ 478 \pm 31 $MeV and $338 \pm 48$~MeV, respectively.
Several other experimental results can be explained by
the existence of the $f_0$ meson resonance~\cite{Augustin:1988ja}\cite{Amsler:1995bz}\cite{Antinori:1995wz}\cite{Ishida:1995xx}\cite{Ishida:1999ms}.
%
%
\section{Evidence for light-quark hybrid exotic mesons}
There are two ways to study hybrid exotic mesons.
One approach is to use lattice QCD calculation.
A summary of current LQCD calculations for the lowest
lying light quark ($u$ or $d$) $J^{PC}= 1^{-+}$ hybrid exotic meson is
listed  in Table~\ref{he_TABLE:ch9}.
These lattice QCD calculations are quenched or partially quenched.
\begin{table}[b]
\caption{  \label{he_TABLE:ch9}
Quenched lattice QCD calculations for the mass of
$u \bar u g$ hybrid exotic meson ($J^{PC}= 1^{-+}$).
}
\begin{center}
\begin{tabular}{|c|c|c|}
\hline
Collaboration(Year) &  Computed Mass(GeV)  &   Reference     \\
\hline
UKQCD (1997)      & $1.87 \pm 0.2    $     &\cite{Lacock:1996ny}  \\
\hline
SESAM (1998)      & $1.90 \pm 0.2    $     &\cite{Lacock:1998be}  \\
\hline
Mei and Luo(2003) & $2.013 \pm 0.026 $     &\cite{Mei:2002ip}     \\
\hline
MILC(2004)        & $1.792 \pm 0.139 $     &\cite{Bernard:2003jd} \\
\hline
\end{tabular}
\end{center}\end{table}
All of this numbers indicate
that the rough mass of hybrid exotic meson is about 1.8-2.1 GeV.

Another approach is to directly perform
experiment~\cite{Carman:2005ps}.
Several high statistics experimental searches for
hybrid exotic mesons have been carried out.
These experiments show credible evidence for the existence of
$J^{PC}= 1^{-+}$ hybrid exotics  $\pi_1(1400)$,
$\pi_1(1600)$, $\pi_1(2000)$.
The $\pi_1(1400)$ was first reported by the VES Collaboration
at IHEP~\cite{Beladidze:1993km} and was confirmed by
the E852 Collaboration at BNL in the
$\pi^{-} p \to \eta\pi^{-}p$ reaction~\cite{Thompson:1997bs}.
This was followed with additional confirmation by the Crystal Barrel
Collaboration at the CERN  LEAR facility in antiproton-neutron
annihilation~\cite{Abele:1998gn}.
In the Particle Data Group (PDG) listings~\cite{Eidelman:2004wy},
the mass of the $\pi_1(1400)$ is $M=1376 \pm 17$ MeV and its width
$\Gamma=300 \pm 40 $ MeV with observed decays to $\pi\eta$.
A second $J^{PC}= 1^{-+}$ exotic hybrid candidate,
the $\pi_1(1600)$, was first observed at BNL E852
in $\pi^{-} p \to \rho p \to \pi^{+}\pi^{-}\pi^{-} p$~\cite{Adams:1998ff}.
Since that time further evidence for the $\pi_1(1600)$
has been provided through its decays into $\eta'\pi$~\cite{Ivanov:2001rv},
$f_1 \pi$~\cite{Kuhn:2004en}, and
$b_1 \pi$~\cite{Lu:2004yn}.
The PDG~\cite{Eidelman:2004wy} lists the mass and width of this
state as $M = 1596 ^{+25}_{-14}$ MeV and
$\Gamma = 312^{+64}_{-24}$ MeV.

Another candidate $J^{PC}= 1^{-+}$  meson is $\pi_1(2000)$,
which has been seen through its decays into
$b_1 \pi$~\cite{Lu:2004yn} and $f_1 \pi$~\cite{Kuhn:2004en}
at BNL E852.
The $J^{PC}$ for $b_1$ is $1^{+-}$, and the $J^{PC}$ for $f_1$ is
$1^{++}$.
This state is the least controversial
of the announced exotics, because its mass is consistent with
theoretical expectations and its decay modes are consistent
with expectations for hybrid mesons.
However the statistical accuracy of the existing data
still needs improvement.

Because the lattice simulation of the heavy quark hybrid exotics
(namely, $b\bar b g$) is difficult,
and searching them from experiment need pretty high energy,
hence there are no papers about this kind of heavy quark hybrid
exotics.
The result of this dissertation will bring a little contribution
in this field.
%
%
\section{Hybrid exotic quarkonium decay}
Recent observations of charmonium states in $B$-meson
decays suggest that charmonium hybrid mesons ($h_c$)~\footnote{
The notation $h_c$ is introduced in Ref.~\cite{Brambilla:2004wf}.
I use his notation in the discussion of this section.
} can be produced in $B$ meson decay~\cite{Close:1997wp,Chiladze:1998ti,Brambilla:2004wf}.
Some of these states are likely to be narrow with subsequent decays
to $J/\psi \, \pi^+\pi^-$, etc.

Recent developments in both theory and experiment lead us to
expect that charmonium hybrids will be produced in $B$ decays.
The partial widths for $B \to c\bar{c} +X$, with $c\bar{c}$ representing
specific final states such as $J/\psi$, $\chi_{c0}$,
$\chi_{c1}$, $\chi_{c2}$, \etc
Possible decay schemes are~\cite{Brambilla:2004wf}:
$h_c(0^{+-},2^{+-}) \to J/\psi + (\pi^+\pi^-,\eta,\eta')$,
$h_c(1^{-+})        \to \eta_c + (\pi^+\pi^-,\eta,\eta')$.
These decay modes can give distinctive signals.
Both $0^{+-}$ and $2^{+-}$ should decay via the
$J/\psi \, \pi\pi$ cascade~\cite{Brambilla:2004wf}.
The $h_c(1^{-+})$ state is expected to be the lightest exotic $c\bar{c}$
hybrid, and therefore in this case the cascade goes to $\eta_c \pi\pi$.

The $X(3872)$ is a narrow state decaying into $\pi^+ \pi^- J/\psi$,
with a mass $M_X \sim 3872\,\mathrm{MeV}$.  Given the observed mass,
in the charmonium region, it is natural
to assert that the $X(3872)$ is itself a charmonium state. However it
is difficult to identify the $X(3872)$ with any of the
expected narrow $c\bar{c}$ mesons,
leading to suggestions that it is  a more exotic
particle~\cite{Acosta:2003zx}\cite{Abazov:2004kp}\cite{Aubert:2004ns}.
%
%
\section{Summary}
We reviewed arguments that, in the heavy quark limit, the decay channel
of  a $J^{PC}=1^{-+}$ spin-exotic hybrid meson will be dominated
by the creation of a flavor singlet light quark-antiquark.
In this dissertation, we explored this decay in lattice QCD
with $N_f=3$ flavors of dynamical
sea quark with masses close to the physical values.
We studied the transition between an excited hybrid gluonic state
(i.e., $\Pi_u$ state)
and a ground state gluonic state (i.e., $\chi_b$ state)
with the production of a flavor-singlet light quark-antiquark pair.
The dominant decay channel should be the $\pi\pi$ channel.
We introduced a phenomenological model for the decay
and directly studied the transition
to two pions on the lattice.
We obtained a little bit smaller transition amplitude for this decay channel.

Our numerical result for the scalar $f_0$ meson is  $am=0.47(7)$
(or $m_{f_0} = (768 \pm 136)$~MeV
for this lattice spacing $a=0.12 \,{\rm  fm}$, the mass of up quark
($am_u = 0.005$), and the mass of strange quark ($am_s = 0.05$).
The energy release from the hybrid meson at $10.833$~GeV to the
$\chi_b$  state at $9.893$~GeV is $0.940$~GeV.
Therefore, it is impossible to decay to a scalar meson heavier than $940$~MeV.
For $b$ quarks the relevant transitions is $H \to \chi_b \,  \pi\pi$.
Our raw lattice result is an overall decay rate of
$3.62(98)$~MeV for the scalar meson.

The evaluation of this transitions amplitude
in this dissertation is a fascinating step for
the phenomenological study of hybrid decays.
Our results are consistent with our expectation that
the spin-exotic hybrid meson states are relatively narrow.
We hope our results will be able to guide the experimental search
for a $\bar b b g$ and $f_0$ meson, etc.

In this dissertation, we performed our research on one ensemble of
lattice gauge configuration.
That means our results are at a single lattice spacing and light quark
mass. As we discussed in Chapter 3. The physical one should be in
continuum limit (lattice spacing $a \to 0$). Hence, we
still work on other different configurations. In the continuum
limit, we get the decay rates that have physical means.

%% file: appA.tex
\chapter{The $PC$ of the hybrid meson}
In this appendix, we want to prove Eq.~(\ref{appb:eq:1}),
and Eq.~(\ref{appb:eq:2}), which we use in Chapter 4 to evaluate
the $PC$  of the hybrid meson.

Here we first introduce the azimuthal angle $\theta$,
the zenith  angle $\psi$ of the axis of the $Q\overline{Q}$
and  $R$, the distance (radius) between $Q$ and $\overline{Q}$
with respect to a fixed system of coordinates $\xi, \eta, \zeta$.
Besides the fixed system of the coordinates $\xi, \eta, \zeta$,
we also introduce a moving system of the coordinates $x, y, z$.
And the two system have the same origin.
The $z$-axis is in the direction of the axis of $Q\overline{Q}$,
and the $x$-axis is lying in the $\xi-\eta$ plane.
In bpdy-fixed system of the coordinates $x, y, z$,
we also introduce the  cylinder coordinates:
radial $\rho$,  angle $\alpha$, and $z$.
The angle $\alpha$ is measured in
xy-plane from the x-axis with $0 \le \alpha < 2\pi$.

The coordinates $\xi_g, \eta_g, \zeta_g$ of the gluon and
the coordinates $x_g, \ y_g, \ z_g$ of the gluon
have following relationship
\bea
x_g &=& - \xi_g \sin\,\psi +  \eta_g \cos\,\psi    \\
y_g &=& - \xi_g \cos\,\theta \cos\,\psi +  \eta_g \cos\,\theta
           \sin\,\psi + \zeta_g \sin\,\theta       \\
z_g &=& - \xi_g \sin\,\theta \cos\,\psi +  \eta_g \sin\,\theta
           \sin\,\psi + \zeta_g \cos\,\theta  .
\label{relatin:eq}
\eea

We know, under the operation of parity,
that the $\theta, \psi$ change according to
\bea
\theta &\rightarrow& \pi - \theta  \\
\psi   &\rightarrow& \pi + \psi   .
\eea
Also the  $\xi,  \eta,  \zeta$ change sign.
From Eqs.~(B.1,B.2,B.3), we obtain that,
under the operation of parity, the $x$ and $z$ do not change sign,
and $y$ change sign. This means that, under the operation of
parity,
\bea
\alpha &\rightarrow&  -\alpha  .
\eea

The wave function of the hybrid meson is, then,
\be
\chi_{n,\Lambda,L,M_L} =
\phi_{n,\Lambda,L}(\rho, \alpha, z)
\Xi_{n,\Lambda,L,M_L}(r)
\Theta_{L,M_L,\Lambda}(\theta)
\Phi_{M_L}(\psi)  ,
\ee
where $L$ is the total angular momentum of the hybrid meson,
$M_L$ is the $\zeta$-component of  $L$,
$\Lambda$ is the $z$-component of the angular momentum of the gluon,
and $n$ stands for the others quantum numbers,
which specify the state of the  hybrid meson,
$\phi_{n,\Lambda,L}(\rho, \alpha, z)$ is the gluon wave function,
and $\Xi_{n,\Lambda,L,M_L}(r)$ is the ``radial part'' of
$Q\overline{Q}$ wave function.
The $\theta$ dependence is given by $\Theta_{L,M_L,\Lambda}(\theta)$.
Finally, $\Phi_{M_L}(\psi)$ gives  the ``$\psi$'' dependence.

According to the solution in Page 299 of Ref.~\cite{Landau:Lif},
we obtain
\bea
\Theta_{L,M_L,\Lambda}(\theta) & = &  (-i)^L
\sqrt{ \frac{(2L\PLUSONE 1)!(L\PLUSONE M_L)!}
     { (L\MINUSONE\Lambda)!(L\PLUSONE\Lambda)!(L\MINUSONE M_L)!} }
\times \frac
{ (1\MINUSONE\cos\,\theta )^{ \frac{\Lambda - M_L}{2}  }  }
{ (1\PLUSONE\cos\,\theta )^{ \frac{\Lambda + M_L}{2}  }  }
\times  \ \ \ \ \ \ \ \ \ \  \nonumber \\
& &\Bigg\{
\left ( \frac{\partial}{\partial \cos\,\theta} \right )^{L-M_L}
(1\MINUSONE\cos\,\theta )^{L-\Lambda}(1\PLUSONE\cos\,\theta )^{L+\Lambda}
\Bigg\} ,
\eea
which we are called the left-handed gluon state, and
\bea
\Theta_{L,M_L,-\Lambda}(\theta) & = &  (-i)^L
\sqrt{ \frac{(2L\PLUSONE 1)!(L\PLUSONE M_L)!}
     { (L\MINUSONE\Lambda)!(L\PLUSONE\Lambda)!(L\MINUSONE M_L)!} }
\times \frac
{ (1\MINUSONE\cos\,\theta )^{ \frac{-\Lambda - M_L}{2} } }
{ (1\PLUSONE\cos\,\theta )^{ \frac{-\Lambda + M_L}{2} } }
\times   \ \ \ \ \ \ \ \ \ \ \nonumber \\
& & \hspace{-0.8cm} \Bigg\{
\left ( \frac{\partial}{\partial \cos\,\theta} \right )^{L-M_L}
(1\MINUSONE\cos\,\theta )^{L+\Lambda}(1\PLUSONE\cos\,\theta )^{L-\Lambda}
\Bigg\} ,
\eea
which we are called the right-handed gluon state.
It is obvious that the left-handed gluon state
and the right-handed state are not eigenstates of parity.
Now if we combine them, we can construct eigenstates of parity.
\bea
&& \Theta_{L,M_L,\Lambda}(\theta) + \epsilon \Theta_{L,M_L,-\Lambda}(\theta) =
(-i)^L \sqrt{ \frac{(2L+1)!(L\PLUSONE M_L)!}{ (L\MINUSONE\Lambda)!(L\PLUSONE\Lambda)!(L-M_L)!} }
\times \Bigg\{  \ \ \ \ \ \ \ \ \ \ \ \ \ \ \nonumber\\
& &  \ \ \ \ \
\ \frac
{ (1\MINUSONE\cos\,\theta )^{ \frac{\Lambda - M_L}{2}  }  }
{ (1\PLUSONE\cos\,\theta )^{ \frac{\Lambda + M_L}{2}  }  }
\left ( \frac{\partial}{\partial \cos\,\theta} \right )^{L-M_L}
(1\MINUSONE\cos\,\theta )^{L\MINUSONE\Lambda}(1\PLUSONE\cos\,\theta )^{L\PLUSONE\Lambda} \ + \nonumber\\
& & \ \ \ \   \epsilon \frac
{ (1\MINUSONE\cos\,\theta )^{ \frac{\Lambda - M_L}{2} } }
{ (1\PLUSONE \cos\,\theta )^{ \frac{\Lambda + M_L}{2} } }
\left( \frac{\partial}{\partial \cos\,\theta} \right )^{ L-M_L }
(1\MINUSONE\cos\,\theta )^{L-\Lambda}
(1\PLUSONE \cos\,\theta )^{L+\Lambda}
\ \Bigg\}   ,
\eea
where $\epsilon = \pm 1$.  \\

Hence, we obtain
\be
P \left\{ \Theta_{L,M_L,\Lambda}(\theta) + \epsilon
\Theta_{L,M_L,-\Lambda}(\theta) \right\}
= \Theta_{L,M_L,\Lambda}(\pi-\theta) + \epsilon
\Theta_{L,M_L,-\Lambda}(\pi-\theta) .
\ee
Now if we consider the identity $ \cos(\pi-\theta) = - \cos\,\theta $,
we obtain
\bea
\hspace{-0.8cm} && \hspace{-0.8cm}
P \left\{ \Theta_{L,M_L,\Lambda}(\theta) + \epsilon
\Theta_{L,M_L,-\Lambda}(\theta) \right\} =
(-i)^L \sqrt{ \frac{(2L\PLUSONE1)!(L\PLUSONE M_L)!}{ (L\MINUSONE\Lambda)!(L\PLUSONE\Lambda)!(L-M_L)!} }
\times \epsilon(-1)^{L-M_L}  \nonumber\\
& & \hspace{-0.8cm}  \ \left \{
\ \epsilon \frac
{ (1\PLUSONE\cos\,\theta )^{ \frac{\Lambda - M_L}{2}  }  }
{ (1\MINUSONE\cos\,\theta )^{ \frac{\Lambda + M_L}{2}  }  }
\left ( \frac{\partial}{\partial \cos\,\theta} \right )^{L-M_L}
(1\PLUSONE\cos\,\theta )^{L - \Lambda}
(1\MINUSONE\cos\,\theta )^{L + \Lambda}  \ +
\right. \nonumber\\
& & \hspace{-0.8cm}  \left.
\ \frac
{ (1\MINUSONE\cos\,\theta )^{ \frac{\Lambda - M_L}{2}  }  }
{ (1\PLUSONE\cos\,\theta )^{ \frac{\Lambda + M_L}{2}  }  }
\left ( \frac{\partial}{\partial \cos\,\theta} \right )^{L-M_L}
(1\PLUSONE\cos\,\theta )^{L - \Lambda}
(1\MINUSONE\cos\,\theta )^{L + \Lambda}
\ \right\} .
\eea
Hence, we obtain
\be
P \left\{
\Theta_{L,M_L,\Lambda}(\theta) + \epsilon
\Theta_{L,M_L,-\Lambda}(\theta) \right\}
=
\epsilon(-1)^{L-M_L} \left\{
\Theta_{L,M_L, \Lambda}(\theta) + \epsilon
\Theta_{L,M_L,-\Lambda}(\theta) \right\} .
\ee
The ``$\psi$'' part of the gluon wave function (namely, $\Phi(\psi)$) is
\be
\Phi_{M_L}(\psi) = \frac{1}{\sqrt{2\pi}}e^{iM_L\psi} .
\ee
It is obvious that
\be
P \Phi_{M_L}(\psi) =  (-1)^{M_L} \Phi_{M_L}(\psi) .
\ee
And the ``$\alpha$'' part of the gluon wave function in
$\phi_{n,\Lambda,L}(\rho, \alpha, z)$ is
\be
\phi_{\Lambda}(\alpha)  = \frac{1}{\sqrt{2\pi}}e^{i\Lambda\alpha} .
\ee
It is obvious that
\be
P \phi_{\Lambda}(\alpha) =  \phi_{-\Lambda}(\alpha) .
\ee

If we consider the intrinsic parity of the meson is -1~\cite{Perkins:1982xb},
the total parity of the hybrid meson is given by
$ P = \epsilon(-1)^{L-M_L}   \times
            (-1)^{M_L}       \times
            (-1)^{1}  $,
that is,
\be
P = \epsilon(-1)^{L + 1} .
\ee

Let us define the $PC$ quantum number of the gluon to be $\eta$,
if we also consider that the symmetry of the spin wave
function under charge-conjugation is $(-1)^{S+1}$~\cite{Perkins:1982xb},
we obtain the charge-conjugation of the hybrid meson, namely
\be
C = \epsilon \eta (-1)^{L+S} .
\ee

%% file: appB.tex
\chapter{Bubble contribution for  {\lowercase{$a_0$}} correlator}
The mass of the $a_0$ meson can be reliably determined on the lattice.
In order to determine the mass of the $a_0$ meson,
we evaluate the $a_0$ correlator
\bea
C(t) &=&
\sum_{\bf x}  \langle
\bar{d}( {\bf x}, t) u( {\bf x}, t)
\bar{u}( {\bf 0}, 0) d( {\bf 0}, 0)
\rangle
\label{cor::appC} \, .
\eea
The extraction of the mass of the $a_0$ meson ($J^P=0^+$ and $I=1$)
is straightforward.
However there exist many multihadron states with $J^P=0^+$ and $I=1$
which can propagate between the source and the sink.
Of special interest in multihadron states is
{\it the intermediate state with two pseudoscalars} $P_1 P_2$ which
we refer to as
{\it the bubble contribution} ($B$)~\cite{Prelovsek:2005rf}.
If the masses of $P_1$ and $P_2$ are small,
the bubble contribution $B$ gives a considerable contribution
to the $a_0$ correlator, and it should be included
in the fit of the lattice correlator in Eq.~(\ref{cor::appC}), that is,
\be
\label{Ctot::appC}
C(t)=Ae^{-m_{\sigma}t}+B(t)   \ ,
\ee
where we omit the unimportant contributions from the excited $a_0$
meson and other high order multihadron intermediate states.

\section{Coupling of a scalar current to pseudoscalar}
\label{appC::C}
Before we embark on the calculation of the bubble contribution,
in this section we first derive the coupling of a point scalar current
$\bar d_r(x)u_r(x)$ to a pair of the pseudoscalar fields at the lowest
energy order of the staggered chiral perturbation theory (S$\chi$PT),
where the subscript $r$ in the expression $u_r(x)$ is
the index of the taste replica for a given quark flavor $u$.
The effective scalar current can be determined from the dependence
of the lattice $QCD$ Lagrangian and the staggered chiral Lagrangian on
the spurion field ${\cal M}$, where ${\cal M}$ is the
staggered quark mass matrix. For $n$  Kogut-Susskind (KS) flavors,
${\cal M}$ is a $4n n_r \times 4n n_r $ matrix.
\bea
{\cal M} = \left( \begin{array}{cccc}
m_{uu} I\otimes I_R&m_{ud} I\otimes I_R&m_{us} I \otimes I_R& \cdots \\*
m_{du} I\otimes I_R&m_{dd} I\otimes I_R&m_{ds} I \otimes I_R& \cdots \\*
m_{su} I\otimes I_R&m_{sd} I\otimes I_R&m_{ss} I \otimes I_R& \cdots \\*
        \vdots & \vdots & \vdots & \ddots \end{array} \right) ,
\eea
where $I$ is a $4 \times 4$ unit matrix, and
$I_R$ is  $ n_r \times n_r $ unit replica matrix.
In short, ${\cal M} =  {\bf m } \otimes I \otimes I_R $, where
\bea
{\bf m} = \left( \begin{array}{cccc}
	 m_{uu}   & m_{ud}    & m_{us}     & \cdots \\*
	 m_{du}   & m_{dd}    & m_{ds}     & \cdots \\*
	 m_{su}   & m_{sd}    & m_{ss}     & \cdots \\*
      \vdots & \vdots & \vdots & \ddots \end{array} \right)
\eea
is the $n\times n$ quark mass matrix.
We know from $QCD$ that the quark current $\bar{d}_r(x)u_r(x)$ is given
by
\be
\DPT  \bar{d}_r(x)u_r(x) =
-\frac{\partial{\cal L}_{QCD}}{ \partial{m_{d_r u_r}(x) } },
\ee
where ${\cal L}_{\rm QCD}$ is the $QCD$ Lagrangian.

Since staggered chiral lagrangian (${\cal L}_{S \chi PT}$)
is an effective equivalent Lagrangian for ${\cal L}_{\rm QCD}$ in
low energy limit, the effective current $\bar u_r(x)u_r(x)$
is obtained from
\be
\DPT  \bar{d}_r(x)u_r(x) =
-\frac{ \partial{\cal L}_{S \chi PT}}{\partial{\cal M}_{d_r u_r}(x)} ,
\ee
where
\be
\label{lagrangianschopt::appC}
{\cal L}_{S \chi PT} = \frac{1}{8} f_{\pi}^2 {\rm Tr}
[\partial^\mu \Sigma \partial_\mu \Sigma^\dag] - \frac{1}{4}
 B_0 f_{\pi}^2{\rm Tr}[{\cal M}^\dag \Sigma+\Sigma^\dag{\cal M}]
\ee
is the staggered chiral Lagrangian~\cite{Aubin:2003mg}.
We omit the high order terms and the terms that are
independent of ${\cal M}$,
$f_{\pi}$ is the tree-level pion decay constant
($f_{\pi} = 131 $ MeV~\cite{Aubin:2003mg}), $B_0$
is the constant with the dimension of the mass~\cite{Bardeen:2001jm},
and $ \DPT\Sigma=\exp \left( \frac{2i\Phi}{f_{\pi}}
\right)$~\cite{Aubin:2003mg}.
The field $\DPT\Phi=\sum_{b=1}^{16} \frac{1}{2}T^b \otimes \phi^b$
is described in terms of the mass eigenstate
field $\phi^b$~\cite{Aubin:2003mg},
where $\phi^b$ is a $3 \times 3$ pseudoscalar matrix with flavor
components $\phi^b_{ f_r f'_{r'} }$ with flavor $f,f'$, the index of
the taste replica $r, r'$,
and taste $b$ which is given by generators
$T^b=\{\xi_5,i\xi_5\xi_\mu,i\xi_\mu\xi_\nu,\xi_\mu,\xi_I\}$~\cite{Aubin:2003mg}.
Hence, $\Phi$ is $ 4 n n_r \times 4n n_r $ pseudoscalar matrix
in S$\chi$PT~\cite{Aubin:2003mg}, and the subscripts $u,d$ denote
its valance flavor component.
Therefore, ${\Sigma}$ is also a $4n n_r \times 4n n_r$ matrix.
The {\rm Tr} is the full $4n n_r \times 4n n_r$ trace.
Therefore, the effective current is~\cite{Bardeen:2001jm}
\be
\bar d_r(x) u_r(x) =
B_0 {\rm Tr_t } [\Phi(x)^2]_{d_r u_r} ,
\label{app_current::appC}
\ee
where the notation ${\rm Tr_t}$ stands for the trace over taste.
%
\section{Bubble Contribution}
\label{appC::BC}
In this section we compute the bubble contribution from two
intermediate states to the $a_0$ correlator in Eq.~(\ref{cor::appC}).
From the discussion in Sec.~\ref{appC::C}, the point scalar current
can be described in terms of the pseudoscalar field
$\Phi$ by using S$\chi$PT~\cite{Bardeen:2001jm}\cite{Prelovsek:2005rf},
\bea
\label{current::appC}
\bar  d_r(x) u_r(x)  &=&  B_0 {\rm Tr_t}[\Phi(x)^2]_{d_r u_r} \\
\bar  u_r(x) d_r(x)  &=&  B_0 {\rm Tr_t}[\Phi(x)^2]_{u_r d_r} ,
\eea
where $\DPT B_0=\frac{M_\pi^2}{2m_q}$ is the coupling
of the point scalar current to the pseudoscalar field $\Phi(x)$,
which can be numerically determined from the lattice simulation.

Now we consider the lattice simulations with staggered KS valence
quarks and three flavors of staggered KS dynamical sea quarks
with $m_u=m_d\not =m_s$. We only consider the correlators where
the valence-quark mass is equal to the $u$ or $d$ dynamical sea-quark mass.
We know that our $a_0$ correlator is estimated by
using the taste-singlet source and taste-singlet sink.
This $a_0$ correlator takes contribution from the
$a_0$ meson and the bubble contribution.
For concreteness, the bubble contribution to the $a_0$ correlator
in a theory with $n_r$ tastes per flavor and  three flavors
(namely, $N_f=2+1$) of KS dynamical sea quarks is~\cite{Aubin:2003mg}
\bea
\label{schpt_0::appC}
B^{S\chi PT}_{a_0}(x) &=&
\frac{B_0^2}{n_r}  \sum_{r,r'=1}^{n_r}
\Bigg\{
\left\langle \bar d_r(x) u_r(x) \ \bar u_{r'}(0) d_{r'}(0)
\right\rangle_{\rm Bubble}  \Biggr\} \nonumber \\
 &=&
\frac{B_0^2}{n_r} \sum_{r,r'=1}^{n_r}  \Bigg\{
\left\langle{\rm Tr_t}[\Phi^2]_{d_r u_r}
            {\rm Tr_t}[\Phi^2]_{u_{r'} d_{r'}}
\right\rangle \Biggr\}  ,
\eea
where the subscript $a_0$ specifies the bubble contribution for $a_0$ meson.
If we consider this identity
$
{\rm Tr_t} \left( T^{a} T^{b} \right) = 4  \delta_{ab}
$,
we arrive at
\bea
B^{S \chi PT}_{a_0}(x) \hspace{-0.3cm} &=& \hspace{-0.3cm}
\frac{B_0^2}{n_r}
\sum_{a=1    \atop b=1   }^{16}
\sum_{i=1    \atop j=1   }^{N_f}
\sum_{r,r'=1 \atop t,t'=1}^{n_r}
\langle
\phi^a_{  d_r    i_t   }(x) \phi^a_{i_t      u_r}(x) \
\phi^{b}_{u_{r'} j_{t'}}(0) \phi^{b}_{j_{t'} d_{r'} }(0)
\rangle  .
\eea
where $a,b,a',b'$ are taste indices, $i,j$ are flavor indices,
and   $r,t,r',t'$ are the   indices of the taste replica.
The Wick contractions result in the products of two propagators
for the pseudoscalar fields.
Therefore, the bubble contribution can be expressed in terms of the
pseudoscalar propagators  $\langle \phi^b \ \phi^b\rangle$.
\bea
B^{S \chi PT}_{a_0}(x)
\hspace{-0.3cm} &=& \hspace{-0.3cm}
\frac{B_0^2}{n_r}
\sum_{b=1}^{16} \sum_{r,r'=1}^{n_r} \sum_{r,r'=1}^{n_r}
\Biggl\{
2\langle \phi^b_{u_r d_t}(x) \ \phi^b_{ d_{t'} u_{r'} }(0) \rangle
 \langle \phi^b_{u_t u_r}(x) \ \phi^b_{ d_{r'} d_{t'} }(0) \rangle  +
 \nonumber \\
 \hspace{-0.3cm}  && \hspace{2.8cm}
2\langle \phi^b_{u_r d_t}(x) \ \phi^b_{ d_{t'} u_{r'} }(0) \rangle
 \langle \phi^b_{u_t u_r}(x) \ \phi^b_{ u_{r'} u_{t'} }(0) \rangle  +
 \nonumber \\
 \hspace{-0.3cm}  && \hspace{2.8cm}
 \langle \phi^b_{u_r s_t}(x) \ \phi^b_{ s_{t'} u_{r'} }(0) \rangle
 \langle \phi^b_{s_t u_r}(x) \ \phi^b_{ u_{r'} s_{t'} }(0) \rangle
\Biggr\}
,
\label{schpt_1:a0:appC}
\eea
where we use charge symmetry to equate correlators with
$u \leftrightarrow d$.

First of all, we sum over the indices of the taste replica in
Eq.~(\ref{schpt_1:a0:appC}).
For illustration, we consider the second term in
Eq.~(\ref{schpt_1:a0:appC}) for taste $b=I,V,A$.
\bea
&&\sum_{r,r'}\sum_{t,t'} \Biggl\{
\langle \phi^b_{u_r u_t}(x) \ \phi^b_{u_{r'} u_{t'} }(0) \rangle
\langle \phi^b_{u_t d_r}(x) \ \phi^b_{d_{t'} u_{r'} }(0) \rangle \Biggr\}
\nonumber  \\
&=& \sum_{r,r'} \Biggl\{
\langle \phi^b_{u_r u_r}(x) \ \phi^b_{u_{r'} u_{r'} }(0) \rangle
\langle \phi^b_{u_r d_r}(x) \ \phi^b_{d_{r'} u_{r'} }(0) \rangle
\Biggr\}_{r \not= r'}  + \nonumber  \\
&&
\sum_{r,t} \Biggl\{
\langle \phi^b_{u_r u_t}(x) \ \phi^b_{u_{r} u_{t} }(0) \rangle
\langle \phi^b_{u_t d_r}(x) \ \phi^b_{d_{t} u_{r} }(0) \rangle
\Biggr\}_{r \not= t}  + \nonumber  \\
&&
\sum_{r,t} \Biggl\{
\langle \phi^b_{u_r u_t}(x) \ \phi^b_{u_{t} u_{r} }(0) \rangle
\langle \phi^b_{u_t d_r}(x) \ \phi^b_{d_{r} u_{t} }(0) \rangle
\Biggr\}_{r \not= t} + \nonumber  \\
&&
\sum_{r} \Biggl\{
\langle \phi^b_{u_r u_r}(x) \ \phi^b_{u_{r} u_{r} }(0) \rangle
\langle \phi^b_{u_t d_r}(x) \ \phi^b_{d_{r} u_{r} }(0) \rangle
\Biggr\}  .
\label{First3::appC}
\eea
The first term and the second term in Eq.~(\ref{First3::appC})
should be zero, and the third term in Eq.~(\ref{First3::appC})
should be
\be
n_r(n_r-1)
\langle \phi^b_{u_r u_r} \ \phi^b_{u_{r} u_{r} } \rangle_{\rm CC} \
\langle \phi^b_{u_r d_r} \ \phi^b_{d_{r} u_{r} } \rangle_{\rm CC} ,
\ee
where the notation CC stands for the connected contribution.
The fourth term in Eq.~(\ref{First3::appC}) should be
\bea
&&n_r \Biggl\{
\langle \phi^b_{u_r u_r}(x) \ \phi^b_{u_{r} u_{r} }(0) \rangle_{\rm CC}
\langle \phi^b_{u_r d_r}(x) \ \phi^b_{d_{r} u_{r} }(0) \rangle_{\rm CC}
\ + \ \nonumber\\ && \hspace{0.7cm}
\langle \phi^b_{u_r u_r}(x) \ \phi^b_{u_{r} u_{r} }(0) \rangle_{\rm DC}
\langle \phi^b_{u_r d_r}(x) \ \phi^b_{d_{r} u_{r} }(0) \rangle_{\rm CC}
\Bigg\} ,
\eea
where the notation DC stands for the disconnected contribution.
Hence, the second term in Eq.~(\ref{First3::appC}) can be changed into
\bea
&& \sum_{r,r'}\sum_{t,t'} \Biggl\{
\langle \phi^b_{u_r u_t}(x) \ \phi^b_{u_{r'} u_{t'}}(0) \rangle
\langle \phi^b_{u_t d_r}(x) \ \phi^b_{d_{t'} u_{r'}}(0) \rangle \Biggr\}
= \nonumber\\
&& \hspace{1.5cm}
n_r^2
\langle \phi^b_{u_r u_r}(x) \ \phi^b_{u_r u_r}(0) \rangle_{\rm CC} \
\langle \phi^b_{u_r d_r}(x) \ \phi^b_{d_r u_r}(0) \rangle_{\rm CC}
+ \nonumber\\  && \hspace{1.5cm} n_r
\langle \phi^b_{u_r u_r}(x) \ \phi^b_{u_r u_r}(0) \rangle_{\rm DC}
\langle \phi^b_{u_r d_r}(x) \ \phi^b_{d_r u_r}(0) \rangle_{\rm CC}
.
\eea
Because there is no disconnected contribution for taste $b \not= I,V,A$,
therefore Eq.~(\ref{First3::appC}) can be written as
\bea
&& \sum_{r,r'}\sum_{t,t'}
\Biggl\{
\langle \phi^b_{u_r u_t}(x) \ \phi^b_{u_{r'} u_{t'} }(0) \rangle
\langle \phi^b_{u_t u_r}(x) \ \phi^b_{u_{t'} u_{r'} }(0) \rangle
\Biggr\} \nonumber\\
&&=
n_r^2
\langle \phi^b_{u_r u_r}(x) \ \phi^b_{u_r u_r}(0) \rangle_{\rm CC}
\langle \phi^b_{u_r d_r}(x) \ \phi^b_{d_r u_r}(0) \rangle_{\rm CC} .
\eea
From the similar procedure, for taste $b=I,V,A$, we obtain
\bea
&&\sum_{r,r'}\sum_{t,t'}
\Biggl\{
\langle \phi^b_{u_r d_t}(x) \ \phi^b_{d_{r'} u_{t'} }(0) \rangle
\langle \phi^b_{u_t u_r}(x) \ \phi^b_{d_{t'} d_{r'} }(0) \rangle
\Biggr\} \nonumber\\
&&=
n_r
\langle \phi^b_{u_r d_r}(x) \ \phi^b_{d_r u_r}(0) \rangle_{\rm CC}
\langle \phi^b_{u_r u_r}(x) \ \phi^b_{d_r d_r}(0) \rangle_{\rm DC} ,
\eea
where we consider the
$\langle \phi^b_{uu} \ \phi^b_{dd}\rangle $
does not include the connected contribution.
For taste $b \not= I, V, A$, this term should be zero.
For any taste $b$, we obtain
\bea
&&\sum_{r,r'}\sum_{t,t'}
\Biggl\{
\langle \phi^b_{u_r s_t}(x) \ \phi^b_{s_{r'} u_{t'} }(0) \rangle
\langle \phi^b_{s_t u_r}(x) \ \phi^b_{u_{t'} s_{r'} }(0) \rangle
\Biggr\} \nonumber\\
&& = n_r^2
\langle \phi^b_{u_r s_r}(x) \ \phi^b_{s_r u_r}(0) \rangle_{\rm CC}
\langle \phi^b_{s_r u_r}(x) \ \phi^b_{u_r s_r}(0) \rangle_{\rm CC} .
\eea

From now on, since there is no summation over the index of the
taste replica. We suppress it.
Thus we can rewrite Eq.~(\ref{schpt_1:a0:appC}) as
\bea
B^{S \chi PT}_{a_0}(x) \hspace{-0.3cm} &=& \hspace{-0.3cm}
n_r B_0^2 \sum_{b=1}^{16} \Biggl\{
2\langle \phi^b_{ud}(x) \ \phi^b_{du}(0) \rangle_{\rm CC} \
 \langle \phi^b_{uu}(x) \ \phi^b_{uu}(0) \rangle_{\rm CC}  +
\nonumber\\ \hspace{-0.3cm} && \hspace{1.5cm}
\langle \phi^b_{us}(x) \ \phi^b_{su}(0) \rangle_{\rm CC}  \
 \langle \phi^b_{su} (x) \ \phi^b_{us}(0) \rangle_{\rm CC}
\Biggr\}  \nonumber\\
\hspace{-0.3cm} &+& \hspace{-0.3cm}
4 B_0^2 \sum_{b=I,V,A}
\Biggl\{
   \langle \phi^b_{uu}(x)  \ \phi^b_{uu}(0) \rangle_{\rm DC} \
   \langle \phi^b_{ud}(x)  \ \phi^b_{du}(0) \rangle_{\rm CC}
 \Biggr\}  ,
\label{schpt_0:a0:appC}
\eea
The propagators $\langle \phi^b \ \phi^b\rangle$  for the
pseudoscalar field $\phi^b$ for various tastes $b$ are studied
in Ref.~\cite{Aubin:2003mg}. Propagators
for all tastes ($I,V,A,T,P$) have connected contributions,
while only tastes $I,~V$ and $A$ have disconnected contributions.
Hence, Eq.~(\ref{schpt_0:a0:appC}) can be rewritten by
plugging in the mesonic propagators, that is,
\be
B^{S\chi PT}_{a_0}(t) = F.T.[~B^{S \chi PT}_{a_0}(p)~]_{ {\bf p} =
{\bf 0} } \ \ \ \ {\rm with} \\
\ee
\bea
\label{schpt_6:a0:appC}
B^{S \chi PT}_{a_0}(p) \hspace{-0.3cm} &=& \hspace{-0.3cm}
\frac{B_0^2}{n_r}  \sum_k  \Biggl\{
-4\frac{1}{ (k+p)^2+M_{U_V}^2 }  \delta_V
\frac{ k^2+M_{S_V}^2 }
{(k^2+M_{U_V}^2)(k^2+M_{\eta_V}^2)(k^2+M_{\eta^\prime V}^2)}
\nonumber\\
&& \hspace{1.1cm} - 4\frac{1}{ (k+p)^2+M_{U_A}^2 } \delta_A
\frac{k^2+M_{S_A}^2}
{(k^2+M_{U_A}^2)(k^2+M_{\eta_A}^2)(k^2+M_{\eta^{\prime} A}^2)}
\nonumber\\
&& \hspace{-0.5cm}  - n_r^2 \sum_{b=1}^{16}
\Biggl[2\frac{1}{(k+p)^2+M_{U_b}^2} \frac{1}{k^2+M_{U_b}^2}+
        \frac{1}{(k+p)^2+M_{K_b}^2} \frac{1}{k^2+M_{K_b}^2}
\Biggr] \nonumber \\
&& \hspace{-0.5cm}
-         2 \frac{1}{(k+p)^2+M_{U_I}^2}\frac{1}{k^2+M_{U_I}^2 }
+\frac{2}{3}\frac{1}{(k+p)^2+M_{U_I}^2}\frac{1}{k^2+M_{\eta I}^2}
\Biggr\}
,
\eea
where $\delta_V = a^2\delta_V^{\prime}$,
      $\delta_A = a^2\delta_A^{\prime}$
and~\cite{Prelovsek:2005rf}\cite{Aubin:2003mg}
\vspace{-0.3cm}
\bea
M_{\eta_I}     &=& \frac{1}{3}M_{U_I}^2 + \frac{2}{3}M_{S_I}^2 \\
M_{\eta V}^2   &=&
\frac{1}{2} \left( M^2_{U_V} + M^2_{S_V} + 3 n_r \delta_{V}
 - Z_V \right) \\
M_{\eta' V}^2  &=&
\frac{1}{2} \left( M^2_{U_V} + M^2_{S_V} + 3 n_r \delta_{V}
 + Z_V \right) .
\eea
Here
\be
Z_V = \sqrt{ (M^2_{S_V}-M^2_{U_V})^2 - 2 n_r \delta_{V}
           (M^2_{S_V}-M^2_{U_V}) + 9 (n_r \delta_V)^2
         } \nonumber .
\ee
For the axial taste we just requires $V \to A$.
The pseudoscalar masses
$M_{U_b} \!\equiv\! M_{\pi_b}\!=\!2B_0m_u+a^2     \Delta_b$,
$M_{S_b} \!\equiv\! M_{ss_b} \!=\!2B_0m_s+a^2     \Delta_b$ and
$M_{K_b} \!\equiv\! M_{us_b} \!=\!B_0(m_u+m_s)+a^2\Delta_b$
have been determined by
MILC simulations~\cite{Bernard:2001av,Aubin:2004wf}.
The taste breaking term $a^2\Delta_b$ is independent of
the flavor~\cite{Aubin:2003mg}
and will disappear in the continuum limit (i.e., $a \to 0$).
We can prove that
\be
\frac{ k^2 \PLUSONE M_{S_V}^2 }
{(k^2 \PLUSONE M_{U_V}^2)(k^2 \PLUSONE M_{\eta V}^2)
 (k^2 \PLUSONE M_{\eta^\prime V}^2)}  =
\frac{1}{ 4 n_r \delta_{V} }  \left\{
\frac{2}{ k^2 \PLUSONE M_{U_V}^2 } -
\frac{ C_{V_\eta   } }{ k^2 \PLUSONE M_{\eta  V}^2} +
\frac{ C_{V_{\eta'}} }{ k^2 \PLUSONE M_{\eta' V}^2}
\right\}   ,
\label{FZW_01:appC}
\ee
where
\vspace{-0.3cm}
\bea
C_{V_\eta}    &=& \frac{(M_{S_V}^2-M_{U_V}^2-n_r\delta_V+Z_V)}{Z_V}\\
C_{V_{\eta'}} &=& \frac{(M_{S_V}^2-M_{U_V}^2-n_r\delta_V-Z_V)}{Z_V} .
\label{FZW_03:appC}
\eea
For the axial taste we just requires $V \to A$.
We simplify Eq.~(\ref{schpt_6:a0:appC}) as
\bea
\label{schpt_8:a0:appC}
\hspace{-0.1cm}B^{S \chi PT}_{a_0}(p) \hspace{-0.3cm} &=& \hspace{-0.3cm}
\frac{B_0^2}{n_r}  \sum_k  \Biggl\{
- \frac{1}{(k+p)^2+M_{U_V}^2}\frac{ 4C_{V_{\eta'}}}{k^2+M_{\eta' V}^2}
- \frac{1}{(k+p)^2+M_{U_A}^2}\frac{ 4C_{A_{\eta'}}}{k^2+M_{\eta' A}^2}
\nonumber \\  &&\hspace{-1.5cm}
-8\frac{1}{(k+p)^2+M_{U_V}^2}\frac{1}{k^2+M_{U_V}^2 }
+ \frac{1}{(k+p)^2+M_{U_V}^2}\frac{ 4C_{V_{\eta}} }{k^2+M_{\eta V }^2}
\nonumber \\
&&\hspace{-1.5cm}
-8\frac{1}{(k+p)^2+M_{U_A}^2}\frac{1}{k^2+M_{U_A}^2 }
+ \frac{1}{(k+p)^2+M_{U_A}^2}\frac{ 4C_{V_{\eta}} }{k^2+M_{\eta A}^2}
\nonumber \\
&& \hspace{-1.5cm} + n_r^2 \sum_{b=1}^{16}
\Biggl[2\frac{1}{(k+p)^2+M_{U_b}^2} \frac{1}{k^2+M_{U_b}^2}+
        \frac{1}{(k+p)^2+M_{K_b}^2} \frac{1}{k^2+M_{K_b}^2}
\Biggr] \nonumber \\
&& \hspace{-1.5cm}
-         2 \frac{1}{(k+p)^2+M_{U_I}^2} \frac{1}{k^2+M_{U_I}^2 }
+\frac{2}{3}\frac{1}{(k+p)^2+M_{U_I}^2} \frac{1}{k^2+M_{\eta I}^2}
\Biggr\} .
\eea
For $1+1+1$ theory, we obtain
\bea
B^{S \chi PT}_{a_0}(t) \hspace{-0.3cm} &=& \hspace{-0.3cm}
\frac{B_0^2 }{ 4 n_r L^3 } {\sum_{ \bf k}}^{'} \hspace{-0.1cm} \left\{
\frac{2}{3} \OTS {U_I} {\eta I} -2 \FZW {U_I} \right.
\nonumber \\ && \hspace{-2.0cm}
+ 4C_{V_{\eta }} \OTS {U_V} {\eta  V}
- 4C_{V_{\eta'}} \OTS {U_V} {\eta' V}
\nonumber \\ && \hspace{-2.0cm}
+ 4C_{A_{\eta }} \OTS {U_A} {\eta  A}
- 4C_{A_{\eta'}} \OTS {U_A} {\eta' A}
\nonumber \\ && \hspace{-2.0cm}
- 8 \FZW {U_V} -8\FZW {U_A}
\nonumber \\ && \hspace{-2.0cm} \left.
+\frac{1}{8} \sum_{b=1}^{16} \FZW {U_b} +
 \frac{1}{16}\sum_{b=1}^{16} \FZW {K_b}
\right\} .
\label{schpt_9:a0:appC}
\eea

%% file: appC.tex
\chapter{Bubble contribution for {\lowercase{$f_0$}} correlator}
The mass of the $f_{0}$ meson can be reliably determined on the lattice.
In order to determine the mass of the $f_{0}$ meson,
we use lattice simulations
to evaluate the scalar $f_{0}$ correlator in Eq.~(\ref{scor:ch8}),
that is,
\bea
C(t) &=& \hspace{0.5cm}
\sum_{\bf x}  \langle  \bar{u}_r( {\bf x}, t) u_r( {\bf x}, t)
         \bar{u}_r( {\bf 0}, 0) u_r( {\bf 0}, 0) \rangle_{\rm CC}
+ \nonumber \\
&& \hspace{-0.15cm} 2n_r \sum_{\bf x}
\langle  \bar{u}_r( {\bf x}, t) u_r( {\bf x}, t)
         \bar{u}_r( {\bf 0}, 0) u_r( {\bf 0}, 0) \rangle_{\rm DC}
\label{cor:aD} ,
\eea
where $r$ and $r'$ is the indices of the taste replica.
The extraction of the mass of the $f_{0}$ meson ($J^P=0^+$ and $I=0$)
is straightforward.
However there are many multihadron states with $J^P=0^+$ and $I=0$
can propagate between the source and sink of scalar $f_{0}$ correlator.
Of special interest is the bubble contribution $B$, which
gives a considerable contribution to the scalar $f_{0}$ correlator,
and it should be included in the fit of the lattice correlator
in Eq.~(\ref{cor:aD}),
\be
\label{Ctot}
C(t)=Ae^{-m_{f_{0}}t}+B(t) ,
\ee
where we omit the unimportant contributions from the excited scalar $f_0$
meson and other high order multihadron intermediate states.
%
%
\section{Bubble contribution}
In this section we compute the bubble contribution
to the scalar $f_{0}$ correlator in Eq.~(\ref{cor:aD}).
From the discussion in Sec.~\ref{appC::C}, the point scalar current
can be described in terms of the pseudoscalar field $\Phi$
by using S$\chi$PT~\cite{Bardeen:2001jm}\cite{Prelovsek:2005rf}
\be
\label{current}
\bar  u_r(x) u_r(x)  =  B_0 {\rm Tr_t}[\Phi(x)^2]_{u_r u_r} , \ \ \
\bar  d_r(x) d_r(x)  =  B_0 {\rm Tr_t}[\Phi(x)^2]_{d_r d_r} ,
\ee
where $\DPT B_0$ is the coupling of the point scalar current
to the pseudoscalar field $\Phi(x)$.

We know that our scalar $f_{0}$ correlator is estimated by
using the taste-singlet source and taste-singlet sink.
It takes contribution from the
scalar $f_{0}$ meson and the bubble contribution.
For concreteness, the bubble contribution in a theory with $n_r$
tastes per flavor and  three flavors of KS dynamical
sea quarks is~\cite{Aubin:2003mg}
\be
B^{S\chi PT}_{f_0}(x) =
\frac{B_0^2}{n_r} \sum_{r,r'=1}^{n_r}  \Bigg\{
\left\langle{\rm Tr_t}[\Phi^2]_{u_ru_r}
            {\rm Tr_t}[\Phi^2]_{u_{r'}u_{r'}}
\right\rangle   +
\left\langle{\rm Tr_t}[\Phi^2]_{u_ru_r}
            {\rm Tr_t}[\Phi^2]_{d_{r'}d_{r'}}
\right\rangle   \Biggr\} ,
\label{schpt_0}
\ee
where $\Phi$ is $12n_r \times 12n_r$ pseudoscalar matrix
in S$\chi$PT~\cite{Aubin:2003mg},  the subscripts  $u,d$ denote
its valance flavor component, and the subscript $f_0$
specifies the bubble contribution for the $f_0$ meson.
If we consider this identity
$
{\rm Tr_t} \left( T^{a} T^{b} \right) = 4  \delta_{ab}
$,
we obtain
\bea
\label{schpt:eq1}
B^{S \chi PT}_{f_0}(x) \hspace{-0.3cm} &=& \hspace{-0.3cm}
\frac{B_0^2}{n_r}
\sum_{a=1}^{16}     \sum_{b=1}^{16}
\sum_{i=1}^{N_f}    \sum_{j=1}^{N_f}
\sum_{r,r'=1}^{n_r} \sum_{t,t'=1}^{n_r}
\langle
\phi^a_{  u_r    i_t   }(x) \phi^a_{i_t      u_r}(x) \
\phi^{b}_{u_{r'} j_{t'}}(0) \phi^{b}_{j_{t'} u_{r'} }(0)
\rangle + \nonumber \\
\hspace{-0.3cm} && \hspace{-0.3cm}
\frac{B_0^2}{n_r}
\sum_{a=1}^{16}      \sum_{ b=1   }^{16}
\sum_{i=1}^{N_f}     \sum_{j=1   }^{N_f}
\sum_{r,r'=1}^{n_r}  \sum_{t,t'=1}^{n_r}
\langle
\phi^a_{  u_r    i_t   }(x) \phi^a_{  i_t    u_r    }(x) \
\phi^{b}_{d_{r'} j_{t'}}(0) \phi^{b}_{j_{t'} d_{r'} }(0)
\rangle .
\eea
where $a, b, a', b'$ are taste indices, and $i,j$ are flavor indices,
and   $r, t, r', t'$ are the   indices of the taste replica.
From similar procedure in Sec.~\ref{appC::BC},
we obtain
\bea
B^{S \chi PT}_{f_0}(x) \hspace{-0.3cm} &=& \hspace{-0.3cm}
\frac{1}{n_r} B_0^2 \sum_{b=1}^{16} \Biggl\{
4 n_r^2 \langle \phi^b_{ud}(x) \ \phi^b_{du}(0) \rangle \
  n_r^2 \langle \phi^b_{du}(x) \ \phi^b_{ud}(0) \rangle  +
\nonumber\\ \hspace{-0.3cm} && \hspace{1.5cm}
\langle \phi^b_{us}(x) \ \phi^b_{su}(0) \rangle  \
\langle \phi^b_{su} (x) \ \phi^b_{us}(0) \rangle \Biggr\}  + \nonumber\\
\hspace{-0.3cm} && \hspace{-0.3cm}
\frac{1}{n_r} B_0^2 \sum_{b=I,V,A} \Biggl\{
4n_r^2 \langle \phi^b_{uu}(x)  \ \phi^b_{uu}(0) \rangle_{\rm DC} \
    \langle \phi^b_{uu}(x)  \ \phi^b_{uu}(0) \rangle_{\rm DC} +
\nonumber\\ \hspace{-0.3cm} && \hspace{1.7cm}
4n_r   \langle \phi^b_{uu}(x)  \ \phi^b_{uu}(0) \rangle_{\rm DC} \
    \langle \phi^b_{uu}(x)  \ \phi^b_{uu}(0) \rangle_{\rm CC}
\Biggr\} ,
\label{schpt_0:appD}
\eea
where we consider the $u,d$ quarks are degenerate in mass.
The bubble contribution  in Eq.~(\ref{schpt_0}) can be rewritten by
plugging in the mesonic propagators,
\be
\label{Bschpt}
B^{S\chi PT}_{f_0}(t) = F.T.[~B^{S \chi PT}_{f_0}(p)~]_{ {\bf p} =
{\bf 0} } \ \ \ \ {\rm with} \\
\ee
\bea
\hspace{-0.3cm}
B^{S \chi PT}_{f_0}(p) \hspace{-0.3cm} &=& \hspace{-0.3cm}
\frac{B_0^2}{n_r} \sum_k  \Biggl\{
- \frac{1}{ (k+p)^2+M_{U_I}^2} \frac{1}{   k^2  +M_{U_I}^2}
+ \frac{1}{9}
\frac{1}{ (k+p)^2+M_{\eta I}^2}\frac{1}{ k^2    +M_{\eta I}^2}
\nonumber \\ && \hspace{-1.5cm}
+C_{V_\eta }^2\frac{1}{(k+p)^2+M_{\eta  V}^2}\frac{1}{k^2+M_{\eta  V}^2}
+C_{V_\eta'}^2\frac{1}{(k+p)^2+M_{\eta' V}^2}\frac{1}{k^2+M_{\eta' V}^2}
\nonumber \\ &&  \hspace{-1.5cm}
+C_{A_\eta }^2\frac{1}{(k+p)^2+M_{\eta  A}^2}\frac{1}{k^2+M_{\eta  A}^2}
+C_{A_\eta'}^2\frac{1}{(k+p)^2+M_{\eta' A}^2}\frac{1}{k^2+M_{\eta' A}^2}
\nonumber \\ &&  \hspace{-1.5cm}
-C_{V_\eta}C_{V_{\eta'}} \left[
\frac{1}{(k+p)^2+M_{\eta V}^2}\frac{1}{k^2+M_{\eta' V}^2}+
\frac{1}{(k+p)^2+M_{\eta' V}^2}\frac{1}{k^2+M_{\eta V}^2} \right]
\nonumber \\ &&  \hspace{-1.5cm}
-C_{A_\eta}C_{A_{\eta'}} \left[
\frac{1}{(k+p)^2+M_{\eta A}^2}\frac{1}{k^2+M_{\eta' A}^2}+
\frac{1}{(k+p)^2+M_{\eta' A}^2}\frac{1}{k^2+M_{\eta A}^2} \right]
\nonumber \\ &&  \hspace{-1.5cm}
-4\frac{1}{ (k+p)^2+M_{U_V}^2 }\frac{1}{ k^2+M_{U_V}^2 }
-4\frac{1}{ (k+p)^2+M_{U_A}^2 }\frac{1}{ k^2+M_{U_A}^2 }
\nonumber \\ &&  \hspace{-1.5cm}
+n_r^2 \sum_{b=1}^{16} \left[
4\frac{1}{(k+p)^2+M_{U_b}^2} \frac{1}{k^2+M_{U_b}^2} +
 \frac{1}{(k+p)^2+M_{K_b}^2} \frac{1}{k^2+M_{K_b}^2} \right]
\Biggr\} ,
\label{schpt_6:appC:CC}
\eea
where $\delta_V$,
      $M_{\eta_I}$,
      $M_{\eta V}^2$,
      $M_{\eta' V}^2$,
      $\delta_A$,
      $M_{\eta A}^2$,
      $M_{\eta' A}^2$,
      $C_{V_\eta}$,
      $C_{V_{\eta'}}$,
      $C_{A_\eta}$,
      $C_{A_{\eta'}}$
are denoted in Appendix B.
The Fourier Transform of Eq.~(\ref{schpt_6:appC:CC}) (1+1+1+1 theory) is
\bea
B^{S \chi PT}_{f_0}(t, {\bf p}) \hspace{-0.3cm}&=&\hspace{-0.3cm}
B_1^2 \sum_{ \bf k}\hspace{-0.18cm} \left\{
\frac{1}{9}
\OTS {\eta  I} {\eta I} \hspace{-0.08cm} - \hspace{-0.08cm}
\OTS {U_I} { U_I}
\right. \nonumber \\ && \hspace{-2.4cm}
+C_{V_{\eta}}^2   \OTS {\eta   V} {\eta V}
+C_{V_{\eta'}}^2  \OTS {\eta'  V} {\eta' V}
\nonumber \\ && \hspace{-2.4cm}
+C_{A_{\eta}}^2   \OTS {\eta  A} {\eta A}
+C_{A_{\eta'}}^2  \OTS {\eta'  A} {\eta' A}
\nonumber \\ && \hspace{-2.4cm}
-C_{V_{\eta}}C_{V_{\eta'}} \left[
\OTS {\eta  V} {\eta' V} + \OTS {\eta'  V} {\eta V}
\right]  \nonumber \\ && \hspace{-2.4cm}
-C_{A_{\eta}}C_{A_{\eta'}}  \left[
\OTS {\eta  A} {\eta' A} + \OTS {\eta'  A} {\eta A}
\right]  \nonumber \\ && \hspace{-2.4cm}
-4\OTS {U_V} {U_V}  -4\OTS {U_A} {U_A}
\nonumber \\ && \hspace{-2.4cm} \left.
\frac{1}{16}\sum_{b=1}^{16} \left[
4\OTS {U_b} {U_b} +  \OTS {K_b} {K_b}
\right] \right\} ,
\eea
where $\DPT B_1^2 = \frac{B_0^2}{4 n_r L^3} $ .

%% file: appD.tex
\chapter{LBO treatment of decay channel for hybrid }
In this dissertation, we are interested in the decay of a hybrid exotic
quarkonium $1^{-+}$ state $H$ to $\chi_b$ state with the emission
of a pair of pseudoscalar mesons in a scalar channel ($\pi\pi$).
To make the problem simple we assume that the scalar meson ($\pi\pi$)
in the final state does not interact with the $\chi_b$ final state.
The lattice calculation is performed by the LBO approximation,
and the wave functions of the initial and final state
are easily described in the rest frame of the heavy quarks.
We call it the ``body-fixed frame.''
The displacement vector ${\bf R}$ of the heavy
quarks specifies the $z$ axis of the body-fixed frame.
The exotic hybrid state (H) is $\Pi_u$ and the final
quarkonium state is $\Sigma_g^+$ state ($\chi_b$).

As discussed in Chapter 4,
the potentials are computed in the static approximation.
The wave function of the hybrid exotic in lab frame is~\footnote{
$L^2$-{\bf normalized } Wigner D-functions
$\tilde{D}^J_{M, M'}(\phi,\theta,\alpha)$ and ordinary
Wigner D-functions $D^J_{M, M'}(\phi, \theta,\alpha)$
have relationship,
$\DPT
\tilde{D}^J_{M, M'}(\phi,\theta,\alpha) =
\sqrt{ \frac{2J+1}{8\pi^2} }  D^J_{M, M'}(\phi, \theta,\alpha) .
$
Hence $$
\int_{0}^{2\pi} d\alpha \int_{0}^{\pi}  \sin \theta d\theta
\int_{0}^{2\pi} d\phi \
\tilde{D}^{J_1 *}_{M_1, M_1'}(\phi,\theta,\alpha)
\tilde{D}^{J_2  }_{M_2, M_2'}(\phi,\theta,\alpha)
 =
\delta_{J_1 J_2}  \delta_{M_1 M_2} \delta_{M_1' M_2'} . $$
}
\be
\Psi_H(R,\phi,\theta,\alpha) =
\rho_H(R) \frac{ \tilde{D}^1_{m,1}(\phi,\theta,\alpha) +
\tilde{D}^1_{m,-1}(\phi,\theta,\alpha) }{ \sqrt{2} } ,
\ee
where we ignore the spin degrees of freedom, $m$ is the
projection of the total angular momentum (excluding spin) on the lab
$z$ axis, and $ \pm 1$ is its projection on the body-fixed axis.
The Euler angle $\alpha$ specifies a rotation about the body-fixed axis
and it is an internal coordinate for the excited flux tube.

In lab frame the state $H$ can be described by
\be
\left| H \RR = \int d^3 {\bf R} d^3 {\bf r} \int d\alpha
\frac{e^{i {\bf P}_H \cdot {\bf r} } } { (2\pi)^{3/2}}
\Psi_H(R,\phi,\theta,\alpha) \,
{\cal C}_{Q}^{\dag}({\bf r_1})
{\cal C}_{\overline Q}^{\dag}({\bf r_2}) \,
A(\alpha)
\left| 0 \RR ,
\label{aG:H:D}
\ee
where ${\bf r_1} = ( {\bf R} - {\bf r} ) /2$,
      ${\bf r_2} = ( {\bf R} + {\bf r} ) /2$, and
\be
A(\alpha) =  \sum_\Lambda \left[
a_{g\Lambda}      \frac{ e^{i\Lambda\alpha} }{ \sqrt{2\pi} } +
a_{g\Lambda}^\dag \frac{ e^{-i\Lambda\alpha} }{ \sqrt{2\pi} }
\right] ,
\ee
here the $a_{g\Lambda}$ is the creation operator of gluon.

By convention the wave function of the $\chi_b$ state in rest frame is
\be
\Psi_{\chi_b}(R,\phi,\theta) =
\rho_{\chi_b}(R)Y_{1 m_b}(\theta,\phi) .
\ee
In lab frame the state $A$ can be described by
\be
\left| A \RR = \int d^3 {\bf R} d^3 {\bf r}
\frac{e^{i {\bf P}_A \cdot {\bf r} } } { (2\pi)^{3/2}}
\Psi_{\chi_b}(R,\phi,\theta) \,
{\cal C}_{Q}^{\dag}({\bf r_1}) {\cal C}_{\overline Q}^{\dag}({\bf r_2}) \,
\left| 0 \RR .
\label{aG:A:D}
\ee

In the body-fixed frame with ${ \bf R}$ fixed the initial hybrid exotic
state has only the internal gluonic degree of freedom with a wave
function $ \cos \alpha / \sqrt{\pi} $
corresponding to the choice $\Lambda = \pm 1$.
We can model this transition in the lab frame through an interaction potential
\bea
V_I^{\rm lab} &=& \int d^3 {\bf R} d^3 {\bf r} \int d^3 {\bf r_s'}
\, {\bar x({\bf R'})}
{\cal C}_{Q}^{\dag}({\bf r_1}) {\cal C}_{\overline Q}^{\dag}({\bf r_2})
{\cal C}_{Q}({\bf r_1}) {\cal C}_{\overline Q}({\bf r_2}) \times \nonumber \\
&&
\bar\psi\psi({\bf r_s + r})
\left[ w_{I,1}({\bf r_s'})  a_{g,1} + w_{I,-1}({\bf r_s'}) a_{g,-1} \right] ,
\label{V_I:model:ade}
\eea
where the primed coordinates are the coordinates of the scalar meson
in the body-fixed frame, $w_I$ is the wave function of the scalar
meson in the body-fixed frame as determined by the dynamics of the
transition, and $x(R)$ controls the transition amplitude.
Since final state $\chi_b$ has $\Lambda = 0$,
the scalar meson must carry off
the nonzero $z$-component of the angular momentum in the body-fixed frame.

The wave function of the scalar meson is a free spherical wave
with an appropriate angular momentum ($L=1$).
Hence, its wave function in rest frame is
\be
\psi_{s}(r_s, \theta_s, \phi_s) = N_s j_1(pr)Y_{1,m_s}(\theta_s, \phi_s) ,
\ee
where $N_s$ is a normalization factor,
$j_1(pr)$ is the spherical Bessel function, and
$Y_{1,m_s}(\theta_s,\phi_s)$ is the spherical harmonic function,
and $p$ is the magnitude of the angular momentum.

In this notation we evaluate the transition amplitude in lab frame by
\bea
\LL A, \pi({\bf p_1})  \pi({\bf p_2}) \left | V_I^{\rm lab} \right| H \RR
\hspace{-0.3cm}&=&\hspace{-0.3cm}
\int d^3 {\bf R}
\int r_s^2 dr_s d\Omega_s d\alpha \,
\Psi_H(R,\phi,\theta,\alpha)
\Psi^*_{\chi_b}(R, \theta, \phi) \nonumber \\
\hspace{-0.3cm}&\times&\hspace{-0.3cm}
\psi_s^*(r_s, \theta_s, \phi_s)
\frac{ \cos (\alpha - \phi_s') }{ \sqrt{2\pi} }
x({\bf R}) w_I( {\bf r_s'})
\nonumber \\
\hspace{-0.3cm}&\times&\hspace{-0.3cm}
\LL \pi({\bf p_1}) \pi({\bf p_2}) \left| \bar\psi\psi(0) \right| 0 \RR
\delta^{(3)}({\bf P}_H - {\bf P}_A  - {\bf p}),
\eea
where $\left| \pi({\bf p_1})  \pi({\bf p_2})  \RR =
        a_{{\bf p}_1}^\dag a_{{\bf p}_2}^\dag\left| 0 \RR $ ( namely,
$a_{{\bf p}_1}^\dag a_{{\bf p}_2}^\dag$ is the creation operator for
	$\pi\pi$),
${ \bf p = p_1 + p_2}$, and the $p$ is the magnitude of the angular
momentum ${\bf p}$.
For brevity, we divide $\LL A, \pi({\bf p_1}) \pi({\bf p_2})
 \left | V_I^{\rm lab} \right| H \RR $ into  two parts, that is,
\be
\LL A, \pi({\bf p_1}) \pi({\bf p_2})
\left | V_I^{\rm lab} \right| H \RR \equiv
x_{\pi\pi}(p_1, p_2)
\delta^{(3)}{ ({\bf P}_H - {\bf P}_A  - {\bf p} ) } ,
\ee
where
\vspace{-0.4cm}
\bea
x_{\pi\pi}(p_1, p_2) \hspace{-0.2cm}&=&\hspace{-0.2cm}
\int d^3 {\bf R}
\int r_s^2 dr_s d\Omega_s d\alpha
\Psi_H(R,\phi,\theta,\alpha)
\Psi^*_{\chi_b}(R, \theta, \phi)
\psi_s^*(r_s, \theta_s, \phi_s) \nonumber \\
\hspace{-0.2cm}&\times&\hspace{-0.2cm}
\frac{ \cos (\alpha - \phi_s') }{ \sqrt{2\pi} }
x({\bf R}) w_I( {\bf r_s'})
\LL \pi({\bf p_1}) \pi({\bf p_2}) \left| \bar\psi\psi(0) \right| 0 \RR
.
\eea

The integration over scalar meson coordinates is carried
out in body-fixed frame.  To this goal we use the rotation property
of the spherical harmonic function
\be
Y_{1,m_s}(\theta_s,\phi_s) = \sum_\Lambda
D^1_{m_s,\Lambda}(\phi,\theta)
Y_{1,\Lambda}(\theta_s^\prime,\phi_s^\prime)
\ee
Because the volume element is rotationally invariant, we have
\bea
x_{\pi\pi}(p_1, p_2) &=& \int R^2 dR d\Omega \int r_s^{\prime 2} dr_s' d\Omega_s'
\sum_\Lambda D^{1*}_{m_s,\Lambda}(\phi,\theta)
x(R) \rho_{\chi_b}(R)  \rho_H(R)
N_s j_1(pr_s^\prime) \nonumber \\
&\times &
Y_{1,\Lambda}(\theta_s', \phi_s')
Y^*_{1 m_b}(\theta, \phi )
x(R) w_I(r_s', \theta_s', \phi_s^\prime)
\LL \pi({\bf p_1}) \pi({\bf p_2}) \left| \bar\psi\psi(0) \right| 0 \RR
\nonumber \\
&\times &
\int d\alpha \frac{ \cos (\alpha - \phi_s') }{ \sqrt{\pi} }
\frac{ \tilde{D}^1_{m,  1}(\phi,\theta,\alpha) +
\tilde{D}^1_{m, -1}(\phi,\theta,\alpha) }{ \sqrt{2} } .
\eea
After integrating out the $\alpha$ and rearranging terms,
we obtain~\footnote{
Here we use the convention that
$
D^{j}_{m,n}(\phi,\theta,\alpha) =
e^{i m \phi } d^{j}_{m,n}(\phi,\theta,\alpha) e^{i n \alpha }
$.
}
\bea
x_{\pi\pi}(p_1, p_2) &=& \int R^2 dR d\Omega \int r_s^{\prime 2} dr_s' d\Omega_s'
\sum_\Lambda D^{1*}_{m_s,\Lambda}(\phi,\theta)
\rho_{\chi_b}(R)  \rho_H(R)
N_s j_1(pr_s^\prime) \nonumber \\
&\times &
Y_{1,\Lambda}(\theta_s', \phi_s') Y^*_{1 m_b}(\phi,\theta)
x(R) w_I(r_s', \theta_s', \phi_s^\prime)
\LL \pi({\bf p_1}) \pi({\bf p_2}) \left| \bar\psi\psi(0) \right| 0 \RR
\nonumber \\
   &\times &  \sqrt{ \frac{3}{4\pi} }
   [ e^{im\phi} d_{m, 1}(\theta) e^{-i\phi_s'} +
     e^{im\phi} d_{m,-1}(\theta) e^{ i\phi_s'} ]  .
\eea
When carrying out the $\phi_s'$ integration and rearranging terms,
we can omit the summation over $\Lambda$. In order to normalize
correctly later, we here still keep the integral over $\phi_s'$,
that is,
\bea
x_{\pi\pi}(p_1,p_2) \hspace{-0.15cm}&=&\hspace{-0.15cm} \int R^2  \rho_{\chi_b}(R) \, x(p,R) \, \rho_H(R)dR
\times
\LL \pi({\bf p_1}) \pi({\bf p_2}) \left| \bar\psi\psi(0) \right| 0 \RR
\nonumber \\
    \hspace{-0.15cm}&\times & \hspace{-0.15cm}
    \int d\Omega
    D^{1*}_{m_s,1}(\phi,\theta)
    Y^*_{1 m_b}(\phi,\theta)  e^{im\phi}
    \sqrt{\frac{3}{4\pi}} [ d_{m, 1}(\theta) - d_{m,-1}(\theta) ]
    ,
\label{appG:gamma6}
\eea
where
\be
x(p,R) = x(R) \int r_s^{\prime 2} dr_s'
\cos\theta_s' d \theta_s'
N_s j_1(pr_s^\prime)  Y_{11}(\theta_s', \phi_s')
w_I(r_s', \theta_s', \phi_s^\prime) .
\ee

We can divide Eq.~(\ref{appG:gamma6}) into three parts, that is,
\be
 x_{\pi\pi}(p_1,p_2) = \int R^2 \rho_H(R) x(p,R) \rho_{\chi_b}(R) dR
\times F(m, m_s, m_b) ,
\ee
where
\be
F(m, m_s, m_b) = \sqrt{ \frac{3}{4\pi} }
\int d\Omega
    D^{1*}_{m_s,1}(\phi,\theta)
    Y^*_{1 m_b}(\phi,\theta)  e^{ im\phi}
    [ d_{m, 1}(\theta) - d_{m,-1}(\theta) ] .
\ee
We choose the right measurement choice of spherical harmonics function
$Y$ and Wigner $D$ function, that is,
\bea
Y_{1\,1}(\phi, \theta) &=& -\sqrt{ \frac{3}{8\pi} } \sin\theta e^{i\phi}  \\
Y_{1\,0}(\phi, \theta) &=&  \sqrt{ \frac{3}{4\pi} } \cos\theta   \\
Y_{1\,-1}(\phi, \theta) &=&  \sqrt{ \frac{3}{8\pi} } \sin\theta e^{-i\phi}  ,
\eea
and
\vspace{-0.3cm}
\bea
D_{1,  1}^{1}(\phi, \theta) = \frac{1+\cos\theta}{2} e^{-i\phi} &&
D_{1, -1}^{1}(\phi, \theta) = \frac{1-\cos\theta}{2} e^{-i\phi} \\
D_{-1, 1}^{1}(\phi, \theta) = \frac{1-\cos\theta}{2} e^{i\phi} &&
D_{-1,-1}^{1}(\phi, \theta) = \frac{1+\cos\theta}{2} e^{i\phi} \\
D_{0,  1}^{1}(\phi, \theta) = \frac{\sin\theta}{\sqrt{2}} &&
D_{0, -1}^{1}(\phi, \theta) =-\frac{\sin\theta}{\sqrt{2}} .
\eea
We can show only when
\be
m = m_s + m_b ,
\ee
$F(m, m_s, m_b)$ is nonzero. Here we list all the cases, that is,

Case 1) $m=1$, $m_s=1$, $m_b=0$: \
$F( m, m_s, m_b) = \frac{1}{2} $

Case 2) $m=1$, $m_s=0$, $m_b=1$: \
$F( m, m_s, m_b) = \frac{1}{2} $

Case 3) $m=0$, $m_s=1$, $m_b=-1$: \
$F( m, m_s, m_b) = \frac{1}{2} $

Case 4) $m=0$, $m_s=0$, $m_b=0$: \
$F( m, m_s, m_b) = 0$

Case 5) $m=0$, $m_s=-1$, $m_b=1$: \
$F( m, m_s, m_b) = \frac{1}{2} $

Case 6) $m=-1$, $m_s=-1$, $m_b=0$: \
$F(  m, m_s, m_b) = \frac{1}{2} $

Case 7) $m=-1$, $m_s=0$,  $m_b=-1$: \
$F( m, m_s, m_b) = \frac{1}{2}  $  .

Finally, we obtain
\be
 x_{\pi\pi}(p_1,p_2) = \frac{1}{2}
\int R^2 \rho_H(R) x(p,R) \rho_{\chi_b}(R) dR
\times \LL \pi({\bf p_1}) \pi({\bf p_2}) \left| \bar\psi\psi(0) \right| 0 \RR.
\label{xpppp1}
\ee
If we introduce the reduced radial wave function
$u_H(R)= R\rho_H(R)$, and the reduced radial wave function
$u_{\chi_b}(R)= R\rho_{\chi_b}(R)$, we rewrite above equation as
\be
x_{\pi\pi}(p_1,p_2) = \frac{1}{2} \int u_H(R) x(p,R) u_{\chi_b}(R) dR
\times \LL \pi({\bf p_1}) \pi({\bf p_2}) \left| \bar\psi\psi(0) \right| 0 \RR.
\label{appf:gamma8}
\ee
If we denote
\be
x(p) = \frac{1}{2}
\int u_H(R) x(p,R) u_{\chi_b}(R) dR ,
\label{GGG:DDD}
\ee
we have
\be
x_{\pi\pi}(p) = x(p_1,p_2)
\LL \pi({\bf p_1}) \pi({\bf p_2}) \left| \bar\psi\psi(0) \right| 0 \RR .
\ee
In Chapter 8, we use Eq.~(\ref{appf:gamma8}) to calculate the $ x_{\pi\pi}(p)$.

The integration is done in the body-fixed frame.
Since the transition amplitude
is computed on the lattice, we replace the continuous integration with
a lattice sum.  The wave function is measured in lattice momentum
space for a lattice of spatial size $a^3L^3$ through the expansion
\be
w_I(x,y,z) =
\sum_{k_x \ne 0, k_y, k_z}w_I(k_x,k_y,k_z)
\sin(k_x x) \cos(k_y y) \cos(k_z z) ,
\ee
In this approximation
\be
x(p,R) =
\frac{1}{L^3} \sum_{k_x \ne 0, k_y, k_z}\sum_{x,y,z} a^3
N_s j_1(pr_s') Y_{11}(\theta_s', \phi_s') \,
x( {\bf k}, R) \,
\sin(k_x x) \cos(k_y y) \cos(k_z z) .
\label{xpr:detar}
\ee
With a sufficiently large lattice volume the overlap sum over coordinates
$(x,y,z)$ recovers the continuum momentum conservation constraint $p =
|{\bf k}|$. Here we use Eq.~(\ref{x.def:ch7}).

Now we discuss how to calculate the above equation numerically.
First we calculate the $N_s$, this number come from the normalized
condition
\be
\sum_{x,y,z} N_s^2 \left |  j_1(p r) Y_{11}(\theta, \phi) \right | ^2 = 1 ,
\ee
where
\vspace{-0.3cm}
\bea
Y_{11}(\theta,\phi) &=&  \frac{ x + i y }{ r}  \\
j_1(p r)            &=&  \frac{sin( p r)}{p^2 r^2}-\frac{ \cos(p r)}{p r} ,
\eea
here $r=\sqrt{ x^2 + y^2 + z^2 }$.
Then we can use Eq.~(\ref{xpr:detar}) to calculate  $x(p,R)$ at
any given momentum $p$.

In practice, we can rewrite Eq.~(\ref{xpppp1}) as
\bea
x_{\pi\pi}(p_1,p_2) \hspace{-0.2cm} &=& \hspace{-0.2cm}
\frac{1}{L^3} \sum_{k_x \ne 0, k_y, k_z}\sum_{x,y,z}  a^3
N_s j_1(pr_s') Y_{11}(\theta_s', \phi_s')    \nonumber \\
\hspace{-0.15cm} &\times & \hspace{-0.15cm}
\tilde x( {\bf k}) \,
\LL \pi({\bf p_1}) \pi({\bf p_2}) \left| \bar\psi\psi(0) \right| 0 \RR \,
\sin(k_x x) \cos(k_y y) \cos(k_z z) ,
\label{xpp:fzw}
\eea
where
\be
\tilde x( {\bf k})  = \frac{1}{2}
\int u_H(R) x( {\bf k}, R)  u_{\chi_b}(R) dR .
\label{WLY:FZW1}
\ee
From Chapter 7, we know,
\be
\LL \pi({\bf p_1}) \pi({\bf p_2}) \left| \bar\psi\psi(0) \right| 0 \RR
= \hat b_{\pi\pi} \,
\frac{1}{ \sqrt{2E_{\pi}({\bf p}_1)} \sqrt{2E_{\pi}({\bf p}_2)} }  ,
\label{WLY:FZW3}
\ee
where $b_{\pi\pi}$ is denoted in Eq.~(\ref{bhat_pipi:chap7}).
For notational simplicity, we define
\be
\hat x( {\bf k})  \equiv
\tilde x( {\bf k} )   \hat b_{\pi\pi}.
\label{WLY:FZW4}
\ee
If we define
\be
\hat x_{\pi\pi}(p)  =
\frac{1}{L^3} \sum_{k_x \ne 0, k_y, k_z}\sum_{x,y,z}  a^3
N_s j_1(pr_s') Y_{11}(\theta_s', \phi_s')  \,
\hat x( {\bf k}) \,
\sin(k_x x) \cos(k_y y) \cos(k_z z) ,
\label{xpp:fzw:wly}
\ee
then
\be
x_{\pi\pi}(p_1,p_2)  = \hat x_{\pi\pi}(p)
\frac{1}{ \sqrt{2E_{\pi}({\bf p}_1)} \sqrt{2E_{\pi}({\bf p}_2)} } .
\ee

%% file: appE.tex
\chapter{three-body phase space}
In this dissertation, we are interested in the decay
\be
   H \rightarrow \chi_b + \pi + \pi \ ,
\ee
where two-pion final state is in an scalar singlet channel.
In Appendix D, we calculated the transition amplitude $x_{\pi\pi}(p)$ for
two-pion production as a function of the momentum $p$ of the two-pion system
in the rest frame of the $H$.

The starting point is the Fermi's Golden rule, that is,
\be
d\Gamma = \frac{1}{ (2\pi)^3 }
{\left| \LL f | V_I |  i \RR \right|}^2
(2\pi)^4 \delta^4(p_f - p_i ) df ,
\label{appg:ga}
\ee
where $\LL f | V_I |  i \RR$ is the transition
amplitude from initial state $i$ to the final state $f$,
$df$ is the phase space factor.

The general rule is that the various factors in the phase space are
determined by the normalization of the states.
From the definition in Eq.~(\ref{aG:H:D}),
we can show  that the initial state $\left|H\RR$ is normalized as
\be
\int d^3 {\bf p_{\it H}}
| {\bf p_{\it H} } \left. \RR \LL \right. {\bf p_{\it H}} | = 1 ,
\ee
and from the definition in Eq.~(\ref{aG:A:D}),
the final state $\left|A\RR$ is normalized as
\be
\int d^3 {\bf p_{\it A}}
| {\bf p_{\it A} } \left. \RR \LL \right. {\bf p_{\it A}} | = 1 ,
\ee
Hence, they should not give a factor in the phase space formula.
There are two other momentum factors in our phase space formula,
called $p_1$ and $p_2$. From the Ref.~\cite{SCALAR:2006}, they are
normalized as
\be
\sum_{ {\bf p}_1 } \sum_{ {\bf p}_2 }
| {\bf p}_1 {\bf p}_2\left. \RR \LL \right.
  {\bf p}_1 {\bf p}_2  | = 1 ,
\ee
They are replaced by $p = p_{12}$,
the total momentum of the two pions and $p_1 = k$,
the momentum of one of the pions.
In the analysis of the bubble diagram the momenta $p$ and $k$ are
quantized in a box and summed.
Hence the phase space factor
\be
df = d^3 {\bf p}_A
\frac{L^3 a^3} {(2\pi)^3 \, 2E_1} d^3 {\bf p}_1
\frac{L^3 a^3} {(2\pi)^3 \, 2E_2} d^3 {\bf p}_2 ,
\ee
where $E_1, E_2$ are the energies of two pions.
Hence, according to Fermi's Golden rule,
we obtain transition rate ($\Gamma$) by
\bea
\Gamma &=& \frac{1}{ (2\pi)^3 }
\int dp
\int
d^3 {\bf p}_A
\frac{L^3 a^3} {(2\pi)^3 \, 2E_1} d^3 {\bf p}_1
\frac{L^3 a^3} {(2\pi)^3 \, 2E_2} d^3 {\bf p}_2
(2\pi)^4 \delta^4(p_A + p_1 + p_2 - p_H) \nonumber \\
&& \delta(| {\bf p}_{12}| - p) \, \hat x_{\pi\pi}(p)^2 ,
\label{appg:gamma1}
\eea
where $p_H$ is the four momentum of initial hybrid exotic state $H$,
$p_A$ is the four momentum of $\chi_b$ final state,
$p_1$ is the four momentum of one pion final state,
$p_2$ is the four momentum of other pion final state,
we denote $p_{12} \equiv p_1 + p_2$, and
where $\hat x_{\pi\pi}(p)$ is denoted in Eq.~(\ref{x_pipi2:chap7}),
which is the transition amplitude from initial state to the final state.

To simplify the integration in Eq.~(\ref{appg:gamma1}),
we introduce the integral over the intermediate four momentum $p_{12}$
and its invariant mass $s \equiv p_{12}^2 = E_{12}^2 - |{\bf p}_{12}|^2$.
We rewrite the Eq.~(~\ref{appg:gamma1}) as
\bea
\Gamma &=& \frac{1}{ (2\pi)^3 } \left(\frac{La}{2\pi}\right)^6
\int dp\, dE_{12} \,
d^3 {\bf p}_A
d^3 {\bf p}_{12}
(2\pi)^4 \delta^4(p_A + p_1 + p_2 - p_H) \nonumber \\
&\times& F_{12}(s) \delta( | {\bf p}_{12} | - p) \,  \hat x_{\pi\pi}(p)^2 ,
\eea
where $d^3 {\bf p}_A =
|{\bf p}_A|^2 d |{\bf p}_A| \sin \theta d\theta d\phi$,
and $F_{12}(s)$ is the two-body phase space factor for two-pion system,
that is,
\be
F_{12}(s) = \int \frac{ d^3 {\bf p}_1 }{2E_1}
                 \frac{ d^3 {\bf p}_2 }{2E_2}
\delta^4(p_{12} - p_1 - p_2) ,
\ee
where $d^3 {\bf p}_1 =
|{\bf p}_1|^2 d |{\bf p}_1| \sin \theta d\theta d\phi$.
This phase space factor is a Lorentz invariant. Thus we can easily
evaluate it in the rest frame of two-pion system, that is,
\bea
   F_{12}(s) &=&  4\pi
      \int \frac{|{\bf p}_1|^2}{2E_1 2E_2} d |{\bf p}_1|
      \delta(E_1 + E_2 - \sqrt{s})  \nonumber \\
 &=&
     \pi \, \sqrt{ \frac{1}{4} - \frac{m_\pi^2}{s} } ,
\eea
where $m_\pi$ is the mass of the pion meson.
Therefore, we obtain
\bea
\Gamma &=&  2\pi^2 \left(\frac{La}{2\pi}\right)^6
       \int dp\, dE_{12}
       d^3 {\bf p}_A
       d^3 {\bf p}_{12}
       \delta^4(p_A + p_1 + p_2 - p_H) \nonumber \\
       &\times&  \sqrt{ \frac{1}{4} - \frac{m_\pi^2}{s} } \,
                 \delta( | {\bf p}_{12} | - p) \, \hat x_{\pi\pi}(p)^2 .
\eea
In the rest frame of the $H$ we integrate above equation, and we obtain
\be
  \Gamma = 8\pi^3 \left(\frac{La}{2\pi}\right)^6 \int dp\, dE_{12} \,
           |{\bf p}_A|^2 d |{\bf p}_A|
    \delta( E_A + E_{12} - M_H )  \,
    \sqrt{ \frac{1}{4} - \frac{m_\pi^2}{s} } \,
    \delta( |{\bf p}_A| - p) \hat x_{\pi\pi}(p)^2 ,
\ee
In the rest frame of the $H$ we integrate over
$|{\bf p}_A| = p$ and $s$ to get
\be
\Gamma = 8\pi^3  \left(\frac{La}{2\pi}\right)^6
\int_0^{p_{\rm max}}  dp
\ p^2 \, \hat x_{\pi\pi}(p)^2 \,
\sqrt{ \frac{1}{4} - \frac{m_\pi^2}{(\Delta M)^2 - p^2 } } ,
\ee
where we approximate $E_A = M_A$, and define $\Delta M = M_H - M_A$.
If we consider $(\Delta M)^2 - p^2 \ge 4 m_\pi^2$, we have
\be
p_{\rm max} = \sqrt{(\Delta M)^2 - 4 m_\pi^2} .
\ee